\documentclass[fleqn,usenatbib]{mnras}

\usepackage{newtxtext,newtxmath}

\usepackage[T1]{fontenc}
\usepackage{ae,aecompl}

\usepackage{graphicx}	
\usepackage{amsmath}	
\usepackage{comment}
\usepackage{tcolorbox}
\usepackage{fancyvrb} 
\usepackage{scrextend}
\usepackage{multirow}
\usepackage{hyperref}
\usepackage[normalem]{ulem}
\usepackage[export]{adjustbox}
\usepackage{soul}
\RequirePackage{fontawesome}

\newcommand{\vir}[1]{``#1''}

\newcommand{\Msun}{\mbox{M$_{\odot}$}}
\newcommand{\ace}{\mbox{$\alpha_\mathrm{CE}$}}
\newcommand{\lce}{\mbox{$\lambda_\mathrm{CE}$}}

\newcommand{\Rsun}{\mbox{R$_{\odot}$}}
\newcommand{\Lsun}{\mbox{L$_{\odot}$}}

\newcommand{\Mz}[2]{\mbox{$M_\mathrm{ZAMS {#1}{#2}}$}}
\newcommand{\sevn}{{\sc sevn}}
\newcommand{\bse}{{\sc bse}}
\newcommand{\mobse}{{\sc mobse}}
\newcommand{\combine}{{\sc combine}}

\newcommand{\parsec}{{\sc parsec}}
\newcommand{\posydon}{{\sc posydon}}
\newcommand{\mist}{{\sc mist}}
\newcommand{\mesa}{{\sc mesa}}
\newcommand{\bec}{{\sc bec}}
\newcommand{\cosmorate}{{\sc Cosmo}$\mathcal{R}${\sc ate}}
\newcommand{\gitlab}[1]{\href{{#1}}{\faGitlab}}
\newcommand{\gitimage}[1]{\href{{#1}}{\faFileImageO}}
\newcommand{\gitbook}[1]{\href{{#1}}{\faBook}}

\definecolor{forest}{rgb}{0.0, 0.5, 0.0}

\newtcolorbox{mybox}{colback=red!5!white,colframe=red!75!black}
\newtcolorbox{rbox}[1]{colback=red!5!white,colframe=red!75!black,fonttitle=\bfseries,title=#1}
\newtcolorbox{gbox}[1]{colback=gray!5!white,colframe=gray!75!black,fonttitle=\bfseries,title=#1}
\usepackage{etoolbox}


\title[Compact object mergers with {{\sc sevn}}]{Compact object mergers: exploring uncertainties from stellar and binary evolution with {{\sc sevn}}}

\author[G. Iorio et al.]{
Giuliano Iorio$^{1,2,3,}$\thanks{E-mail: giuliano.iorio.astro@gmail.com},
Michela Mapelli$^{1,2,3,}$\thanks{E-mail: michela.mapelli@unipd.it}, 
Guglielmo Costa$^{1,2,3}$,
Mario Spera$^{4,5}$, 
\newauthor{}
Gast\'on J. Escobar$^{1,2}$, 
Cecilia Sgalletta$^{2,4}$,
Alessandro A. Trani$^{6,7}$,
Erika Korb$^{1,2}$,
\newauthor{}
Filippo Santoliquido$^{1,2}$,
Marco Dall'Amico$^{1,2}$,
Nicola Gaspari$^{8}$,
Alessandro Bressan$^{4,3}$
\\
$^{1}$Dipartimento di Fisica e Astronomia Galileo Galilei, Universit\`a di  Padova, Vicolo dell'Osservatorio 3, I--35122 Padova, Italy\\
$^{2}$INFN--Padova, Via Marzolo 8, I--35131 Padova, Italy\\
$^{3}$INAF--Padova, Vicolo dell'Osservatorio 5, I--35122 Padova, Italy\\
$^{4}$SISSA, via Bonomea 365, I--34136 Trieste, Italy\\
$^{5}$INFN--Trieste, via Valerio 2, I--34127 Trieste, Italy\\
$^{6}$Department of Earth Science and Astronomy, College of Arts and Sciences, The University of Tokyo, 3-8-1 Komaba, Meguro-ku, Tokyo 153-8902, Japan\\
$^{7}$Okinawa Institute of Science and Technology, 1919-1 Tancha, Onna-son, Okinawa 904-0495, Japan\\
$^{8}$Department of Astrophysics/IMAPP, Radboud University, P.O. Box 9010, 6500 GL, Nijmegen, The Netherlands
}

\date{Accepted XXX. Received YYY; in original form ZZZ}

\pubyear{2022}

\begin{document}
\label{firstpage}
\pagerange{\pageref{firstpage}--\pageref{lastpage}}
\maketitle

\begin{abstract}

Population-synthesis codes are an unique tool to explore the parameter space of massive binary star evolution and binary compact object (BCO) formation.  Most  
population-synthesis codes are based on the same stellar evolution model, limiting our ability to explore the main uncertainties. 
Here, we present the new version of  the code \sevn{}, which  overcomes this issue 
by interpolating the main stellar properties from a set of pre-computed evolutionary tracks. We describe the new interpolation and adaptive time-step algorithms of \sevn{}, and the main upgrades on single and binary evolution. 
With \sevn{}, we evolved $1.2\times10^9$ binaries in the  metallicity range $0.0001\leq Z \leq 0.03$, exploring a number of models for electron-capture, core-collapse and pair-instability supernovae, different assumptions for  common envelope,  stability of  mass transfer, quasi-homogeneous evolution  and stellar tides. We find that stellar evolution has a dramatic impact on the formation of single and binary compact objects. 
Just by slightly changing the overshooting parameter ($\lambda_{\rm ov}=0.4$, 0.5) and the pair-instability model, the maximum mass of a black hole can vary from $\approx{60}$ to $\approx{100}\ \Msun$. Furthermore, the formation channels of BCOs and the merger efficiency we obtain with \sevn{} show significant differences with respect to the results of other population-synthesis codes, even when the same binary-evolution parameters are used.  For example, the  main traditional formation channel of BCOs
is strongly suppressed in our models: at high metallicity ($Z\gtrsim{0.01}$)
only $<20$\% of the merging binary black holes and binary neutron stars form via this
channel, while other authors found fractions $>70$\%. 
\end{abstract}

\begin{keywords}
methods: numerical - gravitational waves - binaries: general - stars:mass-loss - stars: black holes 
\end{keywords}

\section{Introduction} \label{sec:introduction}

Since the first detection in September 2015, the LIGO--Virgo--KAGRA collaboration (LVK) has reported 90 binary compact object (BCO) merger candidates, most of them binary black holes (BBHs,  \citealt{abbottGW150914,abbottO1, abbottastrophysics,abbottO2,abbottO2popandrate,abbottO3a,abbottO3apopandrate,abbottGWTC-2.1,abbottGWTC3}). 
The LVK data have confirmed that BBHs exist, and probed a mass spectrum of black holes (BHs) ranging from a few to $\sim{200}$ M$_\odot$ \citep{abbottastrophysics,abbottO2popandrate,abbottO3apopandrate,abbottGWTC3popandrate}. This result  has revolutionised our knowledge of stellar-sized BHs, complementing electromagnetic  \citep[e.g.,][]{oezel2010,farr2011} and microlensing data \citep[e.g.,][]{wyrzykowski2016}. Some peculiar LVK events even challenge current evolutionary models, indicating the existence of compact objects inside the claimed lower \citep[e.g.,][]{abbottGW190814} and upper mass gap \citep[e.g.,][]{abbottGW190521,abbottGW190521astro,abbottGWTC-2.1}. Finally, the first and so far only multi-messenger detection of a binary neutron star (BNS) merger \citep[e.g.,][]{abbottGW170817,abbottmultimessenger} has confirmed the association of kilonovae and short gamma-ray bursts with mergers of neutron stars (NSs), paving the ground for a novel synergy between gravitational-wave (GW) scientists and astronomers. 

This wealth of new data triggered an intense debate on the formation channels of BCOs (see, e.g., \citealt{mandelfarmer2018} and \citealt{mapelli2021review} for two recent reviews on this topic). One of the main problems of the models is the size of the parameter space: even if we restrict our attention to BCO formation via binary evolution, countless assumptions about the evolution of massive binary stars can have a sizeable impact on the final BCO properties. Hence, numerical models used to probe BCO populations need to be computationally fast, while achieving the highest possible level of accuracy and flexibility. Binary population synthesis codes are certainly the fastest approach to model binary star evolution, from the zero-age main sequence (ZAMS) to the final fate. 
For example, the famous \bse{} code \citep{Hurley00,Hurley02}, which is the common ancestor of most binary population synthesis codes, evolves $\mathcal{O}(10^6)$ binary stars in a couple of hours on a single CPU core. 
For comparison, a modern stellar evolution code requires $\mathcal{O}(10-100)$ CPU hours to integrate the evolution of an individual binary star. The speed of binary population synthesis codes is essential not only 
to model the parameter space of massive binary star evolution, but also to guarantee that they can be interfaced with dynamical codes to study the dynamical formation of BCOs in dense stellar clusters \citep[e.g.,][]{banerjee2010,tanikawa2013,mapelli2013,ziosi2014,rodriguez2015,rodriguez2016,mapelli2016,banerjee2017,banerjee2018,rastello2018,banerjee2019,banerjee2020,dicarlo2019,dicarlo2020b,dicarlo2021,kremer2020,kremer2020b,rastello20,ye2021,wang2020,rastello2021,wang2022}.

A large number of binary population synthesis codes have been developed across the years and most of them have been used to study the formation of BCOs, e.g., 
{\sc binary\_c} \citep{binaryc1,izzard2006,izzard2009,binaryc2}, {\sc bpass} \citep{eldridge2017}, the {\sc Brussels} code \citep{vanbeveren1998,dedonder2004,mennekens2014}, 
{\sc bse-LevelC} \citep{kamlah2022}, {\sc combine} \citep{Combine}, {\sc compas} \citep{Compas}, {\sc cosmic} \citep{Cosmic}, {\sc IBis} \citep{tutukov1996}, {\sc metisse} \citep{Metisse}, {\sc mobse}  \citep{mapelli2017,GiacobboWind}, {\sc posydon} \citep{Posydon}, the {\sc Scenario Machine} \citep{lipunov1996,lipunov2009}, {\sc SeBa} \citep{portegieszwart1996,toonen2012}, {\sc sevn} \citep{Spera19,Mapelli20}, and {\sc startrack} \citep{belczynski2002,startrack}.

While all of them are independent codes, most of them rely on the same model of stellar evolution: the accurate and computationally efficient fitting formulas developed by \cite{Hurley00}, based on the stellar tracks by \cite{Pols98}. These fitting formulas express the main stellar evolution properties (e.g., photospheric radius, core mass, core radius, luminosity) as a function of stellar age, mass ($M$), and metallicity ($Z$, mass fraction of elements heavier than helium). The results of binary population synthesis codes adopting such fitting formulas can differ by the way they model stellar winds, compact-remnant formation and binary evolution, but rely on the same  stellar evolution model. This implies that they can probe only a small portion of the parameter space, which is the physics encoded in the original tracks by \cite{Pols98}. Stellar evolution models have dramatically changed since 1998, including, e.g., new calibrations for core overshooting \citep[e.g.,][]{Claret2018, Costa2019}, updated networks of nuclear reactions \citep[e.g.,][]{Cyburt2010, Sallaska2013}, 
updated opacity tables \citep[e.g.,][]{Marigo2009, Poutanen2017}, and new sets of stellar tracks with rotation 
\citep[e.g.,][]{Brott2011, chieffi2013, Georgy2013, Choi2016, Nguyen22}.
Moreover, the newest stellar evolution models probe a much wider mass and metallicity range \citep[e.g.,][]{Spera17} than the range encompassed by \cite{Hurley00} fitting formulas ($0.5\leq{}M/{\rm M}_\odot\leq{}50$, $0.0001\leq{}Z\leq{}0.03$). 

Driven by the need to include up-to-date stellar evolution and a wider range of masses and metallicities, several binary population synthesis codes adopt an alternative strategy with respect to \cite{Hurley00} fitting formulas. {\sc bpass} \citep{eldridge2008,eldridge2016,eldridge2017} integrates stellar evolution on-the-fly with a custom version of the Cambridge {\sc stars} stellar evolution code \citep{eggleton1971,pols1995,eldridge2004}. To limit the computational time, the primary star (i.e., the most massive star in the binary system) is first evolved with {\sc stars}, while the secondary is evolved with the fitting formulas by \cite{Hurley00}. After the evolution of the primary star is complete, the evolution of the secondary is re-integrated with {\sc stars}.

{\sc combine} \citep{Combine}, {\sc metisse} \citep{Metisse}, {\sc posydon} \citep{Posydon} and {\sc sevn} \citep{Spera15,Spera17,Spera19,Mapelli20}  share the same approach to stellar evolution: they include an algorithm that interpolates the main stellar-evolution properties (mass, radius, core mass and radius, luminosity, etc as a function of time and metallicity) from a number of pre-computed tables. The main advantage is that the interpolation algorithm is more flexible than the fitting formulas: it is sufficient to generate new tables, in order to update the stellar-evolution model. Furthermore, this approach allows to easily compare different stellar-evolution models encoding different physics (e.g., different stellar-evolution codes, different overshooting models, different convection criteria). Among the aforementioned codes, {\sc posydon} is the only one that includes tables of binary star evolution, run with the code {\sc mesa} \citep{Mesa,paxton2013,paxton2015,paxton2018}, while the others 
are based on single star evolution tables.  Including binary-evolution in the look-up tables has the advantage of encoding the response of each star to 
interactions between binary components. This level of model sophistication comes with increased data size: 
the look-up tables for a given metallicity weigh $\mathcal{O}(100)$~MB for single star evolution, 
and $\mathcal{O}(10)$~GB for binary evolution, respectively. Overall, binary population synthesis codes based on look-up tables are a powerful tool to probe the parameter space of BCO formation with up-to-date stellar evolution.

Here, we present a new version of our binary population synthesis code {\sc sevn}, and use it to explore some of the main uncertainties in BCO formation springing from stellar and binary evolution. This paper is organised as follows.  
Section~\ref{sec:sevn} 
describes the main features of  \sevn{}. 
In Section~\ref{sec:simsetup}, we  describe the stellar evolution models used in this work, our initial conditions, 
and  the main parameters/assumptions tested with our simulations.   
 Section~\ref{sec:Results} shows the  properties of BCOs formed in our simulations, their mass spectrum,  merger efficiency,  and local merger rate density. 
In Section~\ref{sec:discussion}, we   discuss  our results and their possible caveats.   Finally, Section~\ref{sec:conclusion}  is a summary of our main results.

\section{Description of \sevn{}} \label{sec:sevn}

\begin{figure}
	\centering
	\includegraphics[trim={2.5cm 2cm 3cm 0cm},width=1.0\columnwidth]{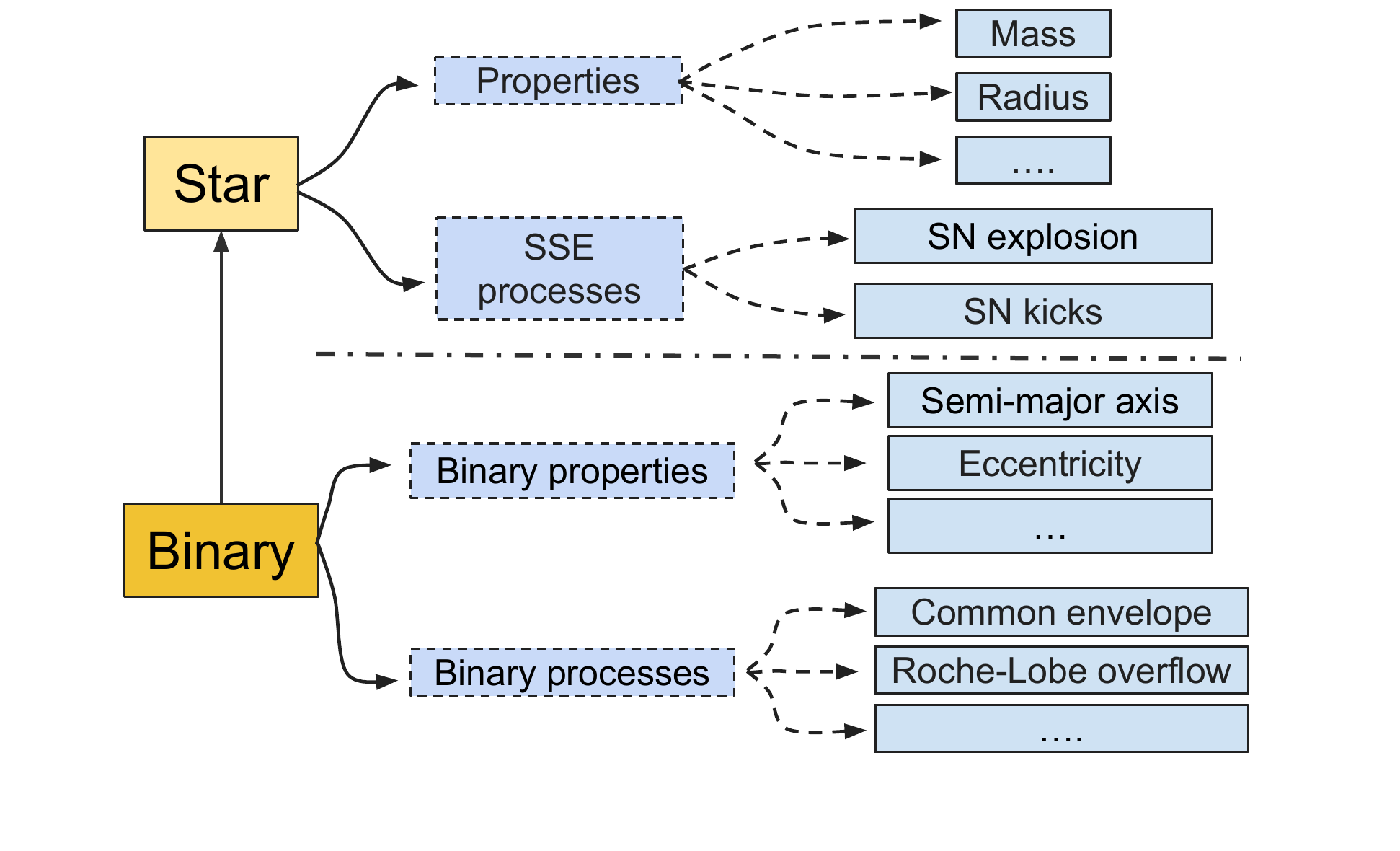}
	\vspace*{-5mm}
	\caption{In \sevn{}, single stars, binary systems, properties and processes are represented with C++ classes. Single stars are characterised by their properties (mass, radius,...) and single stellar evolution processes (supernova explosion type and natal  kicks). Binary stars are characterised by their properties (semi-major axis, eccentricity,..), binary-evolution processes (mass transfer by winds, Roche-lobe overflow, CE, tides,..), and by the two stars component of the binary system.
      \gitlab{https://gitlab.com/iogiul/iorio22_plot/-/tree/v3/SEVN_flowcharts}
 \gitimage{https://gitlab.com/iogiul/iorio22_plot/-/blob/v3/SEVN_flowcharts/SEVNcomponents.pdf}}
	\label{fig:SEVNcomponents} 
\end{figure}

\sevn{} (Stellar EVolution for $N$-body) is a rapid binary population synthesis code,  
 which calculates stellar evolution 
by interpolating pre-computed sets of stellar tracks \citep{Spera15,Spera17,Spera19,Mapelli20}. Binary evolution is implemented by means of 
analytic and semi-analytic prescriptions. 
The main advantage of this strategy is that it makes the implementation more general and flexible:  the stellar evolution models adopted in \sevn{} can easily be changed or updated just by loading a new set of look-up tables. \sevn{} allows to choose the stellar tables at runtime,  without modifying the internal structure of the code or even recompiling it.

The current version of \sevn{} is grounded on the same basic concepts developed for the previous versions \citep[see, e.g.,][]{Spera17,Spera19}, but the code has been completely refactored, improved in many aspects (e.g. time step, modularity),  extended with new functionalities/options, and updated with the latest {\sc parsec} 
stellar evolution tracks \citep{Bressan2012,Chen15,Costa2021,Nguyen22}.  
\sevn{} is written entirely in {\sc C++} (without external dependencies) following the object-oriented programming paradigm. \sevn{} exploits the CPU-parallelisation through   {\sc OpenMP}.
Figure~\ref{fig:SEVNcomponents} shows a schematic representation of the basic \sevn{} components and their relations.

In the following sections, we describe the main  features and options of \sevn{} focusing on the new prescriptions used in this work. 
Additional information about \sevn{} can be found in Appendix~\ref{app:sevn}.
\sevn{} is publicly available at \href{https://gitlab.com/sevncodes/sevn.git}{this link}\footnote{\url{https://gitlab.com/sevncodes/sevn.git}}; the version used in this work is the release \textit{Iorio22}\footnote{\url{https://gitlab.com/sevncodes/sevn/-/releases/iorio22}}.

\subsection{Single star evolution} \label{sec:sse}

In the following sections, we describe the main ingredients used in \sevn{} to integrate stellar evolution from the ZAMS to the formation of the compact remnant. Additional information can be found in Appendix~\ref{app:sevn}.

\subsubsection{Stellar evolution tables} \label{sec:tables}

\begin{table}
\centering
\setlength\tabcolsep{1.5pt} 
\begin{tabular}{lccc}

\multicolumn{4}{c}{\sevn{} tables}                                                  \\ \hline \hline
Table                              & Units                     & Type    & Interpolation              \\ \hline
Time                              & Myr                       & M        & R            \\
Phase$\dagger$                    & Myr$\dagger$              & M        & R            \\
Mass                             & $\Msun{}$                 & M        & LIN            \\
Luminosity                      & $\Lsun{}$                 & M        & LOG           \\
Radius                            & $\Rsun{}$                 & M        & LOG            \\
He-core mass                      & $\Msun{}$                 & M        & LIN            \\
CO-core mass                      & $\Msun{}$                 & M        & LIN            \\
He-core Radius                    & $\Rsun{}$                 & O        & LIN            \\
CO-core Radius                    & $\Rsun{}$                 & O        & LIN            \\
Stellar inertia$\ddagger$                  & $\Msun{} \Rsun^2$         & O        & LOG            \\
Envelope binding energy$\ddagger$         & $\Msun{}^2 \Rsun^{-1} (G^{-1})*$         & O & LOG \\
Surface abundances $\ddagger$        & mass fraction         & O & LIN \\ 
(H,He,C,N,O)        &          &  &  
\\  \hline
\multicolumn{4}{c}{Convective envelope}                                                  \\ \hline
 mass          & normalised to star mass   & O        & LIN            \\
 depth         & normalised to star radius & O        & LIN             \\
 turnover time & yr                        & O        & LIN      
 \\ \hline \hline

\end{tabular}
\caption{Summary of the stellar evolution tables used in \sevn{}. The first column reports the property stored in the table, the second column its units and the third column specifies if a table is mandatory (M) or optional (O). 
\sevn{} includes analytic recipes to  replace the optional tables if they are not available (Appendix~\ref{app:tables}). The fourth column indicates the type of  weights used by \sevn{} during the property interpolation: rational (R), linear (LIN), log (LOG), see Section~\ref{sec:sevninterpolation}. 
$\dagger$ The phase table reports the starting time of each \sevn{} phase (Table~\ref{tab:phases}). 
$\ddagger$Not included in the stellar tracks used in this work (Section~\ref{sec:tracks}). The envelope binding energy is normalised over the gravitational constant $G$ (assumed in solar units and years).   } 
\label{tab:tables}
\end{table}

The \sevn{} stellar-evolution 
tables contain the evolution of the properties of a set of stellar tracks defined by their initial mass $M_\mathrm{ZAMS}$ and metallicity $Z$.  \sevn{} requires, as input, two sets of tables: one for stars that start their life from the hydrogen main sequence (MS; hereafter, H stars), the other  
for stars that are H depleted  (hereafter, pure-He stars).  
Unlike \bse{}, \sevn{} assumes that the stellar models already include wind mass loss. 

Table~\ref{tab:tables} summarises the tables available in \sevn{}.
Each stellar-evolution model comprises (at least) seven  tables grouped by metallicity. The tables for each metallicity are stored in different directories. 
Each table refers to a given stellar property. There are  seven mandatory tables corresponding to the main stellar properties: time,  total stellar mass,  He-core mass,  CO-core mass, stellar radius, bolometric luminosity, and the stellar phase (Section \ref{sec:sevnphase}). 
Each row  in the tables refers to a star with a given  $M_\mathrm{ZAMS}$ and $Z$, each column stores the value of the property  at the time correspondent to the same row and column in the time table. The first column of each row in the mass table identifies the $M_\mathrm{ZAMS}$ of the star.  
The stellar-phase table contains the starting time for the stellar phases (Section \ref{sec:sevnphase}). 
The end of the evolution (i.e., the stellar lifetime) is not reported in the phase table, rather \sevn{} implicitly assumes it is equal to the last value reported in the time tables. 

Additional properties such as the radii of the He and CO cores, the envelope binding energy, and the properties of the convective envelope (mass, extension, eddy turnover timescale) are optional. 
If such tables are not provided (or disabled by the user), \sevn{}  estimates these  properties using alternative analytic approximations (Appendix~\ref{app:tables}).      
These tables are not mandatory because they contain information that is not available in most stellar-evolution tracks, but they are essential to properly model several evolution processes. For example, the properties of the convective envelope allow a more physical identification of the evolutionary phase and can be used to estimate the stability of mass transfer (Section~\ref{sec:rlo}), in addition they also play an  important role in setting the efficiency of stellar tides (Section~\ref{sec:rlo}).
The modular structure of \sevn{} makes it possible to easily introduce new tables to follow the evolution of additional stellar properties. 
\sevn{} does not assume a specific definition for the mass and radius of the He and CO cores. The estimate of such properties depends on the adopted stellar evolution models and/or on the user choice  in the production of the \sevn{} tables (Section \ref{sec:tracks}).

\subsubsection{\sc TrackCruncher}

The most important requirement of the tables is that they must  capture all the main features of the stellar tracks they are generated from, but at the same time they must be as small as possible (up to a few MB each), to make the interpolation fast and to reduce the memory cost.
In order to satisfy these requirements, we 
developed the code {\sc TrackCruncher}, which we use to efficiently generate the tables for \sevn{}. %
This code extracts the properties to store in the \sevn{} tables from a set of stellar tracks, while estimating the starting time of the \sevn{} phases (see Section \ref{sec:sevnphase} and Appendix~\ref{sec:trackstable}). 
In addition, {\sc TrackCruncher} decides which time-steps of the original tracks 
can be omitted in the final tables, in order to reduce the table size. In particular, we store in the final tables only the time-steps of the original tracks that guarantee errors smaller than 2\%  when we perform a linear interpolation to model the evolution of  the stellar properties  (Section \ref{sec:sevninterpolation}).
This track under-sampling  reduces significantly the size of the tables, from  $\mathcal{O}$(1 GB) to $\mathcal{O}$(10 MB). 
For example, the complete set of tables for H stars (pure-He stars) used in this work (see Section \ref{sec:tracks}) occupies only ${\sim} 30$ MB (${\sim} 10$ MB), while the original tracks consume ${\sim} 5$ GB (${\sim} 6$ GB) of disc space. 
This procedure significantly reduces both the storage and runtime memory footprint of \sevn{};  moreover it speeds up single stellar evolution computation (see Section \ref{sec:tstep}).  

 {\sc TrackCruncher} is publicly available at \href{https://gitlab.com/sevncodes/trackcruncher}{this link}\footnote{\url{https://gitlab.com/sevncodes/trackcruncher}}.
 It is optimized to process the outputs of {\sc parsec} \citep{Bressan2012},  {\sc franec} \citep{Limongi2018}, and the {\sc mist}  stellar tracks \citep{Choi2016}, but  can easily be  extended to process the output of other stellar evolution codes. {\sc TrackCruncher} can also be used as a tool to compress and reduce the memory size of stellar tracks. 

The specific description of the stellar tables used in this work can be found in Section \ref{sec:tracks} and Appendix~\ref{sec:trackstable}.

\subsubsection{Stellar phases}  \label{sec:sevnphase}

\setlength{\tabcolsep}{2pt}
\begin{table*}
\begin{tabular}{lcccl}
\hline
\sevn{} Phase & Phase ID  & \sevn{} Remnant subphase & Remnant ID  & \bse{} stellar-type equivalent \\ \hline
Pre-main sequence (PMS) & 0 & -- & 0 & not available \\
Main sequence (MS) & 1 & -- & 0 & 1 if $f_\mathrm{conv}^\dagger<0.8$, else 0 \\
\hline
 Terminal-age main sequence (TAMS) & 2 & --  & 0 & \multirow{2}{*}{\textit{$2$  if $f_\mathrm{conv}^\dagger<0.33$, {\rm else} $3$}} \\
Shell H burning (SHB) & 3 & --  & 0  &  \\
\hline
Core He burning (CHeB) & 4 & --  & 0  & 7 if WR$^\ddagger$, else 4 \\
Terminal-age core He burning  (TCHeB) & 5 & --  & 0 & 7 if WR$^\ddagger$, else: 4  if  $f_\mathrm{conv}^\dagger<0.33$, else 5 \\
Shell He burning (SHeB) & 6 & --  & 0 & 8 if WR$^\ddagger$, else: 4  if  $f_\mathrm{conv}^\dagger<0.33$ else 5 \\ \hline
\multirow{7}{*}{Remnant} & \multirow{7}{*}{7} & He white dwarf (HeWD) & 1 & 10 \\
 &  & CO white  dwarf (COWD)   & 2 & 11 \\
 &  & ONe white dwarf (ONeWD)   & 3 & 12 \\
 &  & neutron star formed via electron capture (ECNS)  & 4 & 13 \\
 &  & neutron star formed via core collapse (CCNS) & 5 & 13 \\
 &  & black hole (BH) & 6 & 14 \\
 &  & no compact remnant (Empty) & -1 & 15 \\ \hline
\end{tabular}
\caption{\sevn{} stellar evolutionary phases (Column~0), identifiers (Column~1) and remnant types (Column~2). Column~3 shows the correspondence to 
\protect\cite{Hurley00,Hurley02} stellar types: 0, low-mass main sequence (MS); 1, main sequence (MS); 2, 
Hertzsprung-gap (HG); 3, first giant branch (GB); 4, core-helium burning (CHeB); 5, early asymptotic giant branch (EAGB);
7, naked-helium MS (HeMS);  8, naked-helium HG (HeHG).  The  \bse{} stellar types 6 (thermally pulsating AGB) and 9 (naked-helium giant branch) do not have a correspondent \sevn{} phase. 
ECNS and CCNS are NSs produced by electron capture and core collapse supernovae, respectively (Section \protect\ref{sec:remform}). $\dagger$ $f_\mathrm{conv}$ is the mass fraction of the convective envelope over the total envelope mass  (total mass in case of MS stars),  $\ddagger$ WR indicates 
Wolf-Rayet (WR) stars, i.e., stars which have a He core  mass larger than 97.9\% of the 
total mass. See Section \protect\ref{sec:sevnphase} for additional details.}
\label{tab:phases}
\end{table*}

\cite{Spera19} found that the interpolation of  stellar evolution properties significantly improves if we use  
the percentage of life of a star instead of the absolute value of the time (Section \ref{sec:sevninterpolation}). 
In order to further refine the 
interpolation, they estimate the percentage of life in  three stellar macro-phases: i) the H phase, in which the star has not developed a He core yet; ii) the He phase, when the star has a He core but not a CO core; iii) the CO phase, when the star has a CO core. 

In the current version of \sevn{}, we refine the definition of macro-phases in \cite{Spera19} by dividing  stellar evolution in seven physically motivated phases.
The phase from time 0 to the ignition of hydrogen burning in the core is the pre-main sequence  (PMS,  phase id~$=0$).
During core-hydrogen burning, the star is in the main sequence (MS, phase id~$=1$) phase until its He core starts to grow (He-core mass $> 0$) and the star enters the terminal-age MS (TAMS, phase id~$=2$). 
The next phase, shell H burning (SHB, phase id~$=3$),  starts when the hydrogen in the core has been completely exhausted and the star is burning hydrogen in a thin shell around the He core.
At the ignition of core helium burning, the star enters the core He burning phase (CHeB, phase id~$=4$), which is followed by  the terminal-age core He burning (TCHeB, phase id~$=5$, CO-core mass $>0$) and the shell He burning (SHeB, phase id~$=6$).  
This last phase starts when  helium   has been completely exhausted in the core. The remnant phase (id~$=7$) begins when the evolution time exceeds the  star's lifetime (see Section \ref{sec:tables}), and the star becomes a compact remnant (Section \ref{sec:remform}).

During its evolution, a star can be stripped of its hydrogen envelope either because of effective stellar winds or due to binary interactions.  If the He-core mass is larger than 97.9\% of the total stellar mass, \sevn{} classifies the star as a Wolf-Rayet (WR) star \citep[e.g.,][]{Bressan2012,Chen15}    
and the star jumps to a new interpolating track  on the pure-He tables (Section \ref{sec:trackchange}).  
In \sevn{}, we do not use special phases for pure-He stars. 
The only difference with respect to hydrogen-rich  stars is that a pure-He star does not go through phases 0--3, but rather starts its life from phase 4 (CHeB). Pure-He stars in \sevn{} are equivalent to the stars defined as naked-He stars in other population synthesis codes derived from \bse{} \citep{Hurley02}. 

During binary evolution, an evolved pure-He star can 
lose its He envelope leaving a naked-CO star. \sevn{} does not have a dedicated phase for such objects, but they are considered compact remnant-like objects and evolve accordingly (Section \ref{sec:sync}). The conversion between \sevn{} stellar phases and {\sc bse} stellar types 
\citep{Hurley00} is summarised in Table \ref{tab:phases}.

\subsubsection{Interpolation} \label{sec:sevninterpolation}

\begin{figure}
	\centering
	\includegraphics[width=1.0\columnwidth]{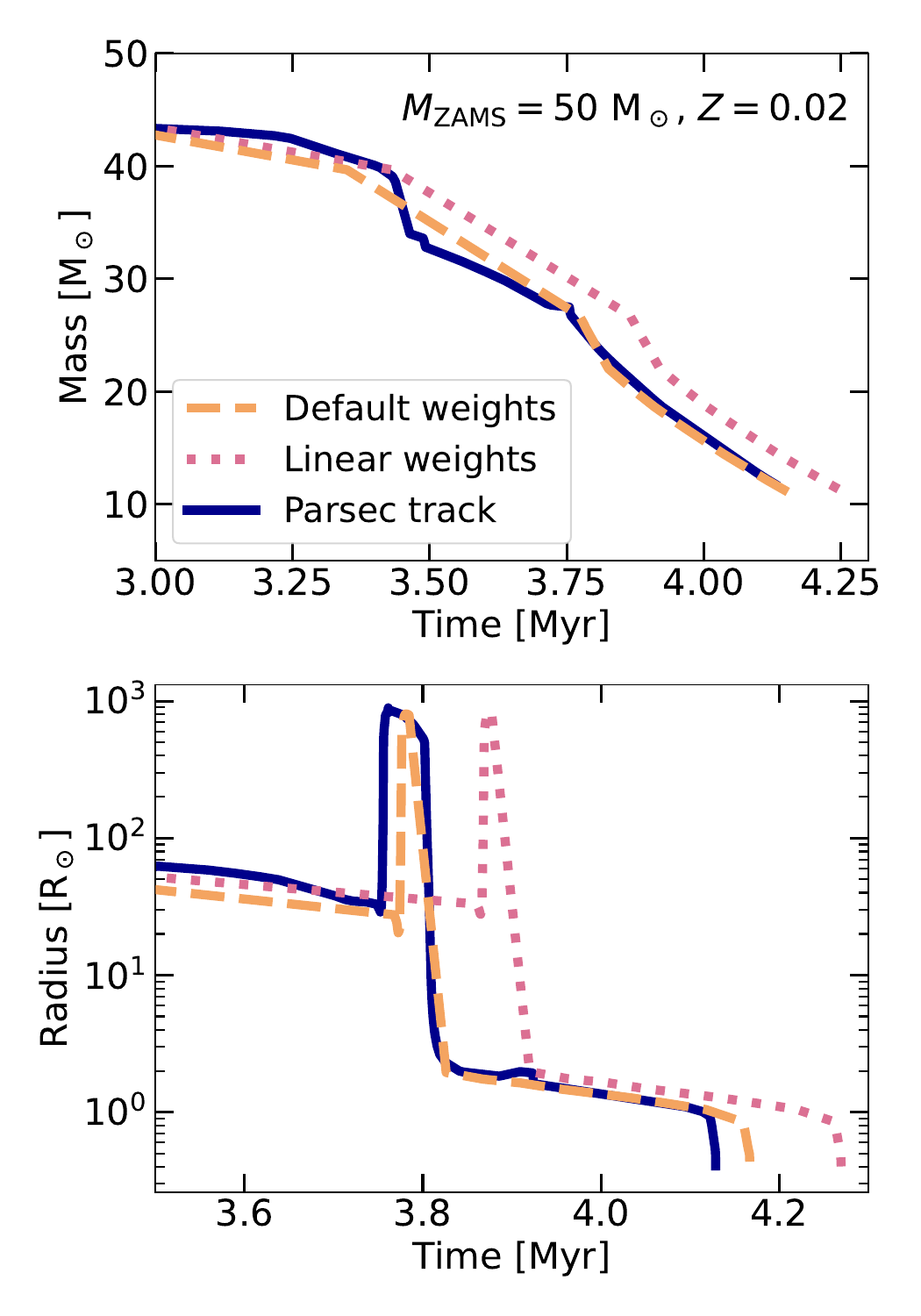}
	\vspace*{-5mm}
	\caption{Time evolution of the mass (top) and radius (bottom) of a star with $M_\mathrm{ZAMS}=50 \ \Msun$ and $Z=0.02$. The blue solid lines refer to a  stellar track obtained with the code \parsec\  (Section~\ref{sec:tracks}), while the other lines show the \sevn{} interpolation using pre-evolved tracks of two stars with  $M_\mathrm{ZAMS,1}=40 \ \Msun$,  $M_\mathrm{ZAMS,2}=60 \ \Msun$, and $Z=0.02$. 
    We obtain the orange dashed line interpolation using the default weights: 
     \textit{linear} weights for the mass (Eq.~\ref{eq:wlinear}), 
     \textit{logarithmic} weights for the radius (Eq.~\ref{eq:wlog}), and  
     \textit{rational} weights for the phase and time (Eq.~\ref{eq:wrational}). 
     In contrast, we obtain the pink dotted  curve using \textit{linear} weights for all the properties. 
     \gitlab{https://gitlab.com/iogiul/iorio22_plot/-/tree/v3/interpolation}
     \gitbook{https://gitlab.com/iogiul/iorio22_plot/-/blob/v3/interpolation/Interpolation_test.ipynb} 
        \gitimage{https://gitlab.com/iogiul/iorio22_plot/-/blob/v3/interpolation/interp_test.pdf}}
\label{fig:interptest} 
\end{figure}

We 
estimate the properties of each star  at a given time via interpolation.  
The method implemented in this version of \sevn{} 
is an improved version with respect to  \cite{Spera19}. 
When a star is initialised, \sevn{} assigns to it four interpolating tracks from the hydrogen or pure-He look-up tables.   These four tracks have two different metallicities ($Z_1$, $Z_2$) and four different ZAMS masses ($M_{\rm ZAMS,1}$, $M_{\rm ZAMS,2}$, $M_{\rm ZAMS,3}$, $M_{\rm ZAMS,4}$, two per metallicity), chosen as  
$M_{\rm ZAMS,1/3}\leq M_{\rm ZAMS}<M_{\rm ZAMS,2/4}$ 
and $Z_1\leq Z<Z_2$
where $M_{\rm ZAMS}$ and $Z$ are the ZAMS mass and the metallicity of the star we want to calculate. 
In case $M_{\rm ZAMS}$ and/or $Z$ are equal to the maximum values in the tables, we use  $M_{\rm ZAMS,1/3}< M_{\rm ZAMS} \leq M_{\rm ZAMS,2/4}$ and $Z_1 < Z \leq Z_2$. 
A given interpolated property $W$ (e.g. the stellar mass) is estimated as follows.
\begin{equation}
W = \frac{Z_2-Z}{Z_2 - Z_1} W_\mathrm{Z,1} + \frac{Z-Z_1}{Z_2 - Z_1}  W_\mathrm{Z,2},
\label{eq:xinterpZ}
\end{equation} where
\begin{equation}
\begin{split}
W_\mathrm{Z,1} & = \beta_1 W_\mathrm{ZAMS,1} + \beta_2 W_\mathrm{ZAMS,2} \\
W_\mathrm{Z,2} & = \beta_3 W_\mathrm{ZAMS,3} + \beta_4 W_\mathrm{ZAMS,4}.
\end{split}
\label{eq:xinterpM}
\end{equation}
In Eq.~\ref{eq:xinterpM},  $W_\mathrm{ZAMS,i}$ indicates the value of the property $W$ in the interpolating tracks with $M_\mathrm{ZAMS,i}$, and $\beta$ are interpolation weights. \sevn{} includes three  different interpolation weights: 
\begin{itemize}
\item \textit{linear}, 
\begin{equation}
\begin{split}
\beta_{1/3} &= \frac{M_\mathrm{ZAMS,2/4} -  M_\mathrm{ZAMS}}{M_\mathrm{ZAMS,2/4} - M_\mathrm{ZAMS,1/3}}, \\
\ \beta_{2/4} &= \frac{M_\mathrm{ZAMS} - M_\mathrm{ZAMS,1/3}}{M_\mathrm{ZAMS,2/4} - M_\mathrm{ZAMS,1/3}};
\label{eq:wlinear}
\end{split}
\end{equation}
\item \textit{logarithmic},
\begin{equation}
\begin{split}
\beta_{1/3} &= \frac{\log M_\mathrm{ZAMS,2/4} -  \log M_\mathrm{ZAMS}}{\log M_\mathrm{ZAMS,2/4} - \log M_\mathrm{ZAMS,1/3}}, \\
\ \beta_{2/4} &= \frac{\log M_\mathrm{ZAMS} - \log M_\mathrm{ZAMS,1/3}}{\log M_\mathrm{ZAMS,2/4} - \log M_\mathrm{ZAMS,1/3}};
\label{eq:wlog}
\end{split}
\end{equation}
\item \textit{rational},
\begin{equation}
\begin{split}
\beta_{1/3} &=  \frac{M_\mathrm{ZAMS,1/3} \left( M_\mathrm{ZAMS,2/4} -  M_\mathrm{ZAMS} \right)}{ M_\mathrm{ZAMS} \left( M_\mathrm{ZAMS,2/4} - M_\mathrm{ZAMS,1/3} \right)}, \\
\ \beta_{2/4} &= \frac{ M_\mathrm{ZAMS,2/4} \left( M_\mathrm{ZAMS} - M_\mathrm{ZAMS,1/3} \right)}{ M_\mathrm{ZAMS} \left( M_\mathrm{ZAMS,2/4} - M_\mathrm{ZAMS,1/3} \right)}.
\label{eq:wrational}
\end{split}
\end{equation}
\end{itemize}
\sevn{} uses   \textit{logarithmic} weights for the properties that are internally stored and interpolated in logarithmic scale, i.e., radius and luminosity. 
\cite{Spera19} introduced the \textit{rational} weights to improve the interpolation. In particular, we found that they drastically improve the estimate of the starting time of the stellar phases and the estimate of the star lifetime.  For all the other properties, \sevn{} uses \textit{linear} weights (Table~\ref{tab:tables}).
Figure~\ref{fig:interptest} clearly shows that the combination of different weights  gives a much more reliable interpolation compared to using only linear weights.

When a star is initialised, \sevn{} uses Eqs.~\ref{eq:xinterpZ} and \ref{eq:xinterpM} to set the starting times of the stellar phases, $t_\mathrm{start,p}$ (see, e.g., Section \ref{sec:sevnphase}), where $W_\mathrm{ZAMS,i}$ represents the phase times from the phase table (Section \ref{sec:tables}). 
We interpolate the stellar lifetime 
in the same way,  assuming that the last element in the \sevn{} time table  sets the stellar lifetime.  
For all the other properties, $W$ has to be estimated at a given time $t$. The corresponding  $W_\mathrm{ZAMS,i}$ in the tables is not estimated at the same absolute time $t$, rather at the same percentage of life in the phase of the interpolated star (Section \ref{sec:sevnphase}):
\begin{equation}
\Theta_\mathrm{p}  = \frac{t-t_\mathrm{start,p}}{t_\mathrm{start,p_{next}}-t_\mathrm{start,p}},
\label{eq:thetap}
\end{equation}
where $t_\mathrm{start,p}$ indicates the starting time of the current phase $p$, and  $t_\mathrm{start,p_\mathrm{next}}$  the starting time of next phase $p_\mathrm{next}$ (Table \ref{tab:phases}). 
Hence, \sevn{} evaluates $W_\mathrm{ZAMS,i}$ at  time 
\begin{equation}
t_\mathrm{i}  = t_\mathrm{start,p,i}  + \Theta_\mathrm{p} \Delta_\mathrm{p,i},
\label{eq:ttrack}
\end{equation}
where $t_\mathrm{start,p,i}$ and $\Delta_\mathrm{p,i}$ are the starting time and the time duration of the current phase for the  interpolating track.
In practice, \sevn{} uses  Eq.~\ref{eq:thetap} to evaluate the times for each of the fourth interpolating tracks. Then, it estimates $W_\mathrm{ZAMS,i}$ in Eq.~\ref{eq:xinterpM} by interpolating (linearly along the time)  the values stored in the tables.

The division into phases guarantees that all the interpolating stars have the same internal structure (e.g., the presence or not of the core) improving significantly the interpolation method and reducing the interpolation errors to a few percent \citep{Spera19}.

\subsubsection{Spin evolution}
\label{sec:spin}

We model the evolution of  stellar rotation through three properties: the fundamental quantity evolved in \sevn{} is the  spin angular momentum $J_\mathrm{spin}$,  then  we derive the angular velocity as 
$\Omega_\mathrm{spin}=J_\mathrm{spin}\,{} I^{-1}$ (where $I$ is the  inertia), and estimate the spin $\omega_\mathrm{spin}$ as the ratio between $\Omega_\mathrm{spin}$ and the critical angular velocity  $\Omega_\mathrm{crit}=\sqrt{G\,{} M\,{} (1.5 \,{}R)^{-3}}$, where $G$ is the gravity constant, $M$ and $R$ are the stellar mass and radius.
 In this work, we estimate stellar inertia following \cite{Hurley02}:
\begin{equation}
	I = 0.1 (M - M_\mathrm{c}) R^{2} + 0.21M_\mathrm{c} R^{2}_\mathrm{c}\;
	\label{eq:inertiahurley},
\end{equation}
where $M_\mathrm{c}$ is the core mass and $R_\mathrm{c}$  the core radius.
The initial rotation of the star is set by  the input value of $\omega_\mathrm{spin}$. 

During the evolution, part of stellar angular momentum is removed through stellar winds and part through  the so-called magnetic braking \citep{Rappaort83}. Following \cite{Hurley02}, we model stellar winds as
\begin{equation}{}
\dot{J}_\mathrm{spin,wind} = \frac{2}{3} \,{} \dot{M}_\mathrm{wind}\,{}R^2,
\label{eq:jspinwind}
\end{equation} where $\dot{M}_\mathrm{wind}$ is the wind mass loss rate, 
and the magnetic braking as
\begin{equation} 
\dot{J}_\mathrm{spin,mb} =  -5.83 \times 10^{-16}\,{}  \frac{M_\mathrm{env}}{M}\,{} \left(\Omega_\mathrm{spin}\,{} R^3\right) \,{} \Msun\,{} {\rm R}_\odot^2 \,{}\mathrm{yr}^{-2}, 
\label{eq:jspinmb}
\end{equation} 
where $M_\mathrm{env}$ is the envelope mass of the star  (the magnetic braking is not active if the star has no core).
In a given time-step, the spin angular momentum is reduced by Eqs.~\ref{eq:jspinwind} and \ref{eq:jspinmb}. 
We impose that $J_{\rm spin}$ cannot become negative.

 After angular momentum, \sevn{} updates angular velocity 
 and spin. 
 If the spin is larger than one (over-critical rotation), the angular momentum is reset to the value for which $\Omega_\mathrm{spin}=\Omega_\mathrm{crit}$. In this work, we do not consider the enhancement of mass loss in stars close to the critical rotation, and we do not stop mass accretion on critically rotating stars. 

The stellar tracks used in this work have been calculated for non-rotating stars. Although inconsistent, this approach is necessary to include spin-dependent 
binary evolution processes (e.g.,  stellar tides,  Section \ref{sec:tides}).  Given the flexibility of \sevn{}, it will be easy to include rotating stellar tracks (e.g., \citealt{Nguyen22}) to investigate the effect of stellar rotation  on  stellar and binary evolution, and compact object formation (e.g., \citealt{Mapelli20,MarchantMoryia21}).

\subsection{Compact remnant formation} \label{sec:remform}

A compact remnant forms when the evolution time exceeds the stellar lifetime. Depending on the final mass of the CO core  ($M_\mathrm{CO,f}$) \sevn{} can trigger the formation of a white dwarf (WD, if the final CO mass is $M_\mathrm{CO,f}<1.38 \ \Msun$), the explosion of an  electron capture supernova (ECSN, $ 1.38 \ \Msun \leq M_\mathrm{CO,f} < 1.44 \ \Msun$) producing an NS (see \citealt{GiacobboEC}, and references therein), or a core-collapse supernova (CCSN, $M_\mathrm{CO,f} \geq 1.44 \ \Msun$) leaving a NS or a BH. 
 
When a WD is formed, its final mass and sub-type are set as follows. If the $M_\mathrm{ZAMS}$ of the current interpolating track is lower than the 
 He-flash threshold mass ($\approx 2 \ \Msun$, Eq.~2 in \citealt{Hurley00}), the WD is an  helium WD (HeWD) and its mass is equal to the final helium mass of the progenitor star, $M_\mathrm{He,f}$. Otherwise, the final mass of  the WD is equal to  $M_\mathrm{CO,f}$ and the compact remnant is a carbon-oxygen WD (COWD) if  $M_\mathrm{He,f}<1.6 \ \Msun$, an oxygen-neon WD (ONeWD)  otherwise (see Section~6 in \citealt{Hurley00}). 
 The radius and luminosity of the WD are set using Eqs.~90 and 91 of \cite{Hurley00} (setting the radius of the NS $R_\mathrm{NS}=11$~km). 
When an ECSN takes place  \citep[e.g.,][]{Kitaura06,Heuvel07}, the star leaves a NS (ECNS, see Table  \ref{tab:phases}).  The mass of the NS depends on the adopted supernova model.

\subsubsection{Core-collapse supernova} \label{sec:snmodel}
In this work, 
we use two core-collapse supernova models, based on the delayed and rapid model by \cite{fryer2012}. These two models differ only by the time at which the shock is revived: $<250$ ms and $>500$ ms for the rapid and delayed model, respectively.
The star directly collapses to a BH if the final carbon-oxygen core mass $M_\mathrm{CO,f} \geq 11 \ \Msun$ (both models), or if $6 \ \Msun \leq M_\mathrm{CO,f}<7 \ \Msun$  (rapid model only). In this case, the mass of the compact remnant  is equal to the pre-supernova mass of the progenitor, $M_\mathrm{f}$, apart from the neutrino mass loss (Section~\ref{sec:neutrino}). In the other cases, the core-collapse supernova explosion is successful and includes a certain amount of fallback. Thus, the final remnant mass depends  on  $M_\mathrm{CO,f}$ (which sets the fallback fraction) and $M_\mathrm{f}$ \citep{fryer2012}.  
Finally, the compact remnant is classified as NS (CCNS, Table \ref{tab:phases}) if the final mass is lower than 3 \Msun, BH otherwise. 

The only difference of our default model between our implementation of the rapid and delayed models and the original models presented by \cite{fryer2012}  consists in the mass function of NSs. In fact, the models by \cite{fryer2012}  fail to reproduce the mass distribution of Galactic BNSs \citep[e.g.,][]{GiacobboCOB,VG2018}. In absence of a successful astrophysical  model for NS masses, we decided to use a toy model as our default choice:  
    we draw the masses of all the NSs (born via ECSNe or CCSNe) from a Gaussian distribution centred at 1.33 \Msun{} with  standard deviation 0.09 \Msun. This model comes from a fit to the Galactic BNS masses \citep{Ozel12,Kiziltan13,oezel2016}.  We impose that the final compact remnant mass cannot be larger than the pre-SN mass of the progenitor star. Hence, the NS mass of an ultra-stripped ECSN is always lower than or equal to its pre-SN CO mass.  
    With this toy model, NSs with mass $>1.6$ M$_\odot$ are rare, which is critical to produce the primary masses of both GW170817 \citep{abbottGW170817} and GW190425 \citep{abbottGW190425}. We set the minimum NS mass to  1.1  \Msun. 
    \sevn{} also includes other core-collapse supernova models, which  are described in Appendix~\ref{app:snmodels}.

The default NS radius is set to $R_{\rm NS}=11$ km \citep{Capano20}, while the bolometric NS luminosity is set using Eq.~93 in \cite{Hurley00}. The BH radius is equal to the Schwarzschild radius, $R_\mathrm{BH}=R_\mathrm{S}=2\,{} G M_\mathrm{BH}/c^2$, where $c$ is the speed of light, while the BH luminosity is set to an arbitrary small value ($10^{-10} \ {\rm L}_\odot$, see Eqs.~95 and 96 in  \citealt{Hurley00}).

\subsubsection{Pair instability and pulsational pair instability} 
\label{sec:pisn}

 Massive stars ($M_{\rm He,f} \gtrsim 32 \ \Msun$, at the end of carbon burning) 
effectively produce    electron-positron pairs in their core. Pair creation lowers the central pressure and causes an hydro-dynamical instability leading to the contraction of the core and explosive ignition of oxygen or even silicon. This  triggers a number of pulses that enhance mass loss (pulsational pair instability, PPI, \citealt{woosley2007,yoshida2016,Woosley2017}). After the pulses, the star re-gains its hydro-static equilibrium and continues its evolution until the final iron core collapse \citep[e.g.,][and references therein]{Woosley2017,Woosley2019}. At even higher core masses ($64\lesssim{}M_{\rm He,f}/\Msun \lesssim 135$, at the end of carbon burning), a powerful single pulse destroys the whole star, leaving no compact remnant (pair instability supernova, PISN, \citealt{barkat1967,ober1983,bond1984,heger2003}). In very high-mass cores ($M_{\rm He,f}\gtrsim 135 \ \Msun$), pair instability triggers the direct collapse of the star.

The new version of \sevn{} includes two  models for PPIs and PISNe: M20 and F19.
M20 is the same model we implemented in the previous  version of \sevn{}  \citep{Mapelli20}. 
This model is based on the fit by \cite{Spera17} to the BH mass obtained with 1D hydrodynamical simulations by \cite{Woosley2017}. 
A star undergoes PPI if the pre-supernova He-core mass, $M_\mathrm{He,f}$, is within 32 and 64 \Msun, while a PISN is triggered for $64 \leq M_\mathrm{He,f}/\Msun \leq 135$. Above $M_\mathrm{He,f} = 135 \ \Msun$, the star directly collapses to a BH, leaving an intermediate-mass BH. 

PISNe leave no compact remnant, while the final mass of the compact remnant after PPI ($M_{\rm PPI}$) is obtained by applying a correction to the BH mass predicted by the adopted core-collapse supernova model ($M_{\rm CCSN}$, Section \ref{sec:snmodel}):

\begin{equation}
    M_\mathrm{PPI}=
    \begin{cases}
    
    \alpha_\mathrm{P} \,{}M_\mathrm{CCSN} & \mathrm{if}\,{}\alpha_\mathrm{P} \,{}M_\mathrm{CCSN}\geq{}4.5\,{}\mathrm{M}_\odot.\\
    0 & \mathrm{if}\,{}\alpha_\mathrm{P} \,{}M_\mathrm{CCSN}<4.5\,{}\mathrm{M}_\odot.
    \end{cases}
\label{eq:PPImass}
\end{equation}

The correction factor $\alpha_\mathrm{P}$ depends on $M_\mathrm{He,f}$ and the pre-supernova mass ratio between the mass of the He core and the total stellar mass  (see Eqs.~4 and 5 in the Appendix~of \citealt{Mapelli20}). 
The correction factor $\alpha_\mathrm{P}$ can take any values from 1 to 0 (a value of 0 corresponds to a  PISN). 
This definition of $\alpha_\mathrm{P}$ allows us to obtain the best fit to the models by \cite{Woosley2017}.  
If $(\alpha_\mathrm{P} \,{}M_\mathrm{CCSN})<4.5\,{}\mathrm{M}_\odot$, we assume that a PISN is triggered and set the mass of the compact remnant to zero. The limit at 4.5 \ \Msun \ is based on 
the least  massive BH formed in the simulations by \cite{Woosley2017}. 

The model F19 
is based on {\sc mesa} simulations of pure-He stars by \cite{farmer2019}. 
They found that the pre-supernova mass of the CO core, $M_\mathrm{CO,f}$, is a robust proxy for the activation of PISNe and PPIs. 
In this model, the star undergoes PPI if $38 \leq M_\mathrm{CO,f}/\Msun  \leq 60$, while the PISN regime begins at $M_\mathrm{CO,f}>60 \ \Msun$.   
The He-mass threshold at which pair instability leads to the direct collapse of a very massive star reported in \cite{farmer2020} is $M_\mathrm{He} \approx{130-135} \ \Msun$ for their fiducial value of the $^{12}{\rm C}(\alpha,\gamma)^{16}$O reaction rate, similar to \cite{Woosley2017}. Hence, we use a threshold $M_\mathrm{He,f}={135} \ \Msun$ for the transition between PISN and direct collapse, for both models F19 and M20.

In both models, we assume that a PISN explosion leaves no compact remnant.  The compact remnant mass in the PPI regime for the model F19  
is estimated as
\begin{equation}
M_\mathrm{PPI}= \min(M_\mathrm{f},M_\mathrm{F19}),
\end{equation}
where  $M_\mathrm{f}$ is the pre-supernova mass of the exploding star and $M_\mathrm{F19}$   is the mass of the BH according to  
Eq.~A1  of \cite{farmer2019}, and depends on $M_\mathrm{CO,f}$ and metallicity.
\cite{farmer2019} simulated only pure-He stars; therefore, here we are implicitly assuming that the first pulse completely removes any hydrogen layer still present in the star. This is a fair assumption, because the binding energy of the envelope in the late evolutionary stages ($\lesssim 10^{48}-10^{49}$~erg, Appendix \ref{app:ebind})  is  lower than the energy liberated during a pulse ($\gtrsim 10^{49}$ erg, e.g., \ \citealt{Woosley2017}).  In all our PPI/PISN models, if the correction for pair instability  produces a zero-mass compact remnant, the remnant is classified as Empty (Table \ref{tab:phases}).

\subsubsection{Neutrino mass loss}
\label{sec:neutrino}

Regardless of the supernova mechanism, the final  mass of the compact remnant  needs to be corrected  to account for neutrino mass loss. We apply the correction proposed by  \cite{Lattimer1989}, in the version discussed by \cite{zevin2020}:
\begin{equation}
M_\mathrm{rem} = \max{\left[\frac{\sqrt{1 + 0.3\,{}M_\mathrm{rem,\,{}bar}} -1 }{0.15},\,{}(M_\mathrm{rem,\,{}bar}-0.5\,{}\mathrm{M}_\odot)\right]},
\label{eq:nmloss}
\end{equation}
where $M_\mathrm{rem}$ and $M_\mathrm{rem,\,{}bar}$ are the gravitational and baryonic mass of the compact remnant, respectively. 

Note that this correction does not apply to the default model for NS masses. In our default model, NS masses are drawn from a Gaussian function that is already a fit to Galactic BNS masses \citep{oezel2016}, hence we do not need to further account for neutrino loss.

\subsubsection{Supernova kicks} \label{sec:snkicks}

After a supernova (ECSN, CCSN), the compact remnant receives a natal kick.
\sevn{} includes several formalisms for the natal kick, as described in Appendix~\ref{app:snkicks}.  In this work,  we use the three following models. 

In the first model (K$\sigma{}265$), the kick magnitude  $V_\mathrm{kick}$ is drawn from a Maxwellian curve with 1D root-mean-square (rms)  $\sigma_\mathrm{kick}$ and the kick direction is drawn from an isotropic distribution. We draw the kick 
assuming an arbitrary Cartesian frame of reference in which the compact remnant is at rest. The default 1D rms, $\sigma_\mathrm{kick}=265 \ \mathrm{km} \,{}\mathrm{s}^{-1}$, is based on the 
proper motions of young Galactic pulsars \citep{H05}. 
In the second model, we test the effect of reducing the kick dispersion by setting $\sigma_\mathrm{kick}=150 \ \mathrm{km} \,{}\mathrm{s}^{-1}$ \citep[K$\sigma{}150$, e.g.,][see Section \protect\ref{sec:models}]{atri2019,broekgaarden2021}. 

In the third model (KGM20), the kick magnitude is  estimated as  
\begin{equation}
V_{\rm kick}=f_{\rm H05}\,{}\frac{\langle{}M_{\rm NS}\rangle{}}{M_{\rm rem}}\,{}\frac{M_{\rm ej}}{\langle{}M_{\rm ej}\rangle},
\end{equation}    
where $f_{\rm H05}$ is a random number drawn from a Maxwellian distribution with  $\sigma_\mathrm{kick}=265 \ \mathrm{km}\,{} \mathrm{s}^{-1}$;  
$\langle{}M_{\rm NS}\rangle{}$ and $\langle{}M_{\rm ej}\rangle$ are the average NS mass and ejecta mass from single stellar evolution, respectively, while
$M_{\rm rem}$ and $M_{\rm ej}$ are the compact object mass and the ejecta mass \citep{GiacobboKick}.  We calibrate the values of $\langle M_\mathrm{ej} \rangle$  using single stellar \sevn{} simulations at $Z=0.02$ and assuming a Kroupa initial mass function (Section~\ref{sec:ic}). 
In this model, ECSNe and stripped (pure-He  pre-supernova stars)/ultra-stripped (naked-CO pre-supernova stars) supernovae  naturally result in smaller kicks with 
respect to non-stripped CCSNe, due to the lower amount of ejected mass \citep{tauris2015,tauris2017}. 
BHs originating from a direct collapse receive zero natal kicks from this mechanism.

In a binary system, natal kicks change the orbital properties, the relative orbital velocity and  the centre of mass of the binary  as described in Appendix~A1 of \cite{Hurley02}.
After the kick, we update the orbital properties of the binary  considering the new relative orbital velocity and the new total mass in the binary. If the semi-major axis is smaller than 0 and/or the eccentricity larger than 1, the binary does not survive the kick.
The centre-of-mass velocity and the orbital properties of the binary system change even without natal kicks (i.e., after WD formation or direct collapse) because of the  mass lost by  the system at the formation of the compact remnant \citep[the so-called Blaauw kick, ][]{blaauw1961}.

\subsection{Binary evolution} \label{sec:bse}

\sevn{} includes the following binary evolution processes: wind mass transfer, Roche-lobe overflow (RLO), common envelope (CE), stellar tides,  
circularisation at the RLO onset, collision at periastron, orbit decay by GW emission, and stellar mergers.
In the next sections, we describe the  formalism used in this work. 

\subsubsection{Wind mass transfer} \label{sec:winds}

\sevn{} assumes that the stellar tracks stored in the tables already include wind mass loss, therefore  wind mass loss is taken into account self-consistently in single stellar evolution. 
In \sevn{}, we also take into account the possibility that some mass and angular momentum lost from a star (the donor) can be accreted by the stellar companion (the accretor). We follow the implementation by \cite{Hurley02}, 
in which the orbit-averaged accretion rate is estimated according to the \cite{Bondi44} mechanism and fast wind approximation (wind velocity larger than orbital velocity). Under such assumptions, the mass accretion rate $\dot{M}_{\rm a}$ is
\begin{equation}
    \dot{M}_\mathrm{a}= -\frac{\alpha_\mathrm{wind}}{\sqrt{1-e^2}} \left( \frac{G M_\mathrm{a}}{V^2_\mathrm{wind}} \right)^2 \frac{ \dot{M}_\mathrm{d} }{2a^2\left( 1 + V^2_\mathrm{f} \right)^{3/2}},
    \label{eq:dMwind}
\end{equation}
where $\dot{M}_\mathrm{d}$ is the wind mass loss rate of the donor star, $a$ the semi-major axis of the binary system,  
\begin{equation}
    V^2_\mathrm{wind} = 2\,{} \beta_\mathrm{wind} \,{}\frac{G\,{} M_\mathrm{d}}{R_\mathrm{eff}}
    \label{eq:Vwind}
\end{equation} is the wind velocity, 
$V^2_\mathrm{f}=G \,{}(M_\mathrm{d}+M_\mathrm{a})\,{} a^{-1} \,{} V^{-2}_\mathrm{wind}$ is the ratio between the characteristic orbital velocity and the wind velocity, and $R_\mathrm{eff}$  is the stellar effective radius, i.e. the minimum between the radius of the star and its Roche lobe (RL) radius (see Section~\ref{sec:rlo}). In the aforementioned equations, $M_{\rm d}$ and $M_{\rm a}$ are the mass of the donor and accretor, respectively. 
In this work, we set the two dimensionless wind parameters $\alpha_{\rm wind}$ and $\beta_{\rm wind}$ to their default values: 
$\alpha_\mathrm{wind}=1.5$, appropriate for Bondi-Hoyle accretion \citep{Hurley02}, and  $\beta_\mathrm{wind}=0.125$, based on observations of cool super-giant stars \citep{Kucinskas98,Hurley02}. 

In eccentric orbits, Eq.~\ref{eq:dMwind} can predict an amount of accreted mass larger than the actual wind mass loss from the donor. Following \cite{Hurley02}, we set $0.8 |\dot{M}_\mathrm{d,wind}|$ as an upper limit for wind mass accretion.

If the accretor is a compact object (BH, NS, or WD), the mass accretion rate is limited by the Eddington limit 
\begin{equation}
    \dot{M}_\mathrm{Edd} =  2.08 \times 10^{-3} \,{}{\rm M}_\odot\,{}{\rm yr}^{-1}\,{}\eta_\mathrm{Edd}\,{} (1+X)^{-1} \,{}\frac{R_\mathrm{a}}{\mathrm{R_\odot}},
    \label{eq:medd}
\end{equation}
where $R_\mathrm{a}$ is the radius of the accretor (in this case, the compact object), and $X=0.760 -3.0\,{} Z$ is the hydrogen mass fraction of the accreted material.  
In this work, we set $\eta_{\rm Edd}=1.0$, enforcing the Eddington limit (see, e.g., \citealt{Briel22} for a study of super-Eddington accretion). 
 Following \cite{Spera19}, we assume that pure-He and naked-CO stars do not accrete any mass  since the winds of these stars are expected to eject a thin envelope on a very short time scale.

The accreted mass brings additional angular momentum to the accretor increasing its spin: 
\begin{equation} 
    \dot{J}_\mathrm{accreted} = \frac{2}{3} R^2_\mathrm{eff}\,{} \dot{M}_\mathrm{a}\,{} \Omega_\mathrm{spin,d},
    \label{eq:dJwind}
\end{equation}
where $\Omega_\mathrm{spin,d}$ is the angular velocity of the donor star. Eq.~\ref{eq:dJwind} is derived assuming that the winds remove a thin shell of matter from the donor star (see Section \ref{sec:spin}). 

Mass exchange by stellar winds causes  a variation of the orbital angular momentum;  the orbital parameters change accordingly \citep{Hurley02}:
\begin{equation}
    \frac{\dot{a}}{a} = -\frac{\dot{M}_\mathrm{d}}{M_\mathrm{a} + M_\mathrm{d}} - \left( \frac{2-e^2}{M_\mathrm{a}} + \frac{1+e^2}{M_\mathrm{a} + M_\mathrm{d}} \right) \frac{\dot{M}_\mathrm{a}}{1-e^2}
    \label{eq:dawind}
\end{equation}
and
\begin{equation}
    \frac{\dot{e}}{e} = -\dot{M}_\mathrm{a} \left[ (M_\mathrm{a} + M_\mathrm{d})^{-1} + 0.5 M_\mathrm{a}^{-1} \right].
    \label{eq:dewind}
\end{equation}
The wind mass loss produces a widening of the orbit; however, the mass accreted onto the companion star mitigates the magnitude of this effect, returning some of the lost angular momentum back to the system (Eq.~\ref{eq:dawind}). In addition, the wind mass accretion reduces the eccentricity, circularising the orbit (Eq.~\ref{eq:dewind}).
These eccentricity variations are negligible compared to those caused by stellar tides (Section \ref{sec:tides}), even during the most intense phases of wind mass loss \citep{Hurley02}.

 \subsubsection{Roche-lobe overflow} \label{sec:rlo}

Assuming circular and synchronous orbits, 
\cite{Eggleton83} derived an 
approximation for the Roche lobe (RL) radius:
\begin{equation}
    R_\mathrm{L}= a \frac{0.49 q^{2/3}}{0.6q^{2/3}
    + \ln \left( 1 + q^{1/3} \right) }
    \label{eq:eqrlo},
\end{equation}
where $q$ is the mass ratio  between the star and its companion. 

In \sevn{}, a Roche lobe overflow (RLO) begins whenever the radius of one of the two stars becomes 
equal to (or larger than) $R_\mathrm{L}$,  
and stops when this  condition is not satisfied anymore, or if the mass transfer leads to a merger or a CE. 
\sevn{} checks for this condition at every time-step. The RLO implementation used in this work is based on \cite{Hurley02}, \cite{Spera19} and \cite{Bouffanais21}.  
\sevn{} makes use of the \bse{} stellar types (Table \ref{tab:phases}) for the implementation of RLO, mass transfer stability, and CE. 
\newline

\noindent
\textit{Stability criterion}

\begin{table}
\begin{center}
\scalebox{1.0}{
\begin{tabular}{lccc}
\cline{2-4}
\multicolumn{1}{l|}{} & \multicolumn{3}{c|}{\sevn{} $q_\mathrm{c}$ option} \\ \hline
\multicolumn{1}{c}{\bse{} type of the donor star} &
  QCBSE &
  QCRS &
  QCBB  \\ \hline
0 (low mass MS)       & 0.695    & 0.695    & 0.695         \\
1 (MS)                & 3.0      & stable   & stable           \\
2 (HG)                & 4.0      & stable   & stable       \\
3/5 (GB/EAGB)        & Eq. \ref{eq:qchw} & Eq. \ref{eq:qchw}  & Eq. \ref{eq:qchw}  \\
4 (CHeB)              & 3.0      & 3.0      & 3.0          \\
7 (HeMS)                & 3.0      & 3.0      & stable            \\
8 (HeHG)             & 0.784    & 0.784    & stable         \\
\textgreater{}10 (WD) & 0.628    & 0.628    & 0.628        \\ \hline
\end{tabular}
}
\caption{Critical mass ratios  as a function of the donor \bse{} stellar type for different \sevn{} options.
 See Table  \ref{tab:phases}  for the further details \bse{} types and their correspondance to \sevn{} phases. 
 The word stable indicates that the mass transfer is always stable.}
\label{tab:qc}
\end{center}
\end{table}

The RLO  changes the mass ratio, the masses and  semi-major axis of the binary system. As a consequence, the RL shrinks or expands (Eq.~\ref{eq:eqrlo}).
If the RL shrinks faster than the donor's radius (or if the RL expands more slowly than the donor's radius) because of the adiabatic response of the star to mass loss, the mass transfer becomes unstable on a dynamical timescale, leading to a stellar merger or a CE configuration. 

The stability of mass transfer can be evaluated by comparing 
the (adiabatic or thermal) response of the donor to mass loss, as expressed by $\zeta= \frac{d \log R}{d \log M}$, 
 to the variation of the RL, $\zeta_{\rm L}= \frac{d \log R_{\rm L}}{d \log M}$ \citep{Webbink85}. 
Stars with radiative envelopes tend to shrink in response to mass loss, while deep convective envelopes tends to maintain the same radius or slightly expand \citep[e.g.,][]{ge2010,ge2015,ge2020a,ge2020b,Klencki21,Temmink2022}. 
In practice, population synthesis codes usually implement a simplified formalism in which the mass transfer stability is evaluated by comparing the mass ratio $q=M_\mathrm{d}/M_\mathrm{a}$ (where $M_{\rm d}$ and $M_{\rm a}$ are the mass of the donor and accretor star, respectively), with some critical value $q_\mathrm{c}$. 
If the mass ratio is larger than $q_\mathrm{c}$, the mass transfer is considered unstable on a dynamical time scale. 
The critical mass ratio is usually assumed to be large ($>2$) for stars with radiative envelopes (e.g., MS stars, stars in the Hertzsprung-gap phase, and pure-He stars), while it is smaller for 
stars with deep convective envelopes \citep[but see][for a significantly different result]{ge2020a,ge2020b}.

In this work, we use three stability options in which
the critical mass ratio depends on the stellar type of the donor: QCBSE, QCRS, and QCBB (Table \ref{tab:phases}). 
The corresponding $q_\mathrm{c}$ values are summarised in Table \ref{tab:qc}. 
 The option QCBSE is the same as the stability criterion used in \bse{} \citep{Hurley02}, \mobse{} \citep{GiacobboCOB,GiacobboEC,GiacobboKick} and \cite{Spera19} (see their Appendix~C2).
In particular for giant stars with deep convective envelopes (\bse{} phases 3, 5),
\begin{equation}
q_\mathrm{c} = 0.362 + \frac{1}{3\left(1-\frac{M_\mathrm{He,d}}{M_\mathrm{d}}\right)},
\label{eq:qchw}
\end{equation}
where $M_\mathrm{He,d}$ is the core helium mass of the donor star.
Eq.~\ref{eq:qchw}  is based on models of condensed polytropes \citep{Webbink88}  and is widely used in population synthesis codes (e.g. \bse{}, \mobse{}).

Our fiducial option QCRS  uses the same $q_\mathrm{c}$ as \cite{Hurley02}, but  mass transfer is assumed to always be stable for donor stars with  radiative envelopes, i.e., stars in the  MS or Hertzsprung-gap (HG) phase (\bse{} phases 1 and 2).

The option QCBB assumes that not only MS and HG donor stars (\bse{} phases 1 and 2), but also  donor pure-He stars (\bse{} phases 7, 8) always undergo stable mass transfer (\citealt{VG2018} used a similar assumption for pure-He stars).   
These differences with respect to the 
QBSE formalism mainly spring from the stellar evolution models used in this work, and will be discussed in Section~\ref{sec:discsse}. 

Additional stability criteria  implemented in \sevn{} are described in  Appendix~\ref{app:rlostab} and summarised in Table \ref{tab:qcother}.
 In addition to the aforementioned mass transfer stability criterion, \sevn{} considers some special cases.  If the RL is smaller than the core radius of the donor star (He core in hydrogen stars and CO core for pure-He stars), the mass transfer is always considered unstable, ignoring the chosen stability criterion. If both the  donor and accretor are helium-rich WDs ({\sc bse} type 10) and the mass transfer is unstable, the accretor explodes as a SNIa, leaving a mass-less remnant.  
In all the other unstable mass transfer cases in  WD binaries, 
the donor is completely swallowed leaving a mass-less compact remnant and no mass is accreted onto the companion. 
If both stars have radius $R\ge{}R_\mathrm{L}$, we assume that the evolution leads either to a CE (when at least one of the two stars has a clear core-envelope separation, corresponding to {\sc bse}  phases 3, 4, 5, 8), or to a stellar merger (for all the other {\sc bse} phases). 
If the object filling the RL is a BH or a NS, the companion must also be a BH or NS. In this case, the system undergoes a compact binary coalescence.

~\newline
\noindent
\textit{Stable Mass transfer} 

In the new version of \sevn{}, we describe the stable mass transfer with a slightly modified formalism with respect to both  \cite{Hurley02} and \cite{Spera19}. Here below, we describe the main differences. 
The mass loss rate depends on how much the donor overfills the RL \citep{Hurley02}:
\begin{equation}
 \dot{M}_\mathrm{d} = -F(M_\mathrm{d})\,{} \left( \ln \frac{R_\mathrm{d}}{R_\mathrm{L}} \right)^3
 \ \Msun \,{}\mathrm{yr}^{-1},
 \label{eq:smt}
\end{equation} 
and the normalisation factor is \footnote{
In \cite{Hurley02} the extra factor for  HG stars is not included and the one for WDs does not include the mass of the donor. However, both are included in the most-updated version of {\sc bse} and {\sc mobse}.}
\begin{multline}
F \left( M_\mathrm{d} \right) =
3 \times 10^{-6} \left( \min \left[ M_\mathrm{d}, M_\mathrm{max,SMT} \right] \right)^2  \times \\
\begin{cases}
\max \left[ \frac{M_\mathrm{env,d}}{M_\mathrm{d}} ,0.01 \right]\quad{} &\text{for HG phase donors (\bse{} phase 2)} \\
10^3 M_\mathrm{d} \left( \       \max \left[ R_\mathrm{d}, 10^{-4} \right] \right)^{-1}\quad{}&\text{for WD donors} \\
1\quad{} &\text{all other cases},
\end{cases}
\end{multline}
where all the quantities are in solar units. 
In this work,  $M_\mathrm{max,SMT}=5 \ \Msun$, as originally reported in \cite{Hurley02}. 
For giant-like stars (i.e., all the stars that developed a core/envelope structure), we limit the mass transfer to the thermal rate (Eq.~60 in \citealt{Hurley02}), while for all the other stellar types (MS stars and WR stars without a CO core) the limit is set by the dynamical rate (Eq.~62 in \citealt{Hurley02}).

The mass accretion rate $\dot{M}_{\rm a}$ 
is simply parameterised as 
\begin{equation}
\dot{M}_\mathrm{a} = \left\{
 \begin{array}{ll}
 \min{(\dot{M}_{\rm Edd},\,{}-f_\mathrm{MT}\,{}\dot{M}_\mathrm{d})} &\mathrm{if}\,{}{\rm the}\,{} \mathrm{accretor}\,{} \mathrm{is}\,{} \mathrm{a}\,{} \mathrm{compact}\,{} \mathrm{object} \\
  - f_\mathrm{MT}\,{} \dot{M}_\mathrm{d}&\mathrm{otherwise}, 
  \end{array}
  \right.
    \label{eq:maccrlo}
\end{equation}
where $\dot{M}_{\rm Edd}$ is the Eddington rate (Eq.~\ref{eq:medd})  and $f_{\rm MT}\in[0,1]$ is the mass accretion efficiency; here, we use $f_{\rm MT}=0.5$. 
Eq.~\ref{eq:maccrlo} contains an important difference with respect to \cite{Hurley02} and \cite{Spera19}: both authors assume that the accretion efficiency depends on the thermal timescale  of the accretor,  thus it can vary from star to star (Eq.~\ref{eq:mtalternative}). 
The advantage of using the simplified approach in Eq.~\ref{eq:maccrlo} is that the parameter $f_\mathrm{MT}$ has a straightforward physical meaning and can be included in parameter exploration  \citep[see, e.g.,][]{Bouffanais21}.

If the accretor is a compact object (WD, NS, or BH), 
 we enforce the Eddington limit (Eq.~\ref{eq:medd}).
Also, we assume that pure-He and naked-CO stars do not accrete  mass during a RLO (Section \ref{sec:winds}). 
Finally, if the accretor is a WD and the accreted material is hydrogen-dominated (e.g., the donor star is not a WR star), a nova explosion is triggered and the actual accreted mass is  reduced by multiplying it for a factor $\epsilon_\mathrm{nova}=0.001$.

We also test another formalism analogous to the treatment of RLO by  \cite{Hurley02} (Section \ref{sec:models}): for stars in the \bse{} phases 1, 2, and 4, Eq.~\ref{eq:maccrlo} is replaced by 
\begin{equation}
 \dot{M}_\mathrm{a}=  - \min \left(1.0, 10 \frac{\tau_\mathrm{\dot{M}}}{\tau_\mathrm{KH,a}} \right) \dot{M}_\mathrm{d}, \ \mathrm{where}
 \ \tau_\mathrm{\dot{M}}=\frac{M_\mathrm{a}}{|\dot{M}_\mathrm{d}|} \\ 
\label{eq:mtalternative}
\end{equation} and $\tau_\mathrm{KH,a}$ is the thermal timescale of the accretor (Eq.~61 in \citealt{Hurley02}). For \bse{} stellar types 3 and 5, this model assumes that the accretor can absorb any transferred material ($f_\mathrm{MT}=1$ in Eq.~\ref{eq:maccrlo}). 
In addition, in a pure-He-pure-He binary, the stars are allowed to accrete mass during RLO  following the prescription in Eq.~\ref{eq:mtalternative}.

~\newline
\noindent
\textit{Orbital variations}

During a non-conservative mass transfer ($f_\mathrm{MT}\neq1$), some angular momentum is lost from the system. We parametrise the angular momentum loss as
\begin{equation}
    \Delta J_\mathrm{orb,lost} = -|\Delta M_\mathrm{loss}|\,{} \gamma_\mathrm{RLO}\,{} a^2 \,{}\sqrt{1-e^2}\,{} \frac{2 \pi}{P},
    \label{eq:djorbrlo}
\end{equation}
where $P$ is the orbital period and $\Delta M_\mathrm{loss}$ is the actual mass lost from the system in a given evolution step, i.e. the difference between the mass lost by the donor and that accreted on the companion. In all our simulations, we assume that mass which is not accreted is isotropically lost from the donor, 
so that $\gamma_\mathrm{RLO}= M^2_\mathrm{d} /(M_{\rm a}+M_{\rm d})^{2}$. 
See Appendix~\ref{app:rloloss} for other available options. 

Apart from the mass lost from the system, we assume that the total binary angular momentum (stellar spins plus orbital angular momentum) is conserved during RLO. Therefore, the spin angular momentum lost by the donor is added to the orbital angular momentum 
\begin{equation}
\Delta J_\mathrm{orb,d} = -\Delta J_\mathrm{spin,d} = -\Delta M_\mathrm{d}\,{} R^2_\mathrm{L}\,{} \Omega_\mathrm{spin,d},
\label{eq:djorbrlodonor}
\end{equation}
where $\Delta M_\mathrm{d}$ is the mass lost by the donor in an evolutionary step and $\Omega_\mathrm{spin,d}$ is the donor angular velocity. 
In contrast, the mass accreted onto the companion removes some orbital angular momentum  and increases the accretor spin:
\begin{equation}
\Delta J_\mathrm{orb,a} = -\Delta J_\mathrm{spin,a} = -\Delta M_\mathrm{a} \,{}\sqrt{G \,{}M_\mathrm{a}\,{} R_\mathrm{acc}}.
\label{eq:djorbrlodisc}
\end{equation}
 The accretion radius, $R_\mathrm{acc}$ is estimated following \cite{Lublow75} and \cite{Ulrich76}. 
The minimum radial distance of the mass stream to the secondary is estimated as \citep{Lublow75}
\begin{equation}
R_\mathrm{min} = 0.0425 \left( q^{-1}+q^{-2} \right)^{0.25}a.
\end{equation}
If $R_\mathrm{min}>R_\mathrm{a}$ (where $R_{\rm a}$ is the radius of the accretor), we assume that the mass is accreted from the inner edge of an accretion disc and $R_\mathrm{acc}=R_\mathrm{a}$. Otherwise, the accretion disc is not formed and the material from the donor hits the accretor in a direct stream. In the latter case, the angular momentum of the transferred material is estimated using the radius at which the disc would have formed if allowed, i.e. $R_\mathrm{acc}=1.7 \,{}R_\mathrm{min}$ \citep{Ulrich76}.

Finally, the variation on the semi-major axis due to the RLO is estimated as 
\begin{equation}
\Delta a = \frac{ (J_\mathrm{orb} + \Delta J_\mathrm{orb,lost} + \Delta J_\mathrm{orb,d} + \Delta J_\mathrm{orb,a})^2\,{} (M_\mathrm{a}+M_\mathrm{d})}{G \,{}(1-e^2)\,{} M^2_\mathrm{d}\,{} {} M^2_\mathrm{a}\,{}   }  - a,
\end{equation}
where the 
masses are considered after the mass exchange 
in the current time-step. 
Accordingly, the stellar spins variations are updated considering Eqs.~\ref{eq:djorbrlodonor}  and  \ref{eq:djorbrlodisc}.

~\newline
\noindent
\textit{Unstable mass transfer}

The outcome of an unstable mass transfer depends on the donor stellar type. 
During an unstable mass transfer, giant like-stars (\bse{} types 3, 4, 5, 8)  undergo a CE evolution (Section \ref{sec:CE}), while stars without a clear envelope/core separation (\bse{} types 0, 1, 7) 
directly merge with their companion  (Section \ref{sec:merger}).
The stars in the HG phase (\bse{} type 2) are peculiar objects in which the  differentiation between  He core and H envelope has not fully developed yet \citep{Ivanova04,dominik2012}. 
It is unclear whether an unstable mass transfer with a HG donor should lead to a CE evolution (optimistic scenario in \citealt{dominik2012}, see also \citealt{VG2018}) or to a  direct merger (pessimistic scenario in \citealt{dominik2012}, see also \citealt{GiacobboCOB}).
In this work, we adopt the pessimistic scenario as default, but we also test the optimistic assumption.

~\newline
\noindent
\textit{Quasi-Homogeneous evolution}

We also test the impact of the quasi-homogeneous evolution (QHE) scenario on the properties of binary compact objects (Section \ref{sec:models}).
In the QHE scenario, a star acquires a significant spin rate due to the accretion of material during a  stable RLO mass transfer. As a consequence, the star remains fully mixed during the MS, burning all the hydrogen into helium \citep{Petrovic05,Cantiello07}. 
\sevn{} implements the QHE  as described in \cite{eldridge11} ad \cite{eldridge12}. 
If this option is enabled, \sevn{} activates the QHE evolution for metal poor ($Z\leq0.004$) MS stars that accrete at least 5\% of their  initial mass through  stable RLO mass transfer and reach a post-accretion mass of at least 10~\Msun{}. 
When a star fulfills the QHE condition,  the  evolution of the radius is frozen. Then, at the end of the MS, the star 
 becomes a pure-He  star 
and the evolutionary phase jumps directly to phase 4 (core He burning, see Table \ref{tab:phases}). 

\subsubsection{Common envelope (CE) evolution} \label{sec:CE}  

The CE phase is a peculiar evolutionary stage of a binary system in which the binary is embedded in the expanded envelope of one or both binary components. The loss of corotation between the binary orbit and the envelope produces drag forces that shrink the orbit, while the CE gains energy and expands  \citep[][and reference therein]{IvanovaCE}. 
The CE evolution described in this section is based on the so-called energy formalism \citep{vandeHeuvel76,Webbink84,Livio88,Iben93}   as described in \cite{Hurley02}.  
This formalism is based on the comparison between the energy needed to unbind the stellar envelope(s) and the orbital energy before and after the CE event. 
The evaluation of the two energy terms depends on two parameters: $\lambda_\mathrm{CE}$ and $\alpha_\mathrm{CE}$.  
The first parameter, $\lambda_\mathrm{CE}$, is a structural parameter that defines the binding energy of the stellar  envelope \citep{Hurley02}, therefore the binding energy of the CE is
\begin{equation}
E_{\rm bind,i}=-G\,{}\left(\frac{M_1\,{}M_{\rm env1}}{\lambda_\mathrm{CE1} R_1}+\frac{M_2\,{}M_{\rm env2}}{\lambda_\mathrm{CE2} R_2}\,{}\right),
\label{eq:Ebind}
\end{equation}
where $M_1$ ($M_2$) is the mass of the primary (secondary) star,  $M_{\rm env1}$ ($M_{\rm env2}$) is the mass of the envelope of the primary (secondary) star, $R_1$ ($R_2$) is the radius of the primary (secondary) star. If the accretor is a compact object or a star without envelope, we set $M_{\rm env2}=0$. If both stars have an envelope, they both lose it when the CE is ejected \citep{Hurley02}. 

In our fiducial model we use the  same formalism for $\lambda_\mathrm{CE}$ as used in \bse{} and described in \cite{Claeys14}\footnote{\cite{Hurley02} assume a constant $\lambda_\mathrm{CE}=0.5$ for all stars (see their Eq.~69). However, in the most updated public version of \bse{}, $\lambda_\mathrm{CE}$ depends on the stellar properties and is estimated following \cite{Claeys14} (see Appendix~\ref{app:lambda} for further details). Eq.~\ref{eq:Ebind} is currently used also in \bse{} and \mobse{}.}. According to this formalism,  $\lambda_\mathrm{CE}$ depends on the mass of the star, its evolutionary phase, the mass of the convective envelope and its radius. Since \cite{Claeys14} do not report a fit for pure-He stars, for such stars we use a constant value of $\lambda_\mathrm{CE}=0.5$. 
In this work, we also  test the $\lambda_\mathrm{CE}$ formalism by \cite{xu2010}, the one  by \cite{Klencki21}, and the constant value $\lambda_\mathrm{CE}=0.1$ as in \cite{Spera19}. More details on the choice of  $\lambda_\mathrm{CE}$ can be found in  Appendix~\ref{app:lambda}.

The parameter $\alpha_\mathrm{CE}$ represents the fraction of orbital energy converted into kinetic energy of the envelope during CE evolution.  
The orbital energy variation during CE is 
\begin{equation}
\Delta E_\mathrm{orb} =   \frac{G M_\mathrm{c,1} M_\mathrm{c,2}}{2} \left( a^{-1}_\mathrm{f} - a^{-1}_\mathrm{i} \right),
\label{eq:eorb}
\end{equation}
where $M_\mathrm{c,1}$ and $M_\mathrm{c,2}$  are the masses of the cores of the two stars, and $a_\mathrm{f}$ ($a_\mathrm{i}$) is the semi-major axis after (before) the CE phase. 
Adopting the same formalism as in \bse{}, we set $E_\mathrm{bind}=0$  and $M_\mathrm{c}=M$ for 
MS stars, pure-He stars without a CO core, naked-CO stars, and compact remnants. 
We thus derive the post-CE separation by imposing $E_\mathrm{bind,i}=\alpha_\mathrm{CE}\,{}\Delta E_\mathrm{orb}$.  
If neither of the stars fills its RL in the post-CE configuration, 
we assume the CE is ejected. Otherwise, the two stars coalesce (Section \ref{sec:merger}). 

Here, we follow the same formalism as \cite{Hurley02}, in which both stars lose their envelope (if they have one) during CE evolution. This assumption is still controversial:  the envelope of the donor star loses co-rotation and then needs to be ejected to allow the survival of the binary system, but the fate of the envelope of the companion star is more uncertain, especially if the companion star is much less evolved than the donor star \citep{ivanova2013}. We will revise this assumption in future work.

For $\alpha_{\rm CE}$, we will adopt values ranging from 0.5 to 5. Values of $\alpha_{\rm CE}>1$ are at odds with the original definition of this parameter.  We consider values of $\alpha_{\rm CE}>1$ to account for the fact that the orbital energy variation is not the only source of energy that contributes to unbind the envelope \citep[e.g.,][and references therein]{Roepke2022}.

\subsubsection{Tides} \label{sec:tides}

Tidal forces between two stars in a binary system tend to synchronise the stellar  and orbital rotation, and circularise the orbit \citep[e.g.,][]{Hut81,Meibom,Justensen21}.
In \sevn{}, we account for the effect of tides on the orbit and stellar rotation following the weak friction analytic models by \cite{Hut81}, as implemented in \cite{Hurley02}. 
The model is based on the spin-orbit coupling caused by the misalignment of the tidal bulges in a star and the perturbing potential generated by the companion. The secular average equations implemented in \sevn{} are:
\begin{multline}
     \dot{a} = -6\,{} k_\mathrm{tides} q \,{}(q+1) \left( \frac{R_\mathrm{eff}}{a} \right)^8 \frac{a}{(1-e^2)^{7.5}} \times \\ \times  \left[ f_1 -(1-e^2)^{2/3}f_1\,{}  \frac{\Omega_\mathrm{spin}}{\Omega_\mathrm{orb}} \right],
     \label{eq:aspin}
\end{multline}

\begin{multline}
     \dot{e} = -27\,{} k_\mathrm{tides} q\,{} (q+1) \left( \frac{R_\mathrm{eff}}{a} \right)^8 \frac{e}{(1-e^2)^{6.5}} \times \\ \times  \left[ f_3 -\frac{11}{18}(1-e^2)^{2/3}f_4\,{}  \frac{\Omega_\mathrm{spin}}{\Omega_\mathrm{orb}} \right],
     \label{eq:espin}
\end{multline}

\begin{multline}
    \dot{J}_\mathrm{spin} = 3 k_\mathrm{tides} \,{}q^2\,{}M\,{}R^2\,{}  \left( \frac{R_\mathrm{eff}}{a} \right)^6  \left(\frac{R_\mathrm{eff}}{R} \right)^2 \frac{\Omega_\mathrm{orb}}{(1-e^2)^{6}} \times \\ \times  \left[ f_2 -(1-e^2)^{2/3}f_5\,{}  \frac{\Omega_\mathrm{spin}}{\Omega_\mathrm{orb}} \right],
    \label{eq:ospin}
\end{multline}
where $q$ is the mass ratio between the perturbing star and the star affected by tides, $\Omega_\mathrm{spin}$ is the stellar angular velocity (see Sec.\ \ref{sec:spin}), $R$ is the stellar radius and
    $R_\mathrm{eff}= \min \left[ R_\mathrm{L}, R \right]$
is the effective radius, i.e. the minimum between the stellar radius and its RL radius (Eq.~\ref{eq:eqrlo}). The effective radius has been introduced to take into account that, during a stable RL mass transfer, the actual radius of the star remain close to its  RL (Section \ref{sec:rlo}). In all the other cases, the effective radius is coincident with the stellar radius. Eqs.~\ref{eq:aspin}--\ref{eq:ospin} have been obtained under the assumption that $R<a$ \citep{Hut81}.   The effective radius ensures this condition since the  (circular) RL is, by definition, always smaller than the semi-major axis (see Sec.\ \ref{sec:rlo}). The factor $R^2_\mathrm{eff} R^{-2}$ in Eq.~\ref{eq:ospin} is a re-scaling factor for the stellar inertia $I$ ($J_\mathrm{spin}=\Omega_\mathrm{spin} I$ and $I \propto R^2$).

In Eqs.~\ref{eq:aspin}, \ref{eq:espin} and \ref{eq:ospin}, $f_1$, $f_2$, $f_3$, $f_4$ and $f_5$ are  polynomial functions of $e^2$, given by \cite{Hut81}. The $k_\mathrm{tides}$ term is the inverse of the timescale of tidal evolution. It is estimated following \cite{zahn75,zahn77} and \cite{Hurley02}\footnote{Eq.~42 in \cite{Hurley02} contains a typo: the ratio $R^2a^{-5}$ should be $R\,{}a^{-2.5}$. The typo 
is explicitly reported and fixed in the \bse{} code documentation in the file evolved2.f.} for radiative envelopes, i.e., 
\begin{equation}
    k_\mathrm{tides} = 3.156\times10^{-5} \left( \frac{M}{\Msun} \right)^{3.34} \left( \frac{R}{\mathrm{R}_\odot} \right) \left( \frac{a}{\mathrm{R}_\odot} \right)^{-2.5} \ \mathrm{yr}^{-1},
    \label{eq:krad}
\end{equation}
and \cite{zahn77}, \cite{rasio96}, and \cite{Hurley02} for convective envelopes:
\begin{equation}
    k_\mathrm{tides} = \frac{2}{21} \left( \frac{\tau_\mathrm{conv}}{\mathrm{yr}} \right)^{-1} \frac{M_\mathrm{conv}}{ M} \min\left\{ 1, \left( \frac{ \pi}{(\Omega_\mathrm{orb}-\Omega_\mathrm{spin}) \tau_\mathrm{conv}} \right)^2 \right\} \ \mathrm{yr}^{-1},
    \label{eq:kconv}
\end{equation}
where $M_\mathrm{conv}$ is the mass of the convective envelope, $\tau_\mathrm{conv}$ is the eddy turnover timescale, i.e. the turnover time of the largest convective cells. 
In this work, the values of $M_\mathrm{conv}$ and $\tau_\mathrm{conv}$ are directly interpolated from the tables (see Section \ref{sec:tables} and Appendix~\ref{app:tables}). The amount of variation of $a$, $e$ and $J_\mathrm{spin}$ is estimated by multiplying Eqs.~\ref{eq:aspin}--\ref{eq:ospin} by the current time-step and adding together the effects of the two stars in the system. We assume that compact remnants (WDs, BHs, NSs) and naked-CO stars (stars stripped of both their hydrogen and helium envelopes) are not affected by tides and  act just as a source of perturbation for the companion star. 

There exists a peculiar stellar rotation, $\Omega_\mathrm{eq}$ ($=\Omega_\mathrm{orb}$ when $e=0$), for which Eq.~\ref{eq:ospin} is 0, i.e. no more angular momentum can be exchanged between the star and the orbit. If necessary, we reduce the effective time-step for tidal process to ensure that both stars are not spun down (or up) past $\Omega_\mathrm{eq}$  \citep{Hurley02}. Tides are particularly effective when there is a large mismatch between $\Omega_\mathrm{eq}$ and $\Omega_\mathrm{spin}$, in tight systems ($R\approx a$), and for large convective envelopes (Eq.~\ref{eq:kconv} gives larger $k_\mathrm{tides}$ compared to Eq.~\ref{eq:krad}).

\subsubsection{Circularization during RLO and  collision at periastron} \label{sec:circ}

Although tides strongly reduce the orbital eccentricity before the onset of a RLO, in some cases the RLO starts with  a non-negligible residual eccentricity ($e\approx0.2-0.5$). Since the  RLO formalism described in Section \ref{sec:rlo} assumes circular orbits, \sevn{} includes an option to completely circularise the orbit at the onset of the RLO. This option is the default and we used it for the results presented in this work.

\sevn{} includes different options to handle orbit circularisation. In this work, we assume that the orbit is circularised at periastron, hence  $a_\mathrm{new}=a_\mathrm{old}\,{}(1-e_\mathrm{old})$ and $e_\mathrm{new}=0$, where $a_\mathrm{old}$ and  $e_\mathrm{old}$ are the semi-major axis and the eccentricity before circularisation \citep[see e.g.][]{VG2018}.  

We also test an alternative formalism in which we circularise the system not only at the onset of RLO, but also whenever  one of the two stars fills its RL at  periastron, i.e,  when $R \geq R_\mathrm{L,per}$ and $R_\mathrm{L,per}$ is estimated using Eq.~\ref{eq:eqrlo} replacing the semi-major axis $a$ with the periastron radius $a\,{}(1-e)$. In this case, we circularise the orbit at periastron and the system starts a RLO episode. 

Other available options in \sevn{}, not used in this work,  assume that  circularisation preserves  the orbital angular momentum, i.e. $a_\mathrm{new}=a_\mathrm{old}\,{}(1-e^{2}_\mathrm{old})$, or  the semi-major axis, i.e. $a_\mathrm{new}=a_\mathrm{old}$. In the latter case, the orbital angular momentum increases after  circularisation.  Finally, it is possible to disable the circularisation, conserving any residual eccentricity during 
the RLO (this assumption is the default in \bse{}). During RLO, the stellar tides, as well the other processes, are still active (Section \ref{sec:sync}). Therefore, the binary can still be circularised during an ongoing  RLO.

During binary evolution, \sevn{} checks if the two stars are in contact at periastron, e.g., if $R_1 + R_2 \leq a (1-e)$. 
If this condition is satisfied, \sevn{} triggers a collision. By default  we disable this check during an ongoing RLO. 
The outcome of the collision is similar to the results of an unstable mass transfer during a RLO (Section \ref{sec:rlo}). If at least one of the two stars has a clear core-envelope separation (\bse{} types $>3$, see Table \ref{tab:phases}) the collision triggers a CE, otherwise a direct stellar merger (Sections \ref{sec:CE} and \ref{sec:merger}).

\subsubsection{Gravitational waves (GWs)} \label{sec:gw}

\sevn{} describes the impact of GW emission on the orbital elements by including the same formalism as \bse{}  \citep{Hurley02}:

\begin{equation}
\dot{a} = - \frac{64G^3 M_1 M_2 \left(M_1 + M_2 \right)}{5c^5 a^3 (1-e^2)^\frac{7}{2}} \left(1+\frac{73}{24}e^2 + \frac{37}{96}e^4\right)\\
\label{eq:gwa}
\end{equation}
\begin{equation}
\dot{e} =- \frac{304G^3 M_1 M_2 \left(M_1 + M_2 \right)}{15c^5 a^4 (1-e^2)^\frac{5}{2}}
\left(1+\frac{121}{304}e^2 \right) e.
\label{eq:gwe}
\end{equation}
The above equations, described in \cite{peters1964}, account for orbital decay and circularisation by GWs. Unlike \bse{} (in which Eqs.~\ref{eq:gwa} and \ref{eq:gwe} are active only when the semi-major axis is $<10$ AU), in \sevn{} they are switched on whenever the GW merger timescale, $t_\mathrm{merge}$, is shorter than the Hubble time. The GW merger timescale is estimated using a  high-precision approximation (Appendix~\ref{app:gwtime}) of the solution of the systems of Eqs.~\ref{eq:gwa} and  \ref{eq:gwe} (errors <0.4\%).

\subsubsection{Stellar mergers} \label{sec:merger}

\begin{table}
\begin{tabular}{lcc}
\hline
Compact object  & Companion          & Merger outcome                                   \\ \hline
BH/NS/WD & H-star/pure-He star & BH/NS/WD (no mass accretion)                     \\
BH       & BH/NS/WD           & BH                                               \\
NS       & NS/WD              & if $M_\mathrm{f}<3 \ \Msun^\dagger$: NS, else: BH        \\
HeWD     & HeWD               & SNIa                                             \\
COWD     & COWD/HeWD          & if $M_\mathrm{f}<1.44 \ \Msun^\ddag$: COWD, else: SNIa \\
ONeWD    & WD   & if $M_\mathrm{f}<1.44 \ \Msun^\ddag$: ONeWD, else: NS  \\ \hline
\end{tabular}
\caption{This Table describe the outcome of a merger between a compact object and its companion, as implemented in \sevn{}. 
A SNIa leaves no compact remnant. 
$\dagger$ Assumed Tolman-Oppenheimer-Volkoff  mass limit for NSs, $\ddag$ assumed Chandrasekhar mass limit for WDs.} 
\label{tab:bcomerger}
\end{table}

When two stars merge, we simply sum their CO cores, He cores and total masses. Further details on merger due to  post-CE coalescence can be found in  Appendix~\ref{app:cemerger}.
The merger product inherits the phase and percentage of life of the most evolved progenitor star.
The most evolved star is the one with the largest \sevn{} phase ID (Table \ref{tab:phases}) or with the largest life percentage if the merging stars are in the same phase.

In \sevn{}, we do not need to define a collision table for the merger between two stars (such as Table~2 of \citealt{Hurley02}), because the interpolation algorithm finds the new post-merger track self-consistently, without the need to define a stellar type for the merger product. \sevn{} makes use of a collision table (Table~\ref{tab:bcomerger})  only to describe outcome of mergers involving compact objects.  
In the case of a merger between a star and a compact object (BH, NS, or WD), we assume that the star is destroyed and no mass is accreted onto the compact object. 
Mergers between WDs can trigger a SNIa explosion leaving no compact object (Table~\ref{tab:bcomerger}). Post-merger ONeWDs exceeding the Chandrasekhar mass limit (1.44 \Msun) become NSs. Similarly, post-merger NSs  more massive than the Tolman-Oppenheimer-Volkoff mass limit (set by default to 3.0 \Msun) become BHs (Section \ref{sec:sync}). 
Apart from the cases leading to a SNIa, the product of a merger between two compact objects is a compact object with the mass equal to the total mass of the pre-merger system. We do not remove the mass lost via GW emission, which is usually $\sim{5}$\% of the total mass of the system \citep[e.g.,][]{jimenez-forteza2017}.  We will add a formalism to take this into account  in the  future versions of \sevn. 

\subsection{The evolution algorithm} \label{sec:evalg}

\subsubsection{Adaptive time-step} \label{sec:tstep}

\sevn{} uses a prediction-correction method to adapt the time-step  accounting for the large physical range of timescales (from a few minutes to several Gyr) typical of stellar and binary evolution.

To decide the time-step, we look at a sub-set of stellar and binary properties (total mass, radius, mass of the He and CO core, semi-major axis, eccentricity, and amount of mass loss during a RLO): if any of them changes too much  during a time-step, we reduce  the time-step and repeat the calculation. 
In practice, we choose a maximum   \emph{relative} variation $\delta_{\rm max}$ 
(0.05 by default) and impose that 
\begin{equation}
     \max_{P \in \,{}\mathrm{properties}}  |\delta P|  \leq   \delta_\mathrm{max}, 
    \label{eq:timestepcomd}
\end{equation}
where  $|\delta P|$ is the absolute value of the  relative property variation.  

\sevn{} predicts the next time-step (${\rm d}t_{\rm next}$) as
\begin{equation}
    {\rm d}t_\mathrm{next} = \min_{P \in \mathrm{properties}} \left( \delta_\mathrm{max} \frac{{\rm d}t_\mathrm{last} }{|\delta P_\mathrm{last}|} \right),
    \label{eq:timestep}
\end{equation}
where ${\rm d}t_\mathrm{last}$ is the last time-step and $\delta P_\mathrm{last}$ is the relative variation of property $P$ during the last time-step, hence $|\delta P_\mathrm{last}|/{\rm d}t_\mathrm{last}$ represents the absolute value of the $\delta P_\mathrm{last}$ time derivative.

After the evolution step (Section \ref{sec:sync}), if the condition in Eq.~\ref{eq:timestepcomd} is not satisfied, a new (smaller) time step is predicted using Eq.~\ref{eq:timestep} and the updated values of $\delta P_\mathrm{last}$ and ${\rm d}t_\mathrm{last}$.
Then, we repeat the evolution  of all the properties with the new predicted time-step until condition~\ref{eq:timestepcomd} is satisfied or until the previous and the new proposed time steps differ by less than 20\%.

We use a special treatment 
when a star approaches a change of phase (including the transformation to a compact remnant). In this case, the prediction-correction method is modified to guarantee that the stellar properties are evaluated just after and before the change of phase.
In practice, if the predicted time-step is large enough to cross the time boundary of the current phase, \sevn{} reduces it so that the next evolution step brings the star/binary 
 $10^{-10}$ Myr before the phase change. Then, the following time-step is set to bring the star/binary 
 $10^{-10}$ Myr beyond the next phase.
This allows us to accurately model  stellar evolution across a phase change. In particular, it is necessary to properly set the stellar properties before a  supernova explosion or WD formation (Section \ref{sec:remform}).

On top of the adaptive method, \sevn{} includes a number of predefined time-step upper limits: 
the evolution time cannot exceed the simulation ending time  or the next output time; the stellar evolution cannot skip more than two points on the tabulated tracks; a minimum number of evaluations ($=10$ by default) for each stellar phase  has to be guaranteed. 
The time-step  distribution in a typical binary evolution model spans 9/10 orders of magnitude, from a few hours to several Myr. 

\begin{figure}
	\centering
	\includegraphics[trim={9cm 1.8cm 9cm 0cm},width=1.0\columnwidth,clip]{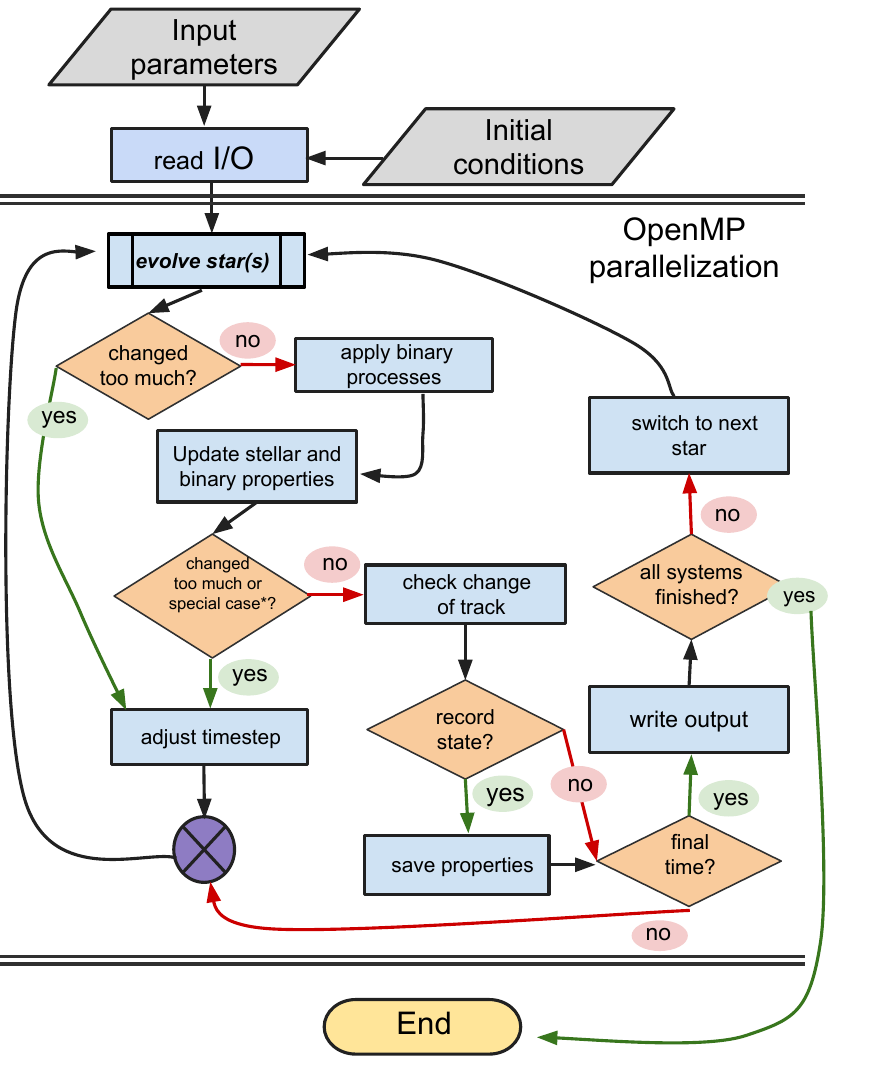}
	\vspace*{-5mm}
	\caption{Schematic representation of the \sevn{} evolution algorithm. The "changed too much"  checks refer to the variation of the stellar and/or binary properties. 
 In the case of single-stellar evolution or in the case of an ionized binary, \sevn{} skips the sections \vir{apply binary processes} and  \vir{update stellar and binary properties}.
 The \vir{special case} check refers to all the  cases in which \sevn{} repeats the evolution to follow a particular binary  evolution process, i.e, CE, merger, and circularistaion at the onset of the RLO (see Section~\ref{sec:sync} for further details).
\gitlab{https://gitlab.com/iogiul/iorio22_plot/-/tree/v3/SEVN_flowcharts}
 \gitimage{https://gitlab.com/iogiul/iorio22_plot/-/blob/v3/SEVN_flowcharts/SEVNflowcharts.pdf}
  }
	\label{fig:SEVNflowchart} 
\end{figure}

\subsubsection{Temporal evolution} \label{sec:sync}

\begin{figure*}
	\centering
	\includegraphics[trim={0cm 0cm 0cm 0cm},width=0.9\textwidth,clip]{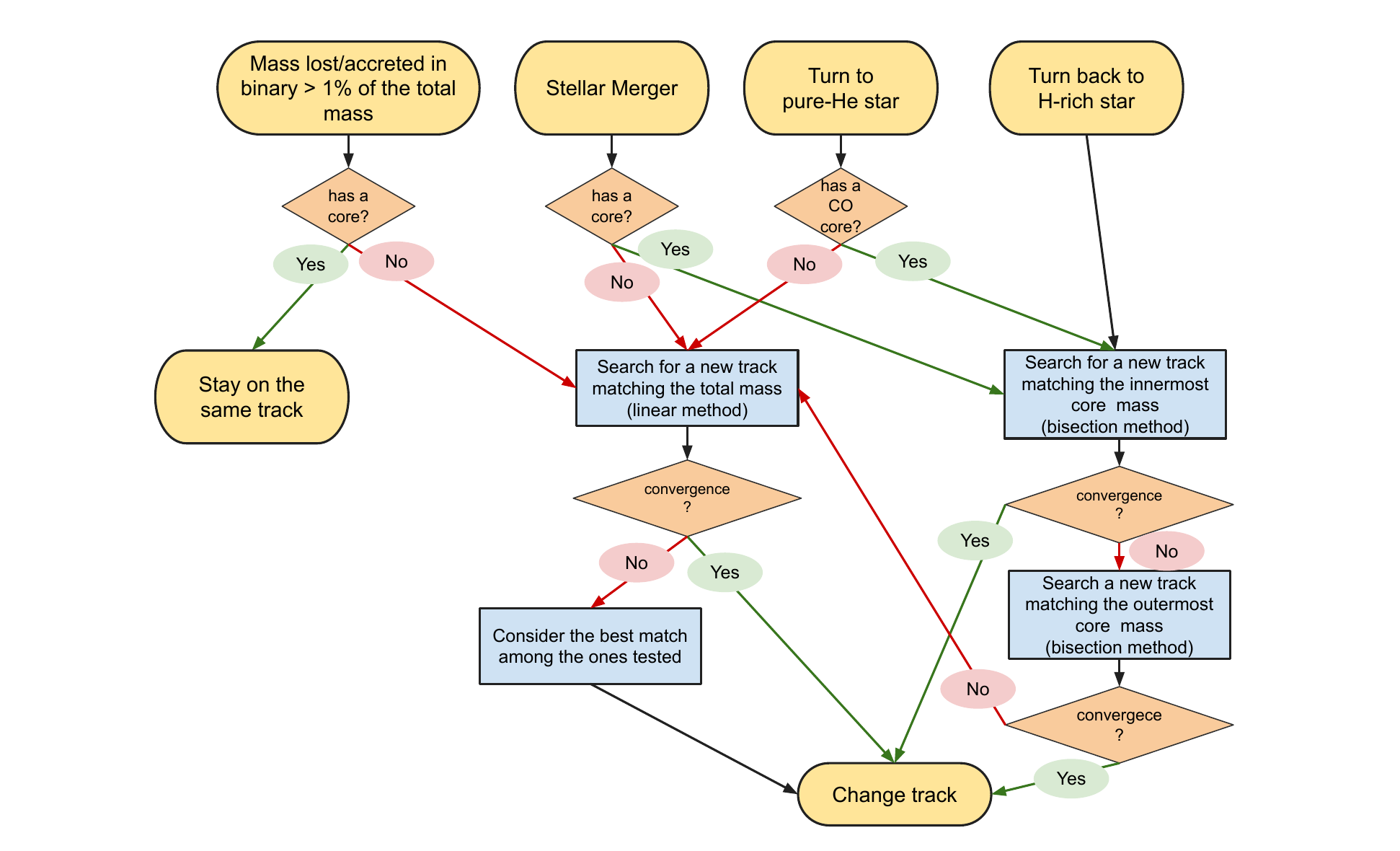}
	\vspace*{-5mm}
	\caption{Schematic representation of the algorithm  \sevn{} uses during a change of stellar track
 (Section~\ref{sec:trackchange}). The elements in the upper row indicate all the cases for which the code searches for a new stellar track: a significant mass loss/mass accretion due to binary interactions, a stellar merger, an H-rich star that loses its envelope turning into a pure-He star, and a pure-He star that accretes a new H envelope turning  back into a H-rich star. 
 In stars with both an He and CO cores, the latter is   the innermost core. In stars with only an He core the innermost and outermost cores coincide. 
\gitlab{https://gitlab.com/iogiul/iorio22_plot/-/tree/v3/SEVN_flowcharts}
 \gitimage{https://gitlab.com/iogiul/iorio22_plot/-/blob/v3/SEVN_flowcharts/SEVNchangetrack.pdf}}
	\label{fig:SEVNchangetrack} 
\end{figure*}

Figure~\ref{fig:SEVNflowchart} summarises the  \sevn{} temporal evolution scheme.  During each time-step, \sevn{} evolves  the two stars independently, then it evaluates and accumulates the property variations, $\Delta P$, caused by  
 each binary-evolution process. The binary 
prescriptions use as input the orbital and stellar properties at the beginning of the evolution step, $P(t_0)$. 

After the integration of the binary-evolution processes, \sevn{} updates each stellar and binary property (Fig.~\ref{fig:SEVNflowchart}). 
In particular, each \emph{binary} property (e.g., semi-major axis, eccentricity)  is updated as $P(t)=P(t_0)+\Delta P$.

Each \emph{stellar evolution} property (e.g., mass of each star) is calculated as $P(t)=P_{\rm s}(t)+\Delta P$, where $P_{\rm s}(t)$ is the value of the property at the end of the time-step as predicted by stellar evolution only. For example, if the property $P(t)$ is the mass of an accretor star during RLO, $P_{\rm s}(t)$ is the mass predicted at the end of the time-step by stellar evolution (accounting for mass loss by winds), while $\Delta P$ is the mass accreted by RLO  and by wind-mass transfer during the time-step. If necessary, the single and binary evolution step is  repeated until the adaptive time-step conditions are satisfied (Section \ref{sec:tstep}). 

\sevn{} evolves the compact remnants passively maintaining their properties constant. 
\sevn{} treats naked-CO stars similar to compact remnants: they evolve passively until they terminate their life and turn into compact remnants.

\sevn{} assumes that the transition from a star to a compact remnant happens at the beginning of the time-step. 
In this case, \sevn{} assigns a mass and a natal kick to the new-born compact object, based on the adopted supernova model. Then, it estimates the next time-step for the updated system.

Similarly, \sevn{} does not use the general adaptive  time-step criterion when  one the following processes takes place: RLO circularisation, merger, or CE. In such cases, \sevn{} uses an arbitrarily small time-step ($\mathrm{d}t_\mathrm{tiny}=10^{-15} \ \mathrm{Myr}$) 
and calculates only the aforementioned process during such time-step. Then, it estimates the new time-step.


At the very end of each evolutionary step, \sevn{} checks if a SNIa must take place. A SNIa is triggered if any of the following conditions is satisfied: i) a HeWD  with mass larger than $0.7 \ \Msun$ has accreted He-rich mass from a WR star, or ii) a COWD 
has accreted at least 0.15 \Msun{} from a WR star. 

Furthermore, \sevn{} checks if any ONeWD (NS) has reached a mass larger than 1.44 \Msun{} (3 \Msun) during the time-step. If this  happens, the ONeWD (NS) becomes a NS (BH). Finally, \sevn{} checks if the stars in the binary need  to jump to a new interpolating track (Section \ref{sec:trackchange}).

\subsubsection{Change of interpolating tracks} \label{sec:trackchange}

During binary evolution, a star can change its mass significantly due to mass loss/accretion, or after a stellar merger. 
In these cases, \sevn{} needs to find a new track, which better matches the current stellar properties.
For stars without a core (MS H-stars or core He burning pure-He stars), \sevn{} moves onto a new evolutionary track every time  the net cumulative mass variations due to binary processes (RLO, wind mass accretion) is larger than 1\% of  the  current star mass.
When a decoupled (He or CO) core is present, its properties drive the evolution of the star \citep[see, e.g.,][Section 7.1]{Hurley00}.  For this reason, we do not allow stars with a He or CO core (H-star with phase $>2$ and pure-He stars with phase $>4$) to change track unless the core mass has changed. 
After a stellar merger, \sevn{} always moves the merger product to a new stellar track. 
When an H-rich star fulfils the WR star condition (He-core mass larger than 97.9\% of the total mass),  the star jumps to a new pure-He track.

\begin{figure*}
	\centering
	\includegraphics[width=1.0\textwidth]{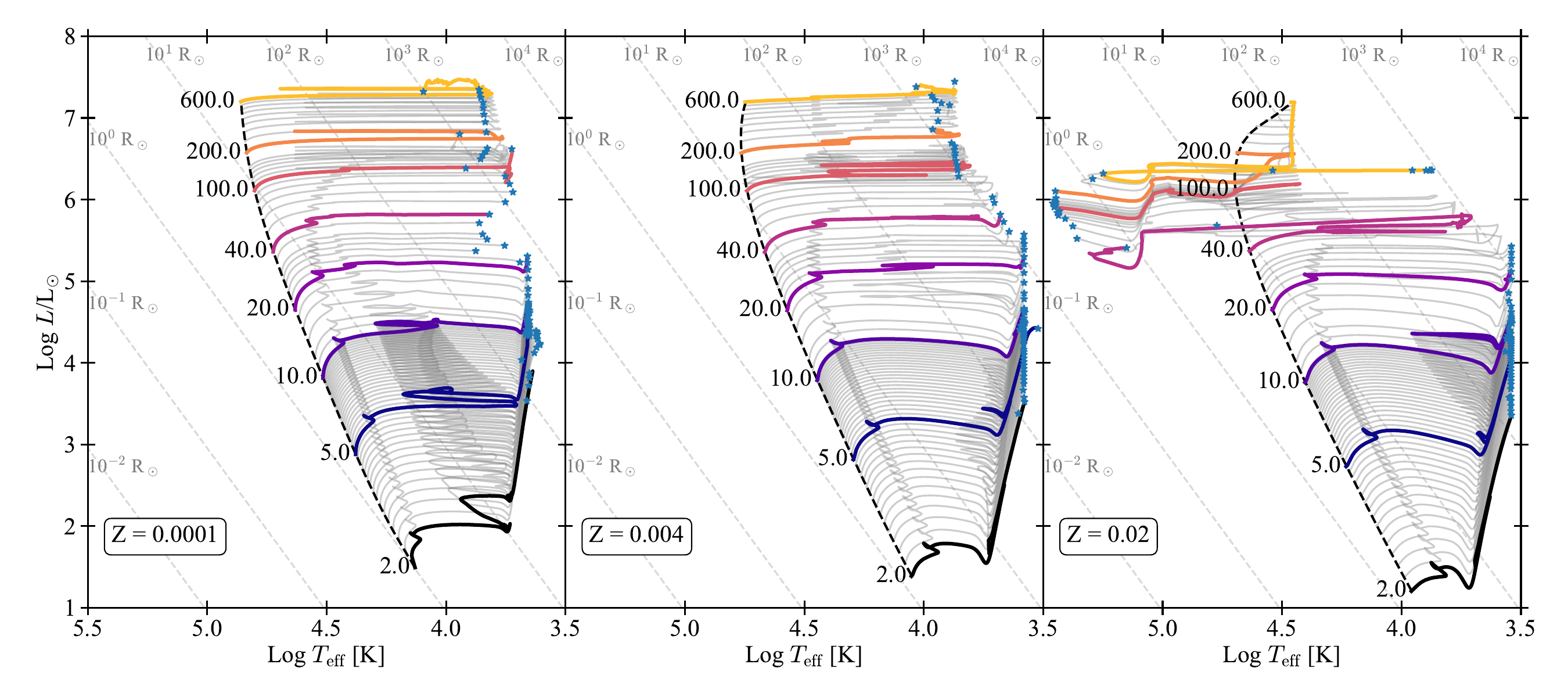}

	\caption{Hertzsprung-Russell (HR) diagram of \textsc{parsec} stellar tracks for three selected metallicities and $\lambda_\mathrm{ov}=0.5$. Different colours indicate selected tracks with initial masses \Mz{} = 2, 5, 10, 20, 40, 100, 200 and 600 \Msun. Solid grey lines show all the other tracks. The blue stars indicate the position of the star at the pre-supernova stage. The dashed black line shows the ZAMS. Diagonal dashed grey lines indicate points in the diagram at a constant radius. 
    \gitlab{https://gitlab.com/iogiul/iorio22_plot/-/tree/v3/parsec_HR}
    \gitimage{https://gitlab.com/iogiul/iorio22_plot/-/blob/v3/parsec_HR/HR.pdf}}
	\label{fig:HRD}
\end{figure*}

When a star moves to a new  track, \sevn{} searches the track 
that best matches the mass (or the mass of the core) of the current star at the same evolutionary stage (\sevn{} phase and percentage of life) and metallicity. 
We define the ZAMS\footnote{For pure-He stars the ZAMS mass is the mass at the beginning of the \sevn{} phase core He burning (Table \ref{tab:phases}).}  mass of such a track as  $M_\mathrm{ZAMS,new}$.
In general, \sevn{} searches  the new track in the H (pure-He) tables for H-rich (pure-He) stars. The only exceptions occur 
when a H-rich star is turned into a pure-He star (in this case, \sevn{} jumps to pure-He tables), and when a pure-He star is transformed back to a H-rich star after a merger (\sevn{} jumps from a pure-He  table to a H-rich table). 

\sevn{} adopts two different strategies to find the best $M_\mathrm{ZAMS,new}$ for stars with or without a core. For stars without a core-envelope separation, 
\sevn{} finds the best $M_\mathrm{ZAMS,new}$ following the method implemented in \citet[][see their Appendix~A2]{Spera19}. Hereafter, we define $M$ as the current mass of the star, $M_\mathrm{p}$ as the mass of the star with ZAMS mass $M_\mathrm{ZAMS}$, estimated at the same phase and percentage of life of the star that is changing track. $M_\mathrm{ZAMS,old}$  is the ZAMS mass of the current  interpolating track. 
Assuming a local linear relation between $M_\mathrm{ZAMS}$ and $M_\mathrm{p}$, we can estimate $M_\mathrm{ZAMS,new}$ using equation
\begin{equation}
M = \frac{M_\mathrm{p,2}-M_\mathrm{p,1}}{M_\mathrm{ZAMS,2} - M_\mathrm{ZAMS,1}}(M_\mathrm{ZAMS,new} - M_\mathrm{ZAMS,1} ) +M_\mathrm{p,1}.
\label{eq:Mlinear}
\end{equation}
As a first guess, we set $M_\mathrm{ZAMS,1}=M_\mathrm{ZAMS,old}$ and $M_\mathrm{ZAMS,2}=M_\mathrm{ZAMS,old}+1.2\delta M$, where $\delta M$ is the cumulative amount of mass loss/accreted due to the binary processes. 
$M_\mathrm{ZAMS,new}$ is accepted as the ZAMS mass of the new interpolating track if 
\begin{equation}
\frac{|M_\mathrm{p,new}-M|}{M} < 0.005,
\label{eq:Mconverge}
\end{equation}
otherwise Eq.~\ref{eq:Mlinear} is iterated replacing $M_\mathrm{ZAMS,1}$ or $M_\mathrm{ZAMS,2}$ with the last estimated $M_\mathrm{ZAMS,new}$. 
The iteration stops when the condition in  Eq.~\ref{eq:Mconverge} is fulfilled, or after 10 steps, or if $M_\mathrm{ZAMS,new}$ is outside the range of the ZAMS mass covered by the stellar tables. If the convergence is not reached, the best $M_\mathrm{ZAMS,new}$ will be the one that gives the minimum value of $|M_\mathrm{p,new}-M|/M$ (it could also be the original $M_\mathrm{ZAMS,old}$). \sevn{} applies this method  also  when H-rich stars without a CO-core turn into pure-He stars (phase $\leq 4$). If the phase is  $<4$,  \sevn{} sets the evolutionary stage of the new track at the beginning of the core-He burning (phase 4). 

For stars with a core,
\sevn{} looks for the best $M_\mathrm{ZAMS,new}$ matching the mass of the innermost core $M_\mathrm{c}$ (He-core for stellar phases 2, 3, 4, and CO-core for phases 5, 6, see Table \ref{tab:phases}). For this purpose, we make use of the bisection method in the ZAMS mass range [$\max(M_\mathrm{c},M_\mathrm{ZAMS,min})$, $M_\mathrm{ZAMS,max}$], where $M_\mathrm{ZAMS,min}$ and $M_\mathrm{ZAMS,max}$ represent the boundaries of the ZAMS mass range covered by the stellar tables (see Sections \ref{sec:tables} and \ref{sec:tracks}). \sevn{} iterates the bisection method until Eq.~\ref{eq:Mconverge} is valid considering the core masses. If the convergence is not reached within 10 steps,  \sevn{} halts the iteration  and the best $M_\mathrm{ZAMS,new}$ is the one that gives the best match to the core mass. Sometimes (e.g. after a merger) the CO core is so massive that no matches can be found. In those cases, \sevn{} applies the same method trying to match the mass of the He core. If the He-core mass is not matched, \sevn{} applies the linear iterative method to match the total mass of the star.  
\sevn{} uses this method also when a pure-He star turns back to an H-rich star after accreting an hydrogen envelope or when a H-rich star with a CO core turns into a pure-He star.

Finally, the star jumps to the new interpolating track with ZAMS mass $M_\mathrm{ZAMS,new}$. \sevn{} updates the four interpolating tracks  and synchronises all the stellar properties with the values of the new interpolating track. The only exceptions are the mass properties (mass, He-core mass, CO-core mass).
If the track-finding methods do not converge (Eq.~\ref{eq:Mconverge}  is not valid), 
the change of track might introduce discontinuities in these properties. 
To avoid this problem, \cite{Spera19} added a formalism that guarantees a continuous temporal evolution. In practice, \sevn{} evolves
the stellar mass and mass of the cores using
\begin{equation}
M_\mathrm{t_1} = M_\mathrm{t_0} \,{}(1 + \delta m), \ \mathrm{where} \   \delta m= \frac{m_\mathrm{t_1}- m_\mathrm{t_0}}{m_\mathrm{t_0}}.
\label{eq:temporal}
\end{equation}
In Eq.~\ref{eq:temporal}, $M_\mathrm{t_1}$ and $M_\mathrm{t_0}$  are the masses of the star (or of the core) estimated at time $t_1$ and $t_0$, while $m_\mathrm{t_1}$ and $m_\mathrm{t_0}$ are the masses obtained from the  interpolating tracks at  time $t_1$ and $t_0$ (see Section \ref{sec:sevninterpolation}).
 Figure \ref{fig:SEVNchangetrack} summarises the algorithm \sevn{} uses to check and handle  a change of track.

\section{Simulation setup} \label{sec:simsetup}

\subsection{\parsec{} Stellar tracks} \label{sec:tracks}

\begin{figure*}
	\centering
	\includegraphics[width=1.0\textwidth]{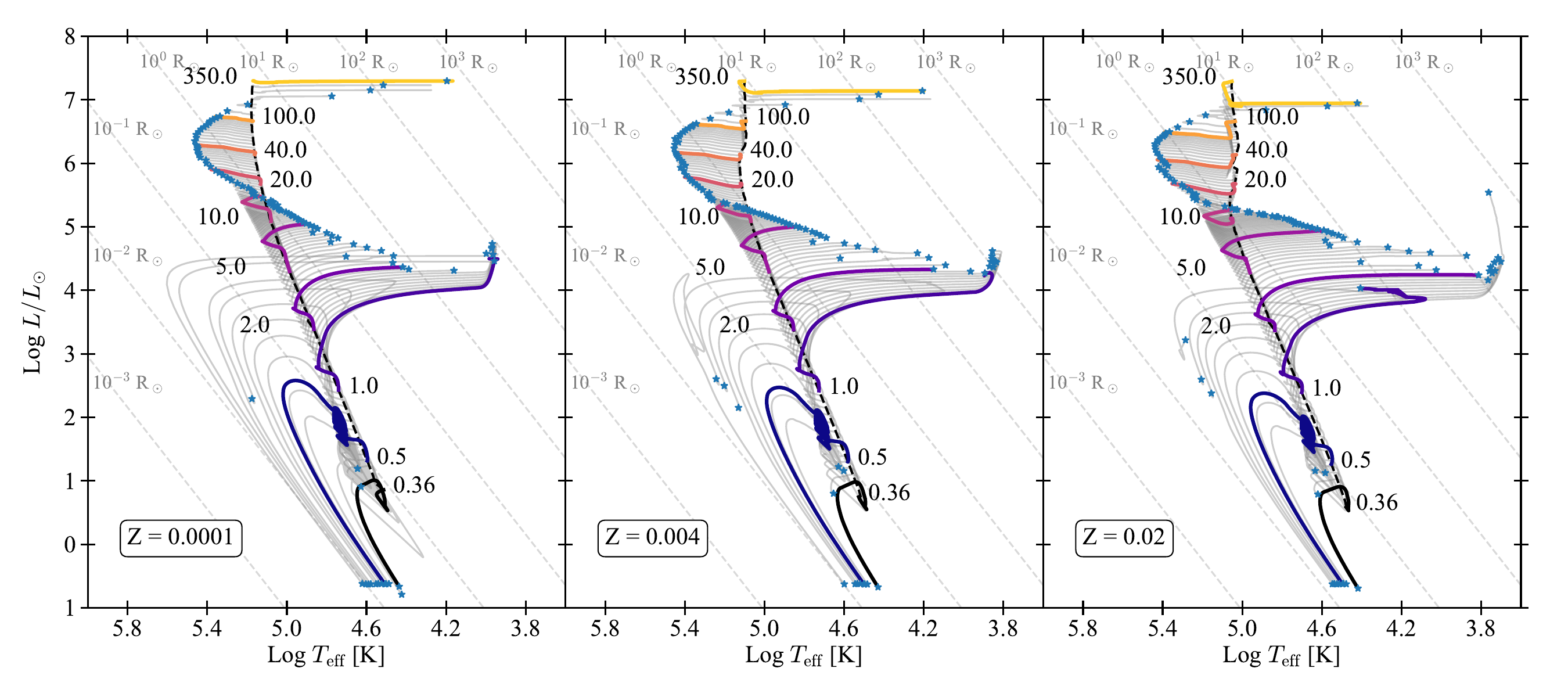}

    \caption{Same as Fig.~\ref{fig:HRD} but for pure-He \textsc{parsec} stellar tracks. 
    \gitlab{https://gitlab.com/iogiul/iorio22_plot/-/tree/v3/parsec_HR}
    \gitimage{https://gitlab.com/iogiul/iorio22_plot/-/blob/v3/parsec_HR/PureHE_HR.pdf}}
	\label{fig:HRDHE}
\end{figure*}

In this work we make use of stellar evolution tracks computed  with the stellar evolutionary code {\sc parsec} \citep{Bressan2012, Costa2019, Costa2021, Nguyen22}. In the following, we briefly describe the input physics assumed and the stellar tracks computed.

For the wind of massive hot stars, we use the mass-loss prescriptions by \citet{Vink2000} and \cite{Vink2001}, which take into account the dependence of the mass-loss on stellar metallicity. We also include the recipes by \cite{Graefener2008} and \cite{Vink2011}, which include the dependence of mass-loss on the Eddington ratio. 
For WR stars, we use prescriptions by \citet{Sander2019}, which 
reproduce the observed Galactic WR type-C (WC) and WR type-O (WO) stars. 
We modified the \cite{Sander2019} recipe, including a metallicity dependence. We refer to \citet{Costa2021} for further details.
For micro-physics, we use a combination of opacity tables from the Opacity Project At Livermore (OPAL)\footnote{\url{http://opalopacity.llnl.gov/}} team \citep{Iglesias1996}, and the \textsc{\ae sopus} tool\footnote{\url{http://stev.oapd.inaf.it/aesopus}} \citep{Marigo2009}, for the regimes of high temperature ($4.2 \leq{} \log{ (T/{\rm K})} \leq{} 8.7$) and low temperature ($3.2 \leq{} \log{(T/{\rm K})}\leq{} 4.1$), respectively. 
We include conductive opacities by \citet{Itoh2008}. 
For the equation of state, we use the \textsc{freeeos}\footnote{\url{http://freeeos.sourceforge.net/}} code version 2.2.1 by Alan W. Irwin, for temperature $\log{(T/{\rm K})} < 8.5$. While for higher temperatures ($\log{ (T/{\rm K})} > 8.5$), we use the code by \cite{Timmes1999}, in which the creation of electron-positron pairs is taken into account.

For internal mixing, we adopt the mixing-length theory \citep[MLT,][]{Bohm-Vitense1958}, with a solar-calibrated MLT parameter $\alpha_\mathrm{MLT} = 1.74$ \citep{Bressan2012}. We use the Schwarzschild criterion \citep{Schwarzschild1958} to define the convective regions, with the core overshooting computed with the ballistic approximation by \cite{bressan1981}.
We computed two different sets of tracks with an overshooting parameter $\lambda_\mathrm{ov}=0.4$ and  0.5. $\lambda_\mathrm{ov}$ is the mean free path of the convective element across the border of the unstable region in units of pressure scale height. 
For the convective envelope, we adopted an undershooting distance $\Lambda_\mathrm{env} = 0.7$ in pressure scale heights. More details on the assumed physics and numerical methodologies can be found in \citet{Bressan2012} and \citet{Costa2021}.

Using the solar-scaled elements mixture by \citet{Caffau2011}, we calculated 13 sets of tracks with a metallicity ranging from $Z = 10^{-4}$ to $4\times10^{-2}$. Each set contains approximately 70 tracks with a mass ranging from 2 to 600 \Msun. For stars in the mass range $2\,{} \Msun\ < \Mz{}{} < 8 \,{}\Msun$, we follow the evolution until the early asymptotic giant branch (E-AGB) phase. 
Stars with an initial mass $\Mz{}{} > 8 \,{}\Msun{}$ are computed  until the advanced core O-burning phase or the beginning of the electron-positron pair instability process.
Figure~\ref{fig:HRD} shows sets of tracks with different metallicities and with the overshooting parameter $\lambda_\mathrm{ov} = 0.5$. 

We also computed new pure-He stellar tracks with {\sc parsec}. 
For pure-He stellar winds, we adopted the prescriptions from \citet{nugis2000}. More details can be found in \citet{Chen15}.
The new sets are computed with the same input physics used for standard stars. The initial composition is set as follows. The hydrogen mass fraction is set to zero ($X = 0$), the helium mass fraction is given by $Y = 1 - Z$, and the metallicity ($Z$) ranges from 10$^{-4}$ to $5\times10^{-2}$. Each set contains 
100 tracks with masses ranging from $\Mz{}{} = 0.36$~\Msun\ to 350~\Msun. Figure~\ref{fig:HRDHE} shows three selected sets of pure-He tracks with different metallicity. These sets of tracks are part of a database that will be described in Costa et al. (in prep.), and will be publicly available in the new \textsc{parsec} Web database repository\footnote{\url{http://stev.oapd.inaf.it/PARSEC}}.

We used  the code {\sc TrackCruncher} (Section \ref{sec:tables}) to produce look-up tables for \sevn{} from the \parsec{} stellar tracks (see Appendix~\ref{app:trackstable} for additional details). 
The \parsec{} tables contain the stellar properties: mass, radius, luminosity, He and CO core mass and radius. 
The He/CO core masses and radii are estimated considering the point at which the H/He mass fraction drops below 0.1\%.
In addition, we produced tables for the properties of the convective envelope (mass, extension, eddy turnover timescale, see Section \ref{sec:tables}). 
For the stellar inertia we use Eq.~\ref{eq:inertiahurley}, while for the binding energy we use Eq.~\ref{eq:Ebind} and  test four different assumptions for the parameter $\lambda_\mathrm{CE}$ (Section~\ref{sec:models}).

\subsubsection{\parsec{} and \mobse{} stellar track  comparison} \label{sec:trackcompare}

\begin{figure*}
	\centering
	\includegraphics[width=1.0\textwidth]{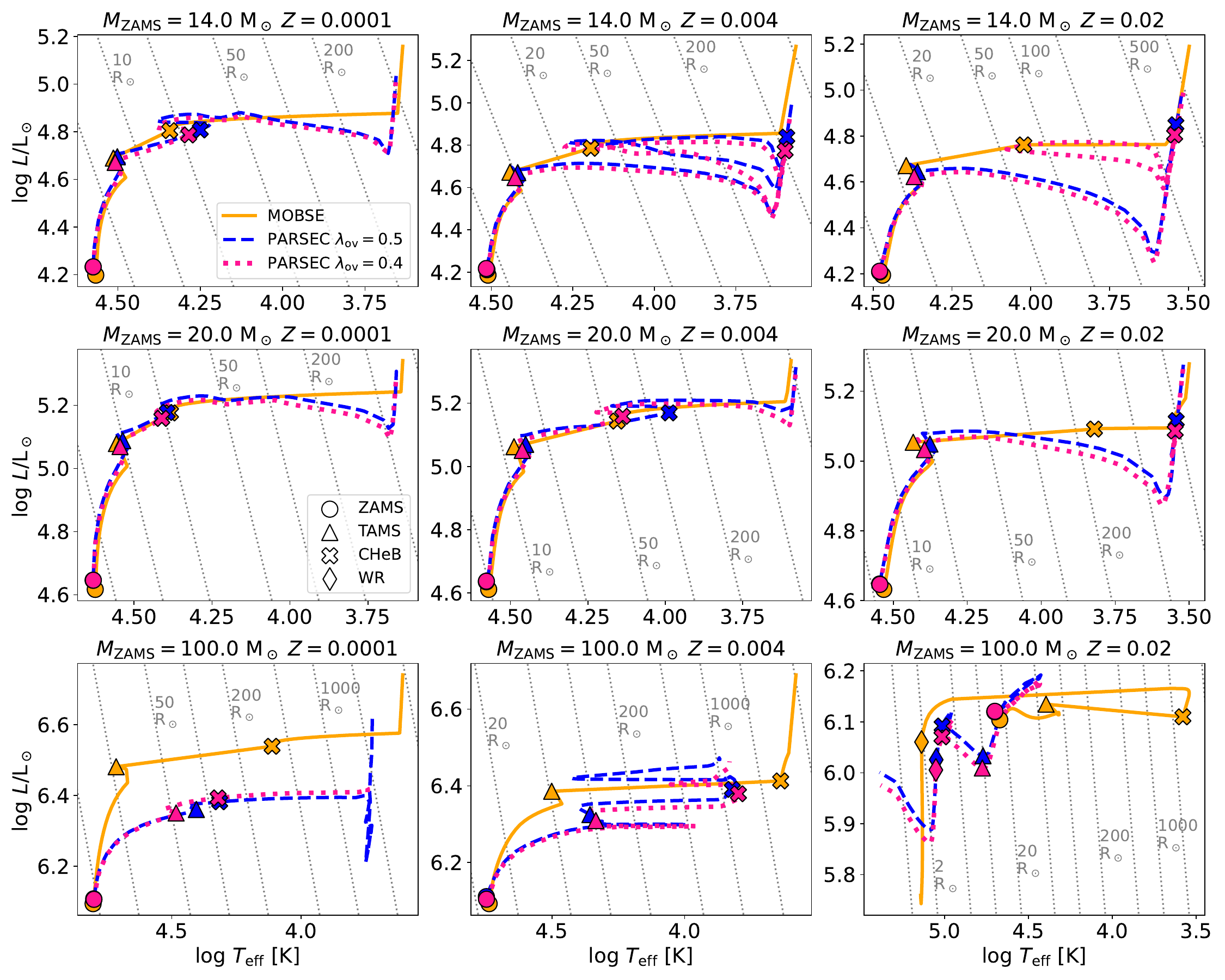}
	\caption{Comparison of stellar tracks in the HR diagram computed by \mobse{} (orange solid lines), and \sevn{} using \parsec{} stellar tables with overshooting parameter $\lambda_\mathrm{ov}=0.5$ (blue dashed lines) and $\lambda_\mathrm{ov}=0.4$ (pink dotted lines). The panels' titles specify the zero-age MS (ZAMS) mass and metallicity of the computed tracks. 
 The markers indicate peculiar phases during the stellar evolution: starting position in the ZAMS (ZAMS, circles); terminal-age MS, i.e. the first time the He-core decouples from the envelope (TAMS, triangles); helium burning ignition in the core (CHeB, crosses); begin of WR star  evolution, i.e. \bse{} phase 7 (WR, diamonds, see Section \ref{sec:sevnphase}). The grey dashed lines indicate points at constant radius: 1, 2, 5, 10, 20, 50, 100, 200, 500, 1000, and 2000 R$_\odot$.  
\sevn{} evolves the WR stars using the pure-He stellar tracks (Section \ref{sec:trackchange}), therefore the PARSEC lines after the diamonds are computed interpolating the pure-He tables (Fig.~\ref{fig:HRDHE}). 
     \gitlab{https://gitlab.com/iogiul/iorio22_plot/-/blob/v3/parsec_mobse_comparison/}
\gitbook{https://gitlab.com/iogiul/iorio22_plot/-/blob/v3/parsec_mobse_comparison/Plot.ipynb}
 \gitimage{https://gitlab.com/iogiul/iorio22_plot/-/blob/v3/parsec_mobse_comparison/TrackCompare.pdf}}
	\label{fig:trackcompare}
\end{figure*}

The stellar evolution implemented in \mobse{} and other \bse{}-like population synthesis codes is based  on the stellar evolution tracks computed by \cite{Pols98}. 
Figure \ref{fig:trackcompare} shows the comparison of the stellar evolution tracks computed with \mobse{}\footnote{\url{https://gitlab.com/micmap/mobse_open}}, and \sevn{} using the \parsec{} tracks for three selected ZAMS masses (14 \Msun{}, NS progenitors; 20 \Msun{}, transition between NS/BH progenitors;   100~\Msun{}, high-mass BH progenitors) at three different metallicities:  $Z=0.0001$, 0.004, and 0.02.

In most cases, the \mobse{} and \sevn{}+\parsec{} stellar tracks show  significant differences, especially for the metal-rich stars.  In the high-mass range of the NS progenitors ($ 14 \lesssim M_\mathrm{ZAMS}/\Msun\lesssim 20 $), the evolution differs substantially after the MS (top panels and middle-right panel in Fig.~\ref{fig:trackcompare}). In particular, in both \parsec{} models, the stars  ignite helium in the red part of the Hertzsprung-Russell (HR) diagram ($T_\mathrm{eff}\approx 3000$~K), while in \mobse{} core He burning begins in a bluer region ($T_\mathrm{eff}\gtrsim 5600$~K) when the stars are still relatively small ($R\lesssim 200 \ \Rsun$).

Figure \ref{fig:trackcompare} shows that the star with $M_\mathrm{ZAMS}=14 \ \Msun$ ignites   helium in an even bluer position in the HR at $T_\mathrm{eff}\approx 10^4$~K,  when it has a radius of $\approx70 \ \Rsun$.
Therefore, in \mobse{}, the NS progenitors tend to interact with their binary companion after or during the core He burning phase. In contrast, when \sevn{} makes use of the \parsec{} tracks, most of the NS progenitors interact before  helium ignition, i.e., during the Hertzsprung gap or giant branch phase (\bse{} types 2 and 3, see Table \ref{tab:phases}). Since most  binary-evolution processes depend on the stellar type  (e.g., RLO, Section \ref{sec:rlo}), these differences have a dramatic impact on the production of BNSs (Sections \ref{sec:bnsform} and \ref{sec:BNSmeff}).

The \parsec{} stellar tracks with different $\lambda_\mathrm{ov}$ values show a  similar evolution in the HR diagram. The largest differences are in the mass range of the NS progenitors at high metallicity. 
For these stars, the tracks with $\lambda_\mathrm{ov}=0.4$ produce a much more extend blue loop (see, e.g.,  the top-right panel in Fig.~\ref{fig:trackcompare}). 
The blue loop is a typical feature of stars in this mass range: at the ignition of core helium burning the star contracts moving to the blue part of the HR diagram, then it expands again at the end of the core He burning toward  the asymptotic giant branch. The $\lambda_\mathrm{ov}=0.4$ and $\lambda_\mathrm{ov}=0.5$ models are similar in many respects, hence we use the term \parsec{} referring to both models, unless specifically noted.

Overall, the \mobse{} stellar tracks reach larger radii during the evolution (up to ten times).  
In particular, high-mass BH progenitors  ($M_\mathrm{ZAMS}\gtrsim 50 \ \Msun$) in \mobse{}  expand up to 2500--10000 \Rsun{}, while in \parsec{}  the maximum radius ranges from $\approx  50 \ \Rsun$ (for $Z\gtrsim0.02$)
to $\approx 2500 \ \Rsun$ (for $Z\lesssim0.001$)
(see, e.g., the lower panels in Fig.~\ref{fig:trackcompare}).
However, in \parsec{} very high-mass ($M_\mathrm{ZAMS}\gtrsim 100 \ \Msun$) metal-poor ($Z\lesssim0.002$) stars reach large radii (up to $\approx 2500 \ \Rsun{}$) during the MS, while in \mobse{} such stars do not expand more than $\approx 50 \ \Rsun{}$ before the  end of the MS (see lower-left panel in Fig.~\ref{fig:trackcompare}). Therefore, in the \sevn{}+\parsec{} simulations 
 massive metal-poor stars tend to interact with their   binary companion during the MS, while in \mobse{} this happens at later evolutionary stages.

In  \mobse{}, high-mass metal-rich stars that become WR stars during the stellar evolution   always expand up to 1000--4000~\Rsun{} before  helium ignition, then they contract  and move toward the blue part of the HR diagram. 
In \parsec{}, only stars with $Z<0.007$ or $M_\mathrm{ZAMS}\lesssim 70 \ \Msun$ expand significantly (up to $\approx1000$  \Rsun) before the WR star phase, the other stars  contract and move to the blue part of the HR diagram already during the evolution in the MS   (see bottom panels in Fig.~\ref{fig:trackcompare}).
As a consequence, 
high-mass metal-rich stars in  \sevn{}+\parsec{} simulations interact less frequently with their binary companion with respect to \mobse{}.

In the mass range of the NS progenitors ($\approx$~8--20~\Msun) at low ($Z<0.001$) and intermediate-high metallicity ($Z>0.003$), the \mobse{}-\parsec{} difference in the maximum stellar radius decreases to $\approx$~0--200~\Rsun{} (see, e.g., the middle-top panel in Fig.~\ref{fig:trackcompare}). There is a small region in the ZAMS mass-metallicity plane ($0.004 \lesssim Z \lesssim 0.008$ and $20 \lesssim M_\mathrm{ZAMS}/\Msun \lesssim 30$), where the \parsec{} stellar tracks reach radii larger than 200--400 \Rsun{} with respect to \mobse{}.

The stellar mass at the end of the star lifetime ($M_\mathrm{f}$) is larger in \parsec{} (up to $\approx{40}$\%) for massive stars ($M_\mathrm{ZAMS} \gtrsim 100 \ \Msun$) and/or stars with high metallicity ($Z>0.008$). At intermediate metallicities ($0.001 < Z < 0.008$), \mobse{} produces larger final masses (up to $\approx25$\%) in the mass range 70--100 \Msun.  The $M_\mathrm{f}$ differences between the two \parsec{} models are within $\approx{10}$\% without a clear trend with  $M_\mathrm{ZAMS}$ and $Z$.  

The final masses of the He and CO cores are similar in the mass range 8--30~\Msun{}. More massive cores  ($\lesssim 30$\%)  are produced by \mobse{} for $M_\mathrm{ZAMS}<8 \ \Msun$ and by   \parsec{}  for $M_\mathrm{ZAMS}>30 \ \Msun$.

At low metallicity ($Z\lesssim0.001$), in the ZAMS mass range 100--150~\Msun, the \parsec{} stellar tracks with $\lambda_\mathrm{ov}=0.5$ end their life with  lighter cores ($\approx 25\%$) with respect to \mobse{} and \parsec{} with $\lambda_\mathrm{ov}=0.4$. This feature, produced by the dredge-up and the envelope undershooting \citep[see,  e.g.,][]{Costa2021}, has a large impact on the mass of the compact remnant when combined with the PISN formalisms (Section \ref{sec:rem_mass_single}). In the rest of the metallicity and ZAMS range, the \parsec{} models with $\lambda_\mathrm{ov}=0.5$ produce slightly ($3\%-5\%$ on average) more massive cores at the end of the evolution with respect to models with $\lambda_\mathrm{ov}=0.4$.

The stellar lifetime in \parsec{} is shorter with respect to \mobse{} up to 25\% for $M_\mathrm{ZAMS}\lesssim 80 \ \Msun$, and up to 40\% for $M_\mathrm{ZAMS}\gtrsim 80 \ \Msun$. 
For $Z<0.01$, the \parsec{} models  with $\lambda_\mathrm{ov}=0.5$ have a slightly longer lifetime ($\approx 5\%$) with respect to models with  $\lambda_\mathrm{ov}=0.4$. This difference increases up to $\approx 15\%$ for massive stars ($M_\mathrm{ZAMS}\gtrsim 200 \ \Msun$).

Using \sevn{}+\parsec{} the ZAMS mass for the  WD/NS transition (Section~\ref{sec:remform}) increases with metallicity from 
$\approx 8 \ \Msun{}$ at $Z=0.0001$  to $\approx 9 \ \Msun{}$ at $Z=0.02$. 
The NS/BH mass transition is at $\approx23 \ \Msun{}$ for the rapid supernova model and at $\approx18\mathrm{-}19 \ \Msun{}$ for the delayed model (see Section~\ref{sec:snmodel}).
In \mobse{}, the WD/NS and the NS/BH mass transitions shift to lower masses:  from $\approx 6 \ \Msun{}$ ($Z=0.0001$) to $\approx 7.5 \  \Msun{}$ ($Z=0.02$) for the WD/NS boundary,  and from  $\approx 20 \ \Msun{}$ ( $\approx 17 \ \Msun{}$)  to $\approx 22 \ \Msun{}$ ( $\approx \ 20 \Msun{}$)  for the NS/BH transition assuming the rapid (delayed) supernova model \citep[see,  e.g., Fig.~1 in][]{GiacobboCOB}. 
Given a 
stellar population following a Kroupa initial mass function (Section~\ref{sec:ic}) and considering only single stellar evolution, \mobse{} produces a larger number of NSs ($\approx 10\mathrm{-}30 \%$) and BHs ($\approx 5\mathrm{-}20 \%$) with respect to \sevn{}+\parsec{}. 

In Appendix~\ref{app:SSE}, we compare the \parsec{} stellar tracks  with the ones from other recent stellar evolution/population synthesis codes (\mist{}, \citealt{Choi2016}; \combine{}, \citealt{Combine}; \posydon{}, \citealt{Posydon}).

\subsection{Setup models}  \label{sec:fiducial} \label{sec:models}

We explore the uncertainties produced by  binary evolution prescriptions using 15 different setup models for the parameters of the \sevn{} simulations.

Unless otherwise specified, we use the \parsec{} stellar tables with $\lambda_\mathrm{ov}=0.5$ for the evolution of H-rich star, and the \parsec{} pure-He tables for the evolution of pure-He stars. 
In addition to the fundamental look-up tables (stellar mass, He and CO core mass, radius, luminosity), we use the stellar tables to evaluate the radial extension of the He and CO cores, and to follow the evolution  of the  convective envelope properties (mass fraction, depth of the convective layers and eddy turnover timescale, see Section~\ref{sec:tables}).

For the fiducial model (F), we set all \sevn{} parameters to their default values (see Sections \ref{sec:sse}  and \ref{sec:bse}). 
We use the rapid supernova model by \cite{fryer2012}, but we draw the NS masses from a Gaussian distribution centred at $M=1.33 \ \Msun$ (Section \ref{sec:snmodel}). We take into account the pair instability and pulsation pair instability using the model M20 by \cite{Mapelli20} (Section \ref{sec:pisn}).  We use the model KGM20  by \cite{GiacobboKick} to draw the natal kicks (Section \ref{sec:snkicks}).
We use the option QCRS (Table \ref{tab:qc}) for the stability of the mass transfer during the RLO, hence the mass transfer is always stable for MS and HG donor stars (\bse{} phases 1 and 2, see Table \ref{tab:phases}), while we follow the \cite{Hurley02} prescriptions in all the other cases. We set the default RLO mass accretion efficiency to $f_{\rm MT}=0.5$ (Eq.~\ref{eq:maccrlo}) as the mean  of the interval, and assume that the mass not accreted during the RLO is lost from the vicinity of the accretor as an isotropic wind (isotropic re-emission option, see Appendix~\ref{app:rloloss}). 
At the onset of RLO, \sevn{} circularises the orbit at  periastron (Section~\ref{sec:circ}).  
During CE, we estimate the envelope binding energy (Eq.~\ref{eq:Ebind}) using the same $\lambda_\mathrm{CE}$ formalism as in \mobse{} and \bse{} (see Appendix~\ref{app:lambda}). 

\begin{table}
\begin{tabular}{lc}
\hline \hline
\multicolumn{1}{c}{Model} & \multicolumn{1}{c}{Parameter variations}                                              \\ \hline
F                         & Fiducial model                                                                        \\ \hline
QCBSE                      & Use QCBSE option for the  RLO mass transfer stability (Table \ref{tab:qc})            \\
QCBB                       & Use QCBB option for the  RLO mass transfer stability  (Table \ref{tab:qc})            \\ 
QHE                       &  Enable quasi-homogeneous evolution during RLO (Section \ref{sec:rlo})                                       \\
RBSE                      & \begin{tabular}[l]{@{}l@{}}Use Eq.~\ref{eq:mtalternative} for mass accretion efficiency during the RLO\\ (same as in \citealt{Hurley02})\end{tabular}   \\ \hline
K$\sigma265$              & Draw supernova kicks from a Maxwellian with $\sigma=265 \ \mathrm{km}{s}^{-1}$               \\
K$\sigma150$              & Draw supernova kicks from a Maxwellian with $\sigma=150 \ \mathrm{km}{s}^{-1}$                \\
F19                       & Use \cite{farmer2019} PISN prescriptions (Section \ref{sec:pisn})  \\
 SND                       & Use the delayed supernova model with a Gaussian distribution \\ & for NS masses  (Section \ref{sec:snmodel})  \\
 \hline
NT                        & Disable tides  (Section \ref{sec:tides})                                                                       \\
NTC                       & \begin{tabular}[c]{@{}c@{}}Disable tides and circularise  when the RLO condition \\ is valid at the pericentre (Section \ref{sec:circ}) \end{tabular}                                                               \\ \hline
LK                        & Use $\lambda_\mathrm{CE}$ by \cite{Klencki21} for CE (Eq.~\ref{eq:Ebind})                         \\
LX                        & Use $\lambda_\mathrm{CE}$ by \cite{XuLi10}  for CE  (Eq.~\ref{eq:Ebind})                        \\
LC                        & Use  $\lambda_\mathrm{CE}=0.1$ for CE  (Eq.~\ref{eq:Ebind})                                                        \\ 
OPT            & QCBSE +  Optimistic CE assumption for HG stars (Section~\ref{sec:CE}) \\
\hline \hline
\end{tabular}
\caption{List of the 15 setup models used in this work to set the \sevn{} single and binary stellar evolution parameters. Column 2 (Parameter variations) describes what we change in each model with respect to the fiducial model.
The fiducial model (F) is described in the main text (Section~\ref{sec:models}).}
 \label{tab:models}
\end{table}

\begin{figure*}
	\centering
	\includegraphics[width=0.9\textwidth]{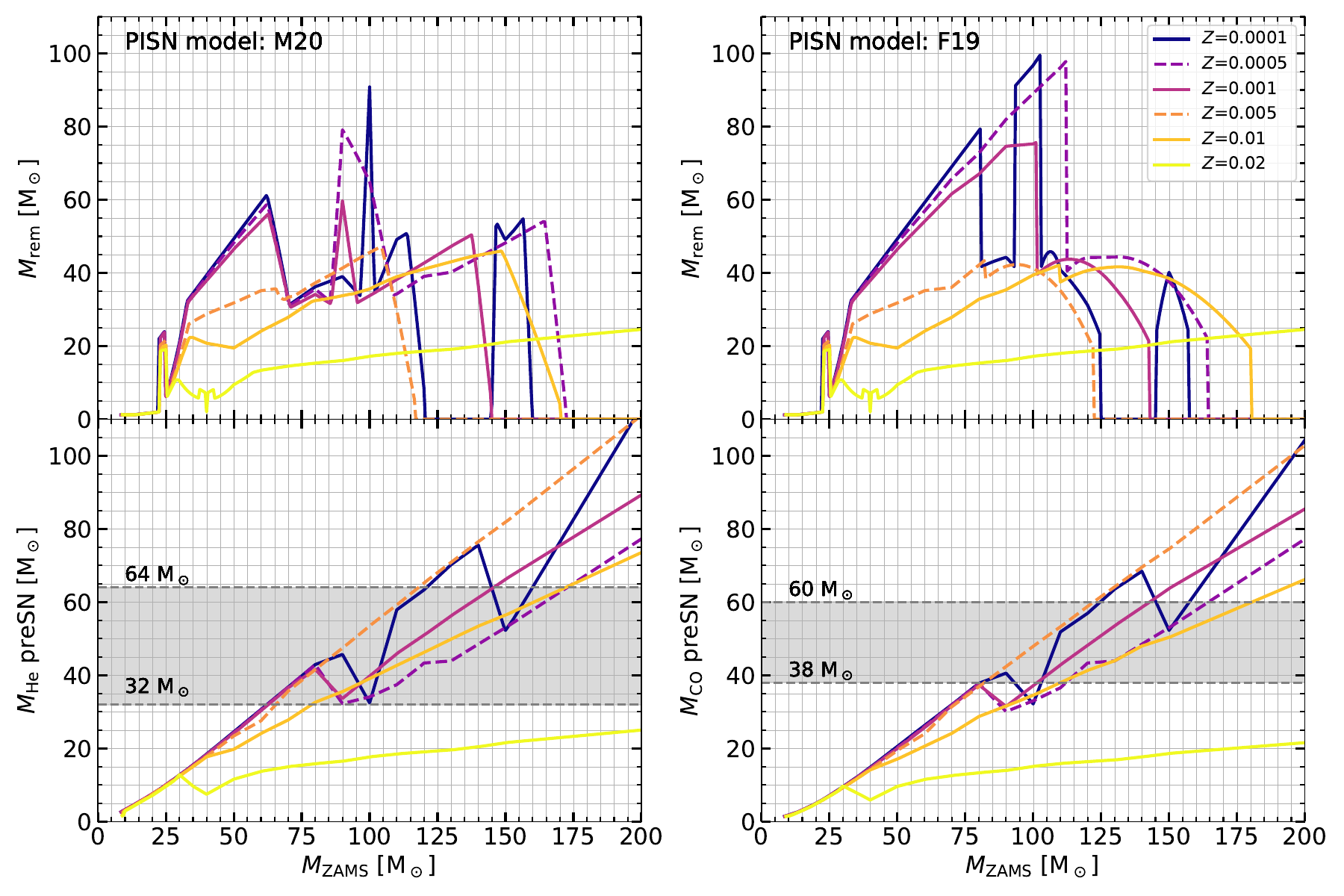}
	\vspace*{-5mm}
	\caption{Compact remnant mass and pre-supernova core masses  from single stellar evolution  as a function of the initial mass $M_\mathrm{ZAMS}$ for the look-up tables obtained from \parsec{} stellar tracks with $\lambda_\mathrm{ov}=0.5$ (Section~\ref{sec:tracks}). 
    The upper-left and upper-right panels show the mass of the compact remnant  considering  the pair-instability model M20 and F19, respectively (Section \ref{sec:pisn}). In both cases, we adopt the rapid supernova model (Section \protect\ref{sec:snmodel}). 
    The lower panels show the pre-supernova mass of the He core, $M_\mathrm{He}$ (left-hand panel), and CO core, $M_\mathrm{CO}$ (right-hand panel). The dashed horizontal lines mark the fundamental mass thresholds for the PISN models.
    In the model  M20, we expect the star to undergo pulsational pair instability (PPI)  between  $32 \ \Msun \leq M_\mathrm{He}\leq 64 \ \Msun$, 
    while for $M_\mathrm{He}>64 \ \Msun$ the star explodes as a PISN leaving no compact remnant. In F19, the PPI and PISN windows start at 
    $M_\mathrm{CO}\geq 38 \ \Msun$ and  $M_\mathrm{CO}>60 \ \Msun$, respectively.
    The different lines indicate different metallicities: $Z$=0.0001 blue solid line, $Z$=0.0005 violet dashed line, $Z=$0.001 violet solid line, $Z$=0.005 orange dashed line, $Z$=0.01 orange solid line,  $Z$=0.02 yellow solid line.  
The  $M_\mathrm{ZAMS}$  of the evolved stars are sampled each 0.5 \Msun{} in the interval  2.5--200 \Msun{}.
\gitlab{https://gitlab.com/iogiul/iorio22_plot/-/tree/v3/PISN}
\gitbook{https://gitlab.com/iogiul/iorio22_plot/-/blob/v3/PISN/Summary_plot.ipynb}
\gitimage{https://gitlab.com/iogiul/iorio22_plot/-/blob/v3/PISN/Mrem_pisn_ov05.pdf}}
\label{fig:mpisn0v05} 
\end{figure*}

Table \ref{tab:models} summarises all the other 14 models and their variations with respect to the fiducial model. 
We test alternative assumptions for the RLO stability with the models QCBSE and  QCBB (see Table \ref{tab:qc}), the model QHE enables the quasi-homogeneous evolution after the RLO mass transfer, while in the model RBSE we set the efficiency of the RLO mass transfer same as in \mobse{} using Eq.~\ref{eq:mtalternative} (most of the time it is equivalent to a conservative mass transfer, i.e. $f_\mathrm{MT}\approx1$ in Eq.~\ref{eq:maccrlo}.). 

The models K$\sigma265$ and  K$\sigma150$ test alternative natal kicks, drawn from a Maxwellian distribution with one-dimensional root-mean square $\sigma_{\rm kick}=265$ and 150 km s$^{-1}$, respectively. In  model F19, we replace the M20 PISN model with the \cite{farmer2019} prescriptions.  In model SND we   explore the delayed supernova model by \cite{fryer2012}, 
drawing the NS masses from a Gaussian distribution.    

We investigate the impact of the stellar tides  disabling them in the model NT. In the model NTC, we disable the tides and  use a less stringent criterion to trigger binary circularisation, enabling it every time the RLO condition is valid at periastron, i.e. using the periastron distance instead of the semi-major axis in Eq.~\ref{eq:eqrlo}  (Section \ref{sec:circ}).  
Furthermore, we test different prescriptions  to evaluate $\lambda_\mathrm{CE}$ during  CE with the model LX  (based on \citealt{XuLi10}), LK (based on \citealt{Klencki21}), and LC (in which $\lambda_\mathrm{CE}=0.1$, see Appendix~\ref{app:lambda}). Finally, model OPT is the closest set-up to the standard \bse{} formalism \citep{Hurley02}: we assume the same mass-transfer stability criteria as \bse{} (QCBSE), and we allow HG donors to survive a CE phase (optimistic CE assumption).

\subsection{Initial conditions} 
\label{sec:ic}

\begin{figure*}
	\centering
	\includegraphics[width=0.9\textwidth]{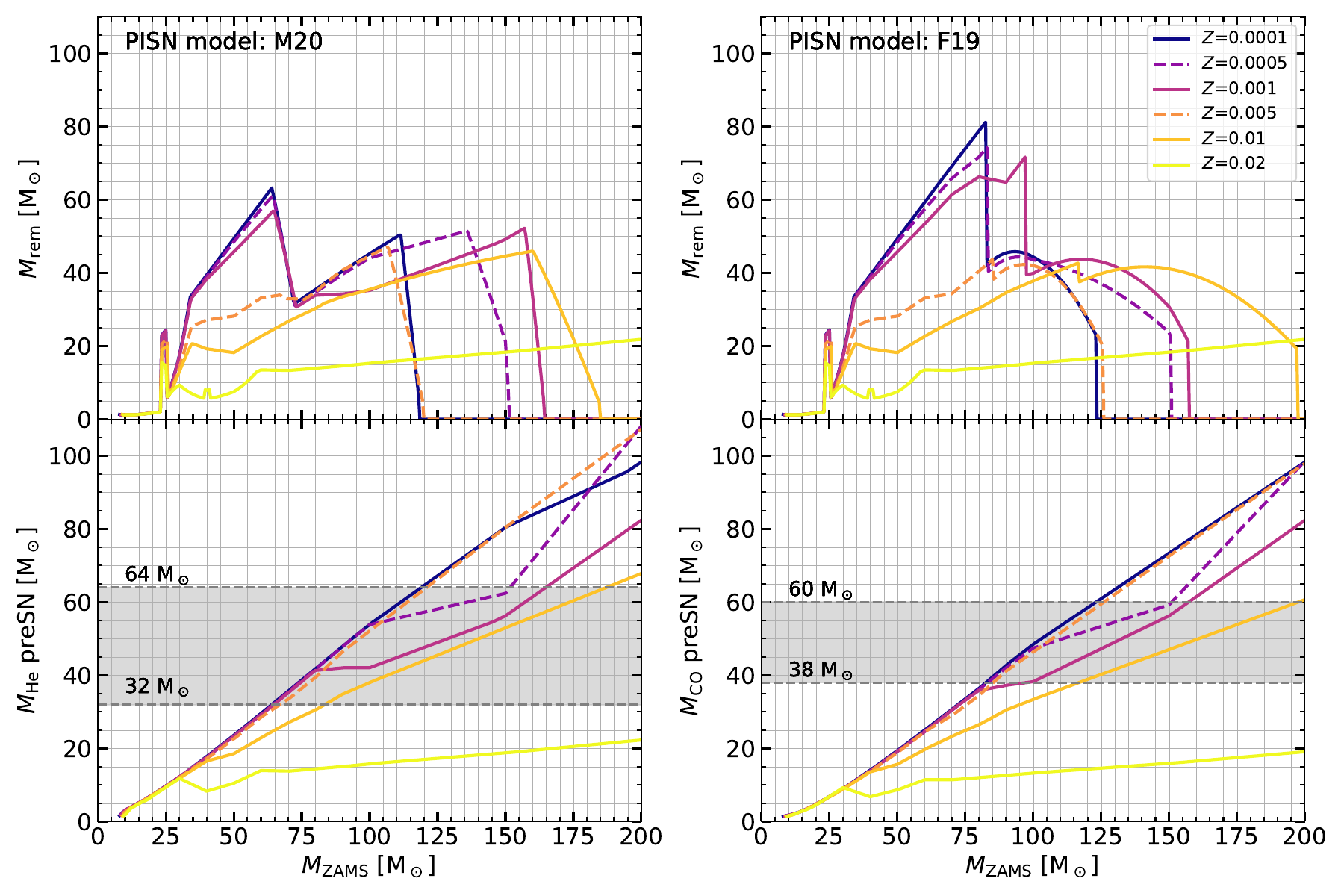}
	\vspace*{-5mm}
	\caption{Same as Fig.~\protect\ref{fig:mpisn0v05}, obtained from \parsec{} stellar tracks with $\lambda_\mathrm{ov}=0.4$ (see Section \ref{sec:tracks}). 
\gitlab{https://gitlab.com/iogiul/iorio22_plot/-/tree/v3/PISN}
\gitbook{https://gitlab.com/iogiul/iorio22_plot/-/blob/v3/PISN/Summary_plot.ipynb}
\gitimage{https://gitlab.com/iogiul/iorio22_plot/-/blob/v3/PISN/Mrem_pisn_ov04.pdf} }
\label{fig:mpisn0v04} 
\end{figure*}

We  randomly draw the initial ZAMS masses of primary stars from a Kroupa initial mass function (IMF,  \citealt{Kroupa2001}) 
\begin{equation}
    \mathrm{pdf}(M_\mathrm{ZAMS,1}) \propto M_\mathrm{ZAMS,1}^{-2.3} \ \ \ \ \ M_\mathrm{ZAMS,1} \in \ [5,150] \ \mathrm{M}_\odot, 
    \label{eq:kroupaimf}
\end{equation}
and the masses of secondary stars 
assuming the distribution of mass ratios from \cite{Sana12}, based on observations of O- and B-type binary stars in open clusters:
\begin{equation}
    \mathrm{pdf}(q) \propto q^{-0.1} \ \ \ \ \ q=\frac{M_\mathrm{ZAMS,2}}{M_\mathrm{ZAMS,1}} \in [q_\mathrm{min},1.0]  \ \mathrm{M}_\odot,
    \label{eq:q}
\end{equation}
with
\begin{equation}
    q_\mathrm{min}= \max \left( \frac{2.2}{M_\mathrm{ZAMS,1}}, 0.1 \right) 
    \label{eq:qmin}
\end{equation}
The lower mass limits for primary stars (5 \Msun) and secondary stars (2.2 \Msun) represent safe boundaries to study  NSs and BHs.  
The upper mass limit (150 \Msun) is a typical mass limit used in the study of NSs and BHs \citep[e.g.][]{GiacobboCOB,Spera19}.  The \parsec{} tracks used in this work reach masses up to 600 \Msun. We investigate this high mass regime  in the follow-up paper \cite{costa2023}. 
We set the initial rotational velocity of the stars to 0.

The initial orbital periods ($P$) and eccentricities ($e$)  have been generated according to the distributions  by \cite{Sana12}:

\begin{equation}
    \mathrm{pdf}(\mathcal{P}) \propto \mathcal{P}^{-0.55} \ \ \ \ \ \mathcal{P} = \log(P/\mathrm{day)} \in [0.15,5.5], 
    \label{eq:pdfP}
\end{equation}

\begin{equation}
    \mathrm{pdf}(e) \propto e^{-0.42} \ \ \ \ \ e \in [0,0.9]. 
    \label{eq:pdfe}
\end{equation}

We generate $10^6$ binary systems and use them as initial conditions in all our simulations (i.e., for different metallicities and different combinations of the main parameters).

The total mass of the simulated binaries is $2.21 \times 10^7 \ \Msun$ corresponding to an effective total mass of $1.74 \times 10^8 \ \Msun$ when taking into account the correction for incomplete IMF sampling due to the mass cuts ($f_{\rm cut}=0.255$)\footnote{We estimate the correction factor 
by applying the mass cuts $M_\mathrm{ZAMS,1}\geq 5 \ \Msun$ and $M_\mathrm{ZAMS,2}\geq 2.2 \ \Msun$  to a  parent  population with the primary mass following a \cite{Kroupa2001} IMF between 0.08 and 150 \Msun{}, and the mass ratio following a \cite{Sana12} distribution between $q_\mathrm{min}=\max (0.08 M^{-1}_\mathrm{ZAMS,1}, 0.1)$ and 1.}

For each of the 15 setup models  (see Table \ref{tab:models}), we ran 60 sets of simulations combining 15 metallicities ($Z=10^{-4}$, $2\times10^{-4}$, $4\times10^{-4}$, $6\times10^{-4}$, $8\times10^{-4}$, $10^{-3}$, $2\times10^{-3}$, $4\times10^{-3}$, $6\times10^{-3}$, $8\times10^{-3}$, $10^{-2}$, $1.4\times10^{-2}$, $1.7\times10^{-2}$, $2\times10^{-2}$, $3\times10^{-2}$) and four values for the $\alpha_\mathrm{CE}$ parameter ($\alpha_\mathrm{CE}=0.5$, 1, 3, 5).
In addition to the 15 models, we generate an extra set of $5\times10^6$ binaries, 
 then we simulate them using the fiducial  model (Section \ref{sec:models}). We use this supplementary dataset to investigate the systematic uncertainties originated by the sampling of the initial conditions. 
In conclusion, we simulate a total of $1.2\times10^9$ binary systems.

We use \sevn{}  to evolve all the binaries until both  stars are compact remnants or, if they collide, until their merger product becomes a compact remnant. 
For a BCO, the orbital decay by GWs is the only active process (Section~\ref{sec:gw}).
Therefore, for each BCO,  we estimate the merger time a posteriori using $t_\mathrm{merge}$ (Eq.~\ref{eq:tcomb}).
\textcolor{black}{ Using  10 threads  
on a server equipped with Intel(R) Xeon(R) Platinum 8168 (2.70 Ghz) CPUs, \sevn{} completed the evolution of each set of $10^6$ binaries in approximately 
1 hour.}

The list of initial conditions, the script used to run SEVN,  and  the simulations outputs  are available in \href{zenodolink}{Zenodo}\footnote{\url{https://doi.org/10.5281/zenodo.7794546}} \citep{paperdataset}.

\section{Results}
\label{sec:Results}

\subsection{Compact remnant mass}\label{sec:rem_mass}

\subsubsection{Single star evolution}\label{sec:rem_mass_single}

\begin{figure*}
	\centering
	\includegraphics[width=0.9\textwidth]{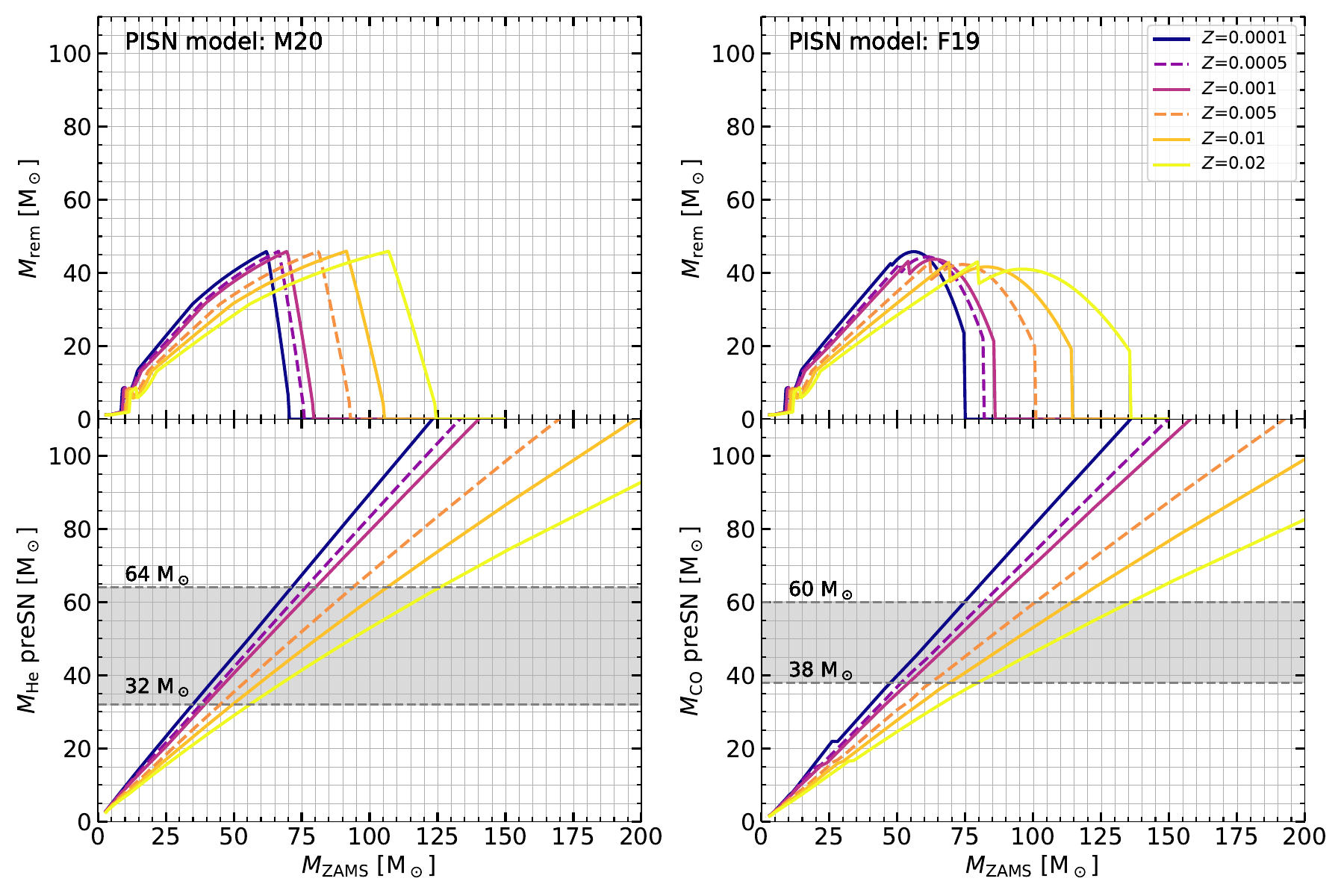}
	\vspace*{-5mm}
	\caption{Same as Fig.~\protect\ref{fig:mpisn0v05}, but for the  \parsec{} pure-He look-up table (see Section \ref{sec:tracks}). $M_\mathrm{ZAMS}$ indicates the initial mass at the beginning of the core He  burning, while the $M_\mathrm{He}$ is equivalent to the pre-supernova stellar mass. 
\gitlab{https://gitlab.com/iogiul/iorio22_plot/-/tree/v3/PISN}
\gitbook{https://gitlab.com/iogiul/iorio22_plot/-/blob/v3/PISN/Summary_plot.ipynb}
\gitimage{https://gitlab.com/iogiul/iorio22_plot/-/blob/v3/PISN/Mrem_pisn_pureHe.pdf}
 }
\label{fig:mpisnHe} 
\end{figure*}

Figures~\ref{fig:mpisn0v05} and  \ref{fig:mpisn0v04}  show the mass spectrum of compact objects that we obtain from single star evolution, by assuming input tables with $\lambda{}_{\rm ov}=0.5$ and 0.4, respectively. 
For each set of evolutionary tables, we show the results of both PISN models, M20 and F19  (Section~\ref{sec:pisn}). These figures show how sensitive the maximum mass of the BH and the PISN window are to the details of stellar evolution \citep[e.g.,][]{farmer2019,farmer2020,Mapelli20,renzo2020,Costa2021,vink2021}. 

In the tables with $\lambda_{\rm ov}=0.5$, several stellar models undergo a dredge-up (e.g., $M_{\rm ZAMS}\approx{100}$ and 150 M$_\odot$ at $Z=10^{-4}$, see Section~\ref{sec:trackcompare}). Because of the dredge-up, the final mass of the He and CO cores of these stars are smaller than those of lower-mass stars, resulting in a non-monotonic trend of both $M_{\rm He,f}$ and $M_{\rm CO,f}$ as a function of $M_{\rm ZAMS}$ (lower panels of Fig.~\ref{fig:mpisn0v05}). If $M_{\rm ZAMS}\approx{100}$~M$_\odot$ and $Z=10^{-4}$, the decrease of $M_{\rm He}$ and $M_{\rm CO}$ caused by the dredge-up allows the stellar models to avoid PPI, producing BHs with mass up to 90 and 100 M$_\odot$ in the M20 and F19 models, respectively. If $M_{\rm ZAMS}\approx{150}$ M$_\odot$ and $Z=10^{-4}$, the star avoids complete disruption by a PISN and collapses to BH after PPI. The details of the mass spectrum rely on the assumed PISN models (M20 and F19), because we do not perform hydrodynamical simulations and should be taken just as indicative trends. Moreover, here we assume that the mass of a BH formed via direct collapse is equal to the total mass of the progenitor star at the onset of core collapse (based on \citealt{fryer2012}). This is an  optimistic assumption, because the residual H-rich envelope is loosely bound and even a small shock triggered by neutrino emission can lead to the ejection of the outer layers \citep[e.g.,][]{fernandez2018,renzo2020b,costa2022}.

In contrast, for $Z<0.02$, $M_{\rm He,f}$ and $M_{\rm CO,f}$ have a perfectly monotonic trend with $M_{\rm ZAMS}$  in the tables with $\lambda_{\rm ov}=0.4$. This results in a much smoother behaviour of $M_{\rm rem}$ versus $M_{\rm ZAMS}$. In this set of tables, the M20 and F19 models lead to a maximum BH mass of $\approx{63}$ and 81~M$_\odot$ (at $Z=10^{-4}$), respectively.

Overall, the F19 model leads to a larger maximum mass, because the PPI regime starts at 
higher stellar masses  with respect to  M20. This result confirms that there are major uncertainties on the lower edge of PISN mass gap from stellar evolution theory \citep[e.g.,][]{farmer2019,farmer2020,renzo2020,Mapelli20,Costa2021,vink2021}.

\cite{farmer2019} do not find such large maximum BH masses, because they simulate only pure-He stars. Figure~\ref{fig:mpisnHe} shows the compact remnant mass, as a function of $M_{\rm ZAMS}$, that we obtain from our pure-He models. Here, the maximum BH mass is $M_{\rm rem}\approx{45}$ M$_\odot$ for both M20 and F19, with very little dependence on $Z$, as already discussed by \cite{farmer2019}.

\begin{figure}
	\centering
	\includegraphics[width=1.0\columnwidth]{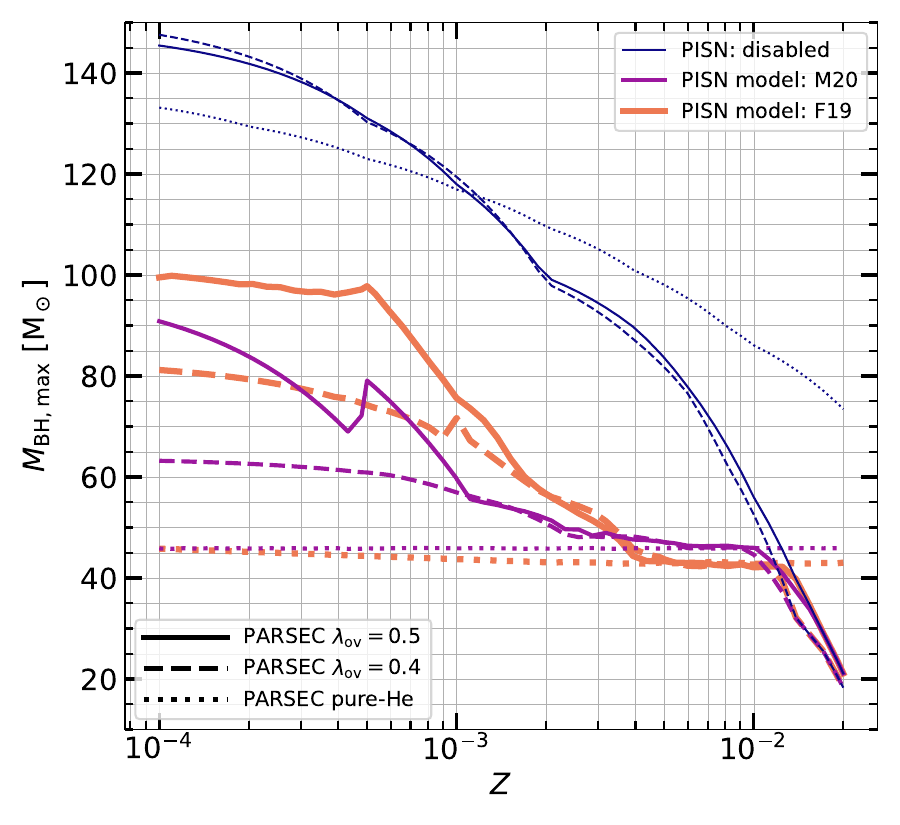}
	\vspace*{-5mm}
	\caption{Maximum BH mass from single stellar evolution as function of metallicity. 
    We estimate the BH mass using the rapid supernova model  (Section \protect\ref{sec:snmodel}) and different models for pair instability. No pair-instability correction: very thin blue lines;  M20 (based on \protect\citealt{Mapelli20}): thin violet lines; F19 (based on \protect\cite{farmer2019}): thick orange lines.  See Section \ref{sec:pisn} for details on pair-instability models.
    The plot shows the results for three different \parsec{} look-up tables: two sets of H-rich stars (from \parsec{} tracks with $\lambda_\mathrm{ov}=0.5$, solid lines, and $\lambda_\mathrm{ov}=0.4$, dashed lines),  and  one for pure-He stars (dotted lines). See Section  \protect\ref{sec:tracks} for additional details on the stellar tracks. We sample the  $M_\mathrm{ZAMS}$ each 0.5 \Msun{} in the interval  2.5--200 \Msun, while the 50 used metallicities are uniformly  sampled in the logarithmic space between $Z=0.0001$ and $Z=0.02$. For pure-He stars, $M_\mathrm{ZAMS}$ indicates the initial mass at the beginning of the core He burning.
    \gitlab{https://gitlab.com/iogiul/iorio22_plot/-/tree/v3/PISN}
\gitbook{https://gitlab.com/iogiul/iorio22_plot/-/blob/v3/PISN/Summary_plot.ipynb}
\gitimage{https://gitlab.com/iogiul/iorio22_plot/-/blob/v3/PISN/MaxBHpisn.pdf}}
\label{fig:maxbh} 
\end{figure}

Figure~\ref{fig:maxbh} shows the maximum BH mass ($M_{\rm BH,\,{}max}$)  that we obtain in our models as a function of metallicity. Here, we do not consider BHs above the upper edge of the PISN mass gap, that we discuss in  \cite{costa2023}. In the H-rich models, $M_{\rm BH,\,{}max}$ increases for decreasing metallicity, because the residual H-rich envelope mass is larger at lower $Z$. In contrast, $M_{\rm BH,\,{}max}$ is almost independent of $Z$ for  pure-He stars.

In the rapid model \citep{fryer2012}, BH progenitors with $M_{\rm CO,f}\in{[6,\,{}7]}$~$\Msun$ (corresponding to a ZAMS mass $M_\mathrm{ZAMS}\approx 25~\Msun$ ) end their life with a direct collapse producing 
large BH masses ($M_\mathrm{rem}\approx 20-25~\Msun$), well visible in Figs.~\ref{fig:mpisn0v05} and \ref{fig:mpisn0v04}. 
At high metallicity ($Z=0.02$),   $M_{\rm CO,f}<7$~\Msun{} 
 between $M_\mathrm{ZAMS}\approx35~\Msun$ and $M_\mathrm{ZAMS}\approx45~\Msun$, and 
$M_{\rm CO,f}<6$~$\Msun$ around  $M_\mathrm{ZAMS}\approx40~\Msun$, resulting in  the fast oscillations of $M_\mathrm{rem}$ 
visible in Figs.~\ref{fig:mpisn0v05} and  \ref{fig:mpisn0v04}.

\subsubsection{Binary evolution}
\label{sec:rem_mass_binary}

\begin{figure*}
	\centering
	\includegraphics[width=2.0\columnwidth]{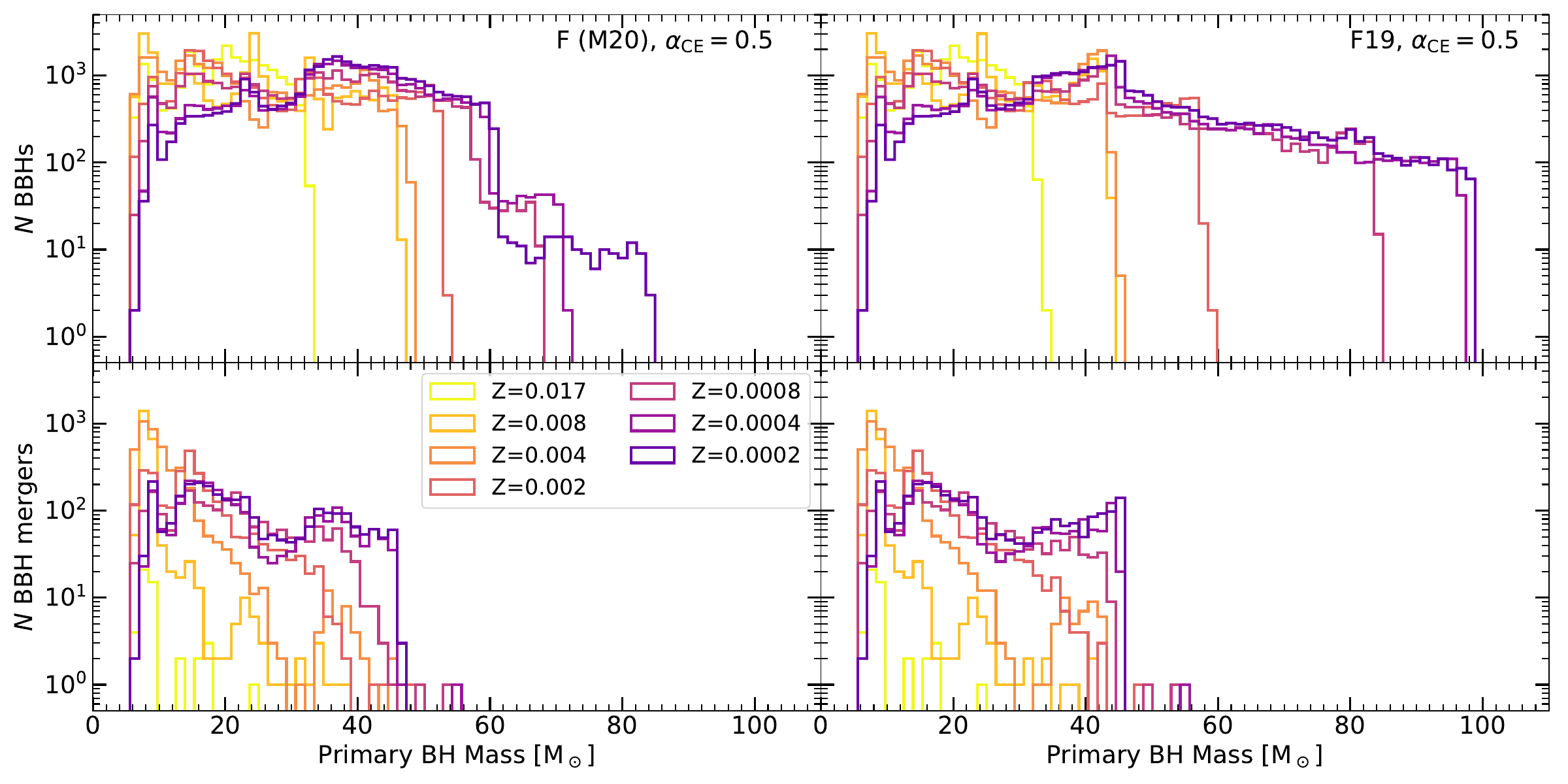}
	 \caption{Upper (Lower) panel: BBHs (BBH mergers) in our  simulations  F (left, pair instability model M20) and F19 (right) with $\alpha_{\rm CE}=0.5$. The colour map refers to the metallicity of the progenitor ($Z=0.017$, 0.008, 0.004, 0.002, 0.0008, 0.0004, and 0.0002). 
    The alternative versions of this plot showing the results for all the \ace{} values, and for all the setup models can be found in the gitlab repository of the paper (\gitlab{https://gitlab.com/iogiul/iorio22_plot/-/tree/v3/Mspectrum}).
     \gitbook{https://gitlab.com/iogiul/iorio22_plot/-/blob/v3/Mspectrum/MassDist.ipynb}
     \gitimage{https://gitlab.com/iogiul/iorio22_plot/-/blob/v3/Mspectrum/primary_mass_all.pdf}
     } 
	 \label{fig:BBHs_BBHmerg}
\end{figure*}

Figure~\ref{fig:BBHs_BBHmerg} shows the distribution of primary BH masses\footnote{Here and in the following, the primary and secondary BH are the most massive and least massive member of a BBH, respectively.}  at the end of our binary-evolution simulations. The upper panel shows all the bound BBHs, while the lower panel shows the sub-sample of BBHs that reach coalescence within the lifetime of the Universe  ($\approx{14}$~Gyr, \citealt{Planck2018}).
We also compare the models M20 (hereafter, fiducial model F)  and F19.

\begin{figure*}
	\centering
	\includegraphics[width=2\columnwidth]{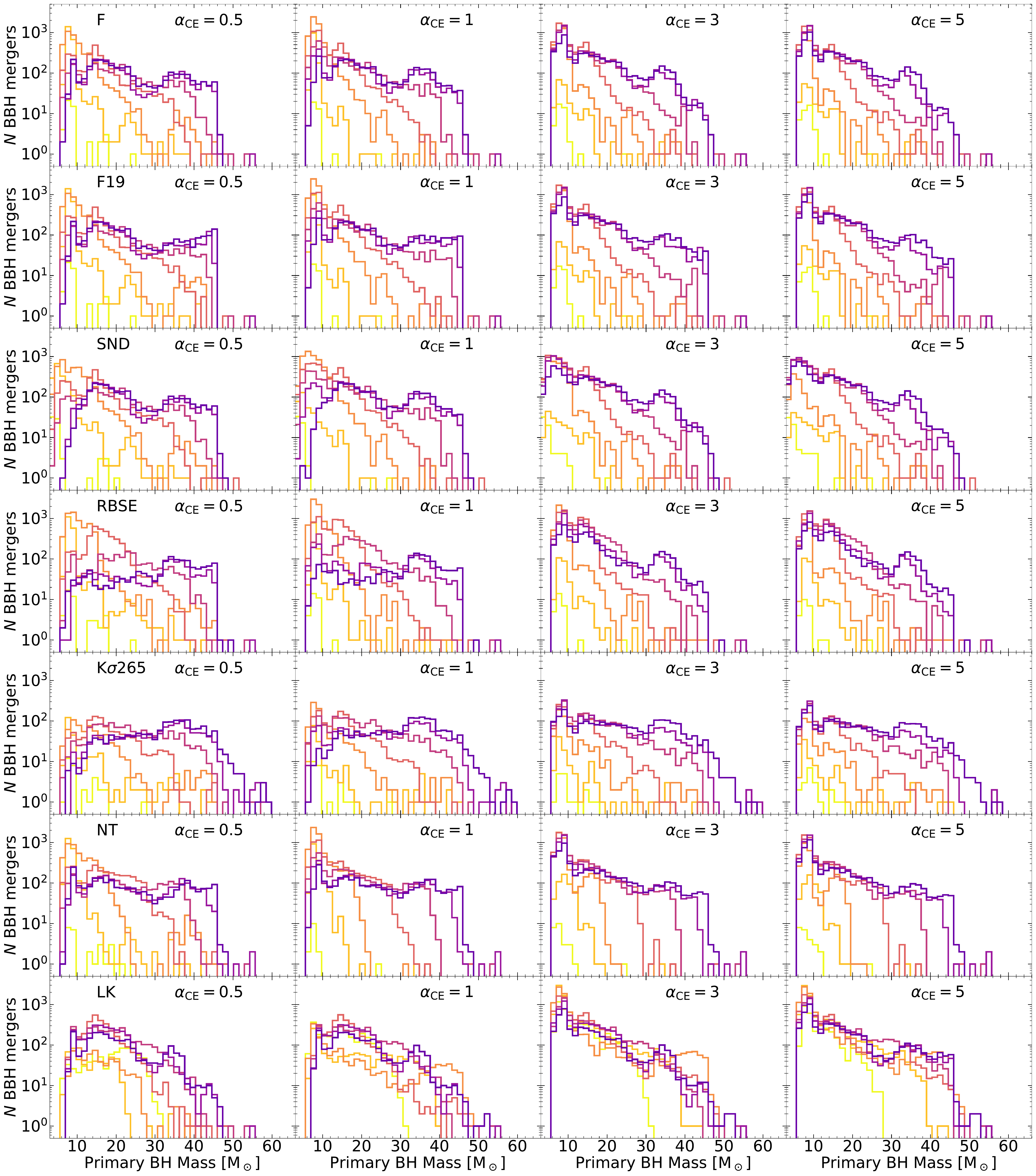}
	\caption{BBH mergers in our simulations. From left to right: $\alpha_{\rm CE}=0.5,$ 1, 3 and 5. The colour map refers to the metallicity $Z$ of the progenitor and is the same as in Fig.~\ref{fig:BBHs_BBHmerg} ($Z=0.017$, 0.008, 0.004, 0.002, 0.0008, 0.0004, and 0.0002). From top to bottom we show the models F, F19, SND, RBSE, K$\sigma{}$265, NT, and LK.
      \gitlab{https://gitlab.com/iogiul/iorio22_plot/-/tree/v3/Mspectrum}
     \gitbook{https://gitlab.com/iogiul/iorio22_plot/-/blob/v3/Mspectrum/MassDist.ipynb}
     \gitimage{https://gitlab.com/iogiul/iorio22_plot/-/blob/v3/Mspectrum/primary_mass_merg.pdf}
 }
	\label{fig:BBHmerg}
\end{figure*}

The maximum BH mass depends on metallicity: higher (lower) mass BHs form from metal-poor (metal-rich) stars because of stellar winds. Binary evolution processes do not change this result, as already reported by many previous studies \citep{dominik2012,mapelli2013,GiacobboCOB,vanson2022}.

The maximum mass of the primary BH in loose BBH systems can be significantly higher than that of the primary BH in BBH mergers. This mainly springs from the assumption that any residual H-rich  envelope collapses to a BH directly if the core-collapse supernova fails. 
In fact, when a binary star is tight enough to evolve into a BBH merger, it undergoes several mass transfer and/or CE phases, which lead to the complete ejection of the stellar envelope. Hence, the two resulting BHs form out of the naked cores of the two progenitor stars, and their mass cannot be $\gg{50}$~M$_\odot$ \citep{GiacobboCOB}.

In contrast, metal-poor single massive stars and massive stars in loose binary systems do not dissipate mass via RLO/CE, retaining a portion of their H-rich  envelope until the onset of core collapse, and can directly collapse to BHs. Hence, the maximum BH mass in loose binary systems is $\approx{80}$ M$_\odot$ ($\approx{100}$ M$_\odot$) in the M20 (F19) scenario.

This effect can contribute to dilute the PISN mass gap, because the genuine edge of the PISN (i.e., the maximum mass of a BH born from a single massive star) starts at $\approx{80-100}$~M$_\odot$, but the maximum mass of BHs in tight binary systems (BBH mergers from isolated binary star evolution) is only $\approx{50}$ M$_\odot$. 

The \textcolor{black}{ third GW transient catalogue  (hereafter, GWTC-3)}  shows that most primary BH masses in BBH mergers are  $\leq{}40$ M$_\odot$ \citep{abbottGWTC3popandrate}. This threshold might indicate that most BBH mergers 
\textcolor{black}{ in GWTC-3} come from isolated binary evolution and suffered from  mass transfer and/or CE. However, in dense star clusters, some of the BHs formed from single stars and loose BBHs might pair up with other BHs and produce merging systems with primary mass up to 80--100 M$_\odot$ \citep[e.g.,][]{mapelli2016,dicarlo2019,dicarlo2020a,dicarlo2020b,banerjee2020,torniamenti2022}. The long tail in the mass spectrum of primary BHs \textcolor{black}{ in GWTC-3}, extending up to $\sim{80}$~M$_\odot$   \citep{abbottGWTC3popandrate} might be populated by such oversized stellar-born  BHs,  rather than by hierarchical mergers \citep[e.g.,][]{miller2002,giersz2015,fragione2019,fragione2020,fragione2020b,kremer2020,mapelli2021,mapelli2022,mehta2022,arcasedda2018b,arcasedda2020,arcasedda2020b} or primordial BHs \citep[e.g.,][]{carr1974,carr2016,bird2016,alihaimoud2017,scelfo2018,deluca2020,deluca2021,franciolini2022,ng2022}.

Figure~\ref{fig:BBHmerg} shows the distribution of primary BH masses in BBH mergers, according to some of the main runs performed in this work. All the considered models show a common trend:
the percentage of low-mass primary BHs increases for larger values of $\alpha_\mathrm{CE}$, especially at low $Z$. In fact, low values of $\alpha_\mathrm{CE}$ tend to facilitate the premature coalescence of a binary system during CE. This suppresses the formation of low-mass BBHs, because their stellar progenitors have relatively small radii and easily merge during CE. In contrast, the efficiency of semi-major axis shrinking drops for large values of $\alpha_\mathrm{CE}$, favouring the survival of both low-mass and high-mass BBHs. The number of high-mass primary BHs  ($>20$ M$_\odot$) increases as $Z$ decreases (especially for $\alpha_{\rm CE}\leq{}1$) because only the most massive BHs merge within the Hubble time at low $Z$ \citep{costa2023}. 

At metallicity $Z\leq{}0.0004$, all our models show a local peak of the primary BH mass distribution at $35-45$ M$_\odot$, reminiscent of the excess found in 
\textcolor{black}{ GWTC-3} \citep{abbottGWTC3popandrate,farah2023,callister2023}. In model F19, the peak is shifted toward larger values ($\approx{45}$ M$_\odot$) than for all of the other models, because of the different treatment of PISNe. Our result suggests that the  peak at $35-40$ M$_\odot$ in the primary BH mass distribution is produced by the interplay between the PISN model and the maximum He core mass of merging BBH progenitors at a given metallicity.

Low-mass BBH mergers are rare in models K$\sigma{}$150 and especially K$\sigma$265 because of their large natal kicks. If natal kicks are large, only the binary systems with the highest binding energy (i.e., the most massive systems) tend to survive. The primary BH mass distribution in the SND models extends to lower values, because the  delayed supernova model produces BHs with mass as low as  3 \Msun{} by construction \citep{fryer2012}. 

Finally, the structure parameter $\lambda{}_{\rm CE}$ has a virtually large impact on the mass spectrum of BBH mergers. Our choice of $\lambda{}_{\rm CE}$ tends to select the typical mass of BBHs merging within the lifetime of the Universe. Hence, a self-consistent choice of  $\lambda{}_{\rm CE}$ is particularly important to capture the BBH mass spectrum \textcolor{black}{ (Sgalletta et al., in prep.)}.

\subsection{Formation channels} \label{sec:channels}

\subsubsection{Classification of formation channels}

In order to discuss the evolutionary paths leading to the formation of merging BCOs, we identify four main formation channels generalising the  classification adopted by \cite{broekgaarden2021}. Channel I includes all the systems that undergo a stable mass transfer before the first compact remnant formation, and later evolve through at least one CE phase. This channel is traditionally considered the most common 
formation channel of BCOs \citep[see,  e.g.,][]{vandeheuvel73,Tauris06,belczynski2018,neijssel2019,Mandel20}. 

Channel II comprises the systems that interact only through at least one stable mass transfer episode \citep[see,  e.g.,][]{Pavlovskii17,vandenheuvel17,GiacobboWind,neijssel2019,Mandel20,Marchant21,gallegosgarcia2021}.

Channel III comprises the systems that trigger at least one CE before the formation of the first compact remnant. Moreover, at the  time of the first compact remnant formation, the system is composed of one H-rich star and one star without H envelope (pure-He or naked-CO star).
The large majority of the systems in this channel pass through a single CE evolution (before the first compact remnant formation) in which the least evolved star has not developed a core yet (single-core CE). This last scenario is equivalent to the definition of channel III in \cite{broekgaarden2021} \cite[see also][]{Schneider2015}.

Channel IV is similar to channel III, but at the time of the first compact remnant formation, both  stars have lost their H envelope. 
The most common evolution route  includes  a single CE evolution (before the first compact remnant formation) in which both stars have a clear core-envelope separation (double-core CE).
 This last scenario is equivalent to the definition of channel IV  in \cite{broekgaarden2021}. This channel is discussed also in other works \citep[e.g.,][]{brown1995,bethe1998,dewi2006,justham2011,VG2018}.

The least frequent, almost negligible, channels include  no interactions during the whole binary evolution (Channel 0) 
and no interactions before the formation of the first compact object (Channel V). Since the binary systems belonging to  channels 0 and V do not interact before the first  supernova kick, such channels   are populated only by systems that receive \emph{lucky} supernova kicks that help to reduce the semi-major axis and/or increase the eccentricity reducing the GW merger time \cite[see, e.g.,][]{broekgaarden2021}.

Table~\ref{tab:channels} summarises the  percentages of merging BCOs formed through the four main channels as a function of $\alpha_\mathrm{CE}$ for the fiducial model (F). 
Figure~\ref{fig:channelsF} shows the formation-channel fractions for the merging BCOs   as a function of  metallicity.  Figure~\ref{fig:channelsFic} shows the cumulative distributions of the primary ZAMS mass, primary compact remnant mass  and initial orbital separation for the  merging BCOs that populate the main formation channels.
Finally, Fig.~\ref{fig:channelsMultimodel}  displays the formation channel  fractions for a sample of alternative  models. 

Table~\ref{tab:channels} indicates that for BBHs and BHNSs higher values of $\alpha_\mathrm{CE}$ favour channels that imply at least one CE episode (channels I, III, and IV). 
This is expected since larger $\alpha_\mathrm{CE}$ values allow more systems to survive CE evolution. 
Merging BNSs cannot form  through stable mass transfer only (channel II), therefore variations of $\alpha_\mathrm{CE}$  change the relative fractions of the other three channels. In particular, channel I becomes progressively dominant with increasing $\alpha_\mathrm{CE}$.

\begin{table}
\centering
\begin{tabular}{cc|cccccc|cccccc|ccccc}
\cline{2-19}
\multicolumn{1}{l|}{}   & &
  \multicolumn{6}{c|}{\begin{tabular}[c]{@{}c@{}}BBHs\\ Channels (\%)\end{tabular}} &
  \multicolumn{6}{c|}{\begin{tabular}[c]{@{}c@{}}BNSs\\ Channels (\%)\end{tabular}} &
  \multicolumn{5}{c}{\begin{tabular}[c]{@{}c@{}}BHNSs\\ Channels (\%)\end{tabular}} \\ \hline
\multicolumn{1}{l|}{$\alpha_\mathrm{CE}$} & & & I    & II   & III & IV  & & & I    & II   & III  & IV &  &  & I    & II  & III  & IV  \\ \hline
0.5                                      & & & 36 & 43 & 2 & 18 & & & 21 & 0 & 17 & 62 & &  & 44 & 31 & 15 & 7 \\
1                                        & & & 39 & 35 & 5 & 21 & & & 49 & 0 & 35 & 16 & &  & 51 & 23 & 18 & 7 \\
3                                        & & & 45 & 27 & 2 & 25 & & & 49 & 0 & 29 & 22 & &  & 52 & 24 & 14 & 10 \\
5                                        & & & 42 & 30 & 2 & 26 & & & 70 & 0 & 14 & 15 & &  & 56 & 23 & 11 & 9 \\ \hline
\end{tabular}
\caption{\textcolor{black}{ Overall percentage (summing up over the  simulated metallicities, Section~\protect\ref{sec:ic}) of  BCOs 
that merge within 14 Gyr in the fiducial model (Section~\protect\ref{sec:models}) formed through a given evolutionary channel: I, II, III, IV (Section~\protect\ref{sec:channels}). }}
\label{tab:channels}
\end{table}

\subsubsection{Formation channels of BBH mergers} \label{sec:bbhform}

Considering the whole merging BBH population (all sampled $\alpha_\mathrm{CE}$ and $Z$) in the fiducial (F) model, the formation channels I and II are the most common ones ($\approx 41\%$ and $\approx 32\%$, respectively)  followed by  channel IV ($\approx 23\%$) and channel III ($\lesssim3\%$).  
In channel I, $\approx99\%$ of the systems undergo just one   CE  after the first compact remnant formation.
 Most of the mass transfer episodes in channel II ($\approx94\%$) cause the complete stripping of the H-rich  envelope of the donor star.
Binaries in channel III go through subsequent stable mass transfer episodes ($\approx25 \%$) or an additional CE  ($\approx70 \%$) after the formation of the first compact remnant, while in channel IV most of the systems ($\approx99\%$) do not experience  any CE after the first compact remnant formation.

Figure~\ref{fig:channelsF} and Table~\ref{tab:channels} indicate that the relative fraction of formation channels  only mildly depend on $\alpha_\mathrm{CE}$. Metallicity has a significant impact on channels I and II, but their cumulative contribution is almost constant up to $Z=0.01$ where  channel IV begins to dominate (Fig.~\ref{fig:channelsF}).

Channel I is mainly ($\approx 98 \%$) populated by  binary systems  
that have the right radius and phase evolution to trigger a stable mass transfer before the first compact remnant formation and a following CE  capable to shrink the orbit enough to produce merging BBHs. 
Since the  relation between the radius and the evolutionary phase varies for different metallicites (see, e.g., the middle panels in Fig.~\ref{fig:trackcompare}), the fraction of channel~I systems wildly depends on $Z$ (see Fig.~\ref{fig:channelsF}).
This formation channel produces light BBHs (primary BH mass  $\lesssim 12 \ \Msun$, see Fig.~\ref{fig:channelsFic}).

High-mass binaries including primary stars with masses within 40--80 \Msun{} produce BBHs preferentially through channel II (Fig.~\ref{fig:channelsFic}).  
Most of such systems are in  tight initial configurations (Fig.~\ref{fig:channelsFic}). Therefore, they are able to interact during the early evolutionary stages in which the stellar envelopes are radiative favouring stable mass transfer that removes the whole stellar envelope.   

The distribution of  channel III BBHs is bimodal: $70 \%$ of the BBHs form from low-mass progenitors, while the others are massive BBHs produced by massive metal-poor progenitors  (Fig.~\ref{fig:channelsFic}).

Channel IV  is  populated by peculiar binaries of  twin stars (mass-ratio $\gtrsim0.9$) that evolve almost synchronously triggering a double-core CE.
For $Z<0.001$, channel IV produces massive BBHs (primary mass up to $45 \ \Msun$, see Fig.~\ref{fig:channelsFic}) with high mass ratio ($q\approx1$). 
At high metallicity ($Z>0.01$), the pure-He stars produced  after  CE ($M\lesssim 15 \ \Msun$, see Fig.~\ref{fig:mpisn0v05}) undergo significant  wind  mass-loss turning into relatively low-mass BHs ($\lesssim 9 \ \Msun$, see Figs.~\ref{fig:channelsF} and  ~\ref{fig:mpisnHe}).

The quasi-homogeneous evolution (model QHE) produces more compact stars after stable RLO mass transfers 
quenching binary interactions. Hence, this model suppresses the channels that depend on stable mass transfer episodes (channels I and II, Fig.~\ref{fig:channelsMultimodel}).
In contrast, the almost conservative mass transfer assumed in the RBSE model (Eq.~\ref{eq:mtalternative}),  favours channel I over channel II (Fig.~\ref{fig:channelsMultimodel}).

Larger natal kicks (K$\sigma$150, K$\sigma$265) 
tend to  randomise  the binary properties  after the supernova kick. As a consequence, the merging BBHs are uniformly distributed among the main formation channels in the whole metallicity range  (Fig.~\ref{fig:channelsMultimodel}).  
Systems that survive large natal kicks  produce binaries with large eccentricities,  
increasing the possibility of triggering a collision at periastron (Fig.~\ref{fig:channelsFic}) and  reducing the GW merger time (Section~\ref{sec:gw}). 
Hence, models K$\sigma$150 and K$\sigma$265 produce the largest fraction of  BBHs from massive binaries ($>100$ \Msun) evolving through channels I, II and III. As a consequence, these models produce also the  largest number of BBHs hosting  massive primary BHs ($>40$ \Msun) among all the tested models  (Fig.~\ref{fig:BBHmerg}).

The model NTC totally suppresses  collisions at periastron, but this does not strongly affect the final results, 
highlighting the relative low importance of such processes for the formation of merging BBHs in our fiducial model.


The higher binding energy predicted by the models LK, LX and LC  (Appendix~\ref{app:ebind}) produces tighter BBHs after CE. As a consequence, channel I becomes accessible to systems with primary stars within the whole ZAMS mass range  (20--150 \Msun{}). 
The inclusion of new systems boosts channel I  especially at high metallicities, producing  massive BBHs (BH primary mass up to $30$ \Msun{}).

The fiducial models with $\alpha_\mathrm{CE}>1$ are qualitatively in agreement with the result by \cite{neijssel2019} (see their Fig.~1).
Our results are consistent with the work by \cite{Combine}, in which the large majority ($> 90 \%$) of  BBH mergers in Galactic-like environments ($Z=0.0088$) form through channel~I (defined as channel C in their Table~C1). 
In \cite{dominik2012}, channel~I represents almost the only way to form merging BBHs both  at solar ($f_\mathrm{CI}\approx 99 \%$) and subsolar ($Z=0.1 Z_\odot$, $f_\mathrm{CI}\approx 93 \%$)  metallicity.
In contrast,  at subsolar metallicity  ($0.0014<Z\lesssim0.002$), almost $50\%$ of our BBHs form through evolution routes alternative to channel~I.

\subsubsection{Formation channels of BNS mergers} \label{sec:bnsform}

\begin{figure*}
	\centering
	\includegraphics[width=2.0\columnwidth]{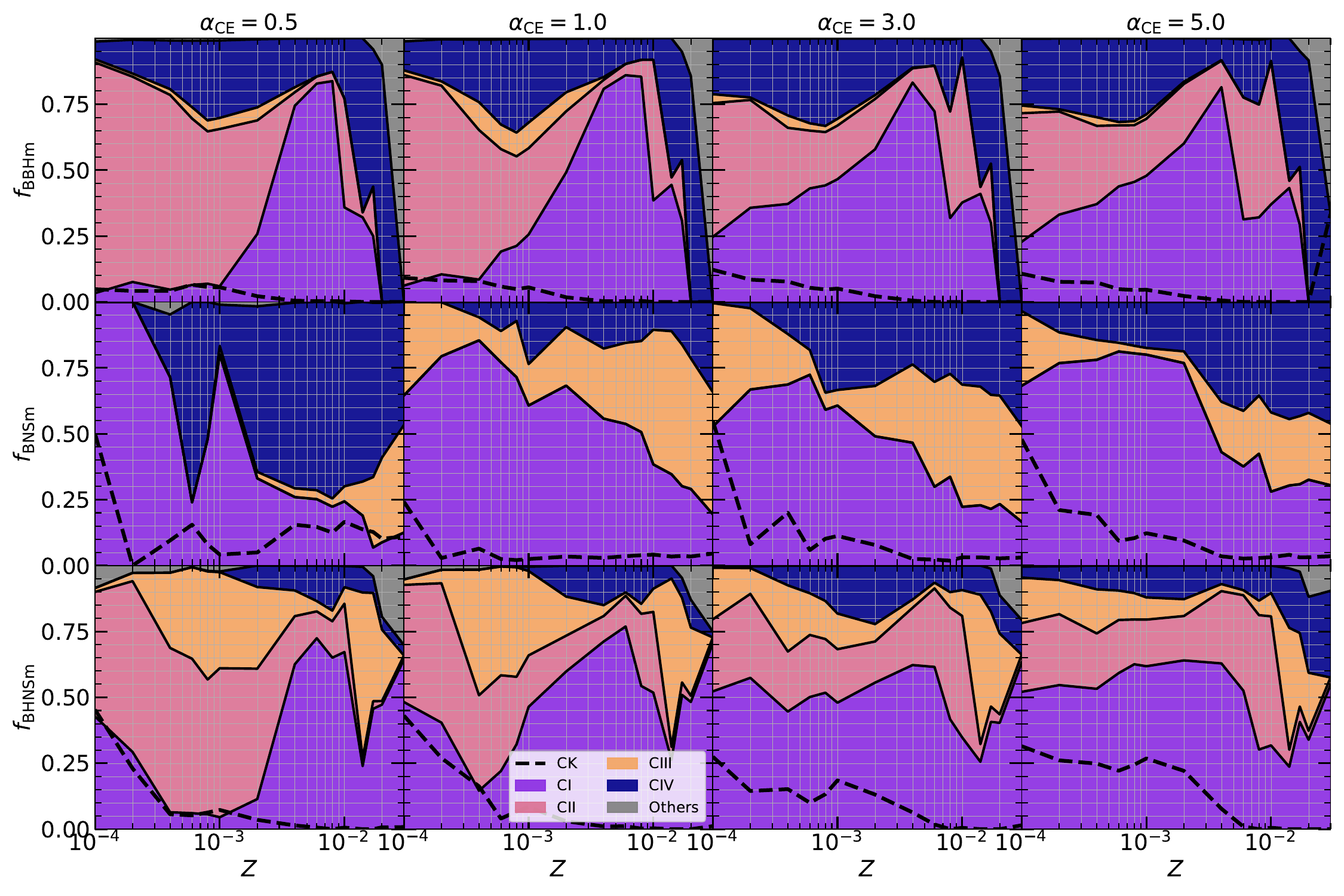}
	\caption{ Relative fraction of systems produced through the four main formation channels  (Section~\protect\ref{sec:channels})  as a function of the metallicity  for  BBHs (upper panels), BNSs (middle panels) and BHNSs (lower panels) that merge within 14 Gyr in our 
     fiducial model (Section~\protect\ref{sec:models}). 
          The panels in each column refer to a different value of the CE efficiency parameter, $\alpha_\mathrm{CE}$, as reported in the titles (Section~\protect\ref{sec:CE}).
          The dashed lines indicate the fraction of systems that undergo at least one CE triggered by a collision at periastron (Section~\protect\ref{sec:circ}). 
          The online repository (\gitlab{https://gitlab.com/iogiul/iorio22_plot/-/tree/v3/formation_channels}) 
          contains  versions of this plot made for the alternative setup models, and alternative versions of this plot made considering the  number of objects rather than the relative fractions.
           \gitbook{https://gitlab.com/iogiul/iorio22_plot/-/blob/v3/formation_channels/Channel_plots.ipynb} 
          \gitimage{https://gitlab.com/iogiul/iorio22_plot/-/blob/v3/formation_channels/plot_channell_aoe_F.pdf}}
	\label{fig:channelsF}
\end{figure*}

Most merging BNSs in the fiducial model form through channel~I ($\approx 59 \%$). The other merging BNS progenitors evolve following  formation channels III ($\approx 22 \%$) and  IV ($\approx 18 \%$). 
In agreement with previous studies \cite[e.g.,][]{GiacobboCOB,Combine,VG2018},
we find that is not possible to produce BNS mergers just through stable mass transfer episodes (channel II).

Since CE evolution is crucial for the formation of BNSs (all the BNS progenitors undergo at least two CE episodes), the relative formation channel  fraction strongly depends on the parameter $\alpha_\mathrm{CE}$ (Table~\ref{tab:channels} and Fig.~\ref{fig:channelsF}). 

Metallicity has a significant impact on the evolution of NS progenitors  (see, e.g., the first two rows in Fig.~\ref{fig:trackcompare}). 
In particular, metal-poor stars tend to interact after core He burning, while  metal-rich  stars interact during the HG or giant-branch phase (Table~\ref{tab:phases}). 
The stellar phase is important to distinguish between stable and unstable mass transfer. 
Moreover, stars with similar radii but in  different  evolution phases can have different 
envelope binding energy favouring or disfavouring CE ejection. 
As a consequence, the relative formation-channel fractions vary significantly with metallicity.

The least massive BNS progenitors evolve through channel I and channel III (see Fig.~\ref{fig:channelsFic}). 
Almost all these binaries ($99\%$) undergo an additional second CE episode when the pure-He secondary expands starting a new unstable RLO (case BB mass transfer, see e.g.\ \citealt{broekgaarden2021}).

Most of the systems  evolving through channel IV (97\%) do not activate a double-core CE, rather they undergo a first stable RLO in which the primary star loses the H-rich  envelope. Later on, the secondary star begins an unstable RLO and 
expels the H-rich  envelope after CE. Before the first NS formation, in almost half of the systems,  the primary star triggers an additional CE turning into a naked-CO star (unstable case BB mass transfer). 
After the first NS formation, the pure-He secondary star triggers an additional  CE episode in 90\% of the cases.  
The minority of binaries that undergo a double-core CE ($\approx 3\%$) contain either massive NS progenitors close to the NS/BH boundary ($\approx 20 \ \Msun$)  or light progenitors ($\approx 11 \Msun$) in an initial wide configuration ($a_\mathrm{ini} \gtrsim 1000 \ \Rsun{}$).


The evolution of the NS progenitors along the HG phase plays an important role in all the three main formation channels, especially at intermediate/high metallicity. 
For $Z>0.001$, the first interaction between the secondary star and the already formed NS begins when the star expands during  the HG phase  (up to $300 \ \Rsun{}$, see Fig.~\ref{fig:trackcompare}). 
In the fiducial  model, stars in the HG phase are always stable and the RLO mass transfer continues until the secondary star changes the \bse{} stellar type. At that point,   the mass transfer becomes unstable due to  large secondary-to-NS mass ratio ($q\gtrsim10$).

In the alternative model QCBSE (Table~\ref{tab:qc}), all the secondary star--NS interactions during the HG phase lead to a direct merger. 
Therefore, the number of BNS progenitors decreases in all the formation channels, but 
the suppression is maximum for channels I and III (Fig.~\ref{fig:channelsMultimodel}).

The model variations of the RBSE and QHE model (at low metallicity) reduce the possibility to start an interaction after the first stable mass transfer reducing the number of channel I BNSs (Fig.~\ref{fig:channelsMultimodel}).

Larger natal kick (models K$\sigma$150 and K$\sigma$265) can easily break the binary after the first NS formation reducing the number of BNSs, except for the tightest ones produced through channel IV (Fig.~\ref{fig:channelsMultimodel}).

Higher envelope binding energies (models LK and LC, see Appendix~\ref{app:ebind})
drastically reduce the number of
BNSs for $\alpha_\mathrm{CE}<1$: such simulations do not produce merging BNSs except for a few peculiar systems at $Z>0.002$. 
Such systems,  evolving through channel IV,  trigger the first CE between a pure-He star  and a partially stripped H-rich star,  then they avoid any interactions after the first NS formation. The channel fractions in model LX are similar to the fiducial model for $\alpha_\mathrm{CE}>1$, and similar to the  LK and LC models in the other cases. 

The relative formation-channel fractions in the other models do not show
significant differences with respect to the fiducial model. 

Both \cite{VG2018} and \cite{Combine} found that formation channel I still dominates ($\gtrsim70 \%$) at high metallicity ($Z=0.014$ and $Z=0.0088$, respectively). In \cite{dominik2012}, the channel I fraction (channel NSNS01 and NSNS03 in their Table~4)  is $\approx 87 \%$  at  $Z=0.02$. 
In contrast, in all our tested models,  the fraction of BNSs formed through channel I  is  always $\lesssim 50 \%$ for $Z>0.008$.
Interestingly, the models in which  the fraction  drops to $\approx 0$  are the most similar ones  (QCBSE and RBSE) to the binary-evolution models by \cite{VG2018} and \cite{dominik2012}.

This large discrepancy derives from two important differences: their optimistic (versus our pessimistic) assumption for CE during the HG phase, and the stellar evolution models. 
We test the optimistic assumption in the OPT model (Fig.~\ref{fig:channelsMultimodel}), and  find that only in the case of $\alpha_\mathrm{CE}=5$ the channel I fraction reaches $\approx 50 \%$ at high metallicity. In  all the other cases, channel I remains subdominant and its fraction even decreases for $\alpha_\mathrm{CE}\leq1$. 
Therefore,  we conclude that the stellar evolution is the main driver of the discrepancy between our channel fractions and those of \cite{dominik2012} and \cite{VG2018}.

Both \cite{dominik2012} and  \cite{VG2018}  used \bse{}-like codes ({\sc startrack} and {\sc compas}), so the difference between their stellar evolution model (based on \citealt{Pols98}) and \parsec{} can be appreciated  in Fig.~\ref{fig:trackcompare} (see also Section~\ref{sec:trackcompare}).
In the mass range of NS progenitors, the \bse{}-like stellar  tracks do not  show a strong dependence on  metallicity and interact mostly after core He burning for  $M_\mathrm{ZAMS}>12 \ \Msun$. 
In contrast, the \parsec{} stellar tracks are markedly different at different metallicity and most of the interactions at $Z>0.001$ are triggered 
during the HG phase leading directly to a merger in the case of  unstable RLO. 
Even considering the optimistic CE model, the binding energies in the HG phase are so high (Appendix~\ref{app:ebind}) that most of the CEs end with a coalescence.

\subsubsection{Formation channels of BHNS mergers} \label{sec:bhnsform}

In the fiducial model (F) 
most 
BHNS mergers form through channel~I ($\approx 51\%$), followed by channel II ($\approx 25 \%$),  III ($\approx 15\%$), and  IV ($\approx 8\%$).
Table~\ref{tab:channels} and Fig.~\ref{fig:channelsF} show that the relative formation-channel fraction remains almost constant for all the values of \ace{}, in the whole metallicity range. The largest differences are found for low  \ace{} values and low/intermediate metallicity, in which  channels I and III are suppressed in favour of channel II, and at high metallicity  where channel II  drops to $\approx 0$\% and channel III rises up to $\approx 30\mathrm{-}50\%$.

\begin{figure}
	\centering
 	\includegraphics[trim={1cm 0 0 3.6cm},width=1.1\columnwidth, clip]{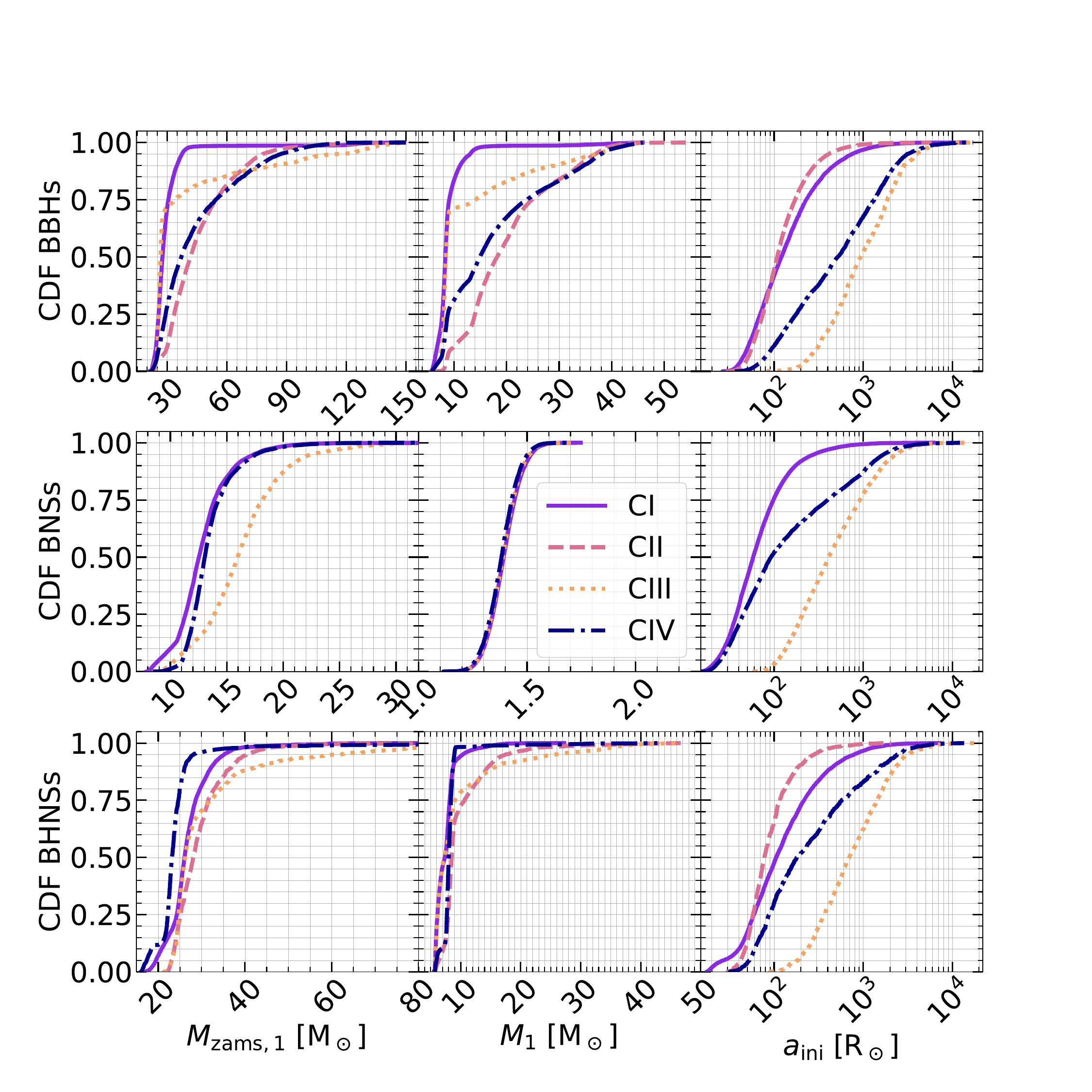}
 	\vspace*{-8mm}
  \caption{
  Cumulative distribution function  for a sample of   properties (left-hand column: ZAMS mass of the primary star; middle column: mass  of the primary compact remnant; right-hand column: initial semi-major axis of  BBHs (upper row), BNSs (central rows) and BHNSs (lower row) that merge within 14 Gyr in the fiducial model  (Section~\ref{sec:models}).  
  The primary is always the most massive object in the binary. Due to binary interactions,  the primary compact remnant can be produced by the secondary star and vice-versa. 
  For each BCO population, we consider all the sampled $\alpha_\mathrm{CE}$ and $Z$ values (Section~\ref{sec:ic}).
  The different lines indicate the four main formation channels (Section~\ref{sec:channels}): I (violet solid), II (pink dashed), III (orange dotted), and IV (blue dot-dashed line).
  The versions of this plot made for the alternative setup models can be found in the online repository (\gitlab{https://gitlab.com/iogiul/iorio22_plot/-/tree/v3/formation_channels}). \gitbook{https://gitlab.com/iogiul/iorio22_plot/-/blob/v3/formation_channels/Plot_cdf.ipynb} 
          \gitimage{https://gitlab.com/iogiul/iorio22_plot/-/blob/v3/formation_channels/summarycdf_channel_F.pdf} }
	\label{fig:channelsFic}
\end{figure}

The most massive BHNS progenitors 
follow 
channel III 
producing the most massive merging BHNSs (Fig.~\ref{fig:channelsFic}). 
Compared to BBHs and BNSs, the  contribution of channel IV  decreases in the whole   \ace{} and  $Z$ range. 
This channel is populated by stars with similar ZAMS mass evolving almost synchronously (see Sections \ref{sec:bbhform} and \ref{sec:bnsform}). 
In the case of BHNS progenitors, this means selecting peculiar systems in a small mass range close to the NS/BH mass boundary ($\approx 22 \ \Msun$, see Fig.~\ref{fig:channelsFic}). Half of the systems trigger  
a double-core CE in the late evolutionary phases (\sevn{} phase 5 or 6, see Table~\ref{tab:phases}). 
The other systems have an initial tighter configuration ($a_\mathrm{ini}\approx 50\mathrm{-}200 \ \Rsun$) and pass through an episode of stable mass transfer before triggering the first CE.

\begin{figure*}
	\centering
	\includegraphics[width=2.0\columnwidth]{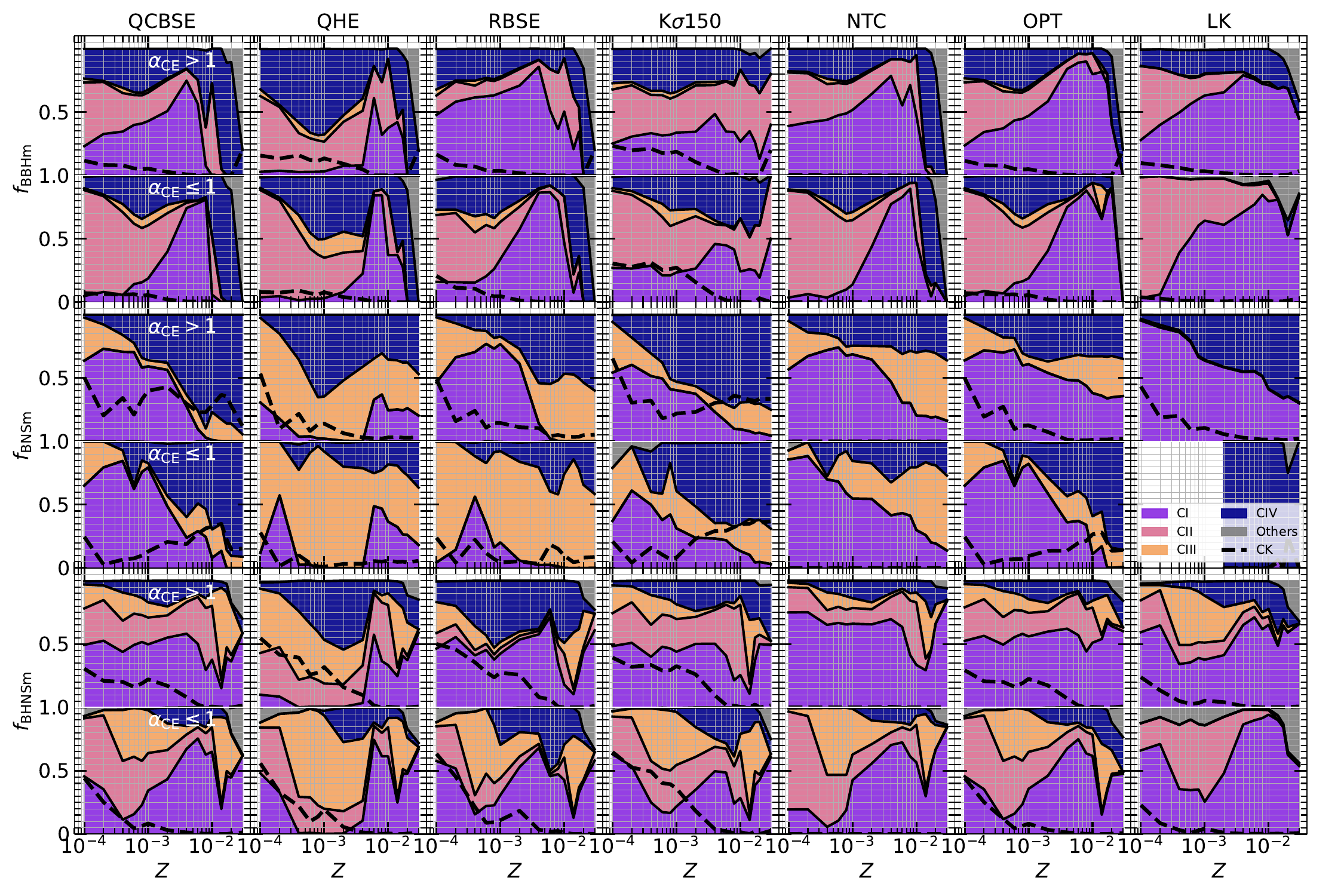}
  	\vspace*{-5mm}
	\caption{Same as Fig. \ref{fig:channelsF}, but showing the relative formation-channel fraction for different  models as indicated in the titles (see Table~\protect\ref{tab:models} and Section~\protect\ref{sec:models}). In each panel, the bottom part of the plot shows the channel fractions considering all  the merging BCOs from the simulations with low-$\alpha_\mathrm{CE}$ values (0.5, 1), while the top part shows the results considering all  the simulations with high-$\alpha_\mathrm{CE}$ values (3, 5). The lack of merging BNSs in the LK simulations for $\alpha_\mathrm{CE}\leq1$ produces the white empty region in the corresponding panel. 
      \gitlab{https://gitlab.com/iogiul/iorio22_plot/-/tree/v3/formation_channels} \gitbook{https://gitlab.com/iogiul/iorio22_plot/-/blob/v3/formation_channels/Channel_plots.ipynb} 
          \gitimage{https://gitlab.com/iogiul/iorio22_plot/-/blob/v3/formation_channels/channel_aoe_multimodel.pdf}}
	\label{fig:channelsMultimodel}
\end{figure*}

Figure~\ref{fig:channelsMultimodel} shows that the variation of simulation parameters does not have a strong impact in the relative channel fraction of merging BHNSs.
The most relevant differences  are present in the model LK (for low \ace{} values), in which the higher binding energies (Appendix~\ref{app:ebind}) totally suppress 
channels III and  IV.  

The results of all our simulations do not agree with the recent results by \cite{broekgaarden2021}, in which almost all the merging BHNSs are formed through  channel I ($86 \%$) and only $8  \%$ of the progenitors evolve through channel II ($4  \%$) and III ($4  \%$).  
In their work, the relative fractions refer to the systems that are detectable by LIGO and Virgo, so they are biased toward binaries with high metallicity  ($Z\gtrsim 0.008$)  hosting massive BHs. In their simulations, such systems form preferentially through channel I. 
In our case, instead, the same  \vir{selection effects}  should boost the percentage of channel III BHNSs, increasing even more the discrepancy (Figs. \ref{fig:channelsF} and \ref{fig:channelsFic}).  
Similarly, our results do not agree with \cite{dominik2012} in which 97\% of the merging BHNSs belong to channel I at $Z=0.02$, while at $Z=0.002$ the channel I fraction decreases to 25\%, and most  of the merging BHNSs ($\approx67\%$) form through channels~III and IV. 
Since the overall number of merging BHNSs in  \cite{broekgaarden2021} is comparable with our results (Fig.~\ref{fig:meff} and Section~\ref{sec:BHNSmeff}), we conclude that the  differences are mostly driven by the different stellar evolution models (both \citealt{dominik2012} and \citealt{broekgaarden2021}   use  \bse{}-like code based on the \citealt{Pols98} stellar tracks). 

Our results are  similar to what found by \cite{Combine}. Assuming $\alpha_\mathrm{CE}=0.5$ and $Z=0.0088$, they estimate that $\approx79\%$ of the merging BHNS belong to channel I, while  the remaining systems evolve through channel II ($\approx19\%$).  
The fraction of merging BHNSs  evolving through stable mass transfers (channel II) is consistent with  our results for $\alpha_\mathrm{CE}=0.5$, but we find that a non-negligible fraction of merging BHNSs ($\approx20\%$) evolve through the CE channels III and IV.  The cause of such difference is  the larger  binding energy of stellar envelopes used in \combine{} (Appendix~\ref{app:SSE}). In our simulation model  with the highest envelope binding energy (LK, see Appendix~\ref{app:ebind}) and $\alpha_\mathrm{CE}\leq1$,  the merging BHNSs form solely through channel I ($\approx90\%$) and  II ($\approx10\%$) at $0.004\leq Z \leq0.01$ (Fig.~\ref{fig:channelsMultimodel}).


\subsection{Merger efficiency} \label{sec:meff}

We define the merger ($\eta$) and formation ($\eta_\mathrm{f}$) efficiency as 
\begin{equation}
\begin{split}
\eta &= \frac{N_\mathrm{BCO}(t_\mathrm{del}<14 \ \mathrm{Gyr})}{M_\mathrm{pop}} \\
\eta_\mathrm{f} &= \frac{N_\mathrm{BCO}}{M_\mathrm{pop}},
\end{split}
\end{equation}
where $N_\mathrm{BCO}$ is the number of  BCOs, $M_\mathrm{pop}$ is the total mass of the simulated stellar population, and $t_\mathrm{del}$ is the delay time, i.e. the time elapsed from the  beginning of the simulation to  the BCO merger. We estimate $M_\mathrm{pop}$ including the correction for the incomplete sample of the IMF (see Section~\ref{sec:ic}), and assuming that half of the population mass is stored in  binaries.

Figure \ref{fig:meff} shows  $\eta_\mathrm{f}$ and $\eta$ as a function of  metallicity for  BBHs, BNSs and BHNSs in our fiducial model.  
Figure~\ref{fig:meff} also compares our results with the merger efficiency found by  \cite{Spera19} using the previous version of \sevn{}  (assuming $\alpha_\mathrm{CE}=1$ and $\lambda_\mathrm{CE}=0.1$),  \cite{GiacobboKick} using \mobse{} ($\alpha_\mathrm{CE}=5$ and $\lambda_\mathrm{CE}$ prescriptions by \citealt{Claeys14}), and \cite{broekgaarden2022} using {\sc COMPAS} ($\alpha_\mathrm{CE}=1$ and $\lambda_\mathrm{CE}$ prescriptions by \citealt{XuLi10}). In Fig. \ref{fig:meffBHBH}, we compare the BBH merger efficiency of the default model with some of the alternative models. Figs.~\ref{fig:meffBNS} and  \ref{fig:meffBHNS} show the same comparison for BNSs and BHNSs.

\subsubsection{BBH merger efficiency} \label{sec:BBHmeff}

\begin{figure*}
	\centering
	\includegraphics[width=1.0\textwidth]{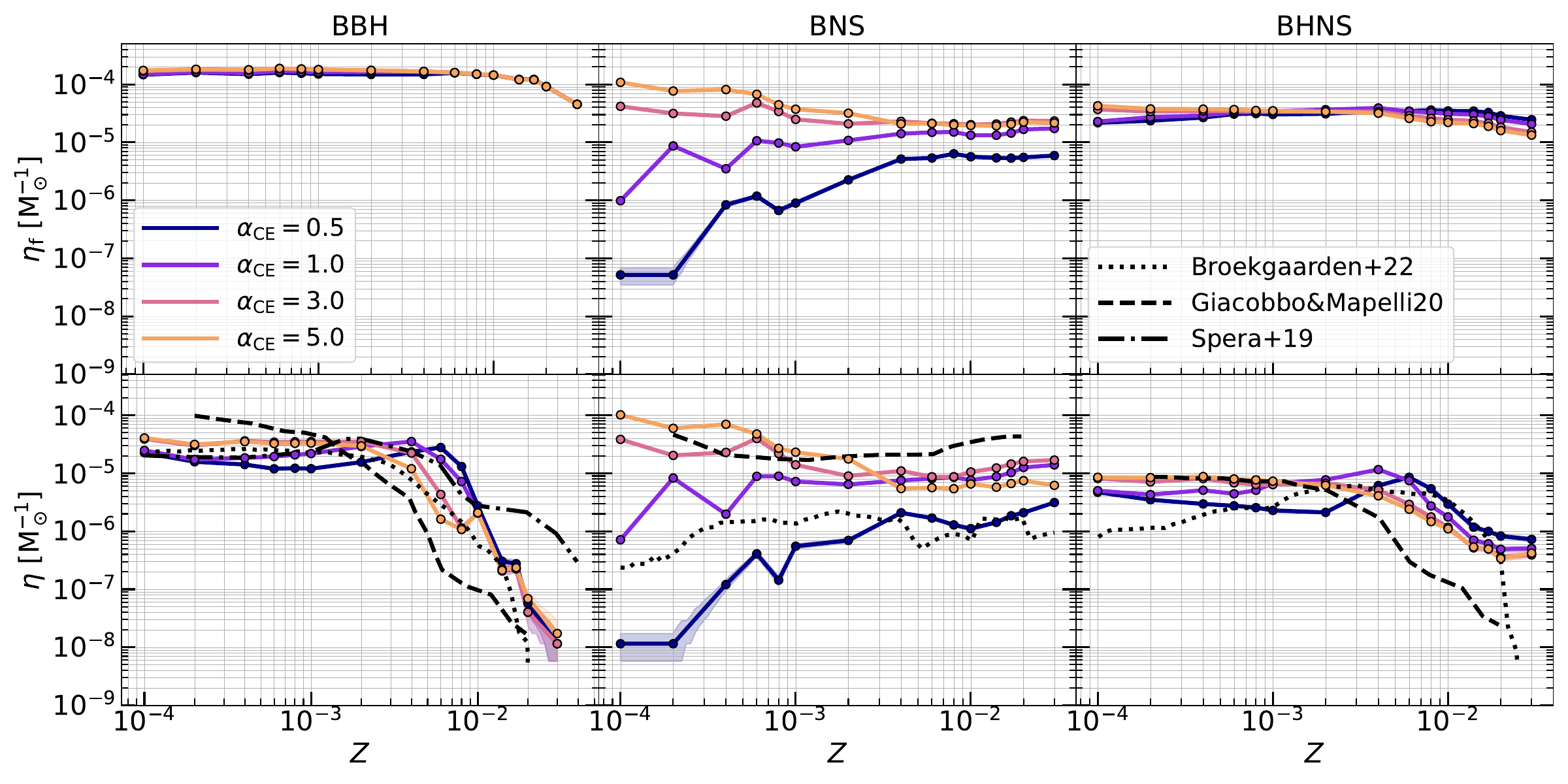}
	\vspace*{-5mm}
	\caption{Formation ($\eta_\mathrm{f}$) and merger ($\eta$) efficiency 
	for BBHs (left-hand panels), BNSs (middle panels) and BHNSs (right-hand panels) as a function of metallicity for the fiducial model (see Section \ref{sec:models}). The total simulated mass has been corrected for the binary fraction (assumed 0.5) and for the partial sampling of the IMF (Section \protect\ref{sec:ic}). The  line  colours show the results obtained assuming different CE efficiency parameters $\alpha_\mathrm{CE}$ (Section \protect\ref{sec:ic}). The shaded areas show the 0.16 and 0.84 percentiles obtained sampling from  Poissonian distributions. The black lines show the merger efficiency by other authors: \protect\cite{broekgaarden2022} (model A, dotted), \protect\cite{GiacobboKick} (model Ej1, dashed), \protect\cite{Spera19} (dot-dashed).
 The versions of this plot made for the alternative setup models can be found in the online repository (\gitlab{https://gitlab.com/iogiul/iorio22_plot/-/tree/v3/merger_efficiency}). 
           \gitbook{https://gitlab.com/iogiul/iorio22_plot/-/blob/v3/merger_efficiency/PLot_paper.ipynb}
           \gitimage{https://gitlab.com/iogiul/iorio22_plot/-/blob/v3/merger_efficiency/mergereff_F.pdf}}
\label{fig:meff} 
\end{figure*}

In the fiducial model, the formation efficiency of BBHs is almost constant at all metallicities and for all the sampled $\alpha_\mathrm{CE}$ ($\eta_\mathrm{f} \approx{10^{-4}} \ \mathrm{M}^{-1}_\odot$),  while the merger efficiency decreases from a few $\times{}10^{-5}$~M$_\odot$ for $Z<0.002$  to $10^{-6}\mathrm{-}10^{-8}$~M$_\odot$ at high metallicity ($Z>0.01$).

The differences between $\eta$ and $\eta_\mathrm{f}$ depend on the different dominant formation channels for BBHs and merging BBHs. Most BBH progenitors ($> 70 \%$) do not interact or interact only via stable mass transfer episodes, hence their final separation is  too large ($>100 \ \Rsun{}$) to make them merge in an Hubble time. 

The increasing importance of stellar winds at high metallicity reduces  $\eta$ for $Z\gtrsim 0.008$. In fact, stars losing a significant amount of mass during the evolution remain more compact (see, e.g., Fig.~\ref{fig:trackcompare}), reducing  binary interactions, and  produce less massive BHs increasing the BBH merger time (Figs.~\ref{fig:mpisn0v05} and~\ref{fig:BBHs_BBHmerg}). Figure~\ref{fig:meff} shows that CE efficiency has a much lower impact on the merger efficiency with respect to the metallicity. The largest differences are at  intermediate metallicities ($0.008<Z<0.004$), where almost 90\% of the BBH progenitors undergo at least one CE episode.

\begin{figure*}
	\centering
	\includegraphics[width=1.0\textwidth]{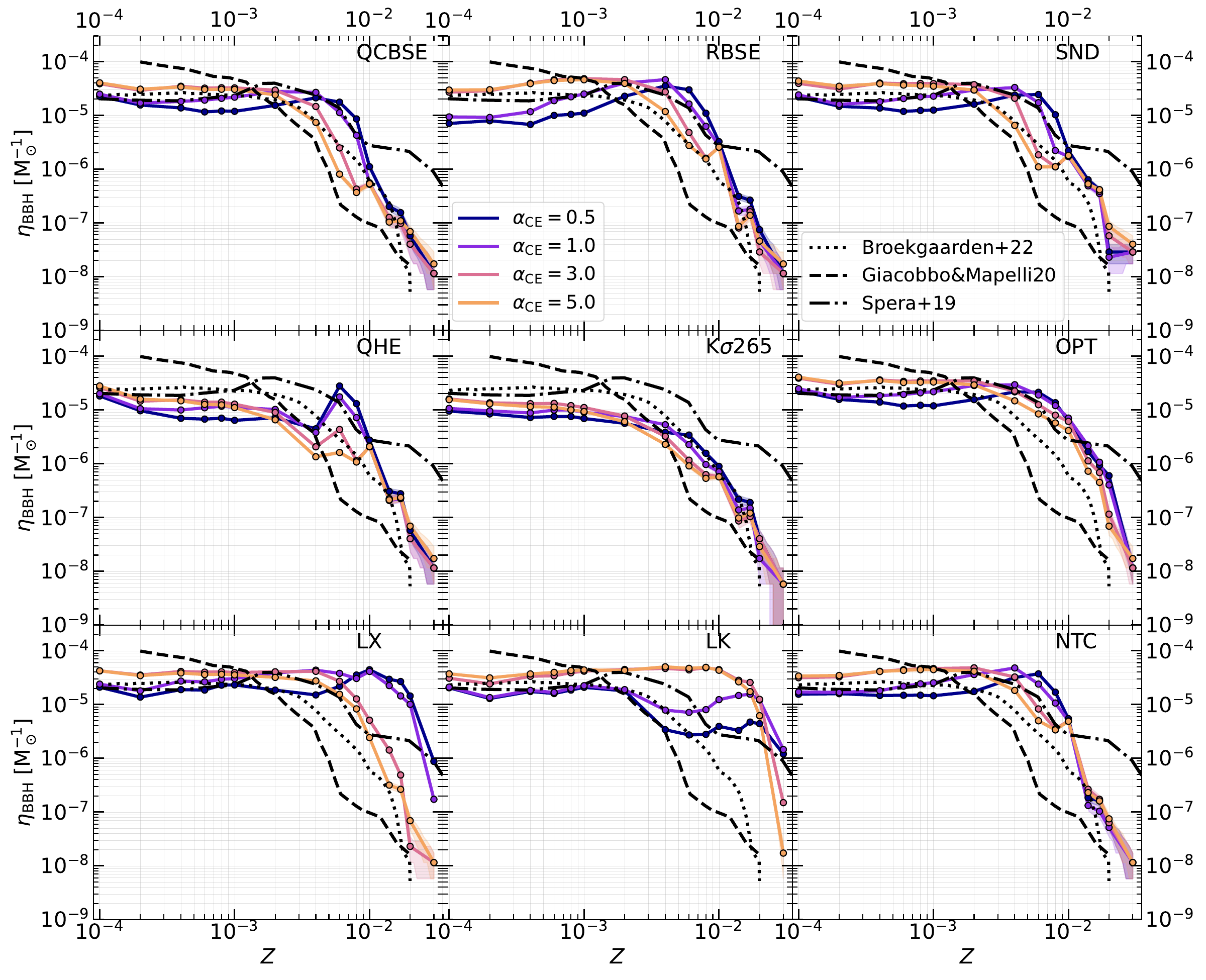}
	\vspace*{-5mm}
	\caption{Same as Fig. \ref{fig:meff}, but showing only the BBH merger efficiency for a selection of further  simulation models as reported by the labels in the top-right corner of each panel (Section \ref{sec:models} and Table~\ref{tab:models}). 
 \gitlab{https://gitlab.com/iogiul/iorio22_plot/-/tree/v3/merger_efficiency}
           \gitbook{https://gitlab.com/iogiul/iorio22_plot/-/blob/v3/merger_efficiency/PLot_paper.ipynb}
           \gitimage{https://gitlab.com/iogiul/iorio22_plot/-/blob/v3/merger_efficiency/multipanel_BHBHm.pdf}
    }
	\label{fig:meffBHBH} 
\end{figure*}

For $Z<0.01$ our results  are in agreement with \cite{Spera19}, especially for $ \alpha_\mathrm{CE}=1$.
At higher metallicity, the  simulations by \cite{Spera19} produce a significantly larger number of BBH mergers. 
This  happens because \cite{Spera19} adopt a constant  value $\lambda_\mathrm{CE}=0.1$, resulting in high binding energies.
 Higher binding energies combined with low $\alpha_\mathrm{CE}$ values let  more massive binaries produce tight BBHs through    channel I.  We find similar results using the LC model in which we also set $\lambda_\mathrm{CE}=0.1$ (Fig.~\ref{fig:meffBHBH}).

The BBH merger efficiency by \cite{GiacobboKick} shows a more steep gradient as a function of metallicity. 
At low metallicity, our simulations produce less BBHs by a factor 3--6. From $Z=0.002$ onward, our BBH merger efficiency becomes 10--100 times larger than what estimated by \cite{GiacobboKick}. 
This trend is present  in 
all our models (Fig.~\ref{fig:meffBHBH}). 
Therefore, 
this difference mostly springs from the different stellar evolution model. 

Our fiducial model with $\alpha_\mathrm{CE}>1$ shows a good agreement with the $\eta$ estimated by \cite{broekgaarden2022}. 
However, the two models are based on many different assumptions  (e.g., different values for $\alpha_\mathrm{CE}$ and $\lambda_\mathrm{CE}$,  different assumptions on the mass transfer stability). 
This comparison highlights how  the effects of binary and  stellar evolution are highly degenerate.

The merger efficiency  drops by up to a factor of 10  in the models QHE, K$\sigma$150  and K$\sigma$265. 
In QHE, the smaller radius of the secondary star reduces the chance of starting a binary interaction, 
while the high supernova kicks in the other two models break a large number of binaries.  
The differences are less evident at high-metallicity, where the quasi-homogeneous evolution is switched off 
and most 
BBH mergers form through   peculiar evolution routes (e.g., channel IV or \emph{lucky} kicks).

Models LX, LC  and LK produce a dramatic increment of BBH mergers  
at high metallicity, because of their high binding energies.
Merging BBHs at low metallicity ($Z\leq{}0.001$) form mainly through  channel II (stable mass transfer) so their number  is not significantly affected by changes in the envelope binding energy. 
Finally, the OPT model produces a factor of 2--10 more BBHs at intermediate and high metallicity.

\subsubsection{BNS merger efficiency} \label{sec:BNSmeff}

Given the low NS mass ($\approx 1.33 \ \Msun$), the only way  for BNS progenitors to survive to supernova kicks is through CE episodes  that shrink the semi-major axis and remove the stellar envelope producing low effective supernova kicks (Section~\ref{sec:snkicks}). Therefore, most of the formed BNSs are tight enough to merge within an Hubble time. As a consequence, the BNS formation and merger efficiency are similar, with the only exception of case $\alpha_{\rm CE}=0.5$ at low metallicity  (Fig.~\ref{fig:meff}). 

Since the formation of BNSs 
passes though at least one CE episode,  
their merger efficiency significantly depends  on $\alpha_\mathrm{CE}$, as already found in other works \cite[see,e. g.][]{VG2018,GiacobboKick,santoliquido2021,broekgaarden2022}. 
The trend of $\eta$ with progenitor's  metallicity also depends on the envelope binding energy, which  is higher for lower metallicity in our models (Appendix~\ref{app:ebind}).
For $\alpha_{\rm CE}=0.5$, we find the largest dependence of $\eta$ on progenitor's metallicity: $\eta$ decreases by 4 orders of magnitude from high to low metallicity. The formation of BNSs is suppressed  at low $Z$ and for $\alpha_\mathrm{CE}\le{}1$, because most CEs end with a premature coalescence. Vice versa, for $\alpha_{\rm CE}\ge{}3$, $\eta$ decreases as the metallicity increases, because larger values of $\alpha_{\rm CE}$ combined with lower binding energies produce wider post-CE systems.

\begin{figure*}
	\centering
	\includegraphics[width=1.0\textwidth]{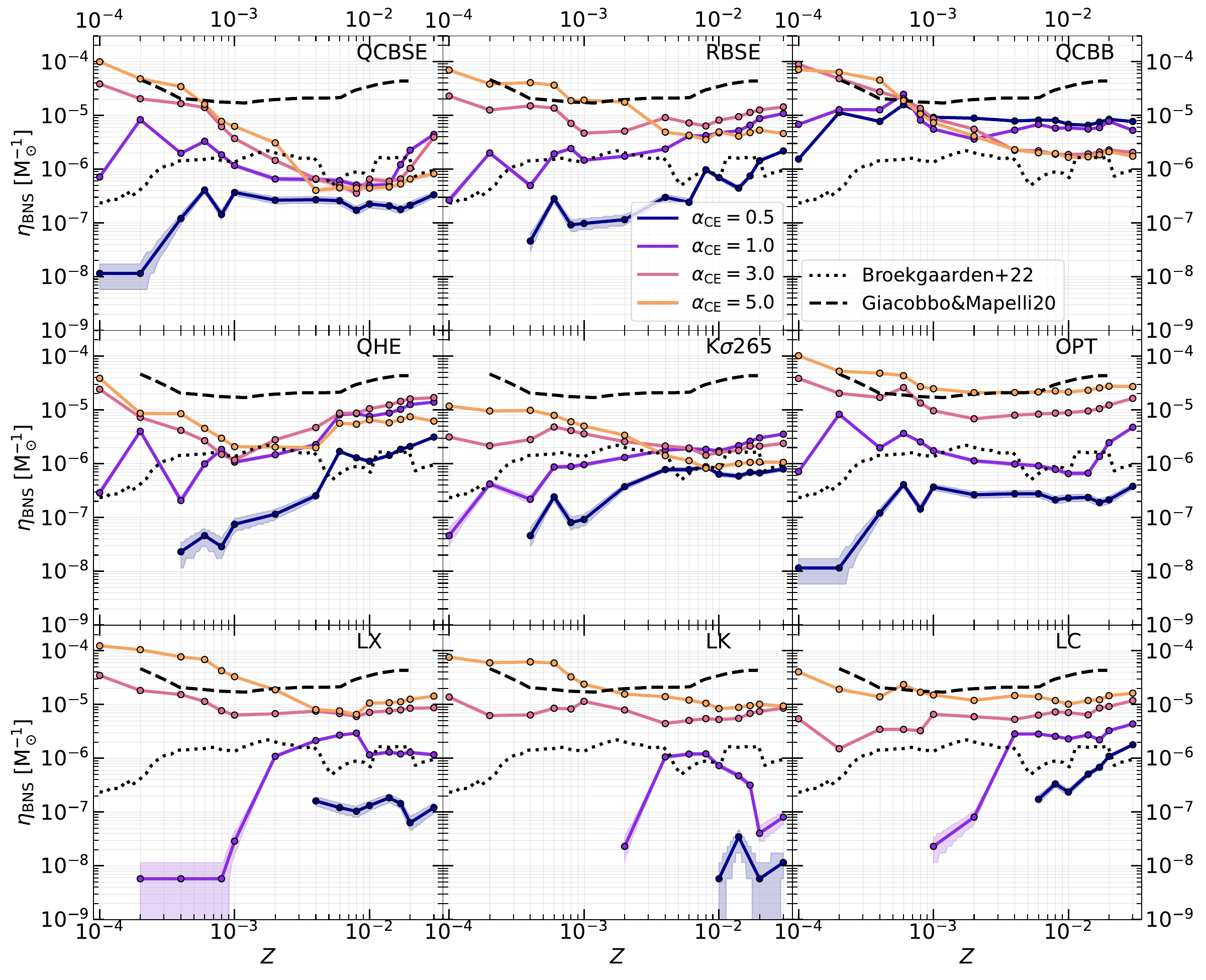}
	\vspace*{-5mm}
	\caption{ 
 \textcolor{black}{ BNS merger efficiency for a selection of   simulation models as specified by the labels in the top-right corner of each panel (Section \ref{sec:models} and Table~\ref{tab:models}). Different colours refer to values of $\alpha_{\rm CE}=0.5,$ 1, 3, 5. The dotted and dashed black lines show the results by \protect\cite{broekgaarden2022} and \protect\cite{GiacobboKick}, respectively.}  
  \gitlab{https://gitlab.com/iogiul/iorio22_plot/-/tree/v3/merger_efficiency}
           \gitbook{https://gitlab.com/iogiul/iorio22_plot/-/blob/v3/merger_efficiency/PLot_paper.ipynb}
           \gitimage{https://gitlab.com/iogiul/iorio22_plot/-/blob/v3/merger_efficiency/multipanel_BHNSm.pdf}
 }
	\label{fig:meffBNS} 
\end{figure*}

The merger efficiency by \cite{GiacobboKick}  shows a flatter metallicity trend for $\ace=5$, while the one by  \cite{broekgaarden2022} is scaled-down by a factor of $\approx{10}$ with respect to our result (assuming $\ace=1$).

Figure~\ref{fig:meffBNS} shows that most of the runs alternative to our fiducial model produce a decrease of the BNS merger efficiency. 
In particular, the enhanced binding energy in   models LX, LK, and LC reduces $\eta$ at low/intermediate metallicities, especially for models with  $\alpha_\mathrm{CE}\leq1$ for which the formation of  BNSs is highly suppressed. 

QCBSE, OPT and QCBB are the most interesting models, since these assume the same mass transfer stability criteria that are usually 
adopted in 
\bse{}-like codes \citep[see, e.g.][]{VG2018,GiacobboKick}. The model QCBSE 
produces a steep metallicity gradient.  
The presence of a metallicity gradient in the merger efficiency has a strong impact on the cosmological evolution of the merger rate density (Section~\ref{sec:MRD}).
 In the OPT model, we also  use the QCBSE option for mass transfer stability;  
 the optimistic CE assumption allows many more systems to survive the CE at high metallicity. 

In model QCBB, mass transfer is always stable if the donor is a pure-He star (case BB mass transfer, see e.g., \citealt{VG2018}).
In simulations with $\alpha_\mathrm{CE}>1$, the configuration of the binaries after the case BB mass transfer is often too wide to produce a merging BNS. Hence, the merger efficiency decreases, especially at high metallicity.
In contrast, for lower $\alpha_\mathrm{CE}$, the BNS progenitors are already in a tight configuration before the case BB mass transfer. 
Avoiding the last CE episode, most of the systems that coalesce in the fiducial model are now able to produce a merging BNS. 
As a consequence, $\eta$ increases and  becomes almost independent of the metallicity.




\subsubsection{BHNS merger efficiency} \label{sec:BHNSmeff}

The formation and merger efficiency of BHNSs is similar to BBHs, although the merger efficiency has a milder dependence on metallicity.
At $Z>0.004$, $\eta$ decreases by one order of magnitude and flattens at $Z>0.01$.
The minimum value of $\eta$ corresponds to the metallicity for which we observe a  suppression of   channel I (Fig.~\ref{fig:channelsF}).

At low metallicity our results agree with the BHNS merger efficiency estimated by \cite{GiacobboKick}, but, similarly to the case of BBHs, their $\eta$ shows a much steeper  trend wih metallicity. 
The results by \cite{broekgaarden2022} are qualitatively in agreement with our results (within a factor of 2--4). Our results and those by  \cite{broekgaarden2022} disagree  only at very high metallicity  ($Z>0.02$),  where our models  substantially differ with respect to the \cite{Pols98} tracks used in  \bse{}-like codes.

As for the other BCOs, the QHE model and the models predicting larger natal kicks reduce the total number of BHNS mergers up to a factor of 10. The models with higher binding energies (LX, LK and LC) allow more metal-rich binaries to shrink enough during  CE, increasing the number of merging BHNSs at $Z>0.01$. 
For low \ace{} values, the significant boost of BHNS mergers at high metallicity produces a rising $\eta$ profile as a function of metallicity.

\begin{figure*}
	\centering
	\includegraphics[width=1.0\textwidth]{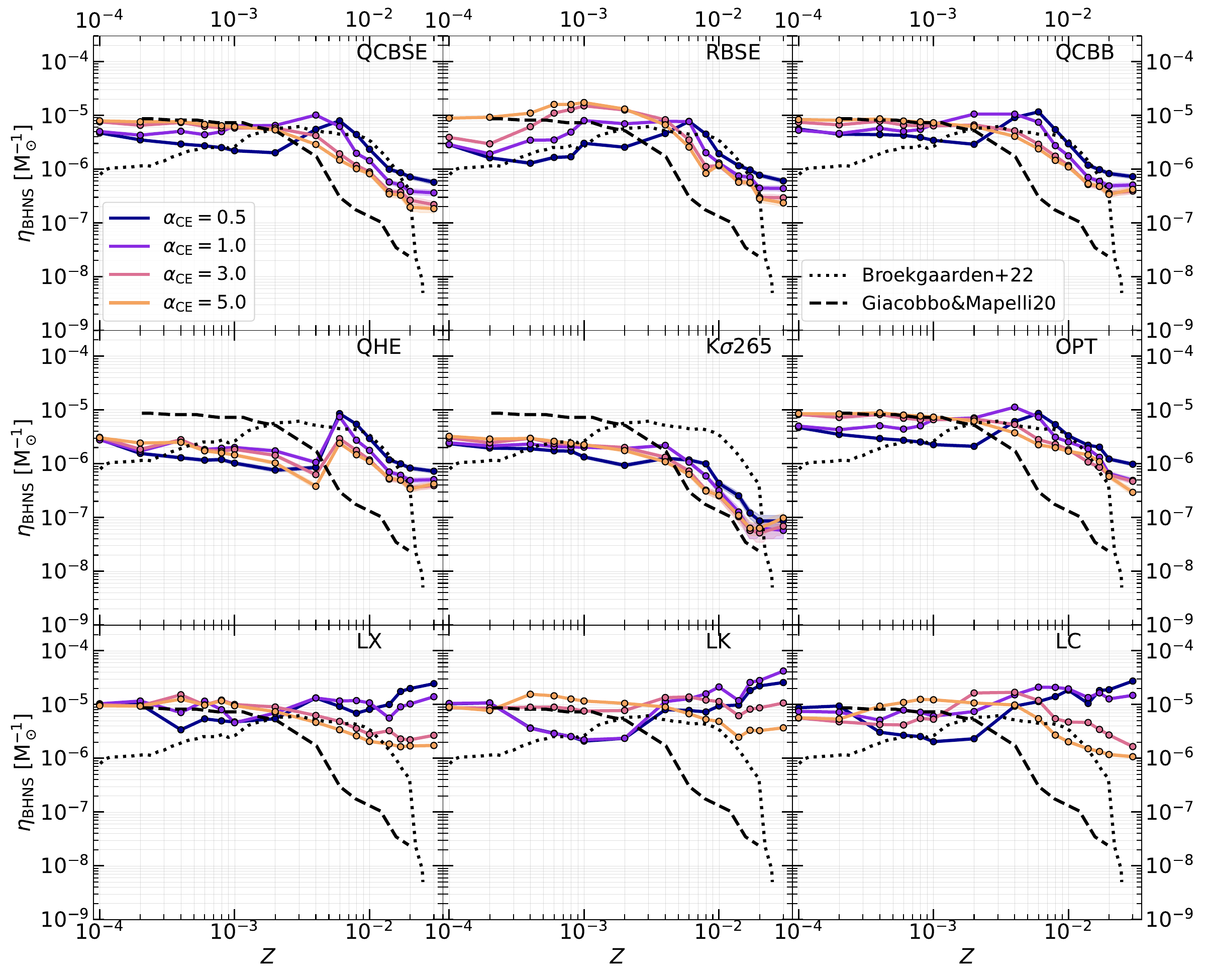}
	\vspace*{-5mm}
	\caption{\textcolor{black}{ Same as Fig. \protect{\ref{fig:meffBNS}} but for BHNSs.}
  \gitlab{https://gitlab.com/iogiul/iorio22_plot/-/tree/v3/merger_efficiency}
           \gitbook{https://gitlab.com/iogiul/iorio22_plot/-/blob/v3/merger_efficiency/PLot_paper.ipynb}
           \gitimage{https://gitlab.com/iogiul/iorio22_plot/-/blob/v3/merger_efficiency/multipanel_NSNSm.pdf}
 }
	\label{fig:meffBHNS} 
\end{figure*}

\subsection{Merger rate density}
\label{sec:MRD}

We estimate the evolution of BCO mergers with redshift by convolving the outputs of \sevn{} with our semi-analytic code \cosmorate{} \citep{santoliquido2020,santoliquido2021}. \cosmorate{}  estimates the merger rate density of compact objects as  
\begin{equation}
\label{eq:mrd}
    \mathcal{R}(z) = \int_{z_{{\rm{max}}}}^{z}\left[\int_{Z_{{\rm{min}}}}^{Z_{{\rm{max}}}} {\rm{SFRD}}(z',Z)\,{} 
    \mathcal{F}(z',z,Z) \,{}{\rm{d}}Z\right]\,{} \frac{{{\rm d}t(z')}}{{\rm{d}}z'}\,{}{\rm{d}}z',
\end{equation}
where 
\begin{equation}
\frac{{\rm{d}}t(z')}{{\rm{d}}z'} = [H_{0}\,{}(1+z')]^{-1}\,{}[(1+z')^3\Omega_{M}+ \Omega_\Lambda]^{-1/2}.
\end{equation}
In the above equation, $H_0$ is the Hubble constant, $\Omega_M$ and $\Omega_\Lambda$ are the matter and energy density, respectively. We adopt the values in \cite{Planck2018}. The term $\mathcal{F}(z',z,Z)$ is given by:
\begin{equation}
\mathcal{F}(z',z,Z) = \frac{1}{M_{{\rm{pop}}}(Z)}\frac{{\rm{d}}\mathcal{N}(z',z, Z)}{{\rm{d}}t(z)},
\end{equation}
where $M_{{\rm{pop}}}(Z)$ is the total initial mass of
the simulated stellar population (including the correction for the
incomplete sample of the IMF, and for the binary fraction),  
and  ${{\rm{d}}\mathcal{N}(z',z, Z)/{\rm{d}}}t(z)$ is the rate of binary compact object mergers forming from stars with initial metallicity $Z$ at redshift $z'$ and merging at $z$, extracted from our \sevn{} catalogues. 
In \cosmorate{},    ${\rm{SFRD}}(z,Z)$  is given by 
\begin{equation}
{\rm{SFRD}}(z',Z) = \psi(z')\,{}p(z',Z),
\end{equation}
where $\psi(z')$ is the cosmic SFR density at formation redshift $z'$, and $p(z',Z)$ is the log-normal distribution of metallicities $Z$ at fixed formation redshift $z'$, with average $\mu(z')$ and spread $\sigma_{Z}$. Here, we take both $\psi{}(z)$ and $\mu{}(z)$ from \cite{madau2017}. 
 Finally, we assume a metallicity spread $\sigma_Z = 0.2$.

\begin{figure}
	\centering
	\includegraphics[width=0.45\textwidth]{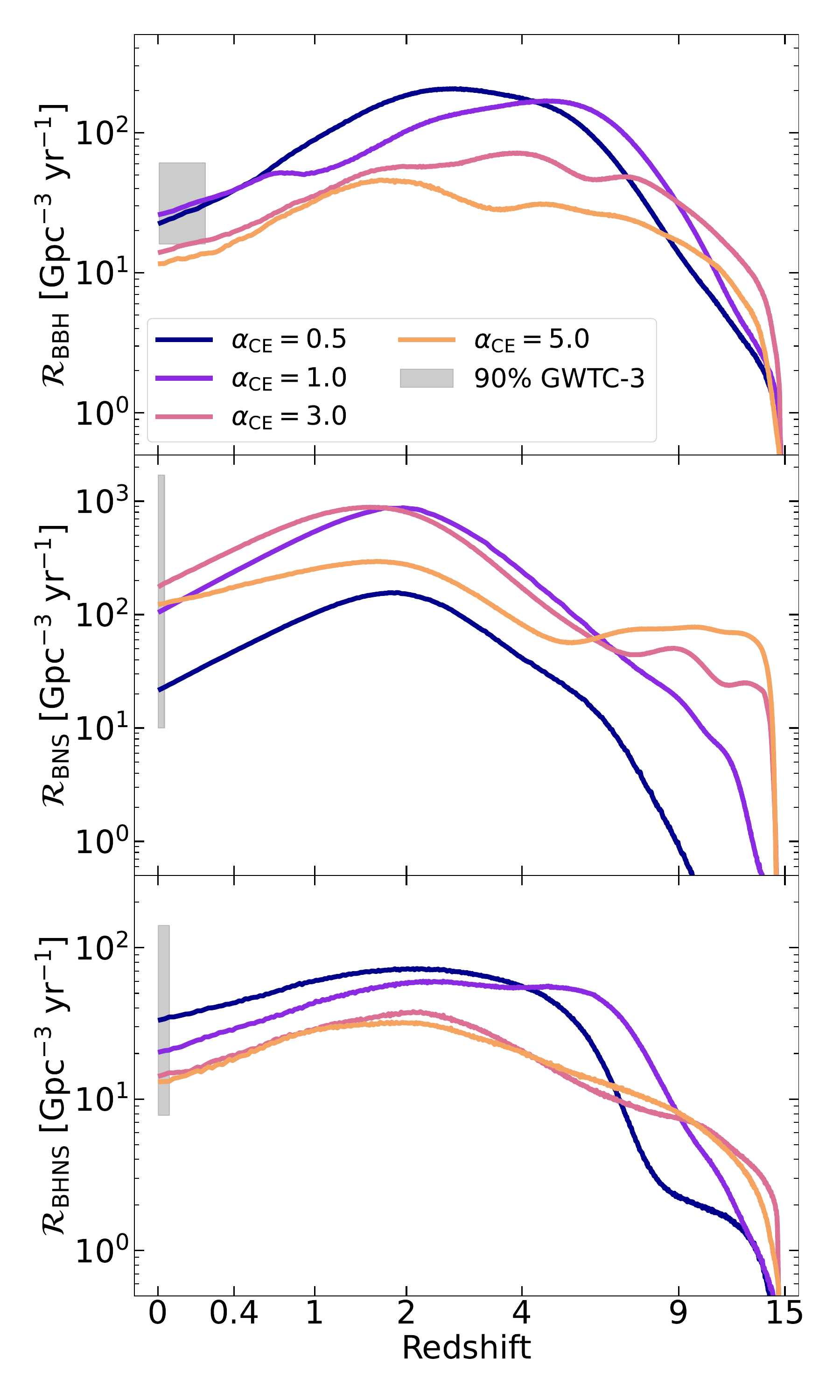}
	\vspace*{-5mm}
	\caption{Merger rate density evolution of BBHs (top), BNSs (middle), and BHNSs (bottom), in the fiducial model (F),  as a function of redshift. The line colours refer to simulations with different $\alpha_\mathrm{CE}$ values as reported in the legend.
The grey shaded area shows the most conservative 90\% credible intervals of the local merger rate density inferred \textcolor{black}{ from GWTC-3} 
\protect\citep{abbottGWTC3popandrate}.
The width of the shaded areas indicates the instrumental horizon obtained by assuming BBHs, BNSs and BHNSs of mass (30, 30), (10, 1.4), and (1.4, 1.4) \Msun{}, respectively.
Alternative versions of this plot referring to the alternative models and alternative metallicity spreads, $\sigma_\mathrm{Z}$ (see main text) can be found in the gitlab repository of the paper (\gitlab{https://gitlab.com/iogiul/iorio22_plot/-/tree/v3/Merger_rate}).
\gitbook{https://gitlab.com/iogiul/iorio22_plot/-/blob/v3/Merger_rate/Plot_paper.ipynb}
\gitimage{https://gitlab.com/iogiul/iorio22_plot/-/blob/v3/Merger_rate/Mrates_s02_F.pdf}} 
	\label{fig:rate_fiducial} 
\end{figure}

\begin{figure}
	\centering
	\includegraphics[width=0.5\textwidth]{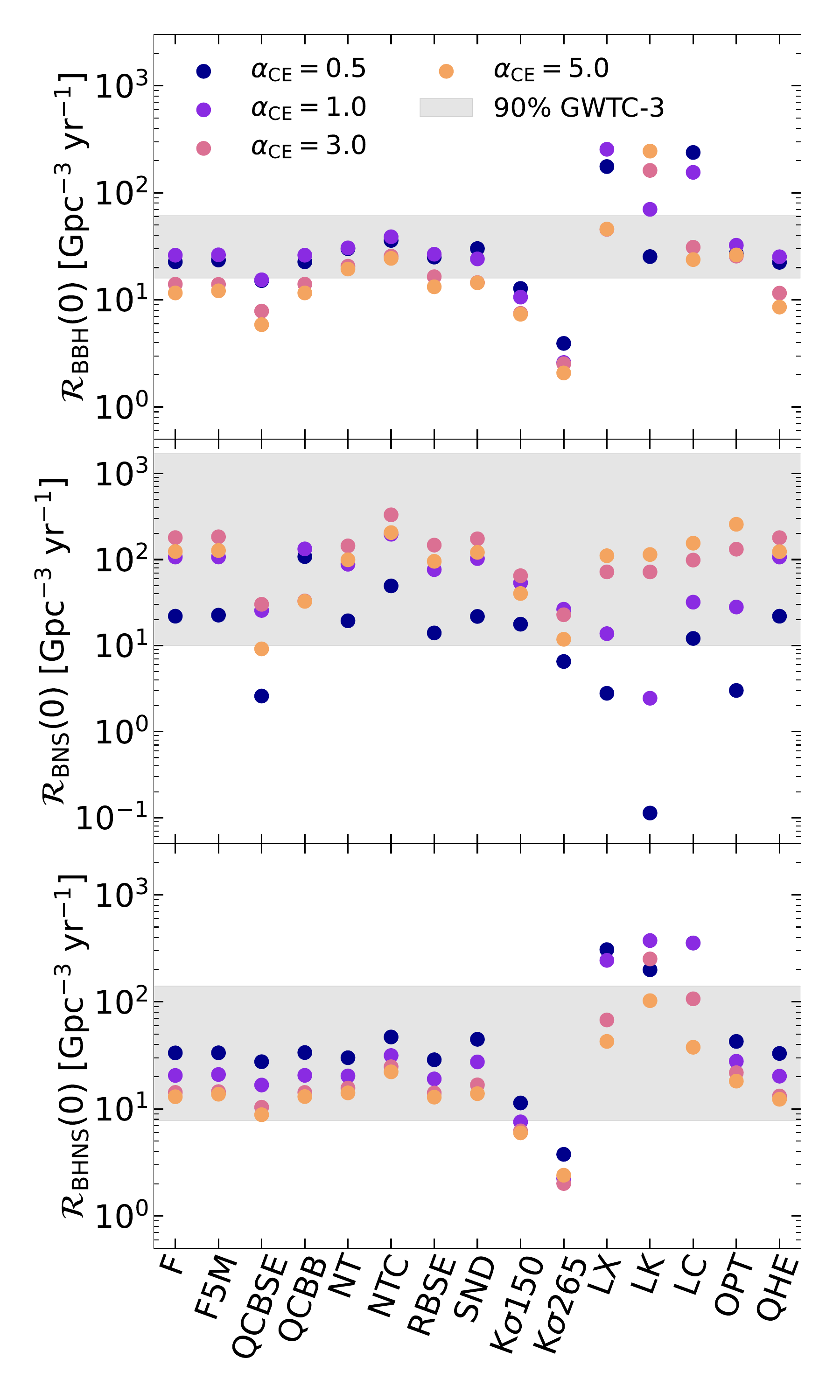}
	\vspace*{-5mm}
	\caption{Merger rate density in the local Universe ($z=0$) of BBHs (top), BNSs (middle), and BHNSs (bottom) for a sub-sample of  models (Table~\ref{tab:models}). Each colour indicates the results of 
	a different $\alpha_\mathrm{CE}$ value. The gray shaded area shows the most conservative 90\% credible intervals  inferred \textcolor{black}{ from GWTC-3} 
    \protect\citep{abbottGWTC3popandrate}.
    Other versions of this plot referring to alternative metallicity spreads $\sigma_\mathrm{Z}$ (see main text) can be found in the gitlab repository of the paper (\gitlab{https://gitlab.com/iogiul/iorio22_plot/-/tree/v3/Merger_rate}).
\gitbook{https://gitlab.com/iogiul/iorio22_plot/-/blob/v3/Merger_rate/Plot_paper.ipynb}
\gitimage{https://gitlab.com/iogiul/iorio22_plot/-/blob/v3/Merger_rate/Mrates0_s0.2.pdf}}
	\label{fig:rate_local} 
\end{figure}

Figure~\ref{fig:rate_fiducial} shows the merger rate density in the comoving frame of BBHs, BNSs, and BHNSs, according to our fiducial model, for the four considered values of $\alpha_{\rm CE}$. For all the considered models, the merger rate density increases as a function of redshift, up to $z\sim{2}$ (or an even higher redshift in the case of BBHs and BHNSs).

The merger rate density of BNSs has a peak for $z\leq{}2$, consistent with the peak of the star formation rate density ($z\approx{2}$, \citealt{madaudickinson2014}) convolved with a short delay time. In contrast, the merger rate density of BHNSs and BBHs peaks at $z>2$, because of the combined effect of star formation rate and metallicity dependence.

The choice of the $\alpha_{\rm CE}$ parameter affects the merger rate density, with an impact of a factor of 10  for BNSs (up to 3 for BHNSs and BBHs).

The results of our fiducial model are within the 90\% credible interval inferred by the \textcolor{black}{ LVK after the third observing run}  \citep{abbottGWTC3popandrate} for 
$\ace \leq 1$ for  BBHs and for all the considered  values of \ace{} for BNSs and BHNSs. Here, we assumed a metallicity spread $\sigma_{\rm Z}=0.2$ which maximises this agreement. For larger metallicity spreads, the models tend to overproduce the merger rate density of BBHs, as already shown by \cite{santoliquido2022}.

Figure~\ref{fig:rate_local} compares the local ($z=0$) merger rate density of several different models run  in this work. We find a factor of 100 difference among different models considered here. In particular, large natal kicks (K$\sigma$265) are associated with the lowest merger rate densities for BBHs and BHNSs.

As already discussed in Section~\ref{sec:BNSmeff}, the combination of the \parsec{} stellar models with the standard criterion for the stability of the mass transfer (QCBSE, see Table~\ref{tab:qc})
drastically reduces the number of BNSs at high metallicity. As a consequence, the model QCBSE produces the lowest local BNS merger rate density.

The models LX, LK, and LC are associated with the highest local merger rate of BBHs and BHNSs, and  the lowest merger rate of BNSs.
In fact, the higher binding energies in such models 
allow more  systems to shrink enough to produce merging  BBHs and BHNSs. 
In contrast, BNS progenitors undergo multiple CE episodes and have  a lower reservoir of binding energy, on average; hence 
they tend to coalesce during CE, especially for ow \ace  values.

In most models with $\sigma_{\rm Z}=0.2$, the local merger rate density of BBHs and BHNSs is $\approx{2-50}$ Gpc$^{-3}$ yr$^{-1}$, while the BNS merger rate density spans from $\approx{3}$ to $\approx{400}$ Gpc$^{-3}$ yr$^{-1}$. Here, we show the results for a fixed value of the median metallicity and  metallicity spread of the Universe: the merger rate density of BBHs and BHNSs are extremely sensitive to this choice \citep[e.g.,][]{chruslinska2019,boco2019,bouffanais2021b,broekgaarden2022,santoliquido2022}.

\section{Discussion} \label{sec:discussion} \label{sec:discsse}

\subsection{Impact of  stellar evolution on BCO properties}

In Section~\ref{sec:trackcompare}, we highlighted the  differences between the \parsec{} stellar tracks used in this work and the ones implemented in \bse{}-like codes  \citep{Pols98}. 
The largest discrepancies are  at high metallicity   and/or for  high-mass stars  (e.g., Fig.~\ref{fig:trackcompare}). \cite{Metisse} showed that different stellar evolution models  can significantly influence the mass spectrum of BHs evolved in isolation \citep[see also][]{Klencki2020}.
In addition, many authors pointed out that the uncertainties in  stellar evolution can have a dramatic impact on 
the mass range in which a star undergoes pair instability 
\citep[e.g.,][]{fields18,Mapelli20,farmer2020,Costa2021,vink2021}.  

In Section~\ref{sec:rem_mass_single} (Fig.~\ref{fig:mpisn0v05}), we showed that  several \parsec{} stellar tracks do not have a monotonic increase of the  core mass as a function of the ZAMS mass due to late dredge-up episodes \cite[][]{Costa2021}. 
As a consequence,  massive metal-poor stars can avoid   PPI ($M_\mathrm{ZAMS}\approx 100 \  \Msun$) or PISN ($M_\mathrm{ZAMS}\approx 150 \  \Msun$) producing massive BHs (up to $\approx 100 \  \Msun{}$), well within the claimed pair instability mass gap (Fig.~\ref{fig:BBHs_BBHmerg}). 
Although such massive BHs cannot merge within an Hubble time  via  isolated binary evolution (Fig.~\ref{fig:BBHmerg} and Section~\ref{sec:rem_mass_binary}), they can have an important role in the formation of massive BCO mergers in dynamically active stellar clusters \citep[see e.g.][]{rastello2018,dicarlo2020a,rastello20,arcasedda21,mapelli2021,rastello2021}.

In Section~\ref{sec:Results}, we show that the details of  stellar evolution  play a fundamental role even during  binary evolution, significantly affecting the properties of BCO mergers. In particular,  using the \parsec{} stellar tracks, we find that the \vir{classic} formation 
channel of BCO mergers (channel I, see Section~\ref{sec:channels})  can be strongly suppressed especially  at high metallicity (Fig.~\ref{fig:channelsF}).

Concerning the merger efficiency, the variations due to differences in the assumed stellar model have an impact as large as that of binary evolution uncertainties 
\citep[e.g., efficiency of  CE,  supernova kicks; see, e.g.,][]{GiacobboKick,santoliquido2021,broekgaarden2022}.

Interestingly, the models for which we find the largest discrepancies  on the formation channel and  merger efficiency (especially regarding the BNSs) with respect to the results  of \bse{}-like codes  are the ones with the most similar assumptions about binary evolution. For example, assuming the QCBSE  stability criterion (Table~\ref{tab:qc}), we obtain a steep metallicity trend for the BNS merger efficiency rather than the almost flat profile usually found in works adopting \bse{}-like codes (Fig.~\ref{fig:meffBNS}).
This implies that the main parameters describing  binary evolution and the underlying  stellar evolution models are highly correlated.
Therefore, any attempt to constrain binary evolution parameters by comparing observations and population-synthesis results could be affected by a  selection bias  of the parameter space.

Our results point out that the investigation of the systematics and uncertainties in stellar evolution  are fundamental  for the  analysis of the  properties BCOs and for the astrophysical interpretation of the results obtained by the LVK. 

The \sevn{} code is designed to explore the parameter space. In fact, it allows to easily test different stellar evolution models using the same exact framework for binary evolution.  In future works, we aim to exploit \sevn{} to   make a more comprehensive comparison of the state-of-the-art stellar evolution models.

\subsection{About CE} \label{sec:discce}

In Section~\ref{sec:Results}, we showed that the parameters related to CE, i.e. \lce{} for the envelope binding energy and \ace{} for CE efficiency have a large impact on the formation of merging BCOs, as already highlighted in many other works \cite[e.g., ][]{dominik2012,VG2018,GiacobboCOB,Combine,GiacobboKick,Klencki21,broekgaarden2021,broekgaarden2022,Vigna2022}.
 Recent works suggest that the models used in binary population synthesis codes may be optimistic regarding CE survival, especially for  massive stars \citep[see,  e.g.,][]{Klencki21,Klencki22}. As a consequence, these codes may overestimate the number of merging BBHs formed through CE \citep[e.g.,][]{Briel22,Marchant21,gallegosgarcia2021}. However, we found that the increase of the binding energy does not always decrease the number of BCO mergers (Section~\ref{sec:bbhform}).  Rather, it allows more (massive) systems to evolve through  channel I (Fig.~\ref{fig:channelsMultimodel}, see also \citealt{Kruckow2016}). 
At high metallicity, higher binding energies  boost the  merger efficiency and the merger rate of BBHs and BHNSs  (Figs. \ref{fig:meffBHBH} and \ref{fig:rate_local}),  allowing the formation of more massive merging BBHs, also influencing the  BH mass spectrum (Fig.~\ref{fig:BBHmerg}).

Variations of \ace{} produce a scatter in the local merger rate density of BNSs up to one order of magnitude. In general, for $\ace<1$, the predicted  merger rates  are just marginally consistent with the one  found by the LVK \textcolor{black}{ after the third observing run} \citep{abbottGWTC3popandrate}.  For BBHs, low \ace{} values  produce a larger number of  mergers. A significant increase in the number of BBH mergers could result in a tension with the local merger rate estimated  \textcolor{black}{ from GWTC-3}  \citep{abbottGWTC3popandrate}  when also the contribution of other formation channels are taken into account (e.g. dynamical formation channel in star clusters).

In contrast, in the lower mass range of  WD binaries, low \ace{} values  seem to be the best match to the observed properties of  post-CE MS--WD systems  \cite[e.g.,][]{zorotovic2010,DeMarco2011,toonen2013,camacho2014}. 

In conclusion, the \ace{}\lce{} model often used in population synthesis codes (but see, e.g., \citealt{Combine} and \citealt{korol2022} for alternative models) could be too simplistic to catch the complex physics of  CE evolution, especially if we assume a constant value for \ace{} throughout the entire stellar mass range.

Recently, there have been many efforts 
to improve the models of  CE \citep[e.g., ][]{Halabi2018,fragos2019,lawsmith2020,Ragoler22,Ryosuke22,Trani22,Vigna2022,DiStefano2023}.
In the future, we aim 
to include and test  additional CE models in \sevn{}. 

\subsection{Other binary evolution processes}

Aside from  CE, the parameters that have a large impact on the formation and merger of  BCOs are the ones regarding supernova kicks and  stability of mass transfer. As expected, large natal kicks reduce the number of merging BCOs and alter their mass spectrum selecting preferentially  massive binaries (Fig.~\ref{fig:BBHmerg}).

The mass transfer stability criterion is the one that mostly correlates with the choice of the stellar evolution model. 
The combination of the \parsec{} stellar tracks with the standard  stability criterion used in \bse{}-like codes (QCBSE, Table~\ref{tab:qc}) produces a suppression of BCO merger efficiency, especially for BNSs (Fig.~\ref{fig:meffBNS}).  Combining the model QCBSE with  the optimistic CE assumption  (model OPT) and large \ace values ($>1$), brings back the efficiency of BCO mergers to the level of the fiducial model.

The quasi-homogeneous evolution reduces binary interactions,  suppressing the number of  BCO mergers at low metallicity (Figs. \ref{fig:meffBHBH}, \ref{fig:meffBNS}, \ref{fig:meffBHNS}), but is thought to be ineffective at high metallicity 
($Z>0.004$). As a consequence it only has  a modest impact onto the local merger rate density  (Fig.~\ref{fig:rate_local}).

The models in which we disable the stellar tides (NT and NTC)    do not significantly alter   the formation channels of BCOs,   their merger efficiency and local merger rate density.  However, Fig.~\ref{fig:BBHmerg} shows that  models without tides  (NT and NTC) produce a flatter  mass spectrum  for  BHs in  BBH mergers.  
These are important results since models of stellar tides  
depend on a large number of parameters and on properties  that are not always available  in  stellar tracks (e.g., stellar rotation and eddy turnover time, Section~\ref{sec:tides}). 
In addition,  recent observations of  binary stars  seem to  challenge the predictions of the classical tide formalism used in  population-synthesis studies, especially regarding dynamic tides \citep{Justensen21,Marcussen2022}. 
Similarly, all the other models we tested do not introduce significant differences in the merger efficiency and local merger rate density, but can alter the features of the mass spectrum of BHs in BCO mergers (see, e.g., the SND and RBSE model in Fig.~\ref{fig:BBHmerg}).

\subsection{Systematics and caveats} \label{sec:caveats}

Recent work highlights the importance of using self-consistent binding energy for the adopted stellar-evolution  models \citep[see e.g.][]{Kruckow2016,Marchant21,Klencki21}. Here, we use binding energy prescriptions that were derived for other stellar models (Appendix~\ref{app:ebind}). However, the four different formalisms we tested
cover a wide range  of binding energies (up to three orders  of magnitude), from low values \citep{Claeys14} to very high ones \citep{Klencki21}.
In a follow-up study (Sgalletta et al., in prep.), we will show the impact of adopting values of the binding energy calculated directly from our stellar-evolution tracks. 
We also aim to investigate the effect of the possible dependence of the envelope binding-energy on the adiabatic mass-loss during CE, as highlighted by \cite{Deloye2010,Ge2010b,Ge2022}.

Although we simulated a large number of binaries, some of the simulations  produce just a few BCOs. In addition, we use the same set of  binaries for all the simulations. 
In order to asses the possible systematic effects due either  to low-number statistic or to the limited sampling of the initial conditions,  we ran a simulation using the fiducial setup (Section~\ref{sec:models}) but with a different set of  $5\times10^6$ binaries. 
The results  of 
these simulations are stored in the  gitlab repository of the paper (\gitlab{https://gitlab.com/iogiul/iorio22_plot}).
We do not find any significant differences with respect to  the fiducial model, except for the merger efficiency  in 
regions of the parameter space in which the simulations produce a low  number of BCOs ($<10$). This happens  for $\alpha_\mathrm{CE}=0.5$ and $Z=0.03$ for BBHs, and  $Z<0.0004$ for BNSs. These differences are within the uncertainties expected for a Poissonian distribution. 

\section{Summary} \label{sec:conclusion}

We presented the new release of the binary 
population-synthesis code \sevn{}.  With respect to its previous versions, \sevn{} has been deeply revised to improve its performance, and to guarantee more flexibility in modelling single and binary star evolution processes: \sevn{} now implements multiple possible options for core-collapse supernovae, pair instability, RLO, CE, natal kicks, stellar tides, and circularisation. The new version of \sevn{} is publicly available at this link \url{https://gitlab.com/sevncodes/sevn.git}, together with an user-guide.

\sevn{} describes stellar evolution by interpolating a set of evolutionary tracks, instead of using the commonly adopted fitting formulas by \cite{Hurley00}. In the new version, we added a completely new set of stellar-evolution tracks run with \parsec{} \citep{Bressan2012,Costa2019} and the {\sc MIST} tracks \citep{Choi2016}. 

We used \sevn{} to investigate the formation and  properties of binary compact objects (BCOs)  exploring a wide portion of the parameter space.   
In the following, we summarise the main results of our analysis.
\begin{itemize}
\item  Stellar evolution plays a fundamental  role in defining the  properties of BCOs, such as their formation channels,  merger efficiency and merger rate density.   
Our results, obtained using \sevn{} with \parsec{} tracks, show  systematic differences with respect the results of \bse{}-like codes  that are  as large as (or even larger than)   the effect of the uncertainties on binary-evolution processes (e.g.,  CE and natal kicks).   
\item We find that there is a degeneracy between the effects of binary-evolution parameters and stellar-evolution models. For example, the classical \bse{}-like stability criterion applied to the \parsec{} tracks induces a strong suppression (more than one order of magnitude) of the BNS merger rate with respect to the results of \bse{}-like codes. 
\item Combining the \parsec{} stellar tracks with the   recent pair-instability prescriptions by  \cite{farmer2019} and \cite{Mapelli20}, 
it is possible to produce  massive BHs (up to $\approx{100}  \ \Msun$),  well within the boundaries of the claimed 
pair-instability mass gap, just through single star evolution.   However, the maximum mass of BHs in BBH mergers is 
$\approx{55} \ \Msun{}$ in all our runs. 
BHs more massive than $\approx{55}$ M$_\odot$ can still merge within the Hubble time, but only if they pair up dynamically with other BHs in dense star clusters and galactic nuclei. 
\item In our simulations, the importance of channel I for BCO formation  (i.e., only stable mass transfer before the first compact remnant formation and then a CE episode)  is strongly suppressed with respect to the large majority of the other works in the literature. In particular, at high metallicity ($Z\gtrsim{0.01}$) only less than $20 \%$ of the merging BBHs and BNSs form via this channel, while other authors found fractions larger than $70 \%$   \citep[ e.g.,][]{dominik2012,GiacobboWind,Combine,VG2018}.
\item The details of  binary circularisation due to  stellar tides do not seem to play an important role for the formation of BCOs. In particular, we obtain very similar results both using the detailed stellar tides formalism by \cite{Hurley02} and 
a simpler model in which the binary is circularised at  periastron at the onset of  RLO.
\item The local merger rate density of our fiducial models  (10--30 Gpc$^{-3}$yr$^{-1}$  for BBHs, 20--200 Gpc$^{-3}$yr$^{-1}$ for BNSs, and 10--40 Gpc$^{-3}$yr$^{-1}$ for BHNSs) is consistent with the most recent estimates by the LVK \textcolor{black}{ (GWTC-3, \citealt{abbottGWTC3popandrate})}. In contrast,  the models  for which the parameters of binary evolution are more similar to the default values of \bse{}-like codes \cite[e.g.,][]{GiacobboKick,santoliquido2021} show  a significant tension with the credible intervals  inferred by the LVK.
\end{itemize}

In conclusion, our work points out  the need to include the uncertainties and systematics  of stellar evolution  in the investigation of the (already large) parameter space  
relevant for the 
formation, evolution and  demography of BCOs. This is  particularly important for the astrophysical interpretation of the results of current and forthcoming GW observatories. 
In this context, \sevn{} represents an unique tool to deeply explore the 
parameter space of BCO formation.



\section*{Acknowledgements}

The authors thank the anonymous referee for suggestions that helped to improve the manuscript. We thank Matthias Kruckow for  kindly sending us  his data to produce Fig.~\ref{fig:combine}. We are grateful to Floor Broekgaarden, Nicola Giacobbo and Tilman Hartwig for their enlightening comments. 
We also thank Manuel Arca Sedda, Alessandro Ballone, Ugo N. Di Carlo, Francesco Iraci, Mario Pasquato, Carole Périgois, Stefano Torniamenti for  the stimulating discussions  during the  development of \sevn{}, and Sara Rastello  for the help in preparing the  public repositories related to this paper and for the  \sevn{} code testing. 
 GC, GI, MM and FS acknowledge financial support from the European Research 
Council for the ERC Consolidator grant DEMOBLACK, under contract no. 
770017.  
EK and MM acknowledge support from PRIN-MIUR~2020 METE, under contract no. 2020KB33TP. AB acknowledges support from PRIN-MIUR 2017 prot. 20173ML3WW 002. 
This research made use of \textsc{NumPy} \citep{Harris20}, \textsc{SciPy} 
\citep{SciPy2020}, \textsc{IPython} \citep{Ipython}. For the plots we used \textsc{Matplotlib} \citep{Hunter2007}.

\section*{Data Availability}
All the data underlying this article are available in  Zenodo  at the link \url{https://doi.org/10.5281/zenodo.7260770} \citep[Version 2.0,][]{paperdataset}.
The codes used in this work are publicly available through gitlab repositories: \sevn{} at \url{https://gitlab.com/sevncodes/sevn.git} (release {\it Iorio22}, \url{https://gitlab.com/sevncodes/sevn/-/releases/iorio22}), {\sc TrackCruncher} at \url{https://gitlab.com/sevncodes/trackcruncher.git}, and {\sc pyblack} at \url{https://gitlab.com/iogiul/pyblack}.
All the Jupyter notebooks used to produce the plots in the paper are available in the gitlab repository \url{https://gitlab.com/iogiul/iorio22_plot.git} (release V3.0, \url{https://gitlab.com/iogiul/iorio22_plot/-/releases/V3.0}). The repository contains also additional plots not showed in this article.
Each plot in the paper reports three icons pointing to specific path of the repository: \faGitlab{} specific folder containing the notebooks, the data and the images, \faBook{} Jupyter notebooks used to make the plot, \faFileImageO{} direct link to the image.


\bibliographystyle{mnras}
\bibliography{main}

\begin{thebibliography}{}
\makeatletter
\relax
\def\mn@urlcharsother{\let\do\@makeother \do\$\do\&\do\#\do\^\do\_\do\%\do\~}
\def\mn@doi{\begingroup\mn@urlcharsother \@ifnextchar [ {\mn@doi@}
  {\mn@doi@[]}}
\def\mn@doi@[#1]#2{\def\@tempa{#1}\ifx\@tempa\@empty \href
  {http://dx.doi.org/#2} {doi:#2}\else \href {http://dx.doi.org/#2} {#1}\fi
  \endgroup}
\def\mn@eprint#1#2{\mn@eprint@#1:#2::\@nil}
\def\mn@eprint@arXiv#1{\href {http://arxiv.org/abs/#1} {{\tt arXiv:#1}}}
\def\mn@eprint@dblp#1{\href {http://dblp.uni-trier.de/rec/bibtex/#1.xml}
  {dblp:#1}}
\def\mn@eprint@#1:#2:#3:#4\@nil{\def\@tempa {#1}\def\@tempb {#2}\def\@tempc
  {#3}\ifx \@tempc \@empty \let \@tempc \@tempb \let \@tempb \@tempa \fi \ifx
  \@tempb \@empty \def\@tempb {arXiv}\fi \@ifundefined
  {mn@eprint@\@tempb}{\@tempb:\@tempc}{\expandafter \expandafter \csname
  mn@eprint@\@tempb\endcsname \expandafter{\@tempc}}}

\bibitem[\protect\citeauthoryear{{Abbott} et~al.,}{{Abbott}
  et~al.}{2016a}]{abbottO1}
{Abbott} B.~P.,  et~al., 2016a, \mn@doi [Physical Review X]
  {10.1103/PhysRevX.6.041015}, \href
  {http://adsabs.harvard.edu/abs/2016PhRvX...6d1015A} {6, 041015}

\bibitem[\protect\citeauthoryear{Abbott et~al.,}{Abbott
  et~al.}{2016b}]{abbottGW150914}
Abbott B.~P.,  et~al., 2016b, \mn@doi [Phys. Rev. Lett.]
  {10.1103/PhysRevLett.116.061102}, 116, 061102

\bibitem[\protect\citeauthoryear{{Abbott} et~al.,}{{Abbott}
  et~al.}{2016c}]{abbottastrophysics}
{Abbott} B.~P.,  et~al., 2016c, \mn@doi [\apjl] {10.3847/2041-8205/818/2/L22},
  \href {http://adsabs.harvard.edu/abs/2016ApJ...818L..22A} {818, L22}

\bibitem[\protect\citeauthoryear{{Abbott} et~al.,}{{Abbott}
  et~al.}{2017a}]{abbottGW170817}
{Abbott} B.~P.,  et~al., 2017a, \mn@doi [Physical Review Letters]
  {10.1103/PhysRevLett.119.161101}, \href
  {http://adsabs.harvard.edu/abs/2017PhRvL.119p1101A} {119, 161101}

\bibitem[\protect\citeauthoryear{{Abbott} et~al.,}{{Abbott}
  et~al.}{2017b}]{abbottmultimessenger}
{Abbott} B.~P.,  et~al., 2017b, \mn@doi [\apjl] {10.3847/2041-8213/aa91c9},
  \href {http://adsabs.harvard.edu/abs/2017ApJ...848L..12A} {848, L12}

\bibitem[\protect\citeauthoryear{{Abbott} et~al.,}{{Abbott}
  et~al.}{2019a}]{abbottO2}
{Abbott} B.~P.,  et~al., 2019a, \mn@doi [Physical Review X]
  {10.1103/PhysRevX.9.031040}, \href
  {https://ui.adsabs.harvard.edu/abs/2019PhRvX...9c1040A} {9, 031040}

\bibitem[\protect\citeauthoryear{{Abbott} et~al.,}{{Abbott}
  et~al.}{2019b}]{abbottO2popandrate}
{Abbott} B.~P.,  et~al., 2019b, \mn@doi [\apjl] {10.3847/2041-8213/ab3800},
  \href {https://ui.adsabs.harvard.edu/abs/2019ApJ...882L..24A} {882, L24}

\bibitem[\protect\citeauthoryear{{Abbott} et~al.,}{{Abbott}
  et~al.}{2020a}]{abbottGW190521}
{Abbott} R.,  et~al., 2020a, \mn@doi [\prl] {10.1103/PhysRevLett.125.101102},
  \href {https://ui.adsabs.harvard.edu/abs/2020PhRvL.125j1102A} {125, 101102}

\bibitem[\protect\citeauthoryear{{Abbott} et~al.,}{{Abbott}
  et~al.}{2020b}]{abbottGW190425}
{Abbott} B.~P.,  et~al., 2020b, \mn@doi [\apjl] {10.3847/2041-8213/ab75f5},
  \href {https://ui.adsabs.harvard.edu/abs/2020ApJ...892L...3A} {892, L3}

\bibitem[\protect\citeauthoryear{{Abbott} et~al.,}{{Abbott}
  et~al.}{2020c}]{abbottGW190814}
{Abbott} R.,  et~al., 2020c, \mn@doi [\apjl] {10.3847/2041-8213/ab960f}, \href
  {https://ui.adsabs.harvard.edu/abs/2020ApJ...896L..44A} {896, L44}

\bibitem[\protect\citeauthoryear{{Abbott} et~al.,}{{Abbott}
  et~al.}{2020d}]{abbottGW190521astro}
{Abbott} R.,  et~al., 2020d, \mn@doi [\apjl] {10.3847/2041-8213/aba493}, \href
  {https://ui.adsabs.harvard.edu/abs/2020ApJ...900L..13A} {900, L13}

\bibitem[\protect\citeauthoryear{{Abbott} et~al.,}{{Abbott}
  et~al.}{2021a}]{abbottGWTC-2.1}
{Abbott} R.,  et~al., 2021a, arXiv e-prints, \href
  {https://ui.adsabs.harvard.edu/abs/2021arXiv210801045T} {p. arXiv:2108.01045}

\bibitem[\protect\citeauthoryear{{Abbott} et~al.,}{{Abbott}
  et~al.}{2021b}]{abbottGWTC3}
{Abbott} R.,  et~al., 2021b, arXiv e-prints, \href
  {https://ui.adsabs.harvard.edu/abs/2021arXiv211103606T} {p. arXiv:2111.03606}

\bibitem[\protect\citeauthoryear{{Abbott} et~al.,}{{Abbott}
  et~al.}{2021c}]{abbottO3a}
{Abbott} R.,  et~al., 2021c, \mn@doi [Physical Review X]
  {10.1103/PhysRevX.11.021053}, \href
  {https://ui.adsabs.harvard.edu/abs/2021PhRvX..11b1053A} {11, 021053}

\bibitem[\protect\citeauthoryear{{Abbott} et~al.,}{{Abbott}
  et~al.}{2021d}]{abbottO3apopandrate}
{Abbott} R.,  et~al., 2021d, \mn@doi [\apjl] {10.3847/2041-8213/abe949}, \href
  {https://ui.adsabs.harvard.edu/abs/2021ApJ...913L...7A} {913, L7}

\bibitem[\protect\citeauthoryear{{Abbott} et~al.,}{{Abbott}
  et~al.}{2023}]{abbottGWTC3popandrate}
{Abbott} R.,  et~al., 2023, \mn@doi [Physical Review X]
  {10.1103/PhysRevX.13.011048}, \href
  {https://ui.adsabs.harvard.edu/abs/2023PhRvX..13a1048A} {13, 011048}

\bibitem[\protect\citeauthoryear{{Aghanim} et~al.,}{{Aghanim}
  et~al.}{2020}]{Planck2018}
{Aghanim} N.,  et~al., 2020, \mn@doi [\aap] {10.1051/0004-6361/201833910},
  \href {https://ui.adsabs.harvard.edu/abs/2020A&A...641A...6P} {641, A6}

\bibitem[\protect\citeauthoryear{{Agrawal}, {Hurley}, {Stevenson}, {Sz{\'e}csi}
   \& {Flynn}}{{Agrawal} et~al.}{2020}]{Metisse}
{Agrawal} P.,  {Hurley} J.,  {Stevenson} S.,  {Sz{\'e}csi} D.,   {Flynn} C.,
  2020, \mn@doi [\mnras] {10.1093/mnras/staa2264}, \href
  {https://ui.adsabs.harvard.edu/abs/2020MNRAS.497.4549A} {497, 4549}

\bibitem[\protect\citeauthoryear{{Ali-Ha{\"\i}moud}, {Kovetz}  \&
  {Kamionkowski}}{{Ali-Ha{\"\i}moud} et~al.}{2017}]{alihaimoud2017}
{Ali-Ha{\"\i}moud} Y.,  {Kovetz} E.~D.,   {Kamionkowski} M.,  2017, \mn@doi
  [\prd] {10.1103/PhysRevD.96.123523}, \href
  {https://ui.adsabs.harvard.edu/abs/2017PhRvD..96l3523A} {96, 123523}

\bibitem[\protect\citeauthoryear{{Arca Sedda}}{{Arca
  Sedda}}{2020}]{arcasedda2020b}
{Arca Sedda} M.,  2020, \mn@doi [\apj] {10.3847/1538-4357/ab723b}, \href
  {https://ui.adsabs.harvard.edu/abs/2020ApJ...891...47A} {891, 47}

\bibitem[\protect\citeauthoryear{{Arca Sedda}, {Mapelli}, {Spera},
  {Benacquista}  \& {Giacobbo}}{{Arca Sedda} et~al.}{2020}]{arcasedda2020}
{Arca Sedda} M.,  {Mapelli} M.,  {Spera} M.,  {Benacquista} M.,   {Giacobbo}
  N.,  2020, \mn@doi [\apj] {10.3847/1538-4357/ab88b2}, \href
  {https://ui.adsabs.harvard.edu/abs/2020ApJ...894..133A} {894, 133}

\bibitem[\protect\citeauthoryear{{Arca Sedda}, {Li}  \& {Kocsis}}{{Arca Sedda}
  et~al.}{2021a}]{arcasedda2018b}
{Arca Sedda} M.,  {Li} G.,   {Kocsis} B.,  2021a, \mn@doi [\aap]
  {10.1051/0004-6361/202038795}, \href
  {https://ui.adsabs.harvard.edu/abs/2021A&A...650A.189A} {650, A189}

\bibitem[\protect\citeauthoryear{{Arca-Sedda}, {Rizzuto}, {Naab}, {Ostriker},
  {Giersz}  \& {Spurzem}}{{Arca-Sedda} et~al.}{2021b}]{arcasedda21}
{Arca-Sedda} M.,  {Rizzuto} F.~P.,  {Naab} T.,  {Ostriker} J.,  {Giersz} M.,
  {Spurzem} R.,  2021b, \mn@doi [\apj] {10.3847/1538-4357/ac1419}, \href
  {https://ui.adsabs.harvard.edu/abs/2021ApJ...920..128A} {920, 128}

\bibitem[\protect\citeauthoryear{{Atri} et~al.,}{{Atri}
  et~al.}{2019}]{atri2019}
{Atri} P.,  et~al., 2019, \mn@doi [\mnras] {10.1093/mnras/stz2335}, \href
  {https://ui.adsabs.harvard.edu/abs/2019MNRAS.489.3116A} {489, 3116}

\bibitem[\protect\citeauthoryear{{Banerjee}}{{Banerjee}}{2017}]{banerjee2017}
{Banerjee} S.,  2017, \mn@doi [\mnras] {10.1093/mnras/stw3392}, \href
  {http://adsabs.harvard.edu/abs/2017MNRAS.467..524B} {467, 524}

\bibitem[\protect\citeauthoryear{{Banerjee}}{{Banerjee}}{2018}]{banerjee2018}
{Banerjee} S.,  2018, \mn@doi [\mnras] {10.1093/mnras/stx2347}, \href
  {http://adsabs.harvard.edu/abs/2018MNRAS.473..909B} {473, 909}

\bibitem[\protect\citeauthoryear{{Banerjee}}{{Banerjee}}{2021}]{banerjee2020}
{Banerjee} S.,  2021, \mn@doi [\mnras] {10.1093/mnras/staa2392}, \href
  {https://ui.adsabs.harvard.edu/abs/2021MNRAS.500.3002B} {500, 3002}

\bibitem[\protect\citeauthoryear{{Banerjee}, {Baumgardt}  \&
  {Kroupa}}{{Banerjee} et~al.}{2010}]{banerjee2010}
{Banerjee} S.,  {Baumgardt} H.,   {Kroupa} P.,  2010, \mn@doi [\mnras]
  {10.1111/j.1365-2966.2009.15880.x}, \href
  {http://adsabs.harvard.edu/abs/2010MNRAS.402..371B} {402, 371}

\bibitem[\protect\citeauthoryear{{Banerjee}, {Belczynski}, {Fryer}, {Berczik},
  {Hurley}, {Spurzem}  \& {Wang}}{{Banerjee} et~al.}{2020}]{banerjee2019}
{Banerjee} S.,  {Belczynski} K.,  {Fryer} C.~L.,  {Berczik} P.,  {Hurley}
  J.~R.,  {Spurzem} R.,   {Wang} L.,  2020, \mn@doi [\aap]
  {10.1051/0004-6361/201935332}, \href
  {https://ui.adsabs.harvard.edu/abs/2020A&A...639A..41B} {639, A41}

\bibitem[\protect\citeauthoryear{{Barkat}, {Rakavy}  \& {Sack}}{{Barkat}
  et~al.}{1967}]{barkat1967}
{Barkat} Z.,  {Rakavy} G.,   {Sack} N.,  1967, \mn@doi [\prl]
  {10.1103/PhysRevLett.18.379}, \href
  {https://ui.adsabs.harvard.edu/abs/1967PhRvL..18..379B} {18, 379}

\bibitem[\protect\citeauthoryear{{Belczynski}, {Kalogera}  \&
  {Bulik}}{{Belczynski} et~al.}{2002}]{belczynski2002}
{Belczynski} K.,  {Kalogera} V.,   {Bulik} T.,  2002, \mn@doi [\apj]
  {10.1086/340304}, \href {http://adsabs.harvard.edu/abs/2002ApJ...572..407B}
  {572, 407}

\bibitem[\protect\citeauthoryear{{Belczynski}, {Kalogera}, {Rasio}, {Taam},
  {Zezas}, {Bulik}, {Maccarone}  \& {Ivanova}}{{Belczynski}
  et~al.}{2008}]{startrack}
{Belczynski} K.,  {Kalogera} V.,  {Rasio} F.~A.,  {Taam} R.~E.,  {Zezas} A.,
  {Bulik} T.,  {Maccarone} T.~J.,   {Ivanova} N.,  2008, \mn@doi [\apjs]
  {10.1086/521026}, \href
  {https://ui.adsabs.harvard.edu/abs/2008ApJS..174..223B} {174, 223}

\bibitem[\protect\citeauthoryear{{Belczynski} et~al.,}{{Belczynski}
  et~al.}{2018}]{belczynski2018}
{Belczynski} K.,  et~al., 2018, \mn@doi [\aap] {10.1051/0004-6361/201732428},
  \href {https://ui.adsabs.harvard.edu/abs/2018A&A...615A..91B} {615, A91}

\bibitem[\protect\citeauthoryear{{Bethe} \& {Brown}}{{Bethe} \&
  {Brown}}{1998}]{bethe1998}
{Bethe} H.~A.,  {Brown} G.~E.,  1998, \mn@doi [\apj] {10.1086/306265}, \href
  {http://adsabs.harvard.edu/abs/1998ApJ...506..780B} {506, 780}

\bibitem[\protect\citeauthoryear{{Bird}, {Cholis}, {Mu{\~n}oz},
  {Ali-Ha{\"i}moud}, {Kamionkowski}, {Kovetz}, {Raccanelli}  \& {Riess}}{{Bird}
  et~al.}{2016}]{bird2016}
{Bird} S.,  {Cholis} I.,  {Mu{\~n}oz} J.~B.,  {Ali-Ha{\"i}moud} Y.,
  {Kamionkowski} M.,  {Kovetz} E.~D.,  {Raccanelli} A.,   {Riess} A.~G.,  2016,
  \mn@doi [Physical Review Letters] {10.1103/PhysRevLett.116.201301}, \href
  {http://adsabs.harvard.edu/abs/2016PhRvL.116t1301B} {116, 201301}

\bibitem[\protect\citeauthoryear{{Blaauw}}{{Blaauw}}{1961}]{blaauw1961}
{Blaauw} A.,  1961, \bain, \href
  {https://ui.adsabs.harvard.edu/abs/1961BAN....15..265B} {15, 265}

\bibitem[\protect\citeauthoryear{{Boco}, {Lapi}, {Goswami}, {Perrotta},
  {Baccigalupi}  \& {Danese}}{{Boco} et~al.}{2019}]{boco2019}
{Boco} L.,  {Lapi} A.,  {Goswami} S.,  {Perrotta} F.,  {Baccigalupi} C.,
  {Danese} L.,  2019, \mn@doi [\apj] {10.3847/1538-4357/ab328e}, \href
  {https://ui.adsabs.harvard.edu/abs/2019ApJ...881..157B} {881, 157}

\bibitem[\protect\citeauthoryear{{B{\"{o}}hm-Vitense}}{{B{\"{o}}hm-Vitense}}{1958}]{Bohm-Vitense1958}
{B{\"{o}}hm-Vitense} E.,  1958, \mn@doi [\zap] {10.1017/CBO9781107415324.004},
  46, 108

\bibitem[\protect\citeauthoryear{{Bond}, {Arnett}  \& {Carr}}{{Bond}
  et~al.}{1984}]{bond1984}
{Bond} J.~R.,  {Arnett} W.~D.,   {Carr} B.~J.,  1984, \mn@doi [\apj]
  {10.1086/162057}, \href
  {https://ui.adsabs.harvard.edu/abs/1984ApJ...280..825B} {280, 825}

\bibitem[\protect\citeauthoryear{{Bondi} \& {Hoyle}}{{Bondi} \&
  {Hoyle}}{1944}]{Bondi44}
{Bondi} H.,  {Hoyle} F.,  1944, \mn@doi [\mnras] {10.1093/mnras/104.5.273},
  \href {https://ui.adsabs.harvard.edu/abs/1944MNRAS.104..273B} {104, 273}

\bibitem[\protect\citeauthoryear{{Bouffanais}, {Mapelli}, {Santoliquido},
  {Giacobbo}, {Iorio}  \& {Costa}}{{Bouffanais} et~al.}{2021a}]{Bouffanais21}
{Bouffanais} Y.,  {Mapelli} M.,  {Santoliquido} F.,  {Giacobbo} N.,  {Iorio}
  G.,   {Costa} G.,  2021a, \mn@doi [\mnras] {10.1093/mnras/stab1589}, \href
  {https://ui.adsabs.harvard.edu/abs/2021MNRAS.505.3873B} {505, 3873}

\bibitem[\protect\citeauthoryear{{Bouffanais}, {Mapelli}, {Santoliquido},
  {Giacobbo}, {Iorio}  \& {Costa}}{{Bouffanais}
  et~al.}{2021b}]{bouffanais2021b}
{Bouffanais} Y.,  {Mapelli} M.,  {Santoliquido} F.,  {Giacobbo} N.,  {Iorio}
  G.,   {Costa} G.,  2021b, \mn@doi [\mnras] {10.1093/mnras/stab1589}, \href
  {https://ui.adsabs.harvard.edu/abs/2021MNRAS.505.3873B} {505, 3873}

\bibitem[\protect\citeauthoryear{{Breivik} et~al.,}{{Breivik}
  et~al.}{2020}]{Cosmic}
{Breivik} K.,  et~al., 2020, \mn@doi [\apj] {10.3847/1538-4357/ab9d85}, \href
  {https://ui.adsabs.harvard.edu/abs/2020ApJ...898...71B} {898, 71}

\bibitem[\protect\citeauthoryear{{Bressan}, {Chiosi}  \& {Bertelli}}{{Bressan}
  et~al.}{1981}]{bressan1981}
{Bressan} A.~G.,  {Chiosi} C.,   {Bertelli} G.,  1981, \aap, \href
  {https://ui.adsabs.harvard.edu/\#abs/1981A&A...102...25B} {102, 25}

\bibitem[\protect\citeauthoryear{{Bressan}, {Marigo}, {Girardi}, {Salasnich},
  {Dal Cero}, {Rubele}  \& {Nanni}}{{Bressan} et~al.}{2012}]{Bressan2012}
{Bressan} A.,  {Marigo} P.,  {Girardi} L.,  {Salasnich} B.,  {Dal Cero} C.,
  {Rubele} S.,   {Nanni} A.,  2012, \mn@doi [\mnras]
  {10.1111/j.1365-2966.2012.21948.x}, \href
  {http://adsabs.harvard.edu/abs/2012MNRAS.427..127B} {427, 127}

\bibitem[\protect\citeauthoryear{{Briel}, {Stevance}  \& {Eldridge}}{{Briel}
  et~al.}{2023}]{Briel22}
{Briel} M.~M.,  {Stevance} H.~F.,   {Eldridge} J.~J.,  2023, \mn@doi [\mnras]
  {10.1093/mnras/stad399}, \href
  {https://ui.adsabs.harvard.edu/abs/2023MNRAS.520.5724B} {520, 5724}

\bibitem[\protect\citeauthoryear{{Broekgaarden} et~al.,}{{Broekgaarden}
  et~al.}{2021}]{broekgaarden2021}
{Broekgaarden} F.~S.,  et~al., 2021, \mn@doi [\mnras] {10.1093/mnras/stab2716},
  \href {https://ui.adsabs.harvard.edu/abs/2021MNRAS.508.5028B} {508, 5028}

\bibitem[\protect\citeauthoryear{{Broekgaarden} et~al.,}{{Broekgaarden}
  et~al.}{2022}]{broekgaarden2022}
{Broekgaarden} F.~S.,  et~al., 2022, \mn@doi [\mnras] {10.1093/mnras/stac1677},
  \href {https://ui.adsabs.harvard.edu/abs/2022MNRAS.516.5737B} {516, 5737}

\bibitem[\protect\citeauthoryear{{Brott} et~al.,}{{Brott}
  et~al.}{2011}]{Brott2011}
{Brott} I.,  et~al., 2011, \mn@doi [\aap] {10.1051/0004-6361/201016114}, \href
  {https://ui.adsabs.harvard.edu/abs/2011A&A...530A.116B} {530, A116}

\bibitem[\protect\citeauthoryear{{Brown}}{{Brown}}{1995}]{brown1995}
{Brown} G.~E.,  1995, \mn@doi [\apj] {10.1086/175268}, \href
  {https://ui.adsabs.harvard.edu/abs/1995ApJ...440..270B} {440, 270}

\bibitem[\protect\citeauthoryear{{Caffau}, {Ludwig}, {Steffen}, {Freytag}  \&
  {Bonifacio}}{{Caffau} et~al.}{2011}]{Caffau2011}
{Caffau} E.,  {Ludwig} H.~G.,  {Steffen} M.,  {Freytag} B.,   {Bonifacio} P.,
  2011, \mn@doi [\solphys] {10.1007/s11207-010-9541-4}, 268, 255

\bibitem[\protect\citeauthoryear{{Callister} \& {Farr}}{{Callister} \&
  {Farr}}{2023}]{callister2023}
{Callister} T.~A.,  {Farr} W.~M.,  2023, \mn@doi [arXiv e-prints]
  {10.48550/arXiv.2302.07289}, \href
  {https://ui.adsabs.harvard.edu/abs/2023arXiv230207289C} {p. arXiv:2302.07289}

\bibitem[\protect\citeauthoryear{{Camacho}, {Torres}, {Garc{\'\i}a-Berro},
  {Zorotovic}, {Schreiber}, {Rebassa-Mansergas}, {Nebot G{\'o}mez-Mor{\'a}n}
  \& {G{\"a}nsicke}}{{Camacho} et~al.}{2014}]{camacho2014}
{Camacho} J.,  {Torres} S.,  {Garc{\'\i}a-Berro} E.,  {Zorotovic} M.,
  {Schreiber} M.~R.,  {Rebassa-Mansergas} A.,  {Nebot G{\'o}mez-Mor{\'a}n} A.,
   {G{\"a}nsicke} B.~T.,  2014, \mn@doi [\aap] {10.1051/0004-6361/201323052},
  \href {https://ui.adsabs.harvard.edu/abs/2014A&A...566A..86C} {566, A86}

\bibitem[\protect\citeauthoryear{{Cantiello}, {Yoon}, {Langer}  \&
  {Livio}}{{Cantiello} et~al.}{2007}]{Cantiello07}
{Cantiello} M.,  {Yoon} S.~C.,  {Langer} N.,   {Livio} M.,  2007, \mn@doi
  [\aap] {10.1051/0004-6361:20077115}, \href
  {https://ui.adsabs.harvard.edu/abs/2007A&A...465L..29C} {465, L29}

\bibitem[\protect\citeauthoryear{{Capano} et~al.,}{{Capano}
  et~al.}{2020}]{Capano20}
{Capano} C.~D.,  et~al., 2020, \mn@doi [Nature Astronomy]
  {10.1038/s41550-020-1014-6}, \href
  {https://ui.adsabs.harvard.edu/abs/2020NatAs...4..625C} {4, 625}

\bibitem[\protect\citeauthoryear{{Carr} \& {Hawking}}{{Carr} \&
  {Hawking}}{1974}]{carr1974}
{Carr} B.~J.,  {Hawking} S.~W.,  1974, \mn@doi [\mnras]
  {10.1093/mnras/168.2.399}, \href
  {https://ui.adsabs.harvard.edu/abs/1974MNRAS.168..399C} {168, 399}

\bibitem[\protect\citeauthoryear{{Carr}, {K{\"u}hnel}  \& {Sandstad}}{{Carr}
  et~al.}{2016}]{carr2016}
{Carr} B.,  {K{\"u}hnel} F.,   {Sandstad} M.,  2016, \mn@doi [\prd]
  {10.1103/PhysRevD.94.083504}, \href
  {http://adsabs.harvard.edu/abs/2016PhRvD..94h3504C} {94, 083504}

\bibitem[\protect\citeauthoryear{{Chen}, {Bressan}, {Girardi}, {Marigo}, {Kong}
   \& {Lanza}}{{Chen} et~al.}{2015}]{Chen15}
{Chen} Y.,  {Bressan} A.,  {Girardi} L.,  {Marigo} P.,  {Kong} X.,   {Lanza}
  A.,  2015, \mn@doi [\mnras] {10.1093/mnras/stv1281}, \href
  {https://ui.adsabs.harvard.edu/abs/2015MNRAS.452.1068C} {452, 1068}

\bibitem[\protect\citeauthoryear{{Chieffi} \& {Limongi}}{{Chieffi} \&
  {Limongi}}{2013}]{chieffi2013}
{Chieffi} A.,  {Limongi} M.,  2013, \mn@doi [\apj]
  {10.1088/0004-637X/764/1/21}, \href
  {http://adsabs.harvard.edu/abs/2013ApJ...764...21C} {764, 21}

\bibitem[\protect\citeauthoryear{{Choi}, {Dotter}, {Conroy}, {Cantiello},
  {Paxton}  \& {Johnson}}{{Choi} et~al.}{2016}]{Choi2016}
{Choi} J.,  {Dotter} A.,  {Conroy} C.,  {Cantiello} M.,  {Paxton} B.,
  {Johnson} B.~D.,  2016, \mn@doi [\apj] {10.3847/0004-637X/823/2/102}, \href
  {https://ui.adsabs.harvard.edu/abs/2016ApJ...823..102C} {823, 102}

\bibitem[\protect\citeauthoryear{{Chruslinska}, {Nelemans}  \&
  {Belczynski}}{{Chruslinska} et~al.}{2019}]{chruslinska2019}
{Chruslinska} M.,  {Nelemans} G.,   {Belczynski} K.,  2019, \mn@doi [\mnras]
  {10.1093/mnras/sty3087}, \href
  {https://ui.adsabs.harvard.edu/abs/2019MNRAS.482.5012C} {482, 5012}

\bibitem[\protect\citeauthoryear{{Claeys}, {Pols}, {Izzard}, {Vink}  \&
  {Verbunt}}{{Claeys} et~al.}{2014}]{Claeys14}
{Claeys} J.~S.~W.,  {Pols} O.~R.,  {Izzard} R.~G.,  {Vink} J.,   {Verbunt}
  F.~W.~M.,  2014, \mn@doi [\aap] {10.1051/0004-6361/201322714}, \href
  {https://ui.adsabs.harvard.edu/abs/2014A&A...563A..83C} {563, A83}

\bibitem[\protect\citeauthoryear{{Claret} \& {Torres}}{{Claret} \&
  {Torres}}{2018}]{Claret2018}
{Claret} A.,  {Torres} G.,  2018, \mn@doi [\apj] {10.3847/1538-4357/aabd35},
  \href {https://ui.adsabs.harvard.edu/abs/2018ApJ...859..100C} {859, 100}

\bibitem[\protect\citeauthoryear{{Costa}, {Girardi}, {Bressan}, {Marigo},
  {Rodrigues}, {Chen}, {Lanza}  \& {Goudfrooij}}{{Costa}
  et~al.}{2019}]{Costa2019}
{Costa} G.,  {Girardi} L.,  {Bressan} A.,  {Marigo} P.,  {Rodrigues} T.~S.,
  {Chen} Y.,  {Lanza} A.,   {Goudfrooij} P.,  2019, \mn@doi [\mnras]
  {10.1093/mnras/stz728}, \href
  {https://ui.adsabs.harvard.edu/abs/2019MNRAS.485.4641C} {485, 4641}

\bibitem[\protect\citeauthoryear{{Costa}, {Bressan}, {Mapelli}, {Marigo},
  {Iorio}  \& {Spera}}{{Costa} et~al.}{2021}]{Costa2021}
{Costa} G.,  {Bressan} A.,  {Mapelli} M.,  {Marigo} P.,  {Iorio} G.,   {Spera}
  M.,  2021, \mn@doi [\mnras] {10.1093/mnras/staa3916}, \href
  {https://ui.adsabs.harvard.edu/abs/2021MNRAS.501.4514C} {501, 4514}

\bibitem[\protect\citeauthoryear{{Costa}, {Ballone}, {Mapelli}  \&
  {Bressan}}{{Costa} et~al.}{2022}]{costa2022}
{Costa} G.,  {Ballone} A.,  {Mapelli} M.,   {Bressan} A.,  2022, \mn@doi
  [\mnras] {10.1093/mnras/stac2222}, \href
  {https://ui.adsabs.harvard.edu/abs/2022MNRAS.516.1072C} {516, 1072}

\bibitem[\protect\citeauthoryear{{Costa}, {Mapelli}, {Iorio}, {Santoliquido},
  {Escobar}, {Klessen}  \& {Bressan}}{{Costa} et~al.}{2023}]{costa2023}
{Costa} G.,  {Mapelli} M.,  {Iorio} G.,  {Santoliquido} F.,  {Escobar} G.~J.,
  {Klessen} R.~S.,   {Bressan} A.,  2023, \mn@doi [arXiv e-prints]
  {10.48550/arXiv.2303.15511}, \href
  {https://ui.adsabs.harvard.edu/abs/2023arXiv230315511C} {p. arXiv:2303.15511}

\bibitem[\protect\citeauthoryear{{Cyburt} et~al.,}{{Cyburt}
  et~al.}{2010}]{Cyburt2010}
{Cyburt} R.~H.,  et~al., 2010, \mn@doi [\apjs] {10.1088/0067-0049/189/1/240},
  \href {https://ui.adsabs.harvard.edu/abs/2010ApJS..189..240C} {189, 240}

\bibitem[\protect\citeauthoryear{{De Donder} \& {Vanbeveren}}{{De Donder} \&
  {Vanbeveren}}{2004}]{dedonder2004}
{De Donder} E.,  {Vanbeveren} D.,  2004, \mn@doi [\nar]
  {10.1016/j.newar.2004.07.001}, \href
  {https://ui.adsabs.harvard.edu/abs/2004NewAR..48..861D} {48, 861}

\bibitem[\protect\citeauthoryear{{De Luca}, {Desjacques}, {Franciolini}, {Pani}
   \& {Riotto}}{{De Luca} et~al.}{2021a}]{deluca2020}
{De Luca} V.,  {Desjacques} V.,  {Franciolini} G.,  {Pani} P.,   {Riotto} A.,
  2021a, \mn@doi [\prl] {10.1103/PhysRevLett.126.051101}, \href
  {https://ui.adsabs.harvard.edu/abs/2021PhRvL.126e1101D} {126, 051101}

\bibitem[\protect\citeauthoryear{{De Luca}, {Franciolini}, {Pani}  \&
  {Riotto}}{{De Luca} et~al.}{2021b}]{deluca2021}
{De Luca} V.,  {Franciolini} G.,  {Pani} P.,   {Riotto} A.,  2021b, \mn@doi
  [\jcap] {10.1088/1475-7516/2021/05/003}, \href
  {https://ui.adsabs.harvard.edu/abs/2021JCAP...05..003D} {2021, 003}

\bibitem[\protect\citeauthoryear{De~Marco, Passy, Moe, Herwig, Mac~Low  \&
  Paxton}{De~Marco et~al.}{2011}]{DeMarco2011}
De~Marco O.,  Passy J.-C.,  Moe M.,  Herwig F.,  Mac~Low M.-M.,   Paxton B.,
  2011, \mn@doi [Monthly Notices of the Royal Astronomical Society]
  {10.1111/j.1365-2966.2010.17891.x}, 411, 2277

\bibitem[\protect\citeauthoryear{{Deloye} \& {Taam}}{{Deloye} \&
  {Taam}}{2010}]{Deloye2010}
{Deloye} C.~J.,  {Taam} R.~E.,  2010, \mn@doi [\apjl]
  {10.1088/2041-8205/719/1/L28}, \href
  {https://ui.adsabs.harvard.edu/abs/2010ApJ...719L..28D} {719, L28}

\bibitem[\protect\citeauthoryear{{Dewi}, {Podsiadlowski}  \& {Sena}}{{Dewi}
  et~al.}{2006}]{dewi2006}
{Dewi} J.~D.~M.,  {Podsiadlowski} P.,   {Sena} A.,  2006, \mn@doi [\mnras]
  {10.1111/j.1365-2966.2006.10233.x}, \href
  {https://ui.adsabs.harvard.edu/abs/2006MNRAS.368.1742D} {368, 1742}

\bibitem[\protect\citeauthoryear{{Di Carlo}, {Giacobbo}, {Mapelli}, {Pasquato},
  {Spera}, {Wang}  \& {Haardt}}{{Di Carlo} et~al.}{2019}]{dicarlo2019}
{Di Carlo} U.~N.,  {Giacobbo} N.,  {Mapelli} M.,  {Pasquato} M.,  {Spera} M.,
  {Wang} L.,   {Haardt} F.,  2019, \mn@doi [\mnras] {10.1093/mnras/stz1453},
  \href {https://ui.adsabs.harvard.edu/abs/2019MNRAS.487.2947D} {487, 2947}

\bibitem[\protect\citeauthoryear{{Di Carlo}, {Mapelli}, {Bouffanais},
  {Giacobbo}, {Santoliquido}, {Bressan}, {Spera}  \& {Haardt}}{{Di Carlo}
  et~al.}{2020a}]{dicarlo2020a}
{Di Carlo} U.~N.,  {Mapelli} M.,  {Bouffanais} Y.,  {Giacobbo} N.,
  {Santoliquido} F.,  {Bressan} A.,  {Spera} M.,   {Haardt} F.,  2020a, \mn@doi
  [\mnras] {10.1093/mnras/staa1997}, \href
  {https://ui.adsabs.harvard.edu/abs/2020MNRAS.497.1043D} {497, 1043}

\bibitem[\protect\citeauthoryear{{Di Carlo} et~al.,}{{Di Carlo}
  et~al.}{2020b}]{dicarlo2020b}
{Di Carlo} U.~N.,  et~al., 2020b, \mn@doi [\mnras] {10.1093/mnras/staa2286},
  \href {https://ui.adsabs.harvard.edu/abs/2020MNRAS.498..495D} {498, 495}

\bibitem[\protect\citeauthoryear{{Di Carlo} et~al.,}{{Di Carlo}
  et~al.}{2021}]{dicarlo2021}
{Di Carlo} U.~N.,  et~al., 2021, \mn@doi [\mnras] {10.1093/mnras/stab2390},
  \href {https://ui.adsabs.harvard.edu/abs/2021MNRAS.507.5132D} {507, 5132}

\bibitem[\protect\citeauthoryear{{Di Stefano}, {Kruckow}, {Gao}, {Neunteufel}
  \& {Kobayashi}}{{Di Stefano} et~al.}{2023}]{DiStefano2023}
{Di Stefano} R.,  {Kruckow} M.~U.,  {Gao} Y.,  {Neunteufel} P.~G.,
  {Kobayashi} C.,  2023, \mn@doi [\apj] {10.3847/1538-4357/acae9b}, \href
  {https://ui.adsabs.harvard.edu/abs/2023ApJ...944...87D} {944, 87}

\bibitem[\protect\citeauthoryear{{Dominik}, {Belczynski}, {Fryer}, {Holz},
  {Berti}, {Bulik}, {Mandel}  \& {O'Shaughnessy}}{{Dominik}
  et~al.}{2012}]{dominik2012}
{Dominik} M.,  {Belczynski} K.,  {Fryer} C.,  {Holz} D.~E.,  {Berti} E.,
  {Bulik} T.,  {Mandel} I.,   {O'Shaughnessy} R.,  2012, \mn@doi [\apj]
  {10.1088/0004-637X/759/1/52}, \href
  {http://adsabs.harvard.edu/abs/2012ApJ...759...52D} {759, 52}

\bibitem[\protect\citeauthoryear{{Eggleton}}{{Eggleton}}{1971}]{eggleton1971}
{Eggleton} P.~P.,  1971, \mn@doi [\mnras] {10.1093/mnras/151.3.351}, \href
  {https://ui.adsabs.harvard.edu/abs/1971MNRAS.151..351E} {151, 351}

\bibitem[\protect\citeauthoryear{{Eggleton}}{{Eggleton}}{1983}]{Eggleton83}
{Eggleton} P.~P.,  1983, \mn@doi [\apj] {10.1086/160960}, \href
  {https://ui.adsabs.harvard.edu/abs/1983ApJ...268..368E} {268, 368}

\bibitem[\protect\citeauthoryear{{Eldridge} \& {Stanway}}{{Eldridge} \&
  {Stanway}}{2012}]{eldridge12}
{Eldridge} J.~J.,  {Stanway} E.~R.,  2012, \mn@doi [\mnras]
  {10.1111/j.1365-2966.2011.19713.x}, \href
  {https://ui.adsabs.harvard.edu/abs/2012MNRAS.419..479E} {419, 479}

\bibitem[\protect\citeauthoryear{{Eldridge} \& {Stanway}}{{Eldridge} \&
  {Stanway}}{2016}]{eldridge2016}
{Eldridge} J.~J.,  {Stanway} E.~R.,  2016, \mn@doi [\mnras]
  {10.1093/mnras/stw1772}, \href
  {https://ui.adsabs.harvard.edu/abs/2016MNRAS.462.3302E} {462, 3302}

\bibitem[\protect\citeauthoryear{{Eldridge} \& {Tout}}{{Eldridge} \&
  {Tout}}{2004}]{eldridge2004}
{Eldridge} J.~J.,  {Tout} C.~A.,  2004, \mn@doi [\mnras]
  {10.1111/j.1365-2966.2004.08041.x}, \href
  {https://ui.adsabs.harvard.edu/abs/2004MNRAS.353...87E} {353, 87}

\bibitem[\protect\citeauthoryear{{Eldridge}, {Izzard}  \& {Tout}}{{Eldridge}
  et~al.}{2008}]{eldridge2008}
{Eldridge} J.~J.,  {Izzard} R.~G.,   {Tout} C.~A.,  2008, \mn@doi [\mnras]
  {10.1111/j.1365-2966.2007.12738.x}, \href
  {https://ui.adsabs.harvard.edu/abs/2008MNRAS.384.1109E} {384, 1109}

\bibitem[\protect\citeauthoryear{{Eldridge}, {Langer}  \& {Tout}}{{Eldridge}
  et~al.}{2011}]{eldridge11}
{Eldridge} J.~J.,  {Langer} N.,   {Tout} C.~A.,  2011, \mn@doi [\mnras]
  {10.1111/j.1365-2966.2011.18650.x}, \href
  {https://ui.adsabs.harvard.edu/abs/2011MNRAS.414.3501E} {414, 3501}

\bibitem[\protect\citeauthoryear{{Eldridge}, {Stanway}, {Xiao}, {McClelland},
  {Taylor}, {Ng}, {Greis}  \& {Bray}}{{Eldridge} et~al.}{2017}]{eldridge2017}
{Eldridge} J.~J.,  {Stanway} E.~R.,  {Xiao} L.,  {McClelland} L.~A.~S.,
  {Taylor} G.,  {Ng} M.,  {Greis} S.~M.~L.,   {Bray} J.~C.,  2017, \mn@doi
  [\pasa] {10.1017/pasa.2017.51}, \href
  {https://ui.adsabs.harvard.edu/abs/2017PASA...34...58E} {34, e058}

\bibitem[\protect\citeauthoryear{{Farah}, {Edelman}, {Zevin}, {Fishbach},
  {Mar{\'\i}a Ezquiaga}, {Farr}  \& {Holz}}{{Farah} et~al.}{2023}]{farah2023}
{Farah} A.~M.,  {Edelman} B.,  {Zevin} M.,  {Fishbach} M.,  {Mar{\'\i}a
  Ezquiaga} J.,  {Farr} B.,   {Holz} D.~E.,  2023, \mn@doi [arXiv e-prints]
  {10.48550/arXiv.2301.00834}, \href
  {https://ui.adsabs.harvard.edu/abs/2023arXiv230100834F} {p. arXiv:2301.00834}

\bibitem[\protect\citeauthoryear{{Farmer}, {Renzo}, {de Mink}, {Marchant}  \&
  {Justham}}{{Farmer} et~al.}{2019}]{farmer2019}
{Farmer} R.,  {Renzo} M.,  {de Mink} S.~E.,  {Marchant} P.,   {Justham} S.,
  2019, \mn@doi [\apj] {10.3847/1538-4357/ab518b}, \href
  {https://ui.adsabs.harvard.edu/abs/2019ApJ...887...53F} {887, 53}

\bibitem[\protect\citeauthoryear{{Farmer}, {Renzo}, {de Mink}, {Fishbach}  \&
  {Justham}}{{Farmer} et~al.}{2020}]{farmer2020}
{Farmer} R.,  {Renzo} M.,  {de Mink} S.~E.,  {Fishbach} M.,   {Justham} S.,
  2020, \mn@doi [\apjl] {10.3847/2041-8213/abbadd}, \href
  {https://ui.adsabs.harvard.edu/abs/2020ApJ...902L..36F} {902, L36}

\bibitem[\protect\citeauthoryear{{Farr}, {Sravan}, {Cantrell}, {Kreidberg},
  {Bailyn}, {Mandel}  \& {Kalogera}}{{Farr} et~al.}{2011}]{farr2011}
{Farr} W.~M.,  {Sravan} N.,  {Cantrell} A.,  {Kreidberg} L.,  {Bailyn} C.~D.,
  {Mandel} I.,   {Kalogera} V.,  2011, \mn@doi [\apj]
  {10.1088/0004-637X/741/2/103}, \href
  {http://adsabs.harvard.edu/abs/2011ApJ...741..103F} {741, 103}

\bibitem[\protect\citeauthoryear{{Fern{\'a}ndez}, {Quataert}, {Kashiyama}  \&
  {Coughlin}}{{Fern{\'a}ndez} et~al.}{2018}]{fernandez2018}
{Fern{\'a}ndez} R.,  {Quataert} E.,  {Kashiyama} K.,   {Coughlin} E.~R.,  2018,
  \mn@doi [\mnras] {10.1093/mnras/sty306}, \href
  {https://ui.adsabs.harvard.edu/abs/2018MNRAS.476.2366F} {476, 2366}

\bibitem[\protect\citeauthoryear{{Fields}, {Timmes}, {Farmer}, {Petermann},
  {Wolf}  \& {Couch}}{{Fields} et~al.}{2018}]{fields18}
{Fields} C.~E.,  {Timmes} F.~X.,  {Farmer} R.,  {Petermann} I.,  {Wolf} W.~M.,
   {Couch} S.~M.,  2018, \mn@doi [\apjs] {10.3847/1538-4365/aaa29b}, \href
  {https://ui.adsabs.harvard.edu/abs/2018ApJS..234...19F} {234, 19}

\bibitem[\protect\citeauthoryear{{Fragione} \& {Loeb}}{{Fragione} \&
  {Loeb}}{2019}]{fragione2019}
{Fragione} G.,  {Loeb} A.,  2019, \mn@doi [\mnras] {10.1093/mnras/stz1131},
  \href {https://ui.adsabs.harvard.edu/abs/2019MNRAS.486.4443F} {486, 4443}

\bibitem[\protect\citeauthoryear{{Fragione} \& {Silk}}{{Fragione} \&
  {Silk}}{2020}]{fragione2020}
{Fragione} G.,  {Silk} J.,  2020, \mn@doi [\mnras] {10.1093/mnras/staa2629},
  \href {https://ui.adsabs.harvard.edu/abs/2020MNRAS.498.4591F} {498, 4591}

\bibitem[\protect\citeauthoryear{{Fragione}, {Loeb}  \& {Rasio}}{{Fragione}
  et~al.}{2020}]{fragione2020b}
{Fragione} G.,  {Loeb} A.,   {Rasio} F.~A.,  2020, \mn@doi [\apjl]
  {10.3847/2041-8213/abbc0a}, \href
  {https://ui.adsabs.harvard.edu/abs/2020ApJ...902L..26F} {902, L26}

\bibitem[\protect\citeauthoryear{Fragos}{Fragos}{2022}]{PosydonZenodo}
Fragos T.,  2022, POSYDON data, \mn@doi{10.5281/zenodo.6655751}, \url
  {https://doi.org/10.5281/zenodo.6655751}

\bibitem[\protect\citeauthoryear{{Fragos}, {Andrews}, {Ramirez-Ruiz}, {Meynet},
  {Kalogera}, {Taam}  \& {Zezas}}{{Fragos} et~al.}{2019}]{fragos2019}
{Fragos} T.,  {Andrews} J.~J.,  {Ramirez-Ruiz} E.,  {Meynet} G.,  {Kalogera}
  V.,  {Taam} R.~E.,   {Zezas} A.,  2019, \mn@doi [\apjl]
  {10.3847/2041-8213/ab40d1}, \href
  {https://ui.adsabs.harvard.edu/abs/2019ApJ...883L..45F} {883, L45}

\bibitem[\protect\citeauthoryear{{Fragos} et~al.,}{{Fragos}
  et~al.}{2023}]{Posydon}
{Fragos} T.,  et~al., 2023, \mn@doi [\apjs] {10.3847/1538-4365/ac90c1}, \href
  {https://ui.adsabs.harvard.edu/abs/2023ApJS..264...45F} {264, 45}

\bibitem[\protect\citeauthoryear{{Franciolini} et~al.,}{{Franciolini}
  et~al.}{2022}]{franciolini2022}
{Franciolini} G.,  et~al., 2022, \mn@doi [\prd] {10.1103/PhysRevD.105.083526},
  \href {https://ui.adsabs.harvard.edu/abs/2022PhRvD.105h3526F} {105, 083526}

\bibitem[\protect\citeauthoryear{{Fryer}, {Belczynski}, {Wiktorowicz},
  {Dominik}, {Kalogera}  \& {Holz}}{{Fryer} et~al.}{2012}]{fryer2012}
{Fryer} C.~L.,  {Belczynski} K.,  {Wiktorowicz} G.,  {Dominik} M.,  {Kalogera}
  V.,   {Holz} D.~E.,  2012, \mn@doi [\apj] {10.1088/0004-637X/749/1/91}, \href
  {http://adsabs.harvard.edu/abs/2012ApJ...749...91F} {749, 91}

\bibitem[\protect\citeauthoryear{{Gallegos-Garcia}, {Berry}, {Marchant}  \&
  {Kalogera}}{{Gallegos-Garcia} et~al.}{2021}]{gallegosgarcia2021}
{Gallegos-Garcia} M.,  {Berry} C. P.~L.,  {Marchant} P.,   {Kalogera} V.,
  2021, \mn@doi [\apj] {10.3847/1538-4357/ac2610}, \href
  {https://ui.adsabs.harvard.edu/abs/2021ApJ...922..110G} {922, 110}

\bibitem[\protect\citeauthoryear{{Ge}, {Webbink}, {Han}  \& {Chen}}{{Ge}
  et~al.}{2010a}]{Ge2010b}
{Ge} H.,  {Webbink} R.~F.,  {Han} Z.,   {Chen} X.,  2010a, \mn@doi [\apss]
  {10.1007/s10509-010-0286-1}, \href
  {https://ui.adsabs.harvard.edu/abs/2010Ap&SS.329..243G} {329, 243}

\bibitem[\protect\citeauthoryear{{Ge}, {Hjellming}, {Webbink}, {Chen}  \&
  {Han}}{{Ge} et~al.}{2010b}]{ge2010}
{Ge} H.,  {Hjellming} M.~S.,  {Webbink} R.~F.,  {Chen} X.,   {Han} Z.,  2010b,
  \mn@doi [\apj] {10.1088/0004-637X/717/2/724}, \href
  {https://ui.adsabs.harvard.edu/abs/2010ApJ...717..724G} {717, 724}

\bibitem[\protect\citeauthoryear{{Ge}, {Webbink}, {Chen}  \& {Han}}{{Ge}
  et~al.}{2015}]{ge2015}
{Ge} H.,  {Webbink} R.~F.,  {Chen} X.,   {Han} Z.,  2015, \mn@doi [\apj]
  {10.1088/0004-637X/812/1/40}, \href
  {https://ui.adsabs.harvard.edu/abs/2015ApJ...812...40G} {812, 40}

\bibitem[\protect\citeauthoryear{{Ge}, {Webbink}  \& {Han}}{{Ge}
  et~al.}{2020a}]{ge2020b}
{Ge} H.,  {Webbink} R.~F.,   {Han} Z.,  2020a, \mn@doi [\apjs]
  {10.3847/1538-4365/ab98f6}, \href
  {https://ui.adsabs.harvard.edu/abs/2020ApJS..249....9G} {249, 9}

\bibitem[\protect\citeauthoryear{{Ge}, {Webbink}, {Chen}  \& {Han}}{{Ge}
  et~al.}{2020b}]{ge2020a}
{Ge} H.,  {Webbink} R.~F.,  {Chen} X.,   {Han} Z.,  2020b, \mn@doi [\apj]
  {10.3847/1538-4357/aba7b7}, \href
  {https://ui.adsabs.harvard.edu/abs/2020ApJ...899..132G} {899, 132}

\bibitem[\protect\citeauthoryear{{Ge} et~al.,}{{Ge} et~al.}{2022}]{Ge2022}
{Ge} H.,  et~al., 2022, \mn@doi [\apj] {10.3847/1538-4357/ac75d3}, \href
  {https://ui.adsabs.harvard.edu/abs/2022ApJ...933..137G} {933, 137}

\bibitem[\protect\citeauthoryear{{Georgy}, {Ekstr{\"o}m}, {Granada}, {Meynet},
  {Mowlavi}, {Eggenberger}  \& {Maeder}}{{Georgy} et~al.}{2013}]{Georgy2013}
{Georgy} C.,  {Ekstr{\"o}m} S.,  {Granada} A.,  {Meynet} G.,  {Mowlavi} N.,
  {Eggenberger} P.,   {Maeder} A.,  2013, \mn@doi [\aap]
  {10.1051/0004-6361/201220558}, \href
  {https://ui.adsabs.harvard.edu/abs/2013A&A...553A..24G} {553, A24}

\bibitem[\protect\citeauthoryear{{Giacobbo} \& {Mapelli}}{{Giacobbo} \&
  {Mapelli}}{2018}]{GiacobboCOB}
{Giacobbo} N.,  {Mapelli} M.,  2018, \mn@doi [\mnras] {10.1093/mnras/sty1999},
  \href {https://ui.adsabs.harvard.edu/abs/2018MNRAS.480.2011G} {480, 2011}

\bibitem[\protect\citeauthoryear{{Giacobbo} \& {Mapelli}}{{Giacobbo} \&
  {Mapelli}}{2019}]{GiacobboEC}
{Giacobbo} N.,  {Mapelli} M.,  2019, \mn@doi [\mnras] {10.1093/mnras/sty2848},
  \href {https://ui.adsabs.harvard.edu/abs/2019MNRAS.482.2234G} {482, 2234}

\bibitem[\protect\citeauthoryear{{Giacobbo} \& {Mapelli}}{{Giacobbo} \&
  {Mapelli}}{2020}]{GiacobboKick}
{Giacobbo} N.,  {Mapelli} M.,  2020, \mn@doi [\apj] {10.3847/1538-4357/ab7335},
  \href {https://ui.adsabs.harvard.edu/abs/2020ApJ...891..141G} {891, 141}

\bibitem[\protect\citeauthoryear{{Giacobbo}, {Mapelli}  \& {Spera}}{{Giacobbo}
  et~al.}{2018}]{GiacobboWind}
{Giacobbo} N.,  {Mapelli} M.,   {Spera} M.,  2018, \mn@doi [\mnras]
  {10.1093/mnras/stx2933}, \href
  {https://ui.adsabs.harvard.edu/abs/2018MNRAS.474.2959G} {474, 2959}

\bibitem[\protect\citeauthoryear{{Giersz}, {Leigh}, {Hypki}, {L{\"u}tzgendorf}
  \& {Askar}}{{Giersz} et~al.}{2015}]{giersz2015}
{Giersz} M.,  {Leigh} N.,  {Hypki} A.,  {L{\"u}tzgendorf} N.,   {Askar} A.,
  2015, \mn@doi [\mnras] {10.1093/mnras/stv2162}, \href
  {http://adsabs.harvard.edu/abs/2015MNRAS.454.3150G} {454, 3150}

\bibitem[\protect\citeauthoryear{{Gr{\"a}fener} \& {Hamann}}{{Gr{\"a}fener} \&
  {Hamann}}{2008}]{Graefener2008}
{Gr{\"a}fener} G.,  {Hamann} W.-R.,  2008, \mn@doi [\aap]
  {10.1051/0004-6361:20066176}, \href
  {http://adsabs.harvard.edu/abs/2008A%26A...482..945G} {482, 945}

\bibitem[\protect\citeauthoryear{{Halabi}, {Izzard}  \& {Tout}}{{Halabi}
  et~al.}{2018}]{Halabi2018}
{Halabi} G.~M.,  {Izzard} R.~G.,   {Tout} C.~A.,  2018, \mn@doi [\mnras]
  {10.1093/mnras/sty2243}, \href
  {https://ui.adsabs.harvard.edu/abs/2018MNRAS.480.5176H} {480, 5176}

\bibitem[\protect\citeauthoryear{Harris et~al.,}{Harris
  et~al.}{2020}]{Harris20}
Harris C.~R.,  et~al., 2020, \mn@doi [Nature] {10.1038/s41586-020-2649-2}, 585,
  357

\bibitem[\protect\citeauthoryear{{Heger}, {Fryer}, {Woosley}, {Langer}  \&
  {Hartmann}}{{Heger} et~al.}{2003}]{heger2003}
{Heger} A.,  {Fryer} C.~L.,  {Woosley} S.~E.,  {Langer} N.,   {Hartmann} D.~H.,
   2003, \mn@doi [\apj] {10.1086/375341}, \href
  {http://adsabs.harvard.edu/abs/2003ApJ...591..288H} {591, 288}

\bibitem[\protect\citeauthoryear{{Hirai} \& {Mandel}}{{Hirai} \&
  {Mandel}}{2022}]{Ryosuke22}
{Hirai} R.,  {Mandel} I.,  2022, \mn@doi [\apjl] {10.3847/2041-8213/ac9519},
  \href {https://ui.adsabs.harvard.edu/abs/2022ApJ...937L..42H} {937, L42}

\bibitem[\protect\citeauthoryear{{Hobbs}, {Lorimer}, {Lyne}  \&
  {Kramer}}{{Hobbs} et~al.}{2005}]{H05}
{Hobbs} G.,  {Lorimer} D.~R.,  {Lyne} A.~G.,   {Kramer} M.,  2005, \mn@doi
  [\mnras] {10.1111/j.1365-2966.2005.09087.x}, \href
  {https://ui.adsabs.harvard.edu/abs/2005MNRAS.360..974H} {360, 974}

\bibitem[\protect\citeauthoryear{Hunter}{Hunter}{2007}]{Hunter2007}
Hunter J.~D.,  2007, \mn@doi [Computing in Science \& Engineering]
  {10.1109/MCSE.2007.55}, 9, 90

\bibitem[\protect\citeauthoryear{{Hurley}, {Pols}  \& {Tout}}{{Hurley}
  et~al.}{2000}]{Hurley00}
{Hurley} J.~R.,  {Pols} O.~R.,   {Tout} C.~A.,  2000, \mn@doi [\mnras]
  {10.1046/j.1365-8711.2000.03426.x}, \href
  {https://ui.adsabs.harvard.edu/abs/2000MNRAS.315..543H} {315, 543}

\bibitem[\protect\citeauthoryear{{Hurley}, {Tout}  \& {Pols}}{{Hurley}
  et~al.}{2002}]{Hurley02}
{Hurley} J.~R.,  {Tout} C.~A.,   {Pols} O.~R.,  2002, \mn@doi [\mnras]
  {10.1046/j.1365-8711.2002.05038.x}, \href
  {https://ui.adsabs.harvard.edu/abs/2002MNRAS.329..897H} {329, 897}

\bibitem[\protect\citeauthoryear{{Hut}}{{Hut}}{1981}]{Hut81}
{Hut} P.,  1981, \aap, \href
  {https://ui.adsabs.harvard.edu/abs/1981A&A....99..126H} {99, 126}

\bibitem[\protect\citeauthoryear{{Iben} \& {Livio}}{{Iben} \&
  {Livio}}{1993}]{Iben93}
{Iben} Icko J.,  {Livio} M.,  1993, \mn@doi [\pasp] {10.1086/133321}, \href
  {https://ui.adsabs.harvard.edu/abs/1993PASP..105.1373I} {105, 1373}

\bibitem[\protect\citeauthoryear{{Iglesias} \& {Rogers}}{{Iglesias} \&
  {Rogers}}{1996}]{Iglesias1996}
{Iglesias} C.~A.,  {Rogers} F.~J.,  1996, \mn@doi [\apj] {10.1086/177381},
  \href {https://ui.adsabs.harvard.edu/abs/1996ApJ...464..943I} {464, 943}

\bibitem[\protect\citeauthoryear{Iorio et~al.,}{Iorio
  et~al.}{2023}]{paperdataset}
Iorio et~al., 2023, {Daset from the paper "Compact object mergers: exploring
  uncertainties from stellar and binary evolution with SEVN"},
  \mn@doi{10.5281/zenodo.7794546}, \url
  {https://doi.org/10.5281/zenodo.7794546}

\bibitem[\protect\citeauthoryear{{Itoh}, {Uchida}, {Sakamoto}, {Kohyama}  \&
  {Nozawa}}{{Itoh} et~al.}{2008}]{Itoh2008}
{Itoh} N.,  {Uchida} S.,  {Sakamoto} Y.,  {Kohyama} Y.,   {Nozawa} S.,  2008,
  \mn@doi [\apj] {10.1086/529367}, \href
  {https://ui.adsabs.harvard.edu/abs/2008ApJ...677..495I} {677, 495}

\bibitem[\protect\citeauthoryear{{Ivanova} \& {Taam}}{{Ivanova} \&
  {Taam}}{2004}]{Ivanova04}
{Ivanova} N.,  {Taam} R.~E.,  2004, \mn@doi [\apj] {10.1086/380561}, \href
  {https://ui.adsabs.harvard.edu/abs/2004ApJ...601.1058I} {601, 1058}

\bibitem[\protect\citeauthoryear{{Ivanova} et~al.,}{{Ivanova}
  et~al.}{2013a}]{IvanovaCE}
{Ivanova} N.,  et~al., 2013a, \mn@doi [\aapr] {10.1007/s00159-013-0059-2},
  \href {https://ui.adsabs.harvard.edu/abs/2013A&ARv..21...59I} {21, 59}

\bibitem[\protect\citeauthoryear{{Ivanova} et~al.,}{{Ivanova}
  et~al.}{2013b}]{ivanova2013}
{Ivanova} N.,  et~al., 2013b, \mn@doi [\aapr] {10.1007/s00159-013-0059-2},
  \href {http://adsabs.harvard.edu/abs/2013A%26ARv..21...59I} {21, 59}

\bibitem[\protect\citeauthoryear{{Izzard}, {Tout}, {Karakas}  \&
  {Pols}}{{Izzard} et~al.}{2004}]{binaryc1}
{Izzard} R.~G.,  {Tout} C.~A.,  {Karakas} A.~I.,   {Pols} O.~R.,  2004, \mn@doi
  [\mnras] {10.1111/j.1365-2966.2004.07446.x}, \href
  {https://ui.adsabs.harvard.edu/abs/2004MNRAS.350..407I} {350, 407}

\bibitem[\protect\citeauthoryear{{Izzard}, {Dray}, {Karakas}, {Lugaro}  \&
  {Tout}}{{Izzard} et~al.}{2006}]{izzard2006}
{Izzard} R.~G.,  {Dray} L.~M.,  {Karakas} A.~I.,  {Lugaro} M.,   {Tout} C.~A.,
  2006, \mn@doi [\aap] {10.1051/0004-6361:20066129}, \href
  {https://ui.adsabs.harvard.edu/abs/2006A&A...460..565I} {460, 565}

\bibitem[\protect\citeauthoryear{{Izzard}, {Glebbeek}, {Stancliffe}  \&
  {Pols}}{{Izzard} et~al.}{2009}]{izzard2009}
{Izzard} R.~G.,  {Glebbeek} E.,  {Stancliffe} R.~J.,   {Pols} O.~R.,  2009,
  \mn@doi [\aap] {10.1051/0004-6361/200912827}, \href
  {https://ui.adsabs.harvard.edu/abs/2009A&A...508.1359I} {508, 1359}

\bibitem[\protect\citeauthoryear{{Izzard}, {Preece}, {Jofre}, {Halabi},
  {Masseron}  \& {Tout}}{{Izzard} et~al.}{2018}]{binaryc2}
{Izzard} R.~G.,  {Preece} H.,  {Jofre} P.,  {Halabi} G.~M.,  {Masseron} T.,
  {Tout} C.~A.,  2018, \mn@doi [\mnras] {10.1093/mnras/stx2355}, \href
  {https://ui.adsabs.harvard.edu/abs/2018MNRAS.473.2984I} {473, 2984}

\bibitem[\protect\citeauthoryear{{Jim{\'e}nez-Forteza}, {Keitel}, {Husa},
  {Hannam}, {Khan}  \& {P{\"u}rrer}}{{Jim{\'e}nez-Forteza}
  et~al.}{2017}]{jimenez-forteza2017}
{Jim{\'e}nez-Forteza} X.,  {Keitel} D.,  {Husa} S.,  {Hannam} M.,  {Khan} S.,
  {P{\"u}rrer} M.,  2017, \mn@doi [\prd] {10.1103/PhysRevD.95.064024}, \href
  {https://ui.adsabs.harvard.edu/abs/2017PhRvD..95f4024J} {95, 064024}

\bibitem[\protect\citeauthoryear{{Justesen} \& {Albrecht}}{{Justesen} \&
  {Albrecht}}{2021}]{Justensen21}
{Justesen} A.~B.,  {Albrecht} S.,  2021, \mn@doi [\apj]
  {10.3847/1538-4357/abefcd}, \href
  {https://ui.adsabs.harvard.edu/abs/2021ApJ...912..123J} {912, 123}

\bibitem[\protect\citeauthoryear{{Justham}, {Podsiadlowski}  \&
  {Han}}{{Justham} et~al.}{2011}]{justham2011}
{Justham} S.,  {Podsiadlowski} P.,   {Han} Z.,  2011, \mn@doi [\mnras]
  {10.1111/j.1365-2966.2010.17497.x}, \href
  {https://ui.adsabs.harvard.edu/abs/2011MNRAS.410..984J} {410, 984}

\bibitem[\protect\citeauthoryear{{Kamlah} et~al.,}{{Kamlah}
  et~al.}{2022}]{kamlah2022}
{Kamlah} A.~W.~H.,  et~al., 2022, \mn@doi [\mnras] {10.1093/mnras/stab3748},
  \href {https://ui.adsabs.harvard.edu/abs/2022MNRAS.511.4060K} {511, 4060}

\bibitem[\protect\citeauthoryear{{Kitaura}, {Janka}  \&
  {Hillebrandt}}{{Kitaura} et~al.}{2006}]{Kitaura06}
{Kitaura} F.~S.,  {Janka} H.~T.,   {Hillebrandt} W.,  2006, \mn@doi [\aap]
  {10.1051/0004-6361:20054703}, \href
  {https://ui.adsabs.harvard.edu/abs/2006A&A...450..345K} {450, 345}

\bibitem[\protect\citeauthoryear{{Kiziltan}, {Kottas}, {De Yoreo}  \&
  {Thorsett}}{{Kiziltan} et~al.}{2013}]{Kiziltan13}
{Kiziltan} B.,  {Kottas} A.,  {De Yoreo} M.,   {Thorsett} S.~E.,  2013, \mn@doi
  [\apj] {10.1088/0004-637X/778/1/66}, \href
  {https://ui.adsabs.harvard.edu/abs/2013ApJ...778...66K} {778, 66}

\bibitem[\protect\citeauthoryear{{Klencki}, {Nelemans}, {Istrate}  \&
  {Pols}}{{Klencki} et~al.}{2020}]{Klencki2020}
{Klencki} J.,  {Nelemans} G.,  {Istrate} A.~G.,   {Pols} O.,  2020, \mn@doi
  [\aap] {10.1051/0004-6361/202037694}, \href
  {https://ui.adsabs.harvard.edu/abs/2020A&A...638A..55K} {638, A55}

\bibitem[\protect\citeauthoryear{{Klencki}, {Nelemans}, {Istrate}  \&
  {Chruslinska}}{{Klencki} et~al.}{2021}]{Klencki21}
{Klencki} J.,  {Nelemans} G.,  {Istrate} A.~G.,   {Chruslinska} M.,  2021,
  \mn@doi [\aap] {10.1051/0004-6361/202038707}, \href
  {https://ui.adsabs.harvard.edu/abs/2021A&A...645A..54K} {645, A54}

\bibitem[\protect\citeauthoryear{{Klencki}, {Istrate}, {Nelemans}  \&
  {Pols}}{{Klencki} et~al.}{2022}]{Klencki22}
{Klencki} J.,  {Istrate} A.,  {Nelemans} G.,   {Pols} O.,  2022, \mn@doi [\aap]
  {10.1051/0004-6361/202142701}, \href
  {https://ui.adsabs.harvard.edu/abs/2022A&A...662A..56K} {662, A56}

\bibitem[\protect\citeauthoryear{{Korol}, {Belokurov}  \& {Toonen}}{{Korol}
  et~al.}{2022}]{korol2022}
{Korol} V.,  {Belokurov} V.,   {Toonen} S.,  2022, \mn@doi [\mnras]
  {10.1093/mnras/stac1686}, \href
  {https://ui.adsabs.harvard.edu/abs/2022MNRAS.515.1228K} {515, 1228}

\bibitem[\protect\citeauthoryear{{Kremer} et~al.,}{{Kremer}
  et~al.}{2020a}]{kremer2020b}
{Kremer} K.,  et~al., 2020a, \mn@doi [\apjs] {10.3847/1538-4365/ab7919}, \href
  {https://ui.adsabs.harvard.edu/abs/2020ApJS..247...48K} {247, 48}

\bibitem[\protect\citeauthoryear{{Kremer} et~al.,}{{Kremer}
  et~al.}{2020b}]{kremer2020}
{Kremer} K.,  et~al., 2020b, \mn@doi [\apj] {10.3847/1538-4357/abb945}, \href
  {https://ui.adsabs.harvard.edu/abs/2020ApJ...903...45K} {903, 45}

\bibitem[\protect\citeauthoryear{{Kroupa}}{{Kroupa}}{2001}]{Kroupa2001}
{Kroupa} P.,  2001, \mn@doi [\mnras] {10.1046/j.1365-8711.2001.04022.x}, \href
  {http://adsabs.harvard.edu/abs/2001MNRAS.322..231K} {322, 231}

\bibitem[\protect\citeauthoryear{{Kruckow}, {Tauris}, {Langer}, {Sz{\'e}csi},
  {Marchant}  \& {Podsiadlowski}}{{Kruckow} et~al.}{2016}]{Kruckow2016}
{Kruckow} M.~U.,  {Tauris} T.~M.,  {Langer} N.,  {Sz{\'e}csi} D.,  {Marchant}
  P.,   {Podsiadlowski} P.,  2016, \mn@doi [\aap]
  {10.1051/0004-6361/201629420}, \href
  {https://ui.adsabs.harvard.edu/abs/2016A&A...596A..58K} {596, A58}

\bibitem[\protect\citeauthoryear{{Kruckow}, {Tauris}, {Langer}, {Kramer}  \&
  {Izzard}}{{Kruckow} et~al.}{2018}]{Combine}
{Kruckow} M.~U.,  {Tauris} T.~M.,  {Langer} N.,  {Kramer} M.,   {Izzard} R.~G.,
   2018, \mn@doi [\mnras] {10.1093/mnras/sty2190}, \href
  {https://ui.adsabs.harvard.edu/abs/2018MNRAS.481.1908K} {481, 1908}

\bibitem[\protect\citeauthoryear{{Ku{\v{c}}inskas}}{{Ku{\v{c}}inskas}}{1998}]{Kucinskas98}
{Ku{\v{c}}inskas} A.,  1998, \apss, \href
  {https://ui.adsabs.harvard.edu/abs/1998Ap&SS.262..127K} {262, 127}

\bibitem[\protect\citeauthoryear{{Lattimer} \& {Yahil}}{{Lattimer} \&
  {Yahil}}{1989}]{Lattimer1989}
{Lattimer} J.~M.,  {Yahil} A.,  1989, \mn@doi [\apj] {10.1086/167404}, \href
  {https://ui.adsabs.harvard.edu/abs/1989ApJ...340..426L} {340, 426}

\bibitem[\protect\citeauthoryear{{Lau}, {Hirai}, {Gonz{\'a}lez-Bol{\'\i}var},
  {Price}, {De Marco}  \& {Mandel}}{{Lau} et~al.}{2022}]{Lau22}
{Lau} M. Y.~M.,  {Hirai} R.,  {Gonz{\'a}lez-Bol{\'\i}var} M.,  {Price} D.~J.,
  {De Marco} O.,   {Mandel} I.,  2022, \mn@doi [\mnras]
  {10.1093/mnras/stac049}, \href
  {https://ui.adsabs.harvard.edu/abs/2022MNRAS.512.5462L} {512, 5462}

\bibitem[\protect\citeauthoryear{{Law-Smith} et~al.,}{{Law-Smith}
  et~al.}{2020}]{lawsmith2020}
{Law-Smith} J. A.~P.,  et~al., 2020, arXiv e-prints, \href
  {https://ui.adsabs.harvard.edu/abs/2020arXiv201106630L} {p. arXiv:2011.06630}

\bibitem[\protect\citeauthoryear{{Limongi} \& {Chieffi}}{{Limongi} \&
  {Chieffi}}{2018}]{Limongi2018}
{Limongi} M.,  {Chieffi} A.,  2018, \mn@doi [\apjs] {10.3847/1538-4365/aacb24},
  \href {https://ui.adsabs.harvard.edu/abs/2018ApJS..237...13L} {237, 13}

\bibitem[\protect\citeauthoryear{{Lipunov}, {Postnov}  \&
  {Prokhorov}}{{Lipunov} et~al.}{1996}]{lipunov1996}
{Lipunov} V.~M.,  {Postnov} K.~A.,   {Prokhorov} M.~E.,  1996, \aap, \href
  {https://ui.adsabs.harvard.edu/abs/1996A&A...310..489L} {310, 489}

\bibitem[\protect\citeauthoryear{{Lipunov}, {Postnov}, {Prokhorov}  \&
  {Bogomazov}}{{Lipunov} et~al.}{2009}]{lipunov2009}
{Lipunov} V.~M.,  {Postnov} K.~A.,  {Prokhorov} M.~E.,   {Bogomazov} A.~I.,
  2009, \mn@doi [Astronomy Reports] {10.1134/S1063772909100047}, \href
  {https://ui.adsabs.harvard.edu/abs/2009ARep...53..915L} {53, 915}

\bibitem[\protect\citeauthoryear{{Livio} \& {Soker}}{{Livio} \&
  {Soker}}{1988}]{Livio88}
{Livio} M.,  {Soker} N.,  1988, \mn@doi [\apj] {10.1086/166419}, \href
  {https://ui.adsabs.harvard.edu/abs/1988ApJ...329..764L} {329, 764}

\bibitem[\protect\citeauthoryear{{Lubow} \& {Shu}}{{Lubow} \&
  {Shu}}{1975}]{Lublow75}
{Lubow} S.~H.,  {Shu} F.~H.,  1975, \mn@doi [\apj] {10.1086/153614}, \href
  {https://ui.adsabs.harvard.edu/abs/1975ApJ...198..383L} {198, 383}

\bibitem[\protect\citeauthoryear{{Madau} \& {Dickinson}}{{Madau} \&
  {Dickinson}}{2014}]{madaudickinson2014}
{Madau} P.,  {Dickinson} M.,  2014, \mn@doi [\araa]
  {10.1146/annurev-astro-081811-125615}, \href
  {http://adsabs.harvard.edu/abs/2014ARA%26A..52..415M} {52, 415}

\bibitem[\protect\citeauthoryear{{Madau} \& {Fragos}}{{Madau} \&
  {Fragos}}{2017}]{madau2017}
{Madau} P.,  {Fragos} T.,  2017, \mn@doi [\apj] {10.3847/1538-4357/aa6af9},
  \href {https://ui.adsabs.harvard.edu/abs/2017ApJ...840...39M} {840, 39}

\bibitem[\protect\citeauthoryear{{Mandel}}{{Mandel}}{2021}]{mandel2021tgw}
{Mandel} I.,  2021, \mn@doi [Research Notes of the American Astronomical
  Society] {10.3847/2515-5172/ac2d35}, \href
  {https://ui.adsabs.harvard.edu/abs/2021RNAAS...5..223M} {5, 223}

\bibitem[\protect\citeauthoryear{{Mandel} \& {Farmer}}{{Mandel} \&
  {Farmer}}{2022}]{mandelfarmer2018}
{Mandel} I.,  {Farmer} A.,  2022, \mn@doi [\physrep]
  {10.1016/j.physrep.2022.01.003}, \href
  {https://ui.adsabs.harvard.edu/abs/2022PhR...955....1M} {955, 1}

\bibitem[\protect\citeauthoryear{{Mandel} \& {Fragos}}{{Mandel} \&
  {Fragos}}{2020}]{Mandel20}
{Mandel} I.,  {Fragos} T.,  2020, \mn@doi [\apjl] {10.3847/2041-8213/ab8e41},
  \href {https://ui.adsabs.harvard.edu/abs/2020ApJ...895L..28M} {895, L28}

\bibitem[\protect\citeauthoryear{{Mapelli}}{{Mapelli}}{2016}]{mapelli2016}
{Mapelli} M.,  2016, \mn@doi [\mnras] {10.1093/mnras/stw869}, \href
  {http://adsabs.harvard.edu/abs/2016MNRAS.459.3432M} {459, 3432}

\bibitem[\protect\citeauthoryear{Mapelli}{Mapelli}{2021}]{mapelli2021review}
Mapelli M.,  2021, Formation Channels of Single and Binary Stellar-Mass Black
  Holes.
Springer Singapore, Singapore, pp 1--65,
  \mn@doi{10.1007/978-981-15-4702-7_16-1}, \url
  {https://doi.org/10.1007/978-981-15-4702-7_16-1}

\bibitem[\protect\citeauthoryear{{Mapelli}, {Zampieri}, {Ripamonti}  \&
  {Bressan}}{{Mapelli} et~al.}{2013}]{mapelli2013}
{Mapelli} M.,  {Zampieri} L.,  {Ripamonti} E.,   {Bressan} A.,  2013, \mn@doi
  [\mnras] {10.1093/mnras/sts500}, \href
  {http://adsabs.harvard.edu/abs/2013MNRAS.429.2298M} {429, 2298}

\bibitem[\protect\citeauthoryear{{Mapelli}, {Giacobbo}, {Ripamonti}  \&
  {Spera}}{{Mapelli} et~al.}{2017}]{mapelli2017}
{Mapelli} M.,  {Giacobbo} N.,  {Ripamonti} E.,   {Spera} M.,  2017, \mn@doi
  [\mnras] {10.1093/mnras/stx2123}, \href
  {http://adsabs.harvard.edu/abs/2017MNRAS.472.2422M} {472, 2422}

\bibitem[\protect\citeauthoryear{{Mapelli}, {Spera}, {Montanari}, {Limongi},
  {Chieffi}, {Giacobbo}, {Bressan}  \& {Bouffanais}}{{Mapelli}
  et~al.}{2020}]{Mapelli20}
{Mapelli} M.,  {Spera} M.,  {Montanari} E.,  {Limongi} M.,  {Chieffi} A.,
  {Giacobbo} N.,  {Bressan} A.,   {Bouffanais} Y.,  2020, \mn@doi [\apj]
  {10.3847/1538-4357/ab584d}, \href
  {https://ui.adsabs.harvard.edu/abs/2020ApJ...888...76M} {888, 76}

\bibitem[\protect\citeauthoryear{{Mapelli} et~al.,}{{Mapelli}
  et~al.}{2021}]{mapelli2021}
{Mapelli} M.,  et~al., 2021, \mn@doi [\mnras] {10.1093/mnras/stab1334}, \href
  {https://ui.adsabs.harvard.edu/abs/2021MNRAS.505..339M} {505, 339}

\bibitem[\protect\citeauthoryear{{Mapelli}, {Bouffanais}, {Santoliquido}, {Arca
  Sedda}  \& {Artale}}{{Mapelli} et~al.}{2022}]{mapelli2022}
{Mapelli} M.,  {Bouffanais} Y.,  {Santoliquido} F.,  {Arca Sedda} M.,
  {Artale} M.~C.,  2022, \mn@doi [\mnras] {10.1093/mnras/stac422}, \href
  {https://ui.adsabs.harvard.edu/abs/2022MNRAS.511.5797M} {511, 5797}

\bibitem[\protect\citeauthoryear{{Marchant} \& {Moriya}}{{Marchant} \&
  {Moriya}}{2020}]{MarchantMoryia21}
{Marchant} P.,  {Moriya} T.~J.,  2020, \mn@doi [\aap]
  {10.1051/0004-6361/202038902}, \href
  {https://ui.adsabs.harvard.edu/abs/2020A&A...640L..18M} {640, L18}

\bibitem[\protect\citeauthoryear{{Marchant}, {Pappas}, {Gallegos-Garcia},
  {Berry}, {Taam}, {Kalogera}  \& {Podsiadlowski}}{{Marchant}
  et~al.}{2021}]{Marchant21}
{Marchant} P.,  {Pappas} K. M.~W.,  {Gallegos-Garcia} M.,  {Berry} C. P.~L.,
  {Taam} R.~E.,  {Kalogera} V.,   {Podsiadlowski} P.,  2021, \mn@doi [\aap]
  {10.1051/0004-6361/202039992}, \href
  {https://ui.adsabs.harvard.edu/abs/2021A&A...650A.107M} {650, A107}

\bibitem[\protect\citeauthoryear{{Marcussen} \& {Albrecht}}{{Marcussen} \&
  {Albrecht}}{2022}]{Marcussen2022}
{Marcussen} M.~L.,  {Albrecht} S.~H.,  2022, \mn@doi [\apj]
  {10.3847/1538-4357/ac75c2}, \href
  {https://ui.adsabs.harvard.edu/abs/2022ApJ...933..227M} {933, 227}

\bibitem[\protect\citeauthoryear{{Marigo} \& {Aringer}}{{Marigo} \&
  {Aringer}}{2009}]{Marigo2009}
{Marigo} P.,  {Aringer} B.,  2009, \mn@doi [\aap]
  {10.1051/0004-6361/200912598}, \href
  {https://ui.adsabs.harvard.edu/abs/2009A&A...508.1539M} {508, 1539}

\bibitem[\protect\citeauthoryear{{Mehta}, {Buonanno}, {Gair}, {Miller},
  {Farag}, {deBoer}, {Wiescher}  \& {Timmes}}{{Mehta} et~al.}{2022}]{mehta2022}
{Mehta} A.~K.,  {Buonanno} A.,  {Gair} J.,  {Miller} M.~C.,  {Farag} E.,
  {deBoer} R.~J.,  {Wiescher} M.,   {Timmes} F.~X.,  2022, \mn@doi [\apj]
  {10.3847/1538-4357/ac3130}, \href
  {https://ui.adsabs.harvard.edu/abs/2022ApJ...924...39M} {924, 39}

\bibitem[\protect\citeauthoryear{{Meibom} \& {Mathieu}}{{Meibom} \&
  {Mathieu}}{2005}]{Meibom}
{Meibom} S.,  {Mathieu} R.~D.,  2005, \mn@doi [\apj] {10.1086/427082}, \href
  {https://ui.adsabs.harvard.edu/abs/2005ApJ...620..970M} {620, 970}

\bibitem[\protect\citeauthoryear{{Mennekens} \& {Vanbeveren}}{{Mennekens} \&
  {Vanbeveren}}{2014}]{mennekens2014}
{Mennekens} N.,  {Vanbeveren} D.,  2014, \mn@doi [\aap]
  {10.1051/0004-6361/201322198}, \href
  {https://ui.adsabs.harvard.edu/abs/2014A&A...564A.134M} {564, A134}

\bibitem[\protect\citeauthoryear{{Miller} \& {Hamilton}}{{Miller} \&
  {Hamilton}}{2002}]{miller2002}
{Miller} M.~C.,  {Hamilton} D.~P.,  2002, \mn@doi [\mnras]
  {10.1046/j.1365-8711.2002.05112.x}, \href
  {http://adsabs.harvard.edu/abs/2002MNRAS.330..232C} {330, 232}

\bibitem[\protect\citeauthoryear{{Neijssel} et~al.,}{{Neijssel}
  et~al.}{2019a}]{neijssel2019}
{Neijssel} C.~J.,  et~al., 2019a, \mn@doi [\mnras] {10.1093/mnras/stz2840},
  \href {https://ui.adsabs.harvard.edu/abs/2019MNRAS.490.3740N} {490, 3740}

\bibitem[\protect\citeauthoryear{{Neijssel} et~al.,}{{Neijssel}
  et~al.}{2019b}]{Neijssel19}
{Neijssel} C.~J.,  et~al., 2019b, \mn@doi [\mnras] {10.1093/mnras/stz2840},
  \href {https://ui.adsabs.harvard.edu/abs/2019MNRAS.490.3740N} {490, 3740}

\bibitem[\protect\citeauthoryear{{Ng}, {Franciolini}, {Berti}, {Pani}, {Riotto}
   \& {Vitale}}{{Ng} et~al.}{2022}]{ng2022}
{Ng} K. K.~Y.,  {Franciolini} G.,  {Berti} E.,  {Pani} P.,  {Riotto} A.,
  {Vitale} S.,  2022, \mn@doi [\apjl] {10.3847/2041-8213/ac7aae}, \href
  {https://ui.adsabs.harvard.edu/abs/2022ApJ...933L..41N} {933, L41}

\bibitem[\protect\citeauthoryear{{Nguyen} et~al.,}{{Nguyen}
  et~al.}{2022}]{Nguyen22}
{Nguyen} C.~T.,  et~al., 2022, \mn@doi [\aap] {10.1051/0004-6361/202244166},
  \href {https://ui.adsabs.harvard.edu/abs/2022A&A...665A.126N} {665, A126}

\bibitem[\protect\citeauthoryear{{Nugis} \& {Lamers}}{{Nugis} \&
  {Lamers}}{2000}]{nugis2000}
{Nugis} T.,  {Lamers} H.~J.~G.~L.~M.,  2000, \aap, \href
  {https://ui.adsabs.harvard.edu/abs/2000A&A...360..227N} {360, 227}

\bibitem[\protect\citeauthoryear{{O'Connor} \& {Ott}}{{O'Connor} \&
  {Ott}}{2011}]{Oconnor11}
{O'Connor} E.,  {Ott} C.~D.,  2011, \mn@doi [\apj]
  {10.1088/0004-637X/730/2/70}, \href
  {https://ui.adsabs.harvard.edu/abs/2011ApJ...730...70O} {730, 70}

\bibitem[\protect\citeauthoryear{{Ober}, {El Eid}  \& {Fricke}}{{Ober}
  et~al.}{1983}]{ober1983}
{Ober} W.~W.,  {El Eid} M.~F.,   {Fricke} K.~J.,  1983, \aap, \href
  {https://ui.adsabs.harvard.edu/abs/1983A&A...119...61O} {119, 61}

\bibitem[\protect\citeauthoryear{{{\"O}zel} \& {Freire}}{{{\"O}zel} \&
  {Freire}}{2016}]{oezel2016}
{{\"O}zel} F.,  {Freire} P.,  2016, \mn@doi [\araa]
  {10.1146/annurev-astro-081915-023322}, \href
  {https://ui.adsabs.harvard.edu/abs/2016ARA&A..54..401O} {54, 401}

\bibitem[\protect\citeauthoryear{{{\"O}zel}, {Psaltis}, {Narayan}  \&
  {McClintock}}{{{\"O}zel} et~al.}{2010}]{oezel2010}
{{\"O}zel} F.,  {Psaltis} D.,  {Narayan} R.,   {McClintock} J.~E.,  2010,
  \mn@doi [\apj] {10.1088/0004-637X/725/2/1918}, \href
  {https://ui.adsabs.harvard.edu/abs/2010ApJ...725.1918O} {725, 1918}

\bibitem[\protect\citeauthoryear{{{\"O}zel}, {Psaltis}, {Narayan}  \& {Santos
  Villarreal}}{{{\"O}zel} et~al.}{2012}]{Ozel12}
{{\"O}zel} F.,  {Psaltis} D.,  {Narayan} R.,   {Santos Villarreal} A.,  2012,
  \mn@doi [\apj] {10.1088/0004-637X/757/1/55}, \href
  {https://ui.adsabs.harvard.edu/abs/2012ApJ...757...55O} {757, 55}

\bibitem[\protect\citeauthoryear{{Patton} \& {Sukhbold}}{{Patton} \&
  {Sukhbold}}{2020}]{Patton2020}
{Patton} R.~A.,  {Sukhbold} T.,  2020, \mn@doi [\mnras]
  {10.1093/mnras/staa3029}, \href
  {https://ui.adsabs.harvard.edu/abs/2020MNRAS.499.2803P} {499, 2803}

\bibitem[\protect\citeauthoryear{{Pavlovskii}, {Ivanova}, {Belczynski}  \&
  {Van}}{{Pavlovskii} et~al.}{2017}]{Pavlovskii17}
{Pavlovskii} K.,  {Ivanova} N.,  {Belczynski} K.,   {Van} K.~X.,  2017, \mn@doi
  [\mnras] {10.1093/mnras/stw2786}, \href
  {https://ui.adsabs.harvard.edu/abs/2017MNRAS.465.2092P} {465, 2092}

\bibitem[\protect\citeauthoryear{{Paxton}, {Bildsten}, {Dotter}, {Herwig},
  {Lesaffre}  \& {Timmes}}{{Paxton} et~al.}{2011}]{Mesa}
{Paxton} B.,  {Bildsten} L.,  {Dotter} A.,  {Herwig} F.,  {Lesaffre} P.,
  {Timmes} F.,  2011, \mn@doi [\apjs] {10.1088/0067-0049/192/1/3}, \href
  {https://ui.adsabs.harvard.edu/abs/2011ApJS..192....3P} {192, 3}

\bibitem[\protect\citeauthoryear{{Paxton} et~al.,}{{Paxton}
  et~al.}{2013}]{paxton2013}
{Paxton} B.,  et~al., 2013, \mn@doi [\apjs] {10.1088/0067-0049/208/1/4}, \href
  {https://ui.adsabs.harvard.edu/abs/2013ApJS..208....4P} {208, 4}

\bibitem[\protect\citeauthoryear{{Paxton} et~al.,}{{Paxton}
  et~al.}{2015}]{paxton2015}
{Paxton} B.,  et~al., 2015, \mn@doi [\apjs] {10.1088/0067-0049/220/1/15}, \href
  {https://ui.adsabs.harvard.edu/abs/2015ApJS..220...15P} {220, 15}

\bibitem[\protect\citeauthoryear{{Paxton} et~al.,}{{Paxton}
  et~al.}{2018}]{paxton2018}
{Paxton} B.,  et~al., 2018, \mn@doi [\apjs] {10.3847/1538-4365/aaa5a8}, \href
  {https://ui.adsabs.harvard.edu/abs/2018ApJS..234...34P} {234, 34}

\bibitem[\protect\citeauthoryear{{Perez} \& {Granger}}{{Perez} \&
  {Granger}}{2007}]{Ipython}
{Perez} F.,  {Granger} B.~E.,  2007, \mn@doi [Computing in Science Engineering]
  {10.1109/MCSE.2007.53}, 9, 21

\bibitem[\protect\citeauthoryear{{Peters}}{{Peters}}{1964a}]{peters1964}
{Peters} P.~C.,  1964a, \mn@doi [Physical Review] {10.1103/PhysRev.136.B1224},
  \href {http://adsabs.harvard.edu/abs/1964PhRv..136.1224P} {136, 1224}

\bibitem[\protect\citeauthoryear{Peters}{Peters}{1964b}]{Peters64}
Peters P.~C.,  1964b, \mn@doi [Phys. Rev.] {10.1103/PhysRev.136.B1224}, 136,
  B1224

\bibitem[\protect\citeauthoryear{{Petrovic}, {Langer}  \& {van der
  Hucht}}{{Petrovic} et~al.}{2005}]{Petrovic05}
{Petrovic} J.,  {Langer} N.,   {van der Hucht} K.~A.,  2005, \mn@doi [\aap]
  {10.1051/0004-6361:20042368}, \href
  {https://ui.adsabs.harvard.edu/abs/2005A&A...435.1013P} {435, 1013}

\bibitem[\protect\citeauthoryear{{Pols}, {Tout}, {Eggleton}  \& {Han}}{{Pols}
  et~al.}{1995}]{pols1995}
{Pols} O.~R.,  {Tout} C.~A.,  {Eggleton} P.~P.,   {Han} Z.,  1995, \mn@doi
  [\mnras] {10.1093/mnras/274.3.964}, \href
  {https://ui.adsabs.harvard.edu/abs/1995MNRAS.274..964P} {274, 964}

\bibitem[\protect\citeauthoryear{{Pols}, {Schr{\"o}der}, {Hurley}, {Tout}  \&
  {Eggleton}}{{Pols} et~al.}{1998}]{Pols98}
{Pols} O.~R.,  {Schr{\"o}der} K.-P.,  {Hurley} J.~R.,  {Tout} C.~A.,
  {Eggleton} P.~P.,  1998, \mn@doi [\mnras] {10.1046/j.1365-8711.1998.01658.x},
  \href {https://ui.adsabs.harvard.edu/abs/1998MNRAS.298..525P} {298, 525}

\bibitem[\protect\citeauthoryear{{Portegies Zwart} \& {Verbunt}}{{Portegies
  Zwart} \& {Verbunt}}{1996}]{portegieszwart1996}
{Portegies Zwart} S.~F.,  {Verbunt} F.,  1996, \aap, \href
  {http://adsabs.harvard.edu/abs/1996A%26A...309..179P} {309, 179}

\bibitem[\protect\citeauthoryear{{Poutanen}}{{Poutanen}}{2017}]{Poutanen2017}
{Poutanen} J.,  2017, \mn@doi [\apj] {10.3847/1538-4357/835/2/119}, \href
  {https://ui.adsabs.harvard.edu/abs/2017ApJ...835..119P} {835, 119}

\bibitem[\protect\citeauthoryear{{Ragoler}, {Bear}, {Schreier}, {Hillel}  \&
  {Soker}}{{Ragoler} et~al.}{2022}]{Ragoler22}
{Ragoler} N.,  {Bear} E.,  {Schreier} R.,  {Hillel} S.,   {Soker} N.,  2022,
  \mn@doi [\mnras] {10.1093/mnras/stac2148}, \href
  {https://ui.adsabs.harvard.edu/abs/2022MNRAS.515.5473R} {515, 5473}

\bibitem[\protect\citeauthoryear{{Rappaport}, {Verbunt}  \& {Joss}}{{Rappaport}
  et~al.}{1983}]{Rappaort83}
{Rappaport} S.,  {Verbunt} F.,   {Joss} P.~C.,  1983, \mn@doi [\apj]
  {10.1086/161569}, \href
  {https://ui.adsabs.harvard.edu/abs/1983ApJ...275..713R} {275, 713}

\bibitem[\protect\citeauthoryear{{Rasio}, {Tout}, {Lubow}  \& {Livio}}{{Rasio}
  et~al.}{1996}]{rasio96}
{Rasio} F.~A.,  {Tout} C.~A.,  {Lubow} S.~H.,   {Livio} M.,  1996, \mn@doi
  [\apj] {10.1086/177941}, \href
  {https://ui.adsabs.harvard.edu/abs/1996ApJ...470.1187R} {470, 1187}

\bibitem[\protect\citeauthoryear{{Rastello}, {Amaro-Seoane}, {Arca-Sedda},
  {Capuzzo-Dolcetta}, {Fragione}  \& {Tosta e Melo}}{{Rastello}
  et~al.}{2019}]{rastello2018}
{Rastello} S.,  {Amaro-Seoane} P.,  {Arca-Sedda} M.,  {Capuzzo-Dolcetta} R.,
  {Fragione} G.,   {Tosta e Melo} I.,  2019, \mn@doi [\mnras]
  {10.1093/mnras/sty3193}, \href
  {https://ui.adsabs.harvard.edu/abs/2019MNRAS.483.1233R} {483, 1233}

\bibitem[\protect\citeauthoryear{{Rastello}, {Mapelli}, {Di Carlo}, {Giacobbo},
  {Santoliquido}, {Spera}, {Ballone}  \& {Iorio}}{{Rastello}
  et~al.}{2020}]{rastello20}
{Rastello} S.,  {Mapelli} M.,  {Di Carlo} U.~N.,  {Giacobbo} N.,
  {Santoliquido} F.,  {Spera} M.,  {Ballone} A.,   {Iorio} G.,  2020, \mn@doi
  [\mnras] {10.1093/mnras/staa2018}, \href
  {https://ui.adsabs.harvard.edu/abs/2020MNRAS.497.1563R} {497, 1563}

\bibitem[\protect\citeauthoryear{{Rastello}, {Mapelli}, {Di Carlo}, {Iorio},
  {Ballone}, {Giacobbo}, {Santoliquido}  \& {Torniamenti}}{{Rastello}
  et~al.}{2021}]{rastello2021}
{Rastello} S.,  {Mapelli} M.,  {Di Carlo} U.~N.,  {Iorio} G.,  {Ballone} A.,
  {Giacobbo} N.,  {Santoliquido} F.,   {Torniamenti} S.,  2021, \mn@doi
  [\mnras] {10.1093/mnras/stab2355}, \href
  {https://ui.adsabs.harvard.edu/abs/2021MNRAS.507.3612R} {507, 3612}

\bibitem[\protect\citeauthoryear{{Renzo}, {Farmer}, {Justham}, {de Mink},
  {G{\"o}tberg}  \& {Marchant}}{{Renzo} et~al.}{2020a}]{renzo2020}
{Renzo} M.,  {Farmer} R.~J.,  {Justham} S.,  {de Mink} S.~E.,  {G{\"o}tberg}
  Y.,   {Marchant} P.,  2020a, \mn@doi [\mnras] {10.1093/mnras/staa549}, \href
  {https://ui.adsabs.harvard.edu/abs/2020MNRAS.493.4333R} {493, 4333}

\bibitem[\protect\citeauthoryear{{Renzo}, {Cantiello}, {Metzger}  \&
  {Jiang}}{{Renzo} et~al.}{2020b}]{renzo2020b}
{Renzo} M.,  {Cantiello} M.,  {Metzger} B.~D.,   {Jiang} Y.~F.,  2020b, \mn@doi
  [\apjl] {10.3847/2041-8213/abc6a6}, \href
  {https://ui.adsabs.harvard.edu/abs/2020ApJ...904L..13R} {904, L13}

\bibitem[\protect\citeauthoryear{{Riley} et~al.,}{{Riley}
  et~al.}{2022}]{Compas}
{Riley} J.,  et~al., 2022, \mn@doi [\apjs] {10.3847/1538-4365/ac416c}, \href
  {https://ui.adsabs.harvard.edu/abs/2022ApJS..258...34R} {258, 34}

\bibitem[\protect\citeauthoryear{{Rodriguez}, {Morscher}, {Pattabiraman},
  {Chatterjee}, {Haster}  \& {Rasio}}{{Rodriguez} et~al.}{2015}]{rodriguez2015}
{Rodriguez} C.~L.,  {Morscher} M.,  {Pattabiraman} B.,  {Chatterjee} S.,
  {Haster} C.-J.,   {Rasio} F.~A.,  2015, \mn@doi [Physical Review Letters]
  {10.1103/PhysRevLett.115.051101}, \href
  {http://adsabs.harvard.edu/abs/2015PhRvL.115e1101R} {115, 051101}

\bibitem[\protect\citeauthoryear{{Rodriguez}, {Chatterjee}  \&
  {Rasio}}{{Rodriguez} et~al.}{2016}]{rodriguez2016}
{Rodriguez} C.~L.,  {Chatterjee} S.,   {Rasio} F.~A.,  2016, \mn@doi [\prd]
  {10.1103/PhysRevD.93.084029}, \href
  {http://adsabs.harvard.edu/abs/2016PhRvD..93h4029R} {93, 084029}

\bibitem[\protect\citeauthoryear{{R{\"o}pke} \& {De Marco}}{{R{\"o}pke} \& {De
  Marco}}{2023}]{Roepke2022}
{R{\"o}pke} F.~K.,  {De Marco} O.,  2023, \mn@doi [Living Reviews in
  Computational Astrophysics] {10.1007/s41115-023-00017-x}, \href
  {https://ui.adsabs.harvard.edu/abs/2023LRCA....9....2R} {9, 2}

\bibitem[\protect\citeauthoryear{{Sallaska}, {Iliadis}, {Champange}, {Goriely},
  {Starrfield}  \& {Timmes}}{{Sallaska} et~al.}{2013}]{Sallaska2013}
{Sallaska} A.~L.,  {Iliadis} C.,  {Champange} A.~E.,  {Goriely} S.,
  {Starrfield} S.,   {Timmes} F.~X.,  2013, \mn@doi [\apjs]
  {10.1088/0067-0049/207/1/18}, \href
  {https://ui.adsabs.harvard.edu/abs/2013ApJS..207...18S} {207, 18}

\bibitem[\protect\citeauthoryear{{Sana} et~al.,}{{Sana} et~al.}{2012}]{Sana12}
{Sana} H.,  et~al., 2012, \mn@doi [Science] {10.1126/science.1223344}, \href
  {https://ui.adsabs.harvard.edu/abs/2012Sci...337..444S} {337, 444}

\bibitem[\protect\citeauthoryear{{Sander}, {Hamann}, {Todt}, {Hainich},
  {Shenar}, {Ramachandran}  \& {Oskinova}}{{Sander} et~al.}{2019}]{Sander2019}
{Sander} A.~A.~C.,  {Hamann} W.~R.,  {Todt} H.,  {Hainich} R.,  {Shenar} T.,
  {Ramachandran} V.,   {Oskinova} L.~M.,  2019, \mn@doi [\aap]
  {10.1051/0004-6361/201833712}, \href
  {https://ui.adsabs.harvard.edu/abs/2019A&A...621A..92S} {621, A92}

\bibitem[\protect\citeauthoryear{{Santoliquido}, {Mapelli}, {Bouffanais},
  {Giacobbo}, {Di Carlo}, {Rastello}, {Artale}  \& {Ballone}}{{Santoliquido}
  et~al.}{2020}]{santoliquido2020}
{Santoliquido} F.,  {Mapelli} M.,  {Bouffanais} Y.,  {Giacobbo} N.,  {Di Carlo}
  U.~N.,  {Rastello} S.,  {Artale} M.~C.,   {Ballone} A.,  2020, \mn@doi [\apj]
  {10.3847/1538-4357/ab9b78}, \href
  {https://ui.adsabs.harvard.edu/abs/2020ApJ...898..152S} {898, 152}

\bibitem[\protect\citeauthoryear{{Santoliquido}, {Mapelli}, {Giacobbo},
  {Bouffanais}  \& {Artale}}{{Santoliquido} et~al.}{2021}]{santoliquido2021}
{Santoliquido} F.,  {Mapelli} M.,  {Giacobbo} N.,  {Bouffanais} Y.,   {Artale}
  M.~C.,  2021, \mn@doi [\mnras] {10.1093/mnras/stab280}, \href
  {https://ui.adsabs.harvard.edu/abs/2021MNRAS.502.4877S} {502, 4877}

\bibitem[\protect\citeauthoryear{{Santoliquido}, {Mapelli}, {Artale}  \&
  {Boco}}{{Santoliquido} et~al.}{2022}]{santoliquido2022}
{Santoliquido} F.,  {Mapelli} M.,  {Artale} M.~C.,   {Boco} L.,  2022, \mn@doi
  [\mnras] {10.1093/mnras/stac2384}, \href
  {https://ui.adsabs.harvard.edu/abs/2022MNRAS.516.3297S} {516, 3297}

\bibitem[\protect\citeauthoryear{{Scelfo}, {Bellomo}, {Raccanelli}, {Matarrese}
   \& {Verde}}{{Scelfo} et~al.}{2018}]{scelfo2018}
{Scelfo} G.,  {Bellomo} N.,  {Raccanelli} A.,  {Matarrese} S.,   {Verde} L.,
  2018, \mn@doi [\jcap] {10.1088/1475-7516/2018/09/039}, \href
  {http://adsabs.harvard.edu/abs/2018JCAP...09..039S} {9, 039}

\bibitem[\protect\citeauthoryear{{Schneider}, {Izzard}, {Langer}  \& {de
  Mink}}{{Schneider} et~al.}{2015}]{Schneider2015}
{Schneider} F.~R.~N.,  {Izzard} R.~G.,  {Langer} N.,   {de Mink} S.~E.,  2015,
  \mn@doi [\apj] {10.1088/0004-637X/805/1/20}, \href
  {https://ui.adsabs.harvard.edu/abs/2015ApJ...805...20S} {805, 20}

\bibitem[\protect\citeauthoryear{{Schwarzschild}}{{Schwarzschild}}{1958}]{Schwarzschild1958}
{Schwarzschild} M.,  1958, {Structure and evolution of the stars.}.
Princeton, Princeton University Press, 1958.

\bibitem[\protect\citeauthoryear{{Shao} \& {Li}}{{Shao} \& {Li}}{2021}]{Shao21}
{Shao} Y.,  {Li} X.-D.,  2021, \mn@doi [\apj] {10.3847/1538-4357/ac173e}, \href
  {https://ui.adsabs.harvard.edu/abs/2021ApJ...920...81S} {920, 81}

\bibitem[\protect\citeauthoryear{{Soberman}, {Phinney}  \& {van den
  Heuvel}}{{Soberman} et~al.}{1997}]{Soberman97}
{Soberman} G.~E.,  {Phinney} E.~S.,   {van den Heuvel} E.~P.~J.,  1997, \aap,
  \href {https://ui.adsabs.harvard.edu/abs/1997A&A...327..620S} {327, 620}

\bibitem[\protect\citeauthoryear{{Spera} \& {Mapelli}}{{Spera} \&
  {Mapelli}}{2017}]{Spera17}
{Spera} M.,  {Mapelli} M.,  2017, \mn@doi [\mnras] {10.1093/mnras/stx1576},
  \href {https://ui.adsabs.harvard.edu/abs/2017MNRAS.470.4739S} {470, 4739}

\bibitem[\protect\citeauthoryear{{Spera}, {Mapelli}  \& {Bressan}}{{Spera}
  et~al.}{2015}]{Spera15}
{Spera} M.,  {Mapelli} M.,   {Bressan} A.,  2015, \mn@doi [\mnras]
  {10.1093/mnras/stv1161}, \href
  {https://ui.adsabs.harvard.edu/abs/2015MNRAS.451.4086S} {451, 4086}

\bibitem[\protect\citeauthoryear{{Spera}, {Mapelli}, {Giacobbo}, {Trani},
  {Bressan}  \& {Costa}}{{Spera} et~al.}{2019}]{Spera19}
{Spera} M.,  {Mapelli} M.,  {Giacobbo} N.,  {Trani} A.~A.,  {Bressan} A.,
  {Costa} G.,  2019, \mn@doi [\mnras] {10.1093/mnras/stz359}, \href
  {https://ui.adsabs.harvard.edu/abs/2019MNRAS.485..889S} {485, 889}

\bibitem[\protect\citeauthoryear{{Sz{\'e}csi}, {Langer}, {Yoon}, {Sanyal}, {de
  Mink}, {Evans}  \& {Dermine}}{{Sz{\'e}csi} et~al.}{2015}]{BEC2015}
{Sz{\'e}csi} D.,  {Langer} N.,  {Yoon} S.-C.,  {Sanyal} D.,  {de Mink} S.,
  {Evans} C.~J.,   {Dermine} T.,  2015, \mn@doi [\aap]
  {10.1051/0004-6361/201526617}, \href
  {https://ui.adsabs.harvard.edu/abs/2015A&A...581A..15S} {581, A15}

\bibitem[\protect\citeauthoryear{{Tanikawa}}{{Tanikawa}}{2013}]{tanikawa2013}
{Tanikawa} A.,  2013, \mn@doi [\mnras] {10.1093/mnras/stt1380}, \href
  {https://ui.adsabs.harvard.edu/abs/2013MNRAS.435.1358T} {435, 1358}

\bibitem[\protect\citeauthoryear{{Tauris} \& {van den Heuvel}}{{Tauris} \& {van
  den Heuvel}}{2006}]{Tauris06}
{Tauris} T.~M.,  {van den Heuvel} E.~P.~J.,  2006, in , Vol.~39, Compact
  stellar X-ray sources.
Cambridge University Press, pp 623--665

\bibitem[\protect\citeauthoryear{{Tauris}, {Langer}  \&
  {Podsiadlowski}}{{Tauris} et~al.}{2015}]{tauris2015}
{Tauris} T.~M.,  {Langer} N.,   {Podsiadlowski} P.,  2015, \mn@doi [\mnras]
  {10.1093/mnras/stv990}, \href
  {http://adsabs.harvard.edu/abs/2015MNRAS.451.2123T} {451, 2123}

\bibitem[\protect\citeauthoryear{{Tauris} et~al.,}{{Tauris}
  et~al.}{2017}]{tauris2017}
{Tauris} T.~M.,  et~al., 2017, \mn@doi [\apj] {10.3847/1538-4357/aa7e89}, \href
  {http://adsabs.harvard.edu/abs/2017ApJ...846..170T} {846, 170}

\bibitem[\protect\citeauthoryear{{Temmink}, {Pols}, {Justham}, {Istrate}  \&
  {Toonen}}{{Temmink} et~al.}{2023}]{Temmink2022}
{Temmink} K.~D.,  {Pols} O.~R.,  {Justham} S.,  {Istrate} A.~G.,   {Toonen} S.,
   2023, \mn@doi [\aap] {10.1051/0004-6361/202244137}, \href
  {https://ui.adsabs.harvard.edu/abs/2023A&A...669A..45T} {669, A45}

\bibitem[\protect\citeauthoryear{{Timmes} \& {Arnett}}{{Timmes} \&
  {Arnett}}{1999}]{Timmes1999}
{Timmes} F.~X.,  {Arnett} D.,  1999, \mn@doi [\apjs] {10.1086/313271}, \href
  {https://ui.adsabs.harvard.edu/abs/1999ApJS..125..277T} {125, 277}

\bibitem[\protect\citeauthoryear{{Toonen} \& {Nelemans}}{{Toonen} \&
  {Nelemans}}{2013}]{toonen2013}
{Toonen} S.,  {Nelemans} G.,  2013, \mn@doi [\aap]
  {10.1051/0004-6361/201321753}, \href
  {https://ui.adsabs.harvard.edu/abs/2013A&A...557A..87T} {557, A87}

\bibitem[\protect\citeauthoryear{{Toonen}, {Nelemans}  \& {Portegies
  Zwart}}{{Toonen} et~al.}{2012}]{toonen2012}
{Toonen} S.,  {Nelemans} G.,   {Portegies Zwart} S.,  2012, \mn@doi [\aap]
  {10.1051/0004-6361/201218966}, \href
  {https://ui.adsabs.harvard.edu/abs/2012A&A...546A..70T} {546, A70}

\bibitem[\protect\citeauthoryear{{Torniamenti}, {Rastello}, {Mapelli}, {Di
  Carlo}, {Ballone}  \& {Pasquato}}{{Torniamenti}
  et~al.}{2022}]{torniamenti2022}
{Torniamenti} S.,  {Rastello} S.,  {Mapelli} M.,  {Di Carlo} U.~N.,  {Ballone}
  A.,   {Pasquato} M.,  2022, \mn@doi [\mnras] {10.1093/mnras/stac2841}, \href
  {https://ui.adsabs.harvard.edu/abs/2022MNRAS.517.2953T} {517, 2953}

\bibitem[\protect\citeauthoryear{{Trani}, {Rieder}, {Tanikawa}, {Iorio},
  {Martini}, {Karelin}, {Glanz}  \& {Portegies Zwart}}{{Trani}
  et~al.}{2022}]{Trani22}
{Trani} A.~A.,  {Rieder} S.,  {Tanikawa} A.,  {Iorio} G.,  {Martini} R.,
  {Karelin} G.,  {Glanz} H.,   {Portegies Zwart} S.,  2022, \mn@doi [\prd]
  {10.1103/PhysRevD.106.043014}, \href
  {https://ui.adsabs.harvard.edu/abs/2022PhRvD.106d3014T} {106, 043014}

\bibitem[\protect\citeauthoryear{{Tucker} \& {Will}}{{Tucker} \&
  {Will}}{2021}]{tucker2021}
{Tucker} A.,  {Will} C.~M.,  2021, \mn@doi [\prd]
  {10.1103/PhysRevD.104.104023}, \href
  {https://ui.adsabs.harvard.edu/abs/2021PhRvD.104j4023T} {104, 104023}

\bibitem[\protect\citeauthoryear{{Tutukov} \& {Yungelson}}{{Tutukov} \&
  {Yungelson}}{1996}]{tutukov1996}
{Tutukov} A.,  {Yungelson} L.,  1996, \mn@doi [\mnras]
  {10.1093/mnras/280.4.1035}, \href
  {https://ui.adsabs.harvard.edu/abs/1996MNRAS.280.1035T} {280, 1035}

\bibitem[\protect\citeauthoryear{{Ulrich} \& {Burger}}{{Ulrich} \&
  {Burger}}{1976}]{Ulrich76}
{Ulrich} R.~K.,  {Burger} H.~L.,  1976, \mn@doi [\apj] {10.1086/154406}, \href
  {https://ui.adsabs.harvard.edu/abs/1976ApJ...206..509U} {206, 509}

\bibitem[\protect\citeauthoryear{{Vanbeveren}, {De Donder}, {Van Bever}, {Van
  Rensbergen}  \& {De Loore}}{{Vanbeveren} et~al.}{1998}]{vanbeveren1998}
{Vanbeveren} D.,  {De Donder} E.,  {Van Bever} J.,  {Van Rensbergen} W.,   {De
  Loore} C.,  1998, \mn@doi [\na] {10.1016/S1384-1076(98)00020-7}, \href
  {https://ui.adsabs.harvard.edu/abs/1998NewA....3..443V} {3, 443}

\bibitem[\protect\citeauthoryear{{Vigna-G{\'o}mez} et~al.,}{{Vigna-G{\'o}mez}
  et~al.}{2018}]{VG2018}
{Vigna-G{\'o}mez} A.,  et~al., 2018, \mn@doi [\mnras] {10.1093/mnras/sty2463},
  \href {https://ui.adsabs.harvard.edu/abs/2018MNRAS.481.4009V} {481, 4009}

\bibitem[\protect\citeauthoryear{{Vigna-G{\'o}mez}, {Wassink}, {Klencki},
  {Istrate}, {Nelemans}  \& {Mandel}}{{Vigna-G{\'o}mez}
  et~al.}{2022}]{Vigna2022}
{Vigna-G{\'o}mez} A.,  {Wassink} M.,  {Klencki} J.,  {Istrate} A.,  {Nelemans}
  G.,   {Mandel} I.,  2022, \mn@doi [\mnras] {10.1093/mnras/stac237}, \href
  {https://ui.adsabs.harvard.edu/abs/2022MNRAS.511.2326V} {511, 2326}

\bibitem[\protect\citeauthoryear{{Vink}, {de Koter}  \& {Lamers}}{{Vink}
  et~al.}{2000}]{Vink2000}
{Vink} J.~S.,  {de Koter} A.,   {Lamers} H.~J.~G.~L.~M.,  2000, \aap, \href
  {https://ui.adsabs.harvard.edu/abs/2000A&A...362..295V} {362, 295}

\bibitem[\protect\citeauthoryear{{Vink}, {de Koter}  \& {Lamers}}{{Vink}
  et~al.}{2001}]{Vink2001}
{Vink} J.~S.,  {de Koter} A.,   {Lamers} H.~J.~G.~L.~M.,  2001, \mn@doi [\aap]
  {10.1051/0004-6361:20010127}, \href
  {http://adsabs.harvard.edu/abs/2001A%26A...369..574V} {369, 574}

\bibitem[\protect\citeauthoryear{{Vink}, {Muijres}, {Anthonisse}, {de Koter},
  {Gr{\"a}fener}  \& {Langer}}{{Vink} et~al.}{2011}]{Vink2011}
{Vink} J.~S.,  {Muijres} L.~E.,  {Anthonisse} B.,  {de Koter} A.,
  {Gr{\"a}fener} G.,   {Langer} N.,  2011, \mn@doi [\aap]
  {10.1051/0004-6361/201116614}, \href
  {http://adsabs.harvard.edu/abs/2011A%26A...531A.132V} {531, A132}

\bibitem[\protect\citeauthoryear{{Vink}, {Higgins}, {Sander}  \&
  {Sabhahit}}{{Vink} et~al.}{2021}]{vink2021}
{Vink} J.~S.,  {Higgins} E.~R.,  {Sander} A. A.~C.,   {Sabhahit} G.~N.,  2021,
  \mn@doi [\mnras] {10.1093/mnras/stab842}, \href
  {https://ui.adsabs.harvard.edu/abs/2021MNRAS.504..146V} {504, 146}

\bibitem[\protect\citeauthoryear{Virtanen et~al.,}{Virtanen
  et~al.}{2020}]{SciPy2020}
Virtanen P.,  et~al., 2020, \mn@doi [Nature Methods]
  {10.1038/s41592-019-0686-2}, \href {https://rdcu.be/b08Wh} {17, 261}

\bibitem[\protect\citeauthoryear{{Vos}, {{\O}stensen}, {Marchant}  \& {Van
  Winckel}}{{Vos} et~al.}{2015}]{Vos15}
{Vos} J.,  {{\O}stensen} R.~H.,  {Marchant} P.,   {Van Winckel} H.,  2015,
  \mn@doi [\aap] {10.1051/0004-6361/201526019}, \href
  {https://ui.adsabs.harvard.edu/abs/2015A&A...579A..49V} {579, A49}

\bibitem[\protect\citeauthoryear{{Wang}}{{Wang}}{2020}]{wang2020}
{Wang} L.,  2020, \mn@doi [\mnras] {10.1093/mnras/stz3179}, \href
  {https://ui.adsabs.harvard.edu/abs/2020MNRAS.491.2413W} {491, 2413}

\bibitem[\protect\citeauthoryear{{Wang}, {Tanikawa}  \& {Fujii}}{{Wang}
  et~al.}{2022}]{wang2022}
{Wang} L.,  {Tanikawa} A.,   {Fujii} M.,  2022, \mn@doi [\mnras]
  {10.1093/mnras/stac2043}, \href
  {https://ui.adsabs.harvard.edu/abs/2022MNRAS.515.5106W} {515, 5106}

\bibitem[\protect\citeauthoryear{{Webbink}}{{Webbink}}{1984}]{Webbink84}
{Webbink} R.~F.,  1984, \mn@doi [\apj] {10.1086/161701}, \href
  {https://ui.adsabs.harvard.edu/abs/1984ApJ...277..355W} {277, 355}

\bibitem[\protect\citeauthoryear{{Webbink}}{{Webbink}}{1985}]{Webbink85}
{Webbink} R.~F.,  1985, in {Pringle} J.~E.,  {Wade} R.~A.,  eds, , Interacting
  Binary Stars.
Cambridge University Press, p.~39

\bibitem[\protect\citeauthoryear{{Webbink}}{{Webbink}}{1988}]{Webbink88}
{Webbink} R.~F.,  1988, in {Mikolajewska} J.,  {Friedjung} M.,  {Kenyon} S.~J.,
    {Viotti} R.,  eds,  Astrophysics and Space Science Library Vol. 145, IAU
  Colloq. 103: The Symbiotic Phenomenon. p.~311,
  \mn@doi{10.1007/978-94-009-2969-2\_69}

\bibitem[\protect\citeauthoryear{{Weiss}, {Hillebrandt}, {Thomas}  \&
  {Ritter}}{{Weiss} et~al.}{2004}]{CoxGiuli}
{Weiss} A.,  {Hillebrandt} W.,  {Thomas} H.~C.,   {Ritter} H.,  2004, {Cox and
  Giuli's Principles of Stellar Structure}

\bibitem[\protect\citeauthoryear{{Will} \& {Maitra}}{{Will} \&
  {Maitra}}{2017}]{will2017}
{Will} C.~M.,  {Maitra} M.,  2017, \mn@doi [\prd] {10.1103/PhysRevD.95.064003},
  \href {https://ui.adsabs.harvard.edu/abs/2017PhRvD..95f4003W} {95, 064003}

\bibitem[\protect\citeauthoryear{{Woosley}}{{Woosley}}{2017}]{Woosley2017}
{Woosley} S.~E.,  2017, \mn@doi [\apj] {10.3847/1538-4357/836/2/244}, \href
  {https://ui.adsabs.harvard.edu/abs/2017ApJ...836..244W} {836, 244}

\bibitem[\protect\citeauthoryear{{Woosley}}{{Woosley}}{2019}]{Woosley2019}
{Woosley} S.~E.,  2019, \mn@doi [\apj] {10.3847/1538-4357/ab1b41}, \href
  {https://ui.adsabs.harvard.edu/abs/2019ApJ...878...49W} {878, 49}

\bibitem[\protect\citeauthoryear{{Woosley}, {Blinnikov}  \& {Heger}}{{Woosley}
  et~al.}{2007}]{woosley2007}
{Woosley} S.~E.,  {Blinnikov} S.,   {Heger} A.,  2007, \mn@doi [\nat]
  {10.1038/nature06333}, \href
  {https://ui.adsabs.harvard.edu/abs/2007Natur.450..390W} {450, 390}

\bibitem[\protect\citeauthoryear{{Woosley}, {Sukhbold}  \& {Janka}}{{Woosley}
  et~al.}{2020}]{Woosley20}
{Woosley} S.~E.,  {Sukhbold} T.,   {Janka} H.~T.,  2020, \mn@doi [\apj]
  {10.3847/1538-4357/ab8cc1}, \href
  {https://ui.adsabs.harvard.edu/abs/2020ApJ...896...56W} {896, 56}

\bibitem[\protect\citeauthoryear{{Wyrzykowski} et~al.,}{{Wyrzykowski}
  et~al.}{2016}]{wyrzykowski2016}
{Wyrzykowski} {\L}.,  et~al., 2016, \mn@doi [\mnras] {10.1093/mnras/stw426},
  \href {https://ui.adsabs.harvard.edu/abs/2016MNRAS.458.3012W} {458, 3012}

\bibitem[\protect\citeauthoryear{{Xu} \& {Li}}{{Xu} \& {Li}}{2010a}]{xu2010}
{Xu} X.-J.,  {Li} X.-D.,  2010a, \mn@doi [\apj] {10.1088/0004-637X/716/1/114},
  \href {http://adsabs.harvard.edu/abs/2010ApJ...716..114X} {716, 114}

\bibitem[\protect\citeauthoryear{{Xu} \& {Li}}{{Xu} \& {Li}}{2010b}]{XuLi10}
{Xu} X.-J.,  {Li} X.-D.,  2010b, \mn@doi [\apj] {10.1088/0004-637X/722/2/1985},
  \href {https://ui.adsabs.harvard.edu/abs/2010ApJ...722.1985X} {722, 1985}

\bibitem[\protect\citeauthoryear{{Ye}, {Kremer}, {Rodriguez}, {Rui},
  {Weatherford}, {Chatterjee}, {Fragione}  \& {Rasio}}{{Ye}
  et~al.}{2022}]{ye2021}
{Ye} C.~S.,  {Kremer} K.,  {Rodriguez} C.~L.,  {Rui} N.~Z.,  {Weatherford}
  N.~C.,  {Chatterjee} S.,  {Fragione} G.,   {Rasio} F.~A.,  2022, \mn@doi
  [\apj] {10.3847/1538-4357/ac5b0b}, \href
  {https://ui.adsabs.harvard.edu/abs/2022ApJ...931...84Y} {931, 84}

\bibitem[\protect\citeauthoryear{{Yoon}, {Woosley}  \& {Langer}}{{Yoon}
  et~al.}{2010}]{Yoon2010}
{Yoon} S.~C.,  {Woosley} S.~E.,   {Langer} N.,  2010, \mn@doi [\apj]
  {10.1088/0004-637X/725/1/940}, \href
  {https://ui.adsabs.harvard.edu/abs/2010ApJ...725..940Y} {725, 940}

\bibitem[\protect\citeauthoryear{{Yoshida}, {Umeda}, {Maeda}  \&
  {Ishii}}{{Yoshida} et~al.}{2016}]{yoshida2016}
{Yoshida} T.,  {Umeda} H.,  {Maeda} K.,   {Ishii} T.,  2016, \mn@doi [\mnras]
  {10.1093/mnras/stv3002}, \href
  {https://ui.adsabs.harvard.edu/abs/2016MNRAS.457..351Y} {457, 351}

\bibitem[\protect\citeauthoryear{{Zahn}}{{Zahn}}{1975}]{zahn75}
{Zahn} J.~P.,  1975, \aap, \href
  {https://ui.adsabs.harvard.edu/abs/1975A&A....41..329Z} {41, 329}

\bibitem[\protect\citeauthoryear{{Zahn}}{{Zahn}}{1977}]{zahn77}
{Zahn} J.~P.,  1977, \aap, \href
  {https://ui.adsabs.harvard.edu/abs/1977A&A....57..383Z} {57, 383}

\bibitem[\protect\citeauthoryear{{Zevin}, {Spera}, {Berry}  \&
  {Kalogera}}{{Zevin} et~al.}{2020}]{zevin2020}
{Zevin} M.,  {Spera} M.,  {Berry} C. P.~L.,   {Kalogera} V.,  2020, \mn@doi
  [\apjl] {10.3847/2041-8213/aba74e}, \href
  {https://ui.adsabs.harvard.edu/abs/2020ApJ...899L...1Z} {899, L1}

\bibitem[\protect\citeauthoryear{{Ziosi}, {Mapelli}, {Branchesi}  \&
  {Tormen}}{{Ziosi} et~al.}{2014}]{ziosi2014}
{Ziosi} B.~M.,  {Mapelli} M.,  {Branchesi} M.,   {Tormen} G.,  2014, \mn@doi
  [\mnras] {10.1093/mnras/stu824}, \href
  {http://adsabs.harvard.edu/abs/2014MNRAS.441.3703Z} {441, 3703}

\bibitem[\protect\citeauthoryear{{Zorotovic}, {Schreiber}, {G{\"a}nsicke}  \&
  {Nebot G{\'o}mez-Mor{\'a}n}}{{Zorotovic} et~al.}{2010}]{zorotovic2010}
{Zorotovic} M.,  {Schreiber} M.~R.,  {G{\"a}nsicke} B.~T.,   {Nebot
  G{\'o}mez-Mor{\'a}n} A.,  2010, \mn@doi [\aap] {10.1051/0004-6361/200913658},
  \href {https://ui.adsabs.harvard.edu/abs/2010A&A...520A..86Z} {520, A86}

\bibitem[\protect\citeauthoryear{{Zwick}, {Capelo}, {Bortolas}, {Mayer}  \&
  {Amaro-Seoane}}{{Zwick} et~al.}{2020}]{Zwick20}
{Zwick} L.,  {Capelo} P.~R.,  {Bortolas} E.,  {Mayer} L.,   {Amaro-Seoane} P.,
  2020, \mn@doi [\mnras] {10.1093/mnras/staa1314}, \href
  {https://ui.adsabs.harvard.edu/abs/2020MNRAS.495.2321Z} {495, 2321}

\bibitem[\protect\citeauthoryear{{Zwick}, {Capelo}, {Bortolas},
  {V{\'a}zquez-Aceves}, {Mayer}  \& {Amaro-Seoane}}{{Zwick}
  et~al.}{2021}]{Zwick21}
{Zwick} L.,  {Capelo} P.~R.,  {Bortolas} E.,  {V{\'a}zquez-Aceves} V.,  {Mayer}
  L.,   {Amaro-Seoane} P.,  2021, \mn@doi [\mnras] {10.1093/mnras/stab1818},
  \href {https://ui.adsabs.harvard.edu/abs/2021MNRAS.506.1007Z} {506, 1007}

\bibitem[\protect\citeauthoryear{{de Mink}, {Langer}, {Izzard}, {Sana}  \& {de
  Koter}}{{de Mink} et~al.}{2013}]{DeMink13}
{de Mink} S.~E.,  {Langer} N.,  {Izzard} R.~G.,  {Sana} H.,   {de Koter} A.,
  2013, \mn@doi [\apj] {10.1088/0004-637X/764/2/166}, \href
  {https://ui.adsabs.harvard.edu/abs/2013ApJ...764..166D} {764, 166}

\bibitem[\protect\citeauthoryear{{van Son}, {de Mink}, {Chru{\'s}li{\'n}ska},
  {Conroy}, {Pakmor}  \& {Hernquist}}{{van Son} et~al.}{2023}]{vanson2022}
{van Son} L.~A.~C.,  {de Mink} S.~E.,  {Chru{\'s}li{\'n}ska} M.,  {Conroy} C.,
  {Pakmor} R.,   {Hernquist} L.,  2023, \mn@doi [\apj]
  {10.3847/1538-4357/acbf51}, \href
  {https://ui.adsabs.harvard.edu/abs/2023ApJ...948..105V} {948, 105}

\bibitem[\protect\citeauthoryear{{van den Heuvel}}{{van den
  Heuvel}}{1976}]{vandeHeuvel76}
{van den Heuvel} E.~P.~J.,  1976, in {Eggleton} P.,  {Mitton} S.,   {Whelan}
  J.,  eds,  Astrophysics and Space Science Library Vol. 73, Structure and
  Evolution of Close Binary Systems. p.~35

\bibitem[\protect\citeauthoryear{{van den Heuvel}}{{van den
  Heuvel}}{2007}]{Heuvel07}
{van den Heuvel} E.~P.~J.,  2007, in {di Salvo} T.,  {Israel} G.~L.,
  {Piersant} L.,  {Burderi} L.,  {Matt} G.,  {Tornambe} A.,   {Menna} M.~T.,
  eds,  American Institute of Physics Conference Series Vol. 924, The
  Multicolored Landscape of Compact Objects and Their Explosive Origins. pp
  598--606 (\mn@eprint {arXiv} {0704.1215}), \mn@doi{10.1063/1.2774916}

\bibitem[\protect\citeauthoryear{{van den Heuvel} \& {De Loore}}{{van den
  Heuvel} \& {De Loore}}{1973}]{vandeheuvel73}
{van den Heuvel} E.~P.~J.,  {De Loore} C.,  1973, \aap, \href
  {https://ui.adsabs.harvard.edu/abs/1973A&A....25..387V} {25, 387}

\bibitem[\protect\citeauthoryear{{van den Heuvel}, {Portegies Zwart}  \& {de
  Mink}}{{van den Heuvel} et~al.}{2017}]{vandenheuvel17}
{van den Heuvel} E.~P.~J.,  {Portegies Zwart} S.~F.,   {de Mink} S.~E.,  2017,
  \mn@doi [\mnras] {10.1093/mnras/stx1430}, \href
  {https://ui.adsabs.harvard.edu/abs/2017MNRAS.471.4256V} {471, 4256}

\makeatother
\end{thebibliography}


\appendix

\section{Additional features of \sevn{}} \label{app:sevn}

\subsection{Alternative to stellar-evolution tables} \label{app:tables}

\subsubsection{Core radius} \label{app:cradii}

The radii of the He and CO core, if not available in the  stellar-evolution tables, are estimated as 
\begin{equation}
R_\mathrm{core} = R_0 \frac{1.1685 M^{4.6}_\mathrm{c}}{M^{4}_\mathrm{c} + 0.162 M^{3}_\mathrm{c} + 0.0065} \ \  \Rsun,
\label{eq:rcore}
\end{equation}
where $M_\mathrm{c}$ is the mass of the He or CO core in \Msun,  $R_0=0.1075$ for the He core and $R_0=0.0415$ for the CO core.
The functional form of  Eq.~\ref{eq:rcore} is the same as used for the radius of naked helium stars in Eq.~78 of  \cite{Hurley00}. We have adapted the coefficient and the normalisation to 
fit the radius of the  He and CO cores in the \parsec{} stellar tracks. 

\subsubsection{Inertia} \label{app:inertia}

\sevn{} implements the following alternative options to estimate the stellar inertia.

\begin{itemize}
\item  Eq.~109 from \cite{Hurley00}
	\begin{equation}
		I = 0.1 (M - M_\mathrm{c}) R^{2} + 0.21M_\mathrm{c} R^{2}_\mathrm{c};
		\label{eq:appinertiahurley}
	\end{equation}

\item  the formalism by \cite{DeMink13}
	\begin{equation}
		I = k M\,{} R^{2}
		\label{eq:inertiademink},
	\end{equation}
	where $k$ depends on the  mass and radius of the star;
\item  the inertia of an homogeneous sphere
	\begin{equation}
		I = \frac{2}{5} M \,{} R^{2};
		\label{eq:inertiahsphere}
	\end{equation}
\item 
the inertia of an homogeneous hollow sphere (modelling the star's envelope) plus an homogeneous sphere (modelling the core):
	\begin{equation}
		I =\frac{2}{5} (M - M_\mathrm{c}) \frac{R^{5} - R^{5}_\mathrm{c}}{R^{3} - R^{3}_\mathrm{c}} +  \frac{2}{5} M R^{2}
		\label{eq:inertiahspherecore}.
	\end{equation}
\end{itemize}

\subsubsection{Convective envelope}

The {\sc parsec} tables also include the main properties of the convective envelope: the mass, radial extension and turnover timescale of the largest convective cells. 

If these are not available, we estimate the mass and  extension of the  convective region following Section 7.1 of \cite{Hurley00}  and  Eqs.~36--40 of \cite{Hurley02}. In practice, we assume that all the MS stars with $M_\mathrm{ZAMS}>1.25 \ \Msun$ and all the pure-He stars have a  radiative envelope. The  MS stars  with $M_\mathrm{ZAMS}\leq1.25 \ \Msun$ begin their evolution with a fully  convective envelope that progressively recedes until the envelope is fully radiative at the end of the  MS. 
Then, the process is reversed during the terminal-age MS phase (TAMS, \sevn{} phase 2, see Table \ref{tab:phases}):  the convective layers grow and the star becomes fully convective at the end of this phase.

The envelope of H-rich stars more evolved than the TAMS phase is assumed to be fully convective.  
\cite{Hurley00} and \cite{Hurley02} 
use the \bse{} type 2 (HG) to set the transition to a fully convective envelope.  There is not a direct correspondence between the \bse{} type HG and the \sevn{} phase TAMS, since it depends on the mass fraction of the convective envelope (Table \ref{tab:phases}), which is not known a priori if the tables are not used. 
As a consequence, in \sevn{} the transition to a fully convective envelope could happen when the effective temperature of the star is  still hot enough to be dominated by the radiative energy transport.  
We will improve this  in future \sevn{} versions; meanwhile we suggest to include the information about convection in the stellar-evolution tables when possible.

There are no analytic approximations for the turnover timescale: this is set to zero if the  tables are not available. 
Therefore, processes that require this quantity have to implement their own alternative to the tables. For example, in the stellar tides (Section \ref{sec:tides}) the turnover timescale is estimated using Eq.~31 in \cite{Hurley02} if it is not available in the tables. 

\subsubsection{Envelope binding energy}  \label{app:lambda} \label{app:ebind}

\begin{figure*}
	\centering
	\includegraphics[trim={3cm 0 3cm 0cm},width=1.0\textwidth]{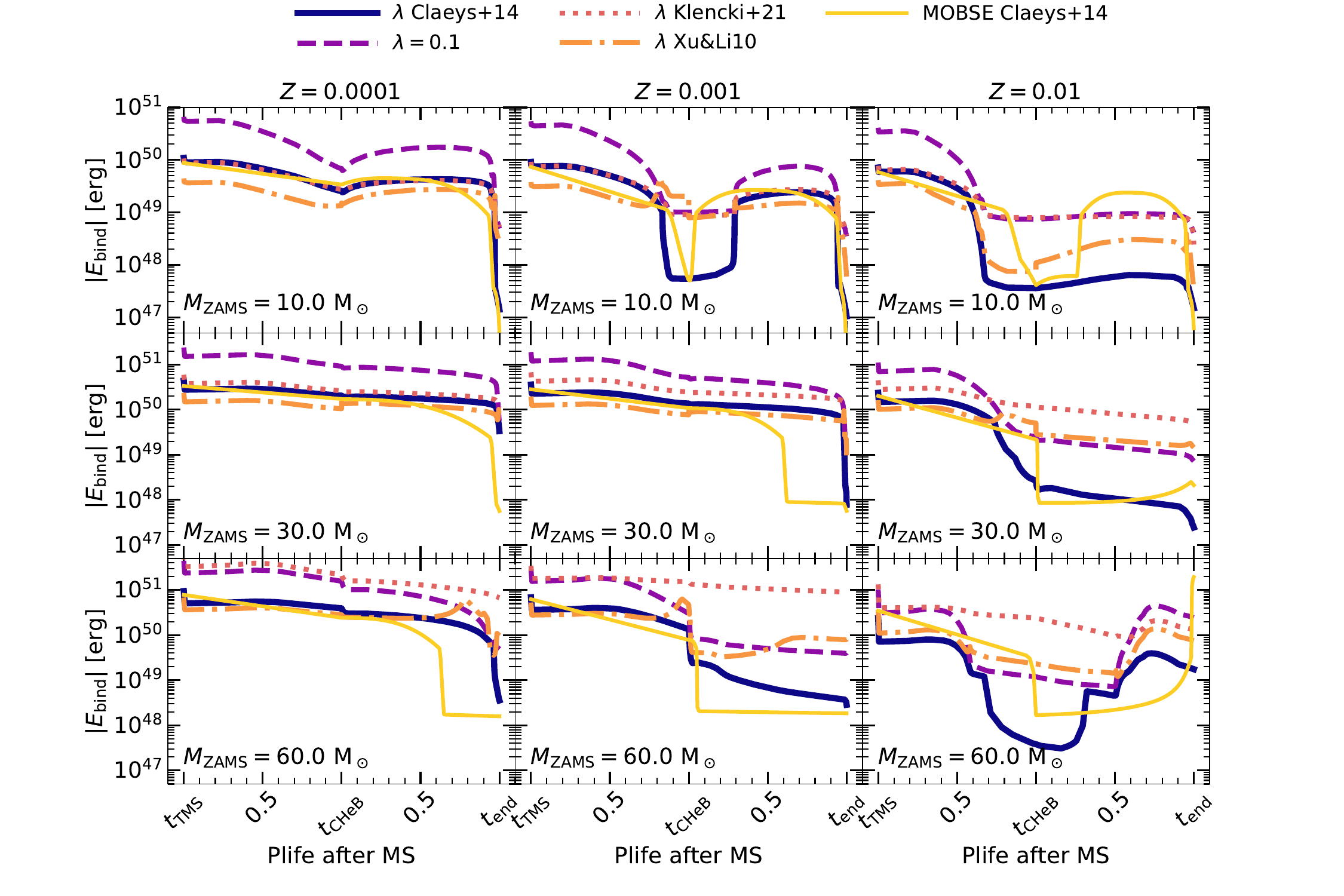}
  	\vspace*{-5mm}
    \caption{Evolution of the envelope binding energy  during  stellar evolution for a sample of stars with $M_\mathrm{ZAMS}=10$ (first~row), 30 (second~row), and $60 \ \Msun{}$ (third~row), and metallicity $Z=0.0001$ (first~column), 0.001 (second~column), 0.01 (third~column). The x-axis indicates the percentage of life after the MS from the TAMS ($t_\mathrm{TAMS}$) up to the CHeB phase ($t_\mathrm{CHeB}$, see Table~\ref{tab:phases}), and from the CHeB  to the 
    moment the star turns into a compact remnant or a pure-He star ($t_\mathrm{end}$). 
    The gold solid thin line shows the binding energy of the \mobse{} stellar models assuming the $\lambda_\mathrm{CE}$ formalism by \protect\cite{Claeys14}. 
     All the other  lines show the binding energy of the \parsec{} stellar models ($\lambda_\mathrm{ov}=0.5$, see Section~\ref{sec:tracks}) assuming different  $\lambda_\mathrm{CE}$ prescriptions: blue solid line following \protect\cite{Claeys14}, violet dashed line using $\lambda_\mathrm{CE}=0.1$, 
     pink dotted line using the prescriptions by \protect\cite{Klencki21}, orange dot-dashed line following \protect\cite{XuLi10}. 
     In all the cases, we estimate the binding energy maximising the fraction of internal and recombination energy (see the main text).
     \gitlab{https://gitlab.com/iogiul/iorio22_plot/-/tree/v3/Ebind}
     \gitbook{https://gitlab.com/iogiul/iorio22_plot/-/blob/v3/Ebind/PlotEbind.ipynb}
     \gitimage{https://gitlab.com/iogiul/iorio22_plot/-/blob/v3/Ebind/Ebind.pdf}
     }
	\label{fig:Ebind}
\end{figure*}

The envelope binding energy is a key quantity to determine the fate of a binary system during a CE phase (Section \ref{sec:CE}). The envelope loss during  CE can be enhanced by taking into account  the internal and recombination energy of the envelope \citep[e.g.,][]{Lau22}, therefore the tables should  contain the effective envelope binding energy, i.e. the gravitational binding energy reduced to take into account the aforementioned additional energy sources.

If the envelope binding energy tables are not available, \sevn{} uses
\begin{equation}
E_\mathrm{bind} = - G \frac{M M_\mathrm{env}}{\lambda_\mathrm{CE} R}.
\label{eq:appebind}
\end{equation}
\sevn{} implements the following options to calculate   the parameter $\lambda_\mathrm{CE}$.

\begin{itemize}
\item{\emph{Constant}}, $\lambda_\mathrm{CE}$ is set to a constant value. It is possible to set different $\lambda_\mathrm{CE}$  values  for H-stars and pure-He stars.
\item{\emph{\bse{}}, $\lambda_\mathrm{CE}$ is estimated as in \bse{} and \mobse{}, i.e. adopting the formalism described by \cite{Claeys14}}. 
Actually, this option of \sevn{} is based on the most updated public version of \bse{} and \mobse{}, which differ from Appendix~A of \cite{Claeys14} for the following aspects.
We replace Eq.~A1 of \cite{Claeys14} with the following equation:
\begin{equation}
\lambda_\mathrm{CE} =
\begin{cases}
2 \lambda_2~~~~~~~~~~~~~~~~~~~~~~~~~~~~~~~~~~~~~~~~~~~~~\mathrm{if} \ f_\mathrm{conv}=0 \\ 
2 \lambda_2 + f^{0.5}_\mathrm{conv} (\lambda_1-2\lambda_2)~~~~~~~~~~~~~~~\mathrm{if} \ 0<f_\mathrm{conv}<1 \\
\lambda_1~~~~~~~~~~~~~~~~~~~~~~~~~~~~~~~~~~~~~~~~~~~~~~~\mathrm{if} \ f_\mathrm{conv}\geq1
\end{cases},
\label{eq:lambdacorr}
\end{equation}
where $f_\mathrm{conv}$ is the mass fraction of the convective envelope with respect to the whole envelope.
In addition, we replace the parameter $\lambda_1$ 
with $2 \lambda_1$ in Eqs.~A6 and A7 of \cite{Claeys14}.
\cite{Claeys14} introduced the parameter $\lambda_\mathrm{ion} \ \in \ [0,1]$ to parametrise the fraction of  internal and recombination energy  
included in the estimate of the binding energy. We use $\lambda_\mathrm{ion}=1$ as default value. 
For pure-He stars, $\lambda_\mathrm{CE}=0.5$. 
\item{\emph{Izzard04}}, same as the \bse{} implementation, but in Eq.~\ref{eq:lambdacorr} we replace $f_\mathrm{conv}$ with the mass of the convective envelope expressed in solar units.
\item{\emph{Xu\&Li10}}, we estimate $\lambda_\mathrm{CE}$ by interpolating on $M_\mathrm{ZAMS}$ the fitting equations by \cite{xu2010} and \cite{XuLi10}. 
Similar to the \bse{} option, it is possible to set the fraction of internal and recombination energy, $\lambda_\mathrm{ion}$, to take into account the estimate of the effective binding energy. We use $\lambda_\mathrm{ion}=1$ as default value. 

For pure-He stars we use the formalism included in {\sc COMPAS}: 
\begin{equation}
\lambda_\mathrm{CE}=0.3 R^{-0.8}_\mathrm{\lambda}, \ \mathrm{with} \ R_\mathrm{\lambda}=\min[120,\max[0.25,R]],
\end{equation}
where $R$ is the stellar radius expressed in solar units. 
\item{\emph{Klencki21}}, we estimate  $\lambda_\mathrm{CE}$ interpolating on $M_\mathrm{ZAMS}$ and $Z$  the fitting formulas by \cite{Klencki21} calibrated on  {\sc MESA} tracks.  
Since \cite{Klencki21} report only H-rich stars, we set  $\lambda_\mathrm{CE}=0.5$ for pure-He stars.
\end{itemize}

We set to 0 the envelope binding energy for all the stars without a core (i.e., \sevn{} phases 0, 1 for H-rich stars and phase 4 for pure-He stars), the naked-CO stars and the compact remnants. 

In Fig.~\ref{fig:Ebind}, we compare the binding energy estimated with different $\lambda_\mathrm{CE}$ prescriptions for the \parsec{} stellar tracks. We also show the binding energy from \mobse{}. 
From the TAMS to the ignition of the core He burning the binding energies from \parsec{} and  \mobse{} (in both cases using the $\lambda_\mathrm{CE}$ prescription by \citealt{Claeys14}) are qualitatively  in agreement. In the later evolutionary phases  the differences are more notable. The prescriptions by \cite{Klencki21} and the constant $\lambda_\mathrm{CE}=0.1$ predict the largest binding energies, while 
the fitting formulas by \cite{XuLi10} yield  values of the binding energy that are generally  intermediate between \cite{Claeys14} and \cite{Klencki21}.

\subsection{Electron-capture and core-collapse supernova models} \label{app:snmodels}


\sevn{} includes the following formalism for electron-capture  and core-collapse supernovae (ECSN and CCSN, respectively).

\begin{itemize}

\item{\emph{Rapid}}, rapid supernova model by \cite{fryer2012}. 

\item{\emph{Delayed}}, delayed supernova model by \cite{fryer2012}. In both the delayed and the rapid model, the final mass depends on the  total and CO-core mass of the star at the onset of core collapse. The mass of the compact remnant of an ECSN is equal to the pre-supernova CO-core mass.

\item{\emph{ Rapid Gaussian}}, same as rapid, but the mass of the NSs (including NSs born from ECSNe) are drawn from a Gaussian distribution (see Section \ref{sec:remform}).
\item{\emph{Delayed Gaussian}}, same as delayed, but the mass of the NSs (including NSs born from ECSNe) are drawn from a Gaussian distribution (see Section \ref{sec:remform}).

\item{\emph{Compactness}}, supernova model based on the compactness parameter, defined as 
\begin{equation}
\xi_\mathrm{2.5}= \frac{2.5}{R(2.5 \ \Msun)/1000\,{} {\rm km}},
\end{equation}
i.e., as the ratio between a characteristic mass (2.5 \Msun) and the radius (in units of 1000 km) enclosing this mass at the onset of the core collapse \citep{Oconnor11}. 
In \sevn{}, we estimate the compactness  using  Eq.~2 in \cite{Mapelli20}.
The compactness can be used to define the final fate of a massive star. In practice, 
it is possible to  define  a compactness threshold $\xi_\mathrm{c}$ so  a compactness value below the threshold  ($\xi_\mathrm{2.5}\leq\xi_\mathrm{c}$) leads to a supernova explosion, while when $\xi_\mathrm{2.5}>\xi_\mathrm{c}$ the star undergoes a  direct collapse (see \citealt{Mapelli20}, and reference therein). By default, $\xi_\mathrm{c}=0.35$. \sevn{} also includes a   stochastic explosion/implosion decision aimed to reproduce the  $\xi_\mathrm{2.5}$ distributions in Fig. 3 of \cite{Patton2020}.
If a supernova explosion is triggered, we always assume that the compact remnant is a NS with mass drawn from a  Gaussian distribution as in the \emph{rapid Gaussian} model (Section \ref{sec:snmodel}). A direct collapse produces a BH  with mass $M_\mathrm{BH}=M_\mathrm{He,f} + 0.9 \left( M_\mathrm{f} -  M_\mathrm{He,f} \right)$, where $M_\mathrm{f}$ and $M_\mathrm{He,f}$ are the pre-supernova total and He-core masses of the star (Eq.~3 in \citealt{Mapelli20}). 

\item{ \emph{Death matrix}}, this model reproduces the results presented in \cite{Woosley20} (see their Fig. 4). For CCSNe, the final remnant mass is obtained by interpolating their Table~2. Compact remnants less massive than 3 \Msun{} are classified as NSs, otherwise as BHs.
The results by \cite{Woosley20} already include the effect of PPI/PISN and neutrino mass loss (Section \ref{sec:pisn}), therefore we do not apply any further correction. 

\item{ \emph{Direct collapse}}, in this model all the CCSNe produce a direct collapse. The mass of the compact remnant is equal to the pre-supernova mass of the star and we do not apply PPI/PISN and neutrino mass loss corrections  (Section \ref{sec:pisn}).

\end{itemize}

\subsection{Kick models} \label{app:snkicks}

In addition to the models  described in Section \ref{sec:snkicks} (K$\sigma265$, K$\sigma150$ and  KGM), \sevn{} includes the following supernova kick models.
\begin{itemize}
\item \emph{KFB}: same formalism as K$\sigma265$ or K$\sigma150$ (see Section \ref{sec:snkicks}), but we correct the module of the kick velocity for the mass fallback during the supernova, i.e. 
$V_\mathrm{kick}=V_\mathrm{M} (1- f_\mathrm{b})$. We  draw  $V_\mathrm{M}$ from a Maxwellian distribution (default 1D rms  $\sigma_\mathrm{kick}=265 \ \mathrm{km}~\mathrm{s}^{-1}$). The fallback fraction, $f_\mathrm{b} \in [0,1]$, is defined as in  \cite{fryer2012} and depends on the supernova model ($f_\mathrm{b}=1$ for direct collapses).  
\item \emph{K0}: all the kicks are set to 0.
\item \emph{KCC15}:  same as \emph{KFB}, but for the CCSNe (including PPIs, see Section \ref{sec:pisn}), we draw $V_\mathrm{M}$ from a Maxwellian curve with $\sigma_\mathrm{kick}=15 \ \mathrm{km}~\mathrm{s}^{-1}$.
\item \emph{KEC15CC265}: same as \emph{KFB}, but for the ECSNe,   we drawn $V_\mathrm{M}$ from a Maxwellian curve with $\sigma_\mathrm{kick}=15 \ \mathrm{km}~\mathrm{s}^{-1}$.
\item \emph{KECUS30}: same as \emph{KFB}, but for the ECSNe and ultra-stripped supernovae,   we draw $V_\mathrm{M}$ from a Maxwellian curve with $\sigma_\mathrm{kick}=30 \ \mathrm{km}~\mathrm{s}^{-1}$. In this model, we define a supernova  as ultra-stripped if the difference between the stellar mass and CO-core mass of the star is lower than 0.1 \Msun{} at the onset of the supernova explosion.
\end{itemize}

\subsection{RLO} 
\subsubsection{Mass transfer stability options} \label{app:rlostab}

\begin{table}
\begin{center}
\scalebox{1.0}{
\begin{tabular}{lccc}
\cline{2-4}
\multicolumn{1}{l|}{} & \multicolumn{3}{c|}{\sevn{} $q_\mathrm{c}$ option} \\ \hline
\multicolumn{1}{c}{\bse{} stellar type Donor} &
  QCH &
  QCCN &
  QCCC \\ \hline
0 (low mass MS)       & 0.695    & 1.717    & 0.695 (1.0)       \\
1 (MS)                & 3.0      & 1.717    & 1.6 (1.0)         \\
2 (HG)                & 4.0      & 3.825    & 4.0 (4.762)       \\
3/5 (GB/AGB)          & Eq. \ref{eq:qch} & Eq. \ref{eq:qchw} & Eq. \ref{eq:qch} (1.15) \\
4 (CHeB)              & 3.0      & 3.0      & 3.0               \\
7 (WR)                & 3.0      & stable   & 3.0               \\
8 (WR-HG)             & 0.784    & stable   & 4.0 (4.762)       \\
\textgreater{}10 (WD) & 0.628    & 0.629    & 3.0 (0.625)       \\ \hline
\end{tabular}
}
\caption{The values in parenthesis for the option QCCC indicate the $q_\mathrm{c}$ when the accretor is a compact remnant (WD, NS, BH). The additional  available option QCSH (not shown in the Table) is the same as QCBSE (see Table \ref{tab:qc})  except  in the case of a BH accretor (see the main text).}
\label{tab:qcother}
\end{center}
\end{table}

Table~\ref{tab:qcother} lists additional critical mass-ratio options implemented in \sevn{} (see Table \ref{tab:qc}).

The option QCH follows exactly the original \cite{Hurley02} implementation, in particular for giant stars with deep convective envelopes (\bse{} type 3, 5)
\begin{equation}
\begin{split}
q_\mathrm{c} &= \frac{ 1.67 -x + 2 \left( \frac{M_\mathrm{He,d}}{M_\mathrm{d}} \right)^5 }{2.13} \\
&\mathrm{with} \ x= 0.30406 +0.0805\zeta + 0.0897\zeta^2 + 0.0878\zeta^3+0.0222\zeta^4 \\ 
&\mathrm{and} \ \zeta= \log \frac{Z}{0.02}, 
\end{split}
\label{eq:qch}
\end{equation}
where  $M_\mathrm{d}$, $M_\mathrm{He,d}$ and $Z$ are the total mass, He-core mass, and  metallicity of the donor star.

The options QCCN and QCCC are taken directly from the code {\sc cosmic} \citep{Cosmic} and based on \cite{Neijssel19}  and \cite{Claeys14}, respectively. 
\sevn{} includes also the option  QCSH based on the work by \cite{Shao21}. It is the same as QCBSE (see Table \ref{tab:qc}), except for  BH accretors. In these case, if the donor-to-accretor mass ratio is lower than 2, the mass transfer is always stable, while if it is larger than $2.1 + 0.8M_\mathrm{a}$ ($M_\mathrm{a}$ is the mass of the accretor) it is always unstable. Between these two cases, the stability condition is checked by comparing  the radius of the donor star, $R_\mathrm{d}$ with 
\begin{equation}
\begin{split}
&R_\mathrm{s} = 6.6 - 26.1\frac{M_\mathrm{d}}{M_\mathrm{a}} + 11.4 \frac{M^2_\mathrm{d}}{M^2_\mathrm{a}}  \ \ \Rsun, \ \ \mathrm{and} \\
&R_\mathrm{u} = -173.8 + 45.5 \frac{M_\mathrm{d}}{\Msun} -0.18 \frac{M^2_\mathrm{d}}{\mathrm{M}^2_\odot} \  \ \Rsun.
\end{split}
\end{equation}
If $R_\mathrm{d}<R_\mathrm{u}$ and $R_\mathrm{d}>R_\mathrm{s}$, the mass transfer is stable, otherwise unstable.

\subsubsection{Angular momentum loss} \label{app:rloloss}

The angular momentum loss during a  non-conservative RLO is parameterised by Eq.~\ref{eq:djorbrlo} and depends on the normalisation parameter  $\gamma_\mathrm{RLO}$. Three different options are available (see \citealt{Soberman97,Tauris06}):
\begin{itemize}
    \item \emph{Jeans mode}, the mass is lost from  the vicinity of the donor star carrying away its specific angular momentum, $\gamma_\mathrm{RLO}= M^2_\mathrm{a} M^{-2}_\mathrm{b}$ ($M_\mathrm{b}$ is the total mass of the binary);
    \item \emph{Isotropric re-emission}, the mass is lost from  the vicinity of the accretor star carrying away its  specific angular momentum, $\gamma_\mathrm{RLO}= M^2_\mathrm{d} M^{-2}_\mathrm{b}$;
    \item \emph{circumbinary disc}, the lost mass settles in a circumbinary disc carrying away a  $\gamma_\mathrm{RLO}$ (real positive number) fraction of the binary angular momentum (see, e.g.,\  \citealt{Vos15}).
\end{itemize}
If the accretion onto a compact object happens at super-Eddington rate, or if there is a nova eruption,  the isotropic re-emission is always used.

\subsection{post-CE coalescence} \label{app:cemerger}

We set the core mass of the coalescence product as the sum of the two stellar cores, $M_\mathrm{c,coal}=M_\mathrm{c,1}+M_\mathrm{c,2}$, while the total mass is estimated as 
\begin{equation}
M_\mathrm{coal} = M_\mathrm{c,coal} + k_\mathrm{CE} M_\mathrm{CE}  + k_\mathrm{NCE} M_\mathrm{NCE},
\label{eq:mmerge}
\end{equation}
where $M_\mathrm{CE}$ is (envelope mass in evolved stars) shared in the CE, while $M_\mathrm{NCE}$ is the mass that is not part of the CE, e.g. the mass of stars in the MS or the mass of pure-He stars without a CO core. 
The factors $k_\mathrm{CE}$ and 
$k_\mathrm{NCE}$ set the mass fraction that remains bound to the star after the CE evolution and the subsequent coalescence. 
They can assume values from 0 to 1 (default value in \sevn{}), or the special value $-1$.  
If both $k_\mathrm{CE}$ and $k_\mathrm{NCE}$  are  set to $-1$, we estimate the final mass of the coalescence product  using  the  method described in \cite{Spera19} (see their section Section 2.3.2). If  $k_\mathrm{NCE}=-1$ and $k_\mathrm{CE}\neq{}-1$ the final mass is obtained using Eq.~77 in \cite{Hurley02} and $k_\mathrm{CE}$ is not considered. Finally, if $k_\mathrm{CE}=-1$ and $k_\mathrm{NCE}\neq{}-1$, we use a re-scaled version of the \cite{Hurley02} implementation in which Eq.~\ref{eq:mmerge} is used and
\begin{equation}
k_\mathrm{CE}= \frac{M_\mathrm{final,Hurley} - M_\mathrm{final,min}}{M_\mathrm{final,max}-M_\mathrm{final,min}},
\end{equation}
where $M_\mathrm{final,Hurley}$ is obtained with Eq.~77 in \cite{Hurley02}, $M_\mathrm{final,min}=M_\mathrm{c,1}+M_\mathrm{c,2}$ and $M_\mathrm{final,max}=M_\mathrm{1}+M_\mathrm{2}$.

\section{From \parsec{} tracks to \sevn{} tables} \label{sec:trackstable}  \label{app:trackstable}

To produce the \sevn{} tables from the \parsec{} stellar tracks (Section \ref{sec:tracks}), we use the code {\sc TrackCruncher} described in Section \ref{sec:tables}.

Firstly, we process each stellar track to set the \sevn{} phase times (Section \ref{sec:sevnphase}). 
Each stellar track is iterated in time until the conditions for starting a given phase are triggered. The correspondent time is used as the starting time of the phase and included in phase  tables (see Appendix~\ref{app:tables}).

The MS (\sevn{} phase 1) starts when the energy production due to the central hydrogen burning (hydrogen burning luminosity) is larger than 60\% of the total luminosity. In addition, the central hydrogen mass fraction must be decreased of at least 1\% with respect to the initial value in the track (at time 0). 

The terminal-age MS  phase begins  when 
the He-core mass is larger than 0. In the \parsec{} stellar tracks used in this work, the He-core  (CO-core) mass is set to 0 until the central hydrogen (helium) mass fraction decreases to  $10^{-3}$.

The shell H burning phase  starts when  the central hydrogen mass fraction  is less than $10^{-8}$.

The core He burning phase  begins when the central helium mass fraction is decreased by at least 1\% with respect to its maximum value.  

The terminal-age core He burning phase starts when the CO-core mass is larger than 0.

The shell He burning phase begins when the central helium mass fraction is lower than $10^{-8}$ and the luminosity produced by C  burning is lower than 20\% of the total luminosity.  

The phases are checked progressively in the order reported above, i.e., a phase cannot be triggered if the previous phase has not been  triggered yet.
{\sc TrackCruncher} rejects all the tracks that do not reach the shell He burning phase.

We assume that the core C burning phase starts  when  its energy output  is larger than or equal to  20\% of the total luminosity.
The subsequent stellar evolution continues on very short time scales ($\lesssim 20$ yr), and the stellar properties required in \sevn{} (e.g., mass, radius, He- and CO-core mass) remain almost constant. For this reason, we do not store in the \sevn{} tables the \parsec{} outputs  after the core C burning, except for the  very last point in the track. 
This allows to reduce the number of points in the table  and to speed-up  single stellar evolution in \sevn{}.

We  also add a check to stop intermediate-mass stars ($M_\mathrm{ZAMS} \lessapprox$ 8-9 \Msun) at the beginning of the asymptotic giant branch (AGB). The late AGB phase is hard to model in detail, and the \parsec{} tracks follow the evolution up to the early AGB.  
To produce more uniform \sevn{} tables in this mass range, we stop the track at the onset of the AGB, i.e., when the central degeneracy parameter, $\eta$, grows to values larger than 15 \citep[][Chapter 3.2]{CoxGiuli}. 
Eventually, we add to the \sevn{} tables the last point of the stellar track and force the star to lose the whole envelope setting the final mass and radius of the star equal to the mass and the radius of the He core.  As a consequence, in \sevn{} the AGB phase will be modelled as a \vir{wind} that reduces linearly the stellar mass from the pre-AGB values to the He-core mass before compact remnant formation. We adopt this pre-processing strategy only for stars that will form a WD, i.e., stars with a maximum CO-core mass lower  than 1.38 \Msun{}  (Section \ref{sec:remform}).

\section{Comparison with other stellar evolution tracks} \label{app:SSE}

\begin{figure*}
	\centering
	\includegraphics[width=1.0\textwidth]{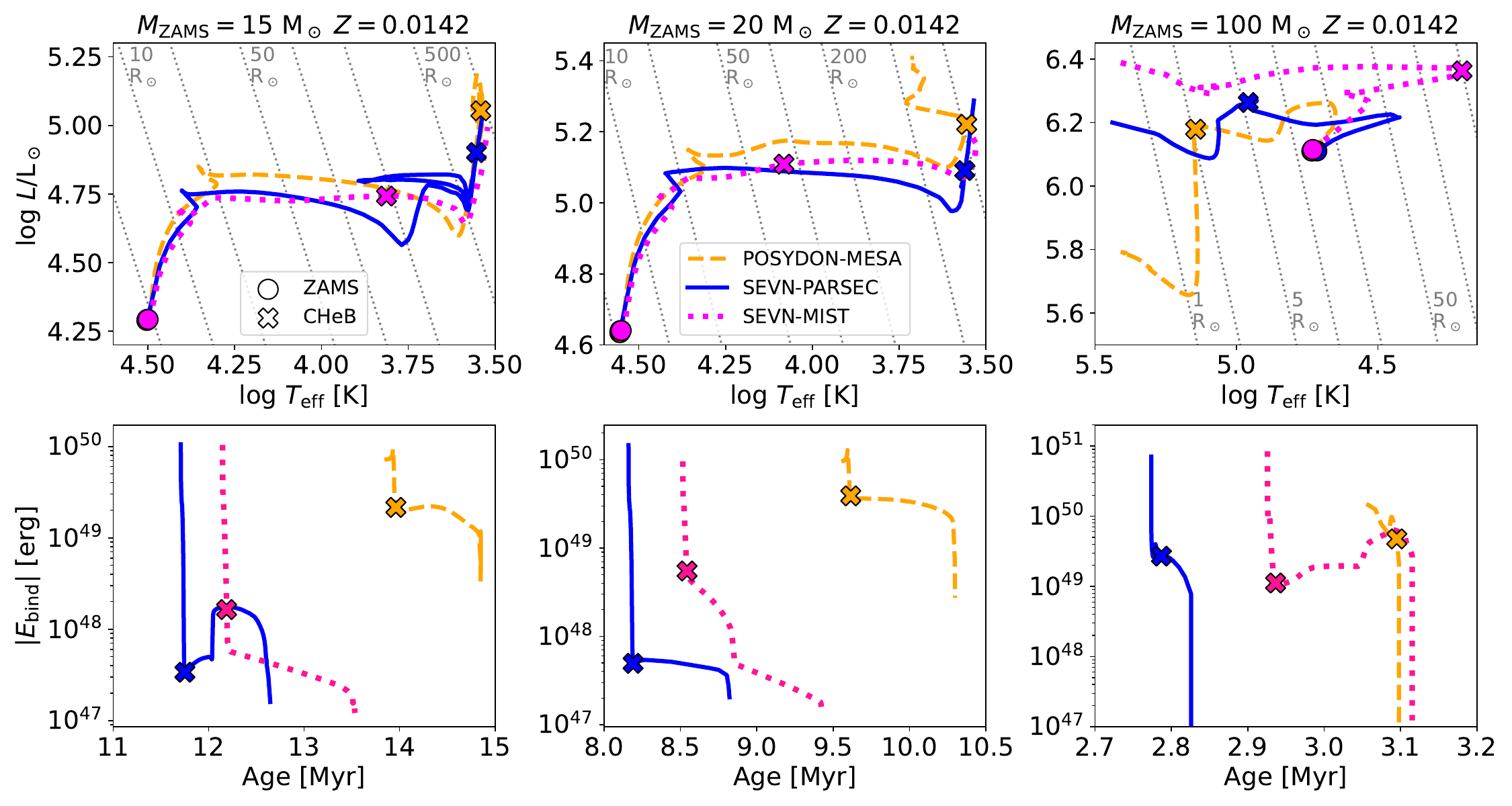}
	\vspace*{-5mm}
	\caption{Comparison among {\sc posydon} using {\sc mesa} tracks (\protect\citealt{Posydon}, orange dashed lines),   \protect\sevn{} using {\sc parsec} tracks (with overshooting parameter $\protect\lambda_\mathrm{ov}=0.5$, blue solid lines), and \protect\sevn{}  using \mist{} tracks  (\citealt{Choi2016}, magenta dotted lines) 
    for stars with $M_{\rm ZAMS}=15$, 20, 100 \protect\Msun{} (from left to right), and metallicity $Z=0.0142$. 
 Upper panels:
    HR diagrams. The grey dashed lines indicate points at constant radius: 1, 2, 5, 10, 20, 50, 100, 200, 500, 1000, and 2000 R$_\odot$.   Lower panels:  evolution of the envelope binding energy estimated following the \protect\cite{Claeys14} formalism for  \protect\sevn{}.(Appendix~\protect\ref{app:lambda}) and taken directly from the tracks for \protect\posydon{}.
    The envelope binding energy is shown only when  the mass of the  helium core  is larger than 0.1 \Msun. 
    The markers indicate the starting position in the ZAMS (circles), and the core He burning (CHeB) ignition  estimated as     described in Appendix~\protect\ref{app:trackstable} (crosses).    \gitlab{https://gitlab.com/iogiul/iorio22_plot/-/tree/v3/posydon_combine_comparison}
    \gitbook{https://gitlab.com/iogiul/iorio22_plot/-/blob/v3/posydon_combine_comparison/Plotnew.ipynb}
    \gitimage{https://gitlab.com/iogiul/iorio22_plot/-/blob/v3/posydon_combine_comparison/compare_posydon.pdf}
    }
	\label{fig:posydon} 
\end{figure*}

\begin{figure*}
	\centering
	\includegraphics[width=1.0\textwidth]{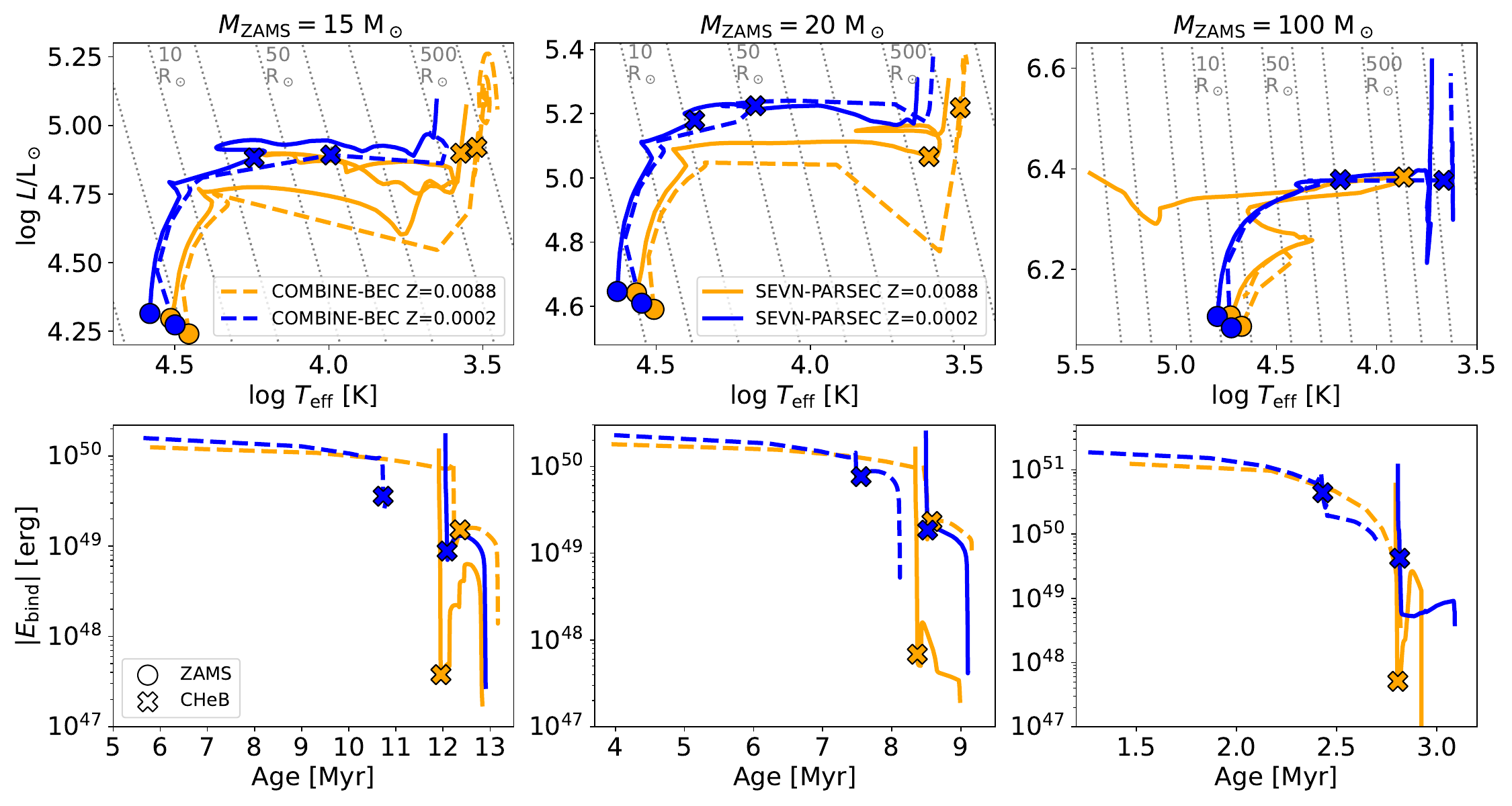}
	\vspace*{-5mm}
	\caption{Same as Fig.~\protect\ref{fig:posydon}, but comparing stars with  metallicity Z=0.0088 \textcolor{black}{(orange)}, Z=0.0002 \textcolor{black}{(blue)} computed by \protect\sevn{} using \protect\parsec{} stellar tables with overshooting parameter $\lambda_\mathrm{ov}=0.5$ (solid lines) , and by \protect\combine{} using \protect\bec{} tracks (\protect\citealt{Combine}, dashed lines). 
 \gitlab{https://gitlab.com/iogiul/iorio22_plot/-/tree/v3/posydon_combine_comparison}
    \gitbook{https://gitlab.com/iogiul/iorio22_plot/-/blob/v3/posydon_combine_comparison/Plotnew.ipynb}
    \gitimage{https://gitlab.com/iogiul/iorio22_plot/-/blob/v3/posydon_combine_comparison/compare_combine.pdf}}
	\label{fig:combine} 
\end{figure*}

Figures~\ref{fig:posydon} and~\ref{fig:combine} 
compare the evolution of stars 
integrated with \sevn{} using the \parsec{} tables ($\lambda_\mathrm{ov}=0.5$, Section~\ref{sec:tracks}) with \sevn{} using the \mist{} tracks \citep{Choi2016}, with \posydon{}\footnote{We use the \posydon{} branch  {\it development} updated to the commit \href{https://github.com/POSYDON-code/POSYDON/tree/80231b4a64cc4bb3bf74f373c4b320edf19ac4f9}{80231b4}, and the MESA stellar tracks from the \cite{PosydonZenodo} 
Zenodo repository (Version 3).}  using \mesa{} tracks \citep{Posydon}, and  with \combine{} \citep{Combine} using \bec{} \citep{Yoon2010,BEC2015} tracks (Kruckow, private communication).
We consider stars with $M_\mathrm{ZAMS}=15,20,100 \ \ \Msun$ corresponding to NS progenitors, to the NS/BH formation boundary, and to high-mass BH progenitors, respectively. We use the metallicity $Z=0.0142$ (the metallicity currently available in \posydon{}) for the comparison with \parsec{}+\mist{} and \posydon, while $Z=0.0088$   and $Z=0.0002$ for the comparison with \combine{} corresponding to their MW-like and IZw18-like models \citep{Brott2011,BEC2015}.

The stars in \sevn{}+\parsec{}, \posydon{} and \combine{} show a similar evolution in the HR diagram. In particular, intermediate-mass stars ($M_\mathrm{ZAMS}\lesssim 20 \ \Msun$) at high-metallicity ignite the core He burning in the red part of the HR diagram ($T_\mathrm{eff}<4000 \ \mathrm{K}$), while in the high-mass high-metallicity range the stars move to the blue before the ignition of the core He burning.

In contrast, stars evolved with \sevn{}+\mist{} show many similarities with the stellar evolution of \bse{}-like codes:  intermediate-mass stars begin  core He burning while they are still hot and relatively compact, and high-metallicity massive stars move to the blue only after the ignition of the core He burning.

Overall, stars in \posydon{} and \mist{} live $\approx10\mathrm{-}20\%$ longer than stars in \parsec{}, while the stellar lifetimes are similar between \parsec{} and \combine{} at $Z=0.0088$, but stars in \combine{} have a shorter ($\approx10\mathrm{-}20\%$) life at $Z=0.0002$. On average, stars in \combine{} and \mist{}  reach larger radii with respect to \parsec{} and \posydon{}.

The envelope binding energy of the stars in \posydon{} and \combine{} is 1--2 orders of magnitude larger than the values estimated in \sevn{} for \parsec{} and \mist{} using the \cite{Claeys14} formalism (Appendix~\ref{app:lambda}). Such difference can have a substantial impact on the production of merging BCOs (see, e.g., Section~\ref{sec:bhnsform} and Section~\ref{sec:discce}).  
The large discrepancy in the starting ages between \parsec{} and \bec{} (lower panels of Fig.~\ref{fig:combine}) depends on the different assumptions used to define 
the He core.
In this work, we consider the He core mass larger than zero  when the central mass fraction of hydrogen is less than 0.001 (Section \ref{sec:tracks}). 
In the \combine{}+\bec{} tracks, the core-envelope boundary  is located at the point in which  the hydrogen mass  fraction is 0.1, but then  the position of the boundary is further refined  based on the average chemical abundances (Kruckow, private communication).
Different definitions of the core-envelope boundary have a strong impact in the  estimate of the envelope binding energy \citep[see e.g.,][]{Kruckow2016}. 

\section{Analytic approximations for the GW merger time} \label{app:gwtime}

\begin{figure}
	\centering
	\includegraphics[width=1.0\columnwidth]{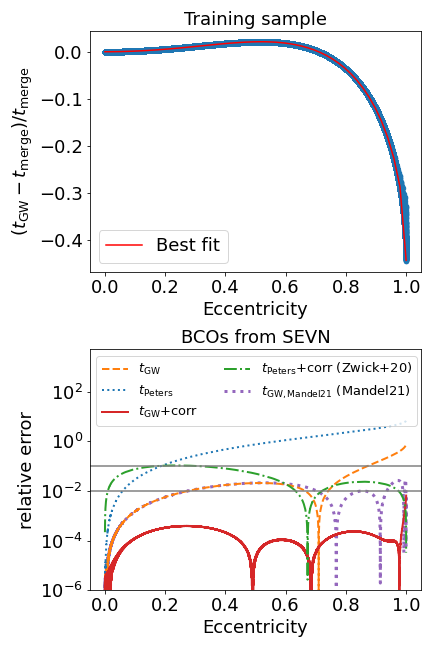}
	\vspace*{-5mm}
	\caption{Upper panel: the blue thick line 
 shows the relative difference between the GW-induced merging time estimated integrating 
 Eqs.~\ref{eq:gwa} and \ref{eq:gwe} and the analytic approximation $t_\mathrm{GW}$ in Eq.~\ref{eq:tgw}. The times have been estimated considering $5\times{}10^5$ BCOs with randomly drawn initial conditions, see text for further details. The red line is an analytic equation fitted to the blue points (Eq.~\protect\ref{eq:fcorr}). Lower panel: Same as the top panel but considering the absolute values of relative errors for a sample of $6\times{}10^4$ BCOs drawn from the fiducial simulations (see Sections \ref{sec:models} and \ref{sec:ic}) and different merging time approximations: $t_\mathrm{peters}$, blue dotted line, Eq.~\ref{eq:tpeters}; $t_\mathrm{GW}$, orange dashed line, Eq.~\protect\ref{eq:tgw}; $t_\mathrm{peters+corr}$ \protect\citep{Zwick20}, green dot-dashed line, Eq.~\protect\ref{eq:tz}; \textcolor{black}{ $t_\mathrm{GW,Mandel21}$ \protect\citep{mandel2021tgw}, purple dotted line, Eq.~\protect\ref{eq:tmandel};}
 $t_\mathrm{GW+corr}$, red solid line, Eq.~\protect\ref{eq:ti}. 
    \gitlab{https://gitlab.com/iogiul/iorio22_plot/-/tree/v3/tmerge}
    \gitbook{https://gitlab.com/iogiul/iorio22_plot/-/blob/v3/tmerge/complete_analysis.ipynb}
    \gitimage{https://gitlab.com/iogiul/iorio22_plot/-/blob/v3/tmerge/tdelay_fit.png}}
	\label{fig:tpeters}
\end{figure}

\begin{table*}
\begin{tabular}{lcccc}
\hline
\multicolumn{1}{c}{Method} & \begin{tabular}[c]{@{}c@{}}Time per system \\ averaged, C++\\ (s)\end{tabular} & \begin{tabular}[c]{@{}c@{}}Time per system \\ averaged, Python\\ (s)\end{tabular} & 
\begin{tabular}[c]{@{}c@{}}Average \\relative error\\ {}\end{tabular} & 
\begin{tabular}[c]{@{}c@{}}Maximum\\ relative error\\ {}\end{tabular} \\ \hline
\multicolumn{1}{l|}{Adaptive 4th order Runge-Kutta} & $1.1 \times 10^{-3}$ & $1.0 \times 10^{-1}$ & benchmark model & benchmark model \\
\multicolumn{1}{l|}{Adaptive Euler} & $3.0 \times 10^{-4}$ & $2.8 \times 10^{-3}$ & $3.8 \times 10^{-3}$ & $2.7 \times 10^{-2}$ \\
\multicolumn{1}{l|}{$t_\mathrm{peters}$ (Eq.~\ref{eq:tpeters})} & $2.9 \times 10^{-8}$ & $4.3 \times 10^{-8}$ & $4.9 \times 10^{-1}$ & $8.7 \times 10^{-1}$ \\
\multicolumn{1}{l|}{$t_\mathrm{GW}$ (Eq.~\ref{eq:tgw})} & $3.0 \times 10^{-8}$ & $3.4 \times 10^{-8}$ & $7.3 \times 10^{-2}$ & $4.3 \times 10^{-1}$ \\
\multicolumn{1}{l|}{$t_\mathrm{peters+corr}$ (\citealt{Zwick20}, Eq.~\ref{eq:tz})} & $4.5 \times 10^{-8}$ & $6.0 \times 10^{-8}$ & $4.5 \times 10^{-2}$ & $1.1 \times 10^{-1}$ \\
\multicolumn{1}{l|}{$t_\mathrm{GW,Mandel21}$ (\citealt{mandel2021tgw}, Eq.~\ref{eq:tmandel})} & $7.9 \times 10^{-8}$ & $8.6 \times 10^{-8}$ & $9.4 \times 10^{-3}$ & $2.8 \times 10^{-2}$ \\
\multicolumn{1}{l|}{$t_\mathrm{GW+corr}$ (Eq.~\ref{eq:ti})} & $8.4 \times 10^{-8}$ & $9.7 \times 10^{-8}$ & $1.7 \times 10^{-4}$ & $5.6 \times 10^{-3}$ \\
\multicolumn{1}{l|}{$t_\mathrm{merge}$ (Eq.~\ref{eq:tcomb})} & $8.4 \times 10^{-8}$ & $1.3 \times 10^{-7}$ & $1.7 \times 10^{-4}$ & $3.3 \times 10^{-3}$ \\ \hline
\end{tabular}
\caption{Performance of different methods to estimate the GW-induced merger time. The first two methods use  a 4th order Runge-Kutta (first row) or Euler (second row)  solver  with an an adaptive time-step scheme.
All the other methods are analytic approximations (further details  are given in the main text). 
The second and third columns contain the average computational time required to estimate the merging time of a single system in \textsc{C++} and \textsc{Python}. 
The fourth and fifth columns contain the average and maximum relative differences of a given method with respect to the merging times estimated with the adaptive 4th order Runge-Kutta scheme.
The values reported in this table have been obtained estimating the merger time for $6\times{}10^4$ BCOs sampled from our fiducial model (Section \ref{sec:models}).
We performed this computation using a serial code and a 3.1 GHz Quad-Core Intel Core i7 processor. The \textsc{Python} script exploits numpy vectorisation. We compiled the \textsc{C++} code 
with the maximum allowed optimisation flag ($-\mathrm{O}3$). The {\sc C++} code used to perform this analysis can be found in the gitlab repository of the paper (\gitlab{https://gitlab.com/iogiul/iorio22_plot/-/tree/v3/tmerge/C\%2B\%2B_code}).}
\label{tab:tgw}
\end{table*}

The GW-induced merger time is estimated integrating 
Eqs.~\ref{eq:gwa} and  \ref{eq:gwe}.  

We test both  performance of an adaptive time-step scheme applied to a 4th order Runge-Kutta and Euler solvers. 
We stop the integration when the semi-major axis becomes smaller than the innermost stable circular orbit (three times the Schwarzchild radius) of the most massive object.
The Runge-Kutta solver offers the most precise evaluation of the merger time at the cost of relatively high computation time, especially using Python (0.1s per integration). In the rest of the Appendix, we consider the merger time estimated with the 4th order Runge-Kutta integration, $t_\mathrm{RK}$, our benchmark to evaluate  the performance of other methods. 
The Euler solver offers a factor of $\approx{3}$ speedup at the cost of an average $\approx 0.4\%$ error and maximum error $\approx 3\%$. 

We can obtain an 
approximation of merging time integrating Eq.~\ref{eq:gwa} by assuming that the eccentricity remains constant during the evolution:
\begin{equation}
t_\mathrm{peters}= \frac{5}{256} \frac{c^5}{G^3} \frac{a^4}{M_1 M_2 (M_1+M_2)} \frac{(1-e^2)^\frac{7}{2}}{\left(1+\frac{73}{24}e^2 + \frac{37}{96}e^4\right)}
\label{eq:tpeters}
\end{equation}
Figure \ref{fig:tpeters} shows that $t_\mathrm{peters}$  quickly diverges from $t_\mathrm{RK}$ for $e>0.1$, in particular, it tends to progressively underestimate the merger time with increasing eccentricity. 
To reduce the time difference, we  remove the part of the denominator depending on the eccentricity in  Eq.~\ref{eq:tpeters}:
\begin{equation}
t_\mathrm{GW}= \frac{5}{256} \frac{c^5}{G^3} \frac{a^4}{M_1 M_2 (M_1+M_2)} (1-e^2)^\frac{7}{2}.
\label{eq:tgw}
\end{equation}
Figure \ref{fig:tpeters} indicates that this simple modification is enough to have a good approximation of the merger time (within a few \%)
up to $e\approx0.8$.

\cite{Zwick20} found that the ratio between $t_\mathrm{peters}$ and the properly integrated merging time depends solely on the eccentricity (see also \citealt{Peters64}), hence they introduced a correction term on $t_\mathrm{peters}$:
\begin{equation}
t_\mathrm{peters+corr}= t_\mathrm{peters} \,{}8^{1-\sqrt{1-e}}.
\label{eq:tz}
\end{equation}
Eq.~\ref{eq:tz} represents a solid improvement with respect to $t_\mathrm{peters}$, especially for very large eccentricities (see Fig. \ref{fig:tpeters}).
However, it gives a less precise approximation with respect to $t_\mathrm{GW}$ for low eccentricities ($e<0.5$).

\textcolor{black}{
Similarly, \cite{mandel2021tgw} proposed an analytic approximation for the GW-induced merger time,
\begin{equation}
t_\mathrm{GW,Mandel21} = t_\mathrm{GW}  (1 + 0.27 e^{10} +0.33e^{20} + 0.2 e^{100}),
\label{eq:tmandel}
\end{equation}
that is accurate to  within 3\% over the entire range of initial eccentricities (Fig. \ref{fig:tpeters}).
}

Based on  \cite{Zwick20} and \cite{mandel2021tgw}, we aim to find a correction term for $t_\mathrm{GW}$.
We produce a training set randomly drawing the initial conditions of $5\times{}10^5$ BCOs. The masses are sampled uniformly between 0.5 and 300 $\mathrm{M}_\odot$, the semi-major axis is sampled uniformly in the logarithmic space between $10^{-1}$ and $10^{10}$ $\mathrm{R}_\odot$. Finally, for half of the sample the eccentricity is drawn uniformly between 0 and 0.95, for the other half between 0.95 and 1.
The upper panel of Fig. \ref{fig:tpeters} shows the relative difference between $t_\mathrm{RK}$ and $t_\mathrm{GW}$.
We fit the relative error curve with an analytic equation deriving the correction term:
\begin{equation}
f_\mathrm{corr}=e^2 \left[ -0.443 + 0.580 \left( 1-e^{3.074} \right)^{
1.105-0.807e+0.193e^2} \right].
\label{eq:fcorr}
\end{equation}

Finally, the GW-induced merger time is approximated as 
\begin{equation}
t_\mathrm{GW+corr}=\frac{ t_\mathrm{GW}  } { 1 + f_\mathrm{corr}(e)}.
\label{eq:ti}
\end{equation}
Eq.~\ref{eq:ti} outperforms all the other tested  approximations, and performs even better than the Euler solver 
(see Table \ref{tab:tgw}). 
For most of the eccentricity range, the relative errors are less than 0.02\% (see Fig. \ref{fig:tpeters}). 
Only for very extreme eccentricities ($e>0.99$), $t_\mathrm{GW+corr}$ begins to produce progressively larger errors, but still contained within $0.6\%$ (maximum relative error at $e=0.99999$). 

The reason for this decrease in precision is evident in the upper panel of Fig. \ref{fig:tpeters}. 
Around $e=0.99$, the relative residuals show an abrupt drop that cannot be properly modelled by the fitting equation.

Figure~\ref{fig:tpeters} shows that the relative error curves of the $t_\mathrm{GW+peters}$ (Eq.~\ref{eq:tz}) and $t_\mathrm{GW+corr}$ (Eq.~\ref{eq:ti}) cross at $e \approx 0.999$. We can exploit the best of the two approximations defining:
\begin{equation}
t_\mathrm{merge} = 
\begin{cases}
t_\mathrm{GW+corr} \ \ \mathrm{for} \ e<0.999 \\ 
t_\mathrm{peters+corr} \ \ \mathrm{for} \ e\geq0.999
\end{cases}
\label{eq:tcomb}
\end{equation}
Eq.~\ref{eq:tcomb} offers a high-precision approximation of the merger time on the whole eccentricity range at the expense of a negligible computation overhead. 
Obviously, all the analytic approximations outperform the adaptive integration in terms of computational time. The speedup is a factor of $10^4\mathrm{-}10^5$ in \textsc{C++} and $10^5\mathrm{-}10^7$ 
in \textsc{Python} (Table \ref{tab:tgw}).

\cite{Zwick20} and  \cite{Zwick21}   introduced additional correction factors to account for post-Newtonian terms \textcolor{black}{ \citep[see also][]{will2017,tucker2021}}. We checked that the corrections are 
negligible for all the BCOs systems tested in our analysis ($5\times{}10^5$ systems with randomly drawn initial conditions and $6\times{}10^4$ systems from the fiducial model, see Sections \ref{sec:models} and \ref{sec:ic}).  Only systems with an initial tight configuration are significantly affected. Such systems merge in a very short time (close to the first periastron passage). Therefore, even if the relative errors could be large, the absolute time difference is negligible for any practical purpose concerning population synthesis studies. 

All the methods discussed in this Appendix are implemented in the function \texttt{estimate\_tgw} contained in the  publicly available \textsc{Python} module \textsc{pyblack}.\footnote{\url{https://gitlab.com/iogiul/pyblack}, use
pip install pyblack to install it.}

\label{lastpage}
\end{document}